\documentclass{uzhthesis}

\usepackage[svgnames,x11names]{xcolor}
\definecolor{lightgray}{HTML}{D5D5D5}
\definecolor{lightergray}{HTML}{EEEEEE}

\usepackage[numbers,sort&compress]{natbib}
\usepackage{amsmath}
\usepackage{amsfonts}
\usepackage{amssymb}
\usepackage{upgreek}
\usepackage{bbm}
\usepackage{slashed}
\usepackage[T1]{fontenc}
\usepackage{multirow}
\usepackage{multibib}\newcites{my}{Mine}\newcites{myt}{Tims}
\usepackage[caption=false]{subfig}

\usepackage{multicol}
\usepackage[printonlyused,nohyperlinks]{acronym}
\makeatletter
    \renewcommand{\ac}[1]{\AC@placelabel{#1}\acs{#1}}
\makeatother
\newcommand{\term}[1]{\emph{\ac{#1}}}
\newcommand{\aterm}[2]{\emph{#1} (#2)}

\usepackage{listings}
\lstdefinelanguage{vim}{
  morekeywords={
  set, let
  map, nmap,
  filetype,
  on, off,
  autocmd,
  Plugin,
  call,
   },
morecomment=[l]{"}, 
morestring=[b]' 
}

\usepackage[colorlinks=true
,urlcolor=blue
,anchorcolor=blue
,citecolor=blue
,filecolor=blue
,linkcolor=blue
,menucolor=blue
,linktocpage=true
,pdfproducer=medialab
]{hyperref}

\usepackage{tikz}
\usetikzlibrary{decorations.pathmorphing,positioning,decorations.pathreplacing,shapes,calc}
\usetikzlibrary{decorations.pathreplacing}
\usetikzlibrary{decorations.markings}
\usetikzlibrary{arrows}

\tikzset{
    photon/.style={decorate, decoration={snake,amplitude=1pt,segment length=6pt}},
    zigzag it/.style={decorate, decoration=zigzag},
    gluon/.style={decorate, draw=black,decoration={coil,amplitude=4pt, segment length=5pt}},
    tightgluon/.style={decorate, draw=black,decoration={coil,amplitude=2pt, segment length=3pt}},   
    fermion/.style={postaction={decorate},
        decoration={markings,mark=at position .55 with {\arrow{>}}}},
    antifermion/.style={postaction={decorate},
        decoration={markings,mark=at position .55 with {\arrow{<}}}},
  on each segment/.style={
    decorate,
    decoration={
      show path construction,
      moveto code={},
      lineto code={
        \path [#1]
        (\tikzinputsegmentfirst) -- (\tikzinputsegmentlast);
      },
      curveto code={
        \path [#1] (\tikzinputsegmentfirst)
        .. controls
        (\tikzinputsegmentsupporta) and (\tikzinputsegmentsupportb)
        ..
        (\tikzinputsegmentlast);
      },
      closepath code={
        \path [#1]
        (\tikzinputsegmentfirst) -- (\tikzinputsegmentlast);
      },
    },
  },
  mid arrow/.style={postaction={decorate,decoration={
        markings,
        mark=at position .5 with {\arrow[#1]{stealth}}
      }}},
 mfermion/.style={postaction={on each segment={mid arrow=black}}}
}

\def\centerarc[#1](#2,#3)(#4:#5:#6)
    { \draw[#1] ({#2+#6*cos(#4)},{#3+#6*sin(#4)}) arc (#4:#5:#6); }

\usepackage{enumitem}
\makeatletter
    \newlist{treelist}{itemize}{5}
    \setlist[treelist]{label=\treelist@label}

    \tikzset{treelist line/.style={thick, line cap=round, rounded corners}}
    \def\treelist@label{%
        \begin{tikzpicture}[remember picture, baseline={([yshift=-.6ex] treelist-bullet-\the\enit@depth.center)}]
            \draw [treelist line] (0, 0) -- node (treelist-bullet-\the\enit@depth) {} ++(.5em, 0);
        \end{tikzpicture}%
        \ifnum\enit@depth>1
            \tikz[remember picture, overlay] \draw [treelist line] (treelist-bullet-\the\numexpr\enit@depth-1\relax.center) |- (treelist-bullet-\the\enit@depth.center);%
        \fi
    }
\makeatother

\newcommand{\mtextsc}[1]{\relax\ifmmode\textsc{#1}\else\ac{#1}\fi}

\newcommand\mcmule{{\sc McMule}}
\def\alphapi{\Big(\frac\alpha\pi\Big)}

\def\I{\mathrm{i}}
\def\D{\mathrm{d}}
\def\E{\mathrm{e}}
\def\tr{\mathrm{tr}}
\newcommand{\sg}[1]{#1_{[{\rm dim}]}}
\newcommand{\s}[2]{#2_{[#1]}}
\newcommand{\sd}[1]{#1_{[d]}}
\newcommand{\sds}[1]{#1_{[d_s]}}
\newcommand{\se}[1]{#1_{[n_\epsilon]}}
\newcommand{\sm}[1]{#1_{[-2\epsilon]}}
\def\neps{n_\epsilon}
\def\Sf{{\rm S}_{[4]}}
\def\QSs{{\rm QS}_{[d_s]}}
\def\QSd{{\rm QS}_{[d]}}
\def\QSe{{\rm QS}_{[n_\epsilon]}}
\def\QSm{{\rm QS}_{[-2\epsilon]}}
\def\cdr{\mtextsc{cdr}}
\def\fdh{\mtextsc{fdh}}
\def\fdf{\mtextsc{fdf}}
\def\hv{\mtextsc{hv}}
\def\dred{\mtextsc{dred}}
\def\dreg{\mtextsc{dreg}}
\def\rs{\mtextsc{rs}}
\def\BM{\mtextsc{bm}}
\def\AC{\mtextsc{ac}}
\def\MS{\overline{\text{MS}}}

\def\mev{{\rm MeV}}
\def\gev{{\rm GeV}}
\newcommand{\Einv}{E\hspace*{-6pt}/}

\newcommand{\Zjet}{Z_{q}}
\newcommand{\Zantijet}{\bar{Z}_{q}}

\newcommand{\den}[1]{\mathcal{D}_{#1}}
\renewcommand{\binom}[2]{\bigg(\genfrac{}{}{0pt}{0}{#1}{#2}\bigg)}
\newcommand{\mbinom}[2]{\bigg(\!\!\bigg(\genfrac{}{}{0pt}{0}{#1}{#2}\bigg)\!\!\bigg)}

\newcommand{\bit}[1]{\D\sigma^{(#1)}}
\newcommand{\bbit}[2]{\D\sigma^{(#1)}_{#2}}
\newcommand{\M}[2]{\mathcal{M}_{#1}^{(#2)}}
\newcommand{\fM}[2]{\mathcal{M}_{#1}^{(#2)f}}

\newcommand\eik{\mathcal{E}}
\newcommand\ieik{\hat{\mathcal{E}}}
\def\xc{\xi_{c}} 
\newcommand{\cdis}[2][c]{\left(\frac{1}{#2}\right)_{\hspace*{-3pt}#1}}

\newcommand{\A}[1]{\mathcal{A}^{(#1)}}
\newcommand{\dZ}[2][i]{\delta Z_{#1}^{(#2)}}

\newcommand{\cA}{{\cal A}}

\def\pref#1{%
 \ifnum0<0#1\relax
   \newcount\foo%
   \foo=0%
   \loop
     \advance\foo +1
     \D\Upsilon_{\the\foo}
   \ifnum\foo<#1
   \repeat
 \else
   \prod_{i=1}^{#1}\D\Upsilon_i
 \fi
  \D \Phi_{n,#1}
}

\def\io{{\rm i0}^+}

\makeatletter
\def\thickhline{%
  \noalign{\ifnum0=`}\fi\hrule \@height \thickarrayrulewidth \futurelet
   \reserved@a\@xthickhline}
\def\@xthickhline{\ifx\reserved@a\thickhline
               \vskip\doublerulesep
               \vskip-\thickarrayrulewidth
             \fi
      \ifnum0=`{\fi}}
\makeatother

\newlength{\thickarrayrulewidth}
\usepackage{chngcntr}
\counterwithout{footnote}{chapter}

\usepackage{ifoddpage}
\usepackage{marginnote}

\makeatletter
  \def\ps@headings{%
      \let\@oddfoot\@empty\let\@evenfoot\@empty
      \def\@evenhead{\thepage\hfil\slshape\leftmark}%
      \def\@oddhead{{\slshape\rightmark}\hfil\thepage}%
      \let\@mkboth\markboth
    \def\chaptermark##1{%
      \markboth {{%
        \ifnum \c@secnumdepth >\m@ne
            \@chapapp\ \thechapter. \ %
        \fi
        ##1}}{}}%
    \def\sectionmark##1{%
      \markright {{%
        \ifnum \c@secnumdepth >\z@
          \thesection. \ %
        \fi
        ##1}}}}
\makeatother

\title{McMule\\[1ex]\huge QED Corrections for Low-Energy Experiments}
\author{Yannick Ulrich}
\from{Deutschland}
\committee{
    \member{Prof. Dr. Adrian Signer (Vorsitz)}
    \member{Prof. Dr. Stefano Pozzorini}
    \member{PD Dr. Michael Spira}
}
\abstractE{We present \mcmule{}, a unified framework for the
calculation of NLO and NNLO corrections to many processes in QED with
massive fermions. This easily extendable program allows users to
calculate an arbitrary observable for any of the processes
implemented. These include various lepton decays as well as certain
low-energy scattering experiments such as $e\mu\to e\mu$ and $\ell
p\to\ell p$ that can be measured to high enough a precision to warrant
QED corrections.
\\
As part of our discussion, we will present a pedagogical introduction
to how these calculations are performed, focusing on technical aspects
supplemented with examples. Our goal is to provide a useful
introduction for those entering the field, covering all aspects
relevant for the practitioner.}

\acknowledgement{First things first: Adrian, I wish to express all my
gratitude to you for suggesting and supervising this wonderful
project. I'm grateful for your support and your supervision but also
for letting me pursue pet projects such as {\tt handyG} and for
involving me in supervisions. Thank you for letting my take care of
our pet mule!
\\\indent
The next Thank-You goes to my collaborators on and around \mcmule{},
Pulak Banerjee, Tim Engel, Christoph Gnendiger, Marco Pruna, and
Adrian Signer. Thank you for working with me and for your many great
ideas that can be found somewhere in these pages. This project would
not have been possible without you. 
\\\indent
Relatedly, I'd like to thank my predecessor, Andrea Visconti, for
patiently explaining the art of two-loop calculations to me and for
sharing his notes with me, even long after he has left academia.
\\\indent
Next, there are those whose supervision I was allowed to actively
join. Thank you, Tim Engel, Nicolas Schalch, Luca Naterop, and Andrea
Gurgone for indulging me during your various projects.
\\\indent
While talking collaborators, I'd like to thank the MUonE theorists and
experimentalists for letting us join your ranks and for your ideas. In
particular, I'd like to thank Massimo Passera for mentioning the
experiment to us over lunch during the PSI 2016 conference, once again
proving that lunch and coffee breaks are the most important parts of
conferences.
\\\indent
Next, I'd like to express my gratitude to my experimental colleagues
who patiently explained to me the many subtleties of experimental
analysis, unwittingly helping to improve \mcmule{}. In alphabetical
order, you are Niklaus Berger, Lukas Gerritzen, Carren Kresse, Alberto
Lusiani, Umberto Marconi, Clara Matteuzzi, Angela Papa, Ann-Katrhin
Perrevoort, Dinko Pocanic, Giada Rutar, Patrick Schwendimann, and
Graziano Venanzoni.
\\\indent
Moving on, I must thank the present and former bachelor, master, and
PhD students of the theory group at PSI for many helpful discussions
and, more importantly, the morning tea breaks. Thank you (in
chronological order) Seraina Glaus, Dario M\"uller, Tim Engel, Lukas
Fritz, Nicolas Schalch, Luca Naterop, David Urwyler, Johannes Lade,
Fiona Kirk, Claudio Manzari, Andrea Gurgone, and Natalie Sch\"ar.
\\\indent
I'm extremely happy to have joined such a wonderful group at PSI.
Thank you for many great lunch and coffee breaks and extracurricular
activities. I'm indebted to Adrian Signer and Michael Spira for
running this amazing group as well as the group's other members, past
and present, for creating such a wonderful working atmosphere. In
addition to our group's students: Emanuele Bagnaschi, Pulak Banerjee
Antonio Coutinho, Andrea Crivellin, Margherita Ghezzi, Chrisoph
Gnendiger, Marco Pruna, Johannes Schlenk, Marc Montull, and Max
Zoller. 
\\\indent
Leaving academia and entering the real world, I'd like to thank the
members of the albrechtstrings orchestra, especially Brigitte,
Juliane, Magnus, Patrick, Stephanie, Ursula, and Sandra.
\\\indent
Last but not least, I'm grateful to my family, both in Northern and
Southern Germany, and of course to my Amazing Group of friends
outside PSI: Nadya, Daniel, and Marko.  Finally, I'd like to thank
John for putting up with me for the last few years. Thanks, mate!
\\\indent
}

\declaration{This thesis is based on the following works to which the
author contributed to directly:
\nocitemy{Pruna:2016spf, Gnendiger:2017pys, Pruna:2017upz,Ulrich:2017adq,
Engel:2018fsb, Engel:2019nfw, Banerjee:2020rww}
\bibliographymy{../muon_ref}{}
\bibliographystylemy{JHEP}
\noindent Hence, text may be copied in verbatim without direct
reference. Further, because the author was closely involved in the
supervision of
\nocitemyt{Engel:2018}
\bibliographymyt{../muon_ref}{}
\bibliographystylemyt{JHEP}
\noindent Sections~\ref{sec:ibp} and~\ref{sec:intergals} follow the
resulting master thesis very closely.
\nocite{Pruna:2016spf, Gnendiger:2017pys, Pruna:2017upz,Ulrich:2017adq,
Engel:2018fsb, Engel:2019nfw,
Banerjee:2020rww,Engel:2018}
}

\begin{document}

\maketitle

\chapter{Relevance of QED}\label{ch:intro}

A naive estimate of the size of radiative corrections in any theory is
generally driven by the size of its coupling. For quantum
electrodynamics (\ac{QED}) this is $\alpha\simeq1/137$, implying that
\ac{QED} corrections can often be safely ignored and are only ever
relevant for experiments with the highest precision. However, this
naive estimate overlooks two aspects. 
\begin{itemize}
    \item
    \ac{QED} corrections can easily become as large as ten percent if they
    include large logarithms of widely different masses and kinematic
    cuts.

    \item
    The other aspect has to do with the experimental precision that
    the theory has to ultimately match or even exceed. Current and
    future experiments will be able to push the precision of event
    rates -- famously far more challenging to measure than shapes --
    to well below the percent level, mandating next-to-leading order
    (\ac{NLO}) or even next-to-next-to-leading order (\ac{NNLO})
    calculations for many processes in \ac{QED}.

\end{itemize}

To facilitate the implementation of many \ac{QED} calculations (10 and
counting up to \ac{NNLO} at the time of this writing) we have
developed a unified framework called \mcmule{} ({\bf M}onte {\bf
c}arlo for {\bf Mu}ons and other {\bf le}ptons). With it, new
processes can be added with relative ease, making \mcmule{} the
defining aspect of the thesis.

In what follows, we list some experiments that in some way or form are
relevant for \mcmule{} even though not all measure processes that can
be calculated with \mcmule{}. Next, we will discuss the implemented
processes sorted by order in perturbation theory.

The thesis-proper begins in Chapter~\ref{ch:qed} with a brief but
mostly standard introduction to \ac{QED}, defining some terminology
that we refer to later. Next, in Chapter~\ref{ch:reg}, we will discuss
different dimensional regularisation schemes with the practitioner in
mind, providing detailed examples. As a next big step, we will discuss
in Chapter~\ref{ch:fks} the infrared (\ac{IR}) subtraction schemes
used by \mcmule{}, the development of which was a corner stone of this
project. We will discuss practical aspects of a two-loop calculation
in Chapter~\ref{ch:twoloop}. For the technically inclined reader, we
will discuss aspects of \mcmule's implementation in
Chapter~\ref{ch:mcmule}.  Most of this will not be relevant for users
of \mcmule{} but serves as a guide on how \mcmule{} could be extended.
Finally, we will review some results obtained by \mcmule{} in
Chapter~\ref{ch:pheno} before finally discussing future developments
in Chapter~\ref{ch:conclusion}.

\section{Relevant experiments}
As mentioned above, experimental progress requires more and more
theory support. While this is of course also true for the \ac{LHC}
experiments that certainly drove the development of technology, we
will focus exclusively on \ac{QED} here.  Still, even though many
experiments have driven this development, an exhaustive list would not
be rewarding here. Instead, we will list some examples, mostly but not
exclusively, focussing on muonic physics that benefit from
fully-differential calculations:

\begin{itemize}

    \item
    Bhabha scattering has been used at various lepton colliders as a
    standard candle for luminosity measurement. Hence, much
    theoretical effort has been devoted to this process. Presently,
    \ac{NNLO} corrections, including leading electron mass effects,
    are known and matched to \aterm{parton shower}{PS}. For a review
    of the state of Bhabha scattering, see for
    example~\cite{Actis:2010gg}.

    \item
    The $g-2$ experiment~\cite{Bennett:2006fi} at Brookhaven, its
    successor at Fermilab~\cite{Grange:2015fou} as well as a novel
    experiment planed at J-PARC~\cite{Saito:2012zz} are precisely
    measuring the anomalous magnetic moment of the muon. This
    observable is thought to be -- due to its high precision -- very
    sensitive to \ac{BSM} and indeed there is a tantalising
    discrepancy between the measurement and the \ac{SM} prediction
    (for example cf.~\cite{Nyffeler:2016gnb}). The theoretical
    prediction is plagued by uncertainties in the \aterm{hadronic vacuum
    polarisation}{HVP} and the hadronic light-by-light
    scattering.

    However, as the \ac{QED} corrections to this process are known to
    the five-loop level~\cite{Aoyama:2017uqe} and $g-2$ is an
    intrinsically inclusive observable, there is nothing further for
    \mcmule{} to directly add to the \ac{QED} calculation of $g-2$.
    Hence, we will refrain from further commenting on the
    determination of the \ac{QED} corrections to $g-2$.
    
    \item
    The proposed MUonE experiment~\cite{Calame:2015fva,
    Abbiendi:2016xup, MUonE:LOI} plans to measure muon-electron
    scattering to high precision in order to independently determine
    the \ac{HVP} contribution to the muon $g-2$ through a novel
    approach.  For this to be competitive with the orthodox
    methodology the relative systematic error needs to be under
    control below $10^{-5}$. Aside from the obvious experimental
    challenges connected to this, the \ac{QED} contributions should be
    known to at least the \ac{NNLO}, level including mass effects and
    matched to \ac{PS}.

    \item
    The P2~\cite{Becker:2018ggl}, PRad~\cite{Xiong:2019umf}, and
    MUSE~\cite{Gilman:2013eiv} experiments are measuring elastic 
    electron-proton and muon-proton scattering, respectively. These
    measurements help to determine the proton radius. However, PRad
    uses M{\o}ller scattering ($ee\to ee$) for normalisation purposes,
    the theory uncertainties of which are a leading systematic.
    
    \item
    The MOLLER experiment~\cite{Benesch:2014bas} and the QWeak
    experiment~\cite{Androic:2018kni} measure the Weinberg angle at
    low $Q^2$ in electron-electron and electron-proton scattering,
    respectively.

    \item
    The MuLan experiment at the Paul Scherrer Institute (\ac{PSI})
    has measured the muon lifetime to 1\,ppm~\cite{Webber:2010zf}.
    This measurement was then, in combination with theoretical
    calculations~\cite{vanRitbergen:1999fi,Anastasiou:2005pn}, used to
    extract the Fermi constant $G_F$.
    
    \item
    The MEG experiment at \ac{PSI}~\cite{Adam:2013mnn} and its
    successor MEG~II~\cite{Baldini:2013ke} are searching for the
    lepton-flavour violating (\ac{LFV}) decay $\mu\to e\gamma$ which
    is predicted by many \ac{BSM} scenarios.  As any observation of
    the \ac{LFV} decay channel would constitute clear evidence of
    \ac{BSM} physics, there is no pressing need for \ac{NLO}
    corrections to this decay mode yet. However, $\mu\to e\gamma$
    becomes indistinguishable from the \term{radiative muon decay}
    $\mu\to e\nu\bar\nu+\gamma$ for small neutrino energies. Hence,
    MEG is searching for a peak on a steeply falling background. It is
    now unsurprising that precise knowledge of this background is
    extremely helpful.

    \item
    The Mu3e experiment at \ac{PSI}~\cite{Perrevoort:2016nuv,
    Blondel:2013ia} is searching for the \ac{LFV} decay $\mu\to eee$.
    This is again difficult to disentangle from the \term{rare muon
    decay} $\mu\to e\nu\bar\nu+ee$ for small neutrino energies.
    Further, Mu3e is sensitive to light but weakly coupled \ac{BSM}
    physics. These potential particles might not appear as a clear
    bump over the falling background but as minute modifications to
    certain differential observables. For these types of analyses,
    radiative correction are essential.

    \item
    The PADME experiment at the INFN National Laboratory of
    Frascati~\cite{Raggi:2014zpa} is searching for annihilation of
    $e^+e^-$ pairs into a photon and a so-called dark photon. As such
    the Standard Model process $ee\to\gamma\gamma$ is of interest for
    PADME.

\end{itemize}

The high experimental accuracy obtained or planned by these
experiments also requires a focussed theory support to make the best
use of their data. This means that from the theoretical side all
relevant processes need to be calculated

\begin{itemize}
    \item
    to the highest order in perturbation theory possible,

    \item
    to be fully-differential, i.e. not just predicting inclusive cross
    section but to instead being able to model the experimental
    situation as closely as possible,

    \item
    to include polarisation effects, should these matter
    experimentally,

    \item
    to include all necessary mass effects wherever possible, and

    \item
    to include resummation where large logarithms are expected.

\end{itemize}

In the following sections we will comment on some of the processes in
\mcmule{}, noticing some practical exceptions to the first point.

Even though \mcmule{} focusses on muonic processes, in some cases
tauonic (eg. $\tau\to e\nu\bar \nu\gamma$) or hadronic (eg. $\ell
p\to\ell p$) processes can be included with only minor changes.

\section{Processes at Leading order}

While leading order (\ac{LO}) calculations are mostly trivial, that
does not necessarily make them futile. In fact, the polarised rare
muon decay $\mu\to e\nu\bar \nu + ee$ was first calculated and made
available to the Mu3e collaboration in a predecessor of the \mcmule{}
framework~\cite{polmatel}. This was required by Mu3e to accurately
simulate their background including polarisation effects which heavily
influence angular distributions. While this was later superseded by a
\ac{NLO} calculation~\cite{Pruna:2016spf,Fael:2016yle}, it was and
still is very helpful for the planning of the Mu3e experiment.

Additionally to their searches for $\mu\to e\gamma$, the MEG
collaboration also looks for the \ac{LFV} decay of a muon into an
electron and a Majoron $J$, a Goldstone boson associated with a
hypothetical spontaneous breaking of lepton
number~\cite{Gelmini:1980re, Chikashige:1980ui} (for a review of the
Majoron in the context of MEG see~\cite{Papa:snf, Ripiccini:2011}, and
reference therein). This particle may decay promptly into
$J\to\gamma\gamma$~\cite{MEGDRMD} resulting in a $\mu\to
e\gamma\gamma$ signature. This becomes indistinguishable from the
double-radiative muon decay $\mu\to e\nu\bar\nu + \gamma \gamma$ if
the neutrinos carry little energy. However, because the process is
heavily suppressed, a \ac{LO} study in \mcmule{} was sufficient to
model the relevant background.

\section{Processes at next-to-leading order}

For many background processes, a \ac{NLO} study is sufficient to meet
the experimental requirements. Notable examples in \mcmule{} are the
radiative ($\mu\to e\nu\bar\nu+\gamma$) and rare ($\mu\to
e\nu\bar\nu+ee$) muon decays. These processes serve as backgrounds to
MEG's and Mu3e's searches for \ac{LFV} decays. As such, especially the
region of low neutrino energy is of particular interest.

\ac{NLO} studies conducted in
\mcmule{}~\cite{Pruna:2016spf,Pruna:2017upz} and
elsewhere~\cite{Fael:2016yle,Fael:2015gua} found relatively large
corrections, reaching up to ten percent in the relevant regimes. In
both cases, the \ac{NLO} correction was driven through large
logarithms that somewhat spoil the perturbative expansion. As we will
see, this is a recurring theme in perturbative calculations in general
and \mcmule{} in particular. However, as in this case the corrections
are largely negative, the \ac{SM} background was generally
overestimated.  Naturally this is preferable as it slightly increases
the actual efficiency.

From a theoretical point of view, an extension to the radiative tau
decay $\tau\to\ell \nu\bar\nu \gamma$ seems natural. This was measured
by {\sc BaBar}~\cite{Oberhof:2015snl,Lees:2015gea}. In the electronic
case ($\ell=e$) the measured \aterm{branching ratio}{BR} was found to
be significantly above the \ac{SM} prediction~\cite{Fael:2015gua}.
Using \mcmule{} we were able to study this discrepancy and found hints
towards a solution~\cite{Pruna:2017upz,Ulrich:2017adq}.

With the high statistics of Belle and its successor, the rare $\tau$
decays $\tau\to\nu\bar\nu \ell ll$ become
accessible~\cite{Sasaki:2017msf}. A \ac{NLO} study that
merges~\cite{Fael:2016yle} with \mcmule{}~\cite{Pruna:2016spf} is
forthcoming~\cite{Fael:tau}.

Finally, we should mention the \ac{NLO} calculation of muon-electron
scattering~\cite{Bardin:1997nc,Kaiser:2010zz} which was revisited
later in the context of the MUonE experiment~\cite{Alacevich:2018vez}
(shortly thereafter confirmed independently by \mcmule{}~\cite{NLOus}
and~\cite{NLOmfmp}) as this allowed the first detailed study of the
situation that will be faced by the MUonE experiment.

\section{Processes at next-to-next-to-leading order}

Even though \ac{NLO} is enough for many background studies, precision
measurements such as the measurement of the Fermi constant
$G_F$~\cite{Webber:2010zf}, the extraction of the Michel parameters by
TWIST~\cite{TWIST:2011aa}, or the planned \ac{HVP} fit by MUonE
require yet higher precision. In these cases we need to turn to
\ac{NNLO}. While \ac{NLO} corrections are essentially solved for
processes involving not too many particles (and no loops at \ac{LO}),
we are far from accomplishing the same feat for \ac{NNLO}. This is
mostly, but not exclusively, due to the lack of two-loop integrals.
Further complication arises from our wish to include mass effects
wherever possible as analytic solutions to integrals with multiple
masses quickly become impossible. In Chapter~\ref{ch:twoloop} we will
comment on this issue and potential shortcuts.

Currently, \mcmule{} implements the conventional muon
decay or \term{Michel decay} $\mu\to\nu\bar\nu e$~\cite{Engel:2019nfw}
and $ee\to\nu\bar\nu$ (which served as a test case) at \ac{NNLO}.
Further, $\mu$-$e$ scattering can be split into gauge invariant
subsets by categorising which fermion radiates (cf.
Section~\ref{sec:colour}). Due to the lightness of the electron,
corrections associated to it are expected to be dominant. These
simpler contributions to $\mu$-$e$ scattering are already implemented
in \mcmule{} at \ac{NNLO}~\cite{Banerjee:2020rww}. The \ac{NNLO}
leptonic corrections to lepton-proton scattering, too, is implemented
because it can be obtained by tweaking $\mu$-$e$ scattering.

\section{Processes at next-to-next-to-next-to-leading order}

While many observables were calculated at \ac{NNLO} for the
\ac{LHC}, only recently a select group of quantities reached
\ac{n3lo} accuracy. Of these, only one -- deep inelastic jet
production \cite{Currie:2018fgr} -- is fully-differential requiring a
subtraction scheme (cf.  Chapter~\ref{ch:fks}).

The dominant contributions to muon-electron scattering would seem like
an ideal candidate to join this select group. It would also be the
first \ac{n3lo} calculation involving massive particles in initial and
final states as well as loops. While this calculation is not yet part
of \mcmule{}, progress is made towards its addition.

\chapter{Introduction to QED}\label{ch:qed}
\setlength\parskip{1em}
Quantum Field Theories (\ac{QFT}) have proven to be immensely
powerful tools to obtain evermore precise theoretical predictions for
the physics at the smallest scale. Usually this is understood in the
framework of the Standard Model (\ac{SM}) of electroweak and strong
interactions. However, we will not be discussing the full \ac{SM}
with its strengths and weaknesses, suffice it to say that, while very
successful, we know that physics beyond the \ac{SM} (\ac{BSM}) must
exist from a variety of evidence. When searching for \ac{BSM}
experimentally, it is crucial to have a precise understanding of the
background due to known physics -- be that the \ac{SM} or one of its
subsets. 

For all processes under consideration here, the background is
dominated by \ac{QED}, a particularly
simple part of the \ac{SM}. This \ac{QFT} is defined through its
Lagrangian\footnote{Through this work, we will use upper Lorentz
indices regardless of whether an object is co- or contravariant. The
summation is still always implicit.}
\begin{subequations}
\begin{align}
\mathcal{L}&= \sum_i \bar\psi_i(\I\gamma^\mu D^\mu-m_i)\psi_i 
  - \frac14 F^{\mu\nu}F^{\mu\nu}\\
           &= \underbrace{
             \sum_i \bar\psi_i(\I\slashed\partial-m_i)\psi_i  
             -\frac14 \big(F^{\mu\nu}\big)^2
            }_\text{free theory}
             -e\sum_i \bar\psi_i\slashed A\psi_i\,,
\label{eq:qedl:split}
\end{align}
where $\psi_i$ are the spinor fields of the leptons and $F^{\mu\nu}$
the electromagnetic field tensor. In the second step we have
introduced some abbreviated notation, most notably the Feynman slash
notation for $\gamma^\mu a^\mu = \slashed{a}$. $D^\mu=\partial^\mu +
\I eA^\mu$ is called the gauge covariant derivative and is a compact
way to describe the interactions of leptons and photons.
\label{eq:qedl}
\end{subequations}

Unfortunately, \ac{QED} -- like all phenomenologically relevant
\ac{QFT}s -- is not exactly solvable. However, the free theory, i.e.
the first two terms of~\eqref{eq:qedl:split} are solvable.  Hence, we
use perturbation theory to expand in the electromagnetic coupling
\begin{align}
\alpha=\frac{e^2}{4\pi} \approx\frac1{137}\,.
\end{align}
This coupling is small enough to serve as an excellent expansion
parameter. Physical quantities like cross sections or decay rates are
now written as
\begin{align}
\sigma =           \sigma^{(0)}
       + \alphapi^1\sigma^{(1)}
       + \alphapi^2\sigma^{(2)}
       + \alphapi^3\sigma^{(3)}
       + \mathcal{O}(\alpha^4)\,,\label{eq:pcount}
\end{align}
where we refer to leading order (\ac{LO}, $\sigma^{(0)}$),
next-to-leading order (\ac{NLO}, $\sigma^{(1)}$) etc.  contributions.

When calculating the contributions $\sigma^{(i)}$ we need to draw all
connected and amputated Feynman diagrams contributing to the same
observable including some fixed number of couplings. Here we
distinguish \term{tree-level diagrams} and \term{loop diagrams}.

Obtaining the \ac{LO} contribution $\sigma^{(0)}$ (which itself can
contain further factors of $\alpha$) is in most cases relatively
straightforward.  Note that $\sigma^{(0)}$ could already contain
loops, i.e. a \term{loop-induced} process. We do not consider this
case here.  Instead, we assume that the first order is always given
through a number of tree-level diagrams. Hence, we can use the number
of loops and the order in perturbation theory interchangeably.

Once we have the matrix element, we need to integrate over the phase
space to obtain a cross section or decay rate. At this stage, experimental
subtleties enter. Modelling these as closely as possible may require
us to include complicated cuts, making analytic integration over the
phase space quickly infeasible. Hence, we will do the integration
numerically. To facilitate the cuts, we define the so-called
\term{measurement function}~\cite{Kunszt:1992tn}. This function takes
as arguments the four-momenta of all particles involved in the
reaction and returns the experimentally measured quantity. The
measurement function has to fulfil certain criteria. We will comment
below on properties it has to fulfil beyond \ac{LO}. But even at
\ac{LO}, an example for an invalid function would be to ask for a
number of photons without also specifying the minimum energy of these
photons.  We call a calculation that can implement any measurement
function without renewed effort \term{fully differential}.

We encounter our first loop diagram in $\sigma^{(1)}$. Because the
momenta of the particles in the loop is not fixed through the momenta
of the external particles, we have to integrate over them.
Unfortunately, these \term{loop integrals} can be divergent for large
momenta (ultraviolet, \ac{UV}) or soft or collinear momenta
(infrared, \ac{IR}).  Hence, the first thing we need to do is to
\term{regularise} these divergences. This is usually done by shifting
the dimension of space-time away from 4 to $d=4-2\epsilon$
(dimensional regularisation, \dreg). Both \ac{IR} and \ac{UV}
singularities now appear as poles in $1/\epsilon$. We will explain how
to do this formally and mathematically consistent in
Chapter~\ref{ch:reg}.

The loop integrals required to solve practical processes tend to be
rather complicated. This complexity obviously increases the more loops
are included. Further, the problem is also made more complicated
through the inclusion of more external particles (with potentially
different masses) as this increases the number of relevant or
\term{active scales} $\mu_i$ that enter in the actual loop integrals.
This is in contrast to other scales (inactive scales) that do not enter
loop integrals like the mass of spectator particles.

Further background information on these topics can be found in various
textbooks such as~\cite{Schwartz:2013pla,Peskin:1995ev,Grozin:2005yg}.

\section{Renormalisation}
\label{sec:renorm}
When computing scattering amplitudes with the
Lagrangian~\eqref{eq:qedl} beyond leading order, we encounter \ac{UV}
singularities that are indicative of our ignorance of the physics at
very high scales. These \ac{UV} singularities are dealt with through
\term{renormalisation}. The main idea is to express scattering
amplitudes in terms of renormalised fields and renormalised
parameters, rather than their bare counterparts, s.t. no \ac{UV}
singularities are present. If to all orders in perturbation theory all
\ac{UV} singularities can systematically be absorbed by a finite
number of \term{renormalisation constants} $Z_i$, we call the theory
\term{renormalisable}. It can be shown that \ac{QED} as well as the
full \ac{SM} are renormalisable.

At this stage we will start using $\psi_{0,i}$, $A_0^\mu$ and
$m_{0,i}$ for the bare quantities of~\eqref{eq:qedl}. The variables
$\psi_i$ etc. shall henceforth be reserved for the renormalised
quantities. In Section~\ref{sec:renschemes} we will be more specific
what is meant by that.

Relating the bare quantities $\psi_{0,i}$, $A_0^\mu$ and $m_{0,i}$ of
the Lagrangian~\eqref{eq:qedl} to the renormalised ones\footnote{In
the notation of~\cite{Grozin:2005yg} $Z_2=Z_\psi$ and
$Z_3=Z_A=Z_\alpha^{-1}$.}
\begin{subequations}
\label{eq:renorm}
\begin{align}
\psi_{0,i} = Z_{2,i}^{1/2}\psi_i\,,
\quad
m_{0,i} = Z_{m,i} m_i
\quad\text{and}\quad
A_0^\mu=Z_3^{1/2}A^\mu\,,
\end{align}
we obtain
\begin{align}
\mathcal{L}&= 
     \sum_i Z_{2,i}\ \bar\psi_i(\I\slashed\partial-Z_{m,i} m_i)\psi_i
         -\frac14 Z_3\ \big(F^{\mu\nu}\big)^2
         -\sum_i Z_2Z_3^{1/2}\ e_0\bar\psi_i\slashed A\psi
\,.
\end{align}
$Z_2$ and $Z_3$ are called the wave-function renormalisation factors,
whereas $Z_m$ is the mass renormalisation. We also need to renormalise
the coupling $e_0$. This is usually expressed in terms of the
vertex-renormalisation factor $Z_1$ as
\begin{align}
e_0 = e \frac{Z_1}{Z_2 Z_3^{1/2}} = Z_3^{-1/2}\ e\,.
\label{eq:ward}
\end{align}
In the last step we have used that to all orders in \ac{QED} $Z_1 =
Z_2$, due to the \term{Ward identity}.
\end{subequations}

In~\eqref{eq:qedl} we have omitted the gauge-fixing terms, containing
the gauge parameter, usually called $\xi$. We will always set this
term to $\xi=1$, i.e. perform all calculations in Feynman gauge. In
general, $\xi$ has to be renormalised as well. However, it can be
shown that, as long as one only considers on-shell scattering
amplitudes or renormalisation constants, this does not matter at any
order in \ac{QED}~\cite{Landau:1955zz, Johnson:1959zz, Fukuda:1978jy}
(also cf.~\cite{Melnikov:2000zc} showing that this ceases to be true
in \ac{QCD} at the three-loop level).

\subsection{Renormalisation schemes}\label{sec:renschemes}
In \dreg{}, the \ac{UV} poles are manifest as poles
$1/\epsilon_\text{UV}$, where we temporarily use the \ac{UV} label to
distinguish \ac{UV} from \ac{IR} poles. At $n$ loops, the highest
\ac{UV} pole is of order $1/\epsilon^n_\text{UV}$. The \ac{UV} part
of the $Z_i=1+\delta Z_i$ is uniquely fixed by the requirement that
all \ac{UV} singularities are absorbed. At one-loop accuracy they are
\begin{align}
\begin{split}
Z_1=Z_2 &= 1+\frac{\alpha}{4\pi}
  \frac{-1}{2\epsilon_\text{UV}} +
  \mathcal{O}\big(\epsilon_\text{UV}^0,\alpha^2\big)\,,
\\
Z_m &= 1+\frac{\alpha}{4\pi}
  \frac{-3}{2\epsilon_\text{UV}} +
  \mathcal{O}\big(\epsilon_\text{UV}^0,\alpha^2\big)\,,\\
Z_3 &= 1+\frac\alpha{4\pi}\frac{\beta_0}{\epsilon_\text{UV}} +
  \mathcal{O}\big(\epsilon_\text{UV}^0,\alpha^2\big)\,,
\end{split}
\label{eq:renconstuv}
\end{align}
where $\beta_0=-4/3 N_F$ in a theory with $N_F$ flavours. However,
there is quite some freedom in choosing a \term{renormalisation
scheme}, i.e. prescription how to fix the terms of the renormalisation
factors that are UV finite. Note that, to the loop order given in
\eqref{eq:renconstuv}, it does not matter whether $\alpha$ has been
renormalised or not, as the difference would be
$\mathcal{O}(\alpha^2)$.

For most choices of the renormalisation scheme, the renormalised
parameters $\alpha$ and $m$ start to exhibit a behaviour known as
\term{running}. These parameters become dependent on the
\term{renormalisation scale} $\mu$, the scale at which the \ac{UV}
subtraction is made.  In particular, this is encountered for the
coupling whose scale dependence is governed by the $\beta$ function
as
\begin{align}
\frac{\partial\alpha(\mu)}{\partial\log\mu} = 2\beta(\alpha(\mu))
\quad
\text{with}
\quad\beta(\alpha)
= -\alpha\Bigg(
    \frac{\alpha}{4\pi}\ \beta_0+ \mathcal{O}(\alpha^2)
\Bigg)\,.
\label{eq:betafunc}
\end{align}
This is a first example of what is called a \aterm{renormalisation
group equation}{RGE}. By choosing the renormalisation scale at the
appropriate scale of the experiment, $Q^2$, one avoids large
logarithms $\log Q^2/\mu^2$ that arise when integrating
\eqref{eq:betafunc}.

The most common renormalisation schemes are the \emph{\ac{msbar}
scheme}, where the finite terms vanish up to some common factors, and
the \aterm{on-shell scheme}{OS}. The latter will be the default in
this project, s.t. for example $m$ and $\alpha$ refer to the \ac{OS}
mass and coupling. Hence, the \ac{OS} scheme deserves some further
elaboration.

The \ac{OS} scheme is constructed to most faithfully reproduce the
classical limit for the input parameters at $Q^2=0$ without the
parameters ever experiencing running. For example, this means that the
electron mass really is $m_e\simeq 0.511\,{\rm MeV}$. To achieve this,
let us consider the one-loop corrections to the fermion propagator
as (following~\cite{Grozin:2005yg})
\begin{align}
\Sigma(p) = m_0\,\Sigma_1(p^2) + (\slashed p-m_0)\Sigma_2(p^2)\,.
\label{eq:defsigma}
\end{align}
To get the physical propagator $S$ from the bare propagator
$S_0=1/(\slashed{p}-m_0)$ we have to sum an infinite number of
$\Sigma$
\begin{align}\begin{split}
S(p) &= S_0(p) + S_0\Sigma(p)S_0 + S_0\Sigma(p)S_0\Sigma(p)S_0 + \cdots
\\
&= \frac1{S_0^{-1}-\Sigma}
= \frac1{\slashed p - m_0 - \Sigma}\,.
\end{split}\end{align}
We now want to describe this in the renormalised quantities, i.e.
\begin{align}
S(p) \frac1{\slashed p - m_0 - \Sigma}
\stackrel!= \frac{Z_2}{\slashed p-m} + \text{regular}\,,
\label{eq:renconm}
\end{align}
where regular refers to terms that do not contribute to the pole as
$\slashed p\to m$.  The \ac{OS} mass of the electron is now just
defined as the pole of the propagator. In principle we could just
plug~\eqref{eq:defsigma} into~\eqref{eq:renconm} and obtain
\begin{align}
Z_m = 1-\Sigma(m)
\quad\text{and}\quad
Z_2 = 1+\frac{\D\Sigma}{\D\slashed{p}}\Big|_{p^2=m^2}\,.
\end{align}
However, calculating $\Sigma'(m)$ can be cumbersome, especially
beyond the one-loop level. Hence, we follow the method set out
by~\cite{Broadhurst:1991fy}: we begin by writing down the perturbative
expansion of $\Sigma$, $Z_m$, and $Z_2$ with the most general
dependence of $p^2$ and $m_0$ allowed by the loop integration
\begin{align}
\begin{split}
Z_2 &= 1 + \sum_{n=1}^{\infty}
    \bigg(\frac{\alpha_0}{m^{2\epsilon}}\bigg)^n F(n)\,,
\\
Z_m &= 1 + \sum_{n=1}^{\infty}
    \bigg(\frac{\alpha_0}{m^{2\epsilon}}\bigg)^n M(n)\,,
\\
\Sigma &=\phantom{1+}\sum_{n=1}^\infty 
    \bigg(\frac{\alpha_0}{(p^2)^{\epsilon}}\bigg)^n\Big(
                m_0\Sigma_1^{(n)}\big(\tfrac{m_0^2}{p^2}\big) 
+ (\slashed p-m_0) \Sigma_2^{(n)}\big(\tfrac{m_0^2}{p^2}\big) 
    \Big)\,,
\end{split}
\end{align}
where everything is expressed in the bare coupling $\alpha_0$.  This
is now what we plug into \eqref{eq:renconm} with $m=Z_m^{-1}m_0$,
expanding to the desired order in $\alpha$. At one-loop accuracy
\begin{align}
M(1) = -\Sigma_1^{(1)}(1)
\quad\text{and}\quad
F(1) = \Sigma_2^{(1)}(1) - 2\Sigma_1^{\prime(1)}(1) - 2\epsilon\Sigma_1^{(1)}(1)\,.
\end{align}
It turns out that this way we still need to calculate $\Sigma_i(1)$
and $\Sigma_i'(1)$ but we are allowed to set $p^2=m_0^2$ \emph{before}
the loop integration. 

For the photon field -- and by extension the coupling -- we proceed
similarly, finding
\begin{align}
Z_3 = \frac1{1-\Pi(0)}\,,
\end{align}
where $\Pi$ is the usual photon self energy, defined through
\begin{align}
\Pi^{\mu\nu}(p) = (p^2g^{\mu\nu} - p^\mu p^\nu)\Pi(p^2)\,.
\label{eq:pimunu}
\end{align}
For a theory with only one massive fermion $\Pi(0)$ depends only on
the mass of this flavour. One can easily calculate
that~\cite{Grozin:2005yg}
\begin{align}
\Pi^{(1)}(0) =-\frac43\frac{e_0^2}{(4\pi)^{d/2}}\ {m^{-2\epsilon}}\ 
    \Gamma(\epsilon)\,.
\end{align}
This way, we have a relation between the \ac{msbar} coupling $\bar\alpha$
and the \ac{OS} coupling $\alpha$ at one-loop accuracy
\begin{align}
\bar\alpha(\mu) = \alpha\bigg(1+\frac43\frac\alpha{4\pi} 
    \log\frac{\mu^2}{m^2}\bigg)\,.
\end{align}

In principle we are free to renormalise the masses and coupling in any
scheme we wish. For the fermion masses, we will always choose the
on-shell scheme. This mass is scale independent and corresponds
directly to the measured value of the lepton masses. Our standard
choice for the coupling is also the on-shell scheme. In this scheme
the coupling is scale independent and corresponds to the measured
value $\alpha\sim 1/137$ in the Thomson limit. However, we
occasionally work with $\bar\alpha$, the coupling in the
\ac{msbar}-scheme. As mentioned above, this coupling depends on the
renormalisation scale $\mu$. If we consider processes at high energies
$Q$ (compared to the fermion masses) this scheme can be useful, as
setting $\mu\sim Q$ allows to resum large logarithms.

All renormalisation constants required up to two-loop accuracy can be
found, expressed in the bare coupling, in Appendix~\ref{ch:fdhconst}.

\subsection{Practical renormalisation}\label{sec:renorm:practical}

In order to obtain scattering matrix elements at a particular order in
perturbation theory, we start by computing all connected and amputated
Feynman diagrams to the required order. Amputated means we do not
include diagrams with self-energy insertions on external
lines. According to the \ac{LSZ} reduction formula, such contributions
are properly included by multiplying the unrenormalised amplitude by
$\sqrt{Z_i}$ for each external line, where $Z_i$ is the wave-function
renormalisation factor in the on-shell scheme. This results in the
renormalised scattering amplitude, but still expressed in terms of the
bare coupling, masses, and gauge parameter. To absorb all \ac{UV}
singularities the bare parameters have to be expressed in terms of the
corresponding renormalised parameters.

Renormalisation beyond one-loop has certain subtleties, most of which
can be explained by pure counting of powers of the coupling $\alpha$.
At the one-loop level, the renormalisation constants $\dZ1$ always
just multiply a tree-level amplitude $\A0$. This ceases to be
sufficient at the two-loop level. Now, additionally to the product of
two-loop renormalisation constants $\dZ2$ with the tree-level
amplitude $\A0$, we need to include one-loop renormalisation of the
one-loop amplitude $\A1\times\dZ1$. Further, the two-loop
renormalisation constants $\dZ2$ themselves need to be renormalised
using constants $\dZ1$. This is called \term{sub-renormalisation}.

Particular attention has to be given to the fermion-mass
renormalisation. Replacing $m_{0,i} = Z_{m,i} m_i = m_i + \delta
m_i^{(1)} + \ldots$ in the lower-order amplitudes and expanding in
$\alpha$ produces all mass counterterms, also those on external lines.
However, the latter have already been taken into account by the
\ac{LSZ} reduction. Hence, in practical calculations it is
advantageous to perform mass renormalisation by explicitly computing
Feynman diagrams with mass counterterms $\delta m_i^{(l)}$ on internal
lines only.

Hence, we arrive at the following practical procedure for two-loop
renormalisation:
\begin{enumerate}

    \item
    For every massive external particle, add the wave function
    renormalisation for heavy fermions $Z_h$
    \begin{align}
    \Big(\frac12\ \dZ[h]2 - \frac18\ \big(\dZ[h]1\big)^2\Big) \times\A0
    \quad\text{and}\quad
    \frac12\ \dZ[h]1 \times\A1\,.
    \end{align} 
    
    \item
    For every massless external fermion, add $\tfrac12\dZ[l]2\times\A0$,
    keeping in mind that these contributions are induced through terms
    proportional to the number of heavy flavours. This means that
    $Z_l=1$ to all orders in theories without at least one massive
    flavour.

    \item
    For every external photon, we have to add the corresponding
    $\tfrac12\dZ[3]1 + ...$ as above.

    \item
    Perform the mass renormalisation of the fermions, i.e. add
    counterterm diagrams obtained through the substitution
    \begin{align}
        \frac\I{\slashed{p}+m} \to 
        \frac\I{\slashed{p}+m} \delta m^{(l)} \frac\I{\slashed{p}+m}
    \end{align}
    for internal fermion lines at the amplitude level. We need $l=2$
    for tree-level diagrams and $l=1$ for one-loop diagrams, as well
    as double insertions with $l=1$ for tree-level diagrams. Note that
    this does not correspond to replacing $m_0 = Z_{m} m$ and
    expanding again in $\alpha$ at the matrix element level, as this
    would lead to the double counting of the mass renormalisation of
    external lines as discussed above.

    \item
    Perform the coupling renormalisation by shifting $\alpha_0\to
    (1+\dZ[\alpha]1 + \dZ[\alpha]2) \times \alpha$ and sorting terms
    according to the now renormalised coupling, dropping every term
    with too high a power in $\alpha$.

    If we have no internal photons at \ac{LO}, i.e. the number of
    external photons coincides with the number of \ac{QED} vertices,
    this step and Step 2 above cancel exactly thanks to the Ward
    identity, meaning neither is necessary (cf. \eqref{eq:ward}).

\end{enumerate}

\section{Effective theories and the muon decay}\label{sec:eft}
A recurring theme of this project is the muon decay as an example
process of high phenomenological relevance. However, the muon does not
decay in pure \ac{QED} as the only \term{weak-isospin} changing
particle in the \ac{SM} is the $W$-boson. The amplitude for $\mu(p)\to
e(q) \nu_\mu(q_3) \bar\nu_e(q_4)$ in the \ac{SM} can be written as
\begin{align}
\cA =
  \frac{\I g}{\sqrt2}
\ \Big[ \bar u_{\nu_\mu}(q_3)\gamma^\alpha P_L u_\mu(p) \Big]
\ \frac{-\I}{q_W^2-m_W^2}
      \Bigg(g^{\alpha\beta}-\frac{q_W^\alpha q_W^\beta}{m_W^2}\Bigg)
  \frac{\I g}{\sqrt2}
\ \Big[ \bar u_{e}(q)\gamma^\beta P_L u_{\nu_e}(q_4) \Big]\,,
\end{align}
with the $W$ coupling $g$ and the usual left-handed projector
$P_L=\tfrac12(1-\gamma_5)$.  While it is of course possible to perform
all calculations, including radiative corrections, in the full
\ac{SM}, that is often unnecessary.  Because the $W$ momentum
$q_W=p-q_3\sim m_\mu$ is much smaller than its mass $m_W$, the $W$
propagator simplifies to
\begin{align}
\frac{-\I}{q_W^2-m_W^2}
    \Bigg(g^{\alpha\beta}-\frac{q_W^\alpha q_W^\beta}{m_W^2}\Bigg)
 = \frac{\I}{m_W^2} g^{\alpha\beta} +
 \mathcal{O}\Big(\frac{q_W^2}{m_W^2}\Big)\,,
\end{align}
resulting in
\begin{align}
\cA =
-\I\frac{g^2}{2m_W^2}
\ \Big[ \bar u_{\nu_\mu}(q_3)\gamma^\alpha P_L u_\mu(p) \Big]
\ \Big[ \bar u_{e}(q)\gamma^\alpha P_L u_{\nu_e}(q_4) \Big]
+\mathcal{O}\Big(\frac{q_W^2}{m_W^2}\Big)\,.
\end{align}
Further, because of the large $W$ mass, radiative corrections due to
the $W$ are also suppressed by $\mathcal{O}(m_\mu^2/m_W^2)$. Hence,
instead of introducing a propagating $W$ boson, we
augment~\eqref{eq:qedl} by
\begin{align}
\mathcal{L}
= \mathcal{L}_{\text{QED}} - \frac{4\, G_F}{\sqrt2} 
\left( \bar\psi_{\nu_\mu} \gamma^\mu P_L \psi_\mu \right)
\left( \bar\psi_e \gamma^\mu P_L \psi_{\nu_e} \right)\,.
\label{eq:lfermi}
\end{align}
Here, we have introduced a \term{dimension-six operator} with a
dimensionful coupling $G_F$. At energies far below $m_W$, the exchange
of a $W$ boson is described well by \eqref{eq:lfermi}. This is a first
example of an \aterm{effective field theory}{EFT}. We have encoded the
high-energy dynamics of the $W$ into a so-called \term{Wilson
coefficient} $G_F$.  The relation of $G_F$ with parameters of the full
\ac{SM} is found through a \term{matching calculation} by calculating
a process both in the full \ac{SM} and in the \ac{EFT} and then fixing
$G_F$ s.t. in the expansion of the \ac{EFT}, i.e.  $m_W\to\infty$,
both agree. In our case we find at LO
\begin{align}
\frac{G_F}{\sqrt2} = \frac{g^2}{8m_W^2} = \frac1{2v^2}\,,
\end{align}
where $v$ is the vacuum expectation value of the Higgs field in the
Standard Model.

There is one more simplification to be done in~\eqref{eq:lfermi}.
Since we cannot measure the neutrinos it is unfortunate that they take
such a prominent role in the calculation. Instead, we would prefer
everything related to neutrinos to factorise. Fortunately, there exist
so-called \term{Fierz identities} to re-arrange spinor bilinears such
as the ones in~\eqref{eq:lfermi}. In our case we find
\begin{align}
\mathcal{L}
= \mathcal{L}_{\text{QED}} - \frac{4\, G_F}{\sqrt2} 
\left( \bar\psi_e \gamma^\mu P_L \psi_\mu \right)  
\left( \bar\psi_{\nu_\mu} \gamma^\mu P_L \psi_{\nu_e} \right)\, .
\label{eq:fierzed}
\end{align}

Because \eqref{eq:fierzed} is the theory we will be using to calculate
radiative corrections to the muon decay, we have to face the issue
that in the strict meaning of the word, \eqref{eq:fierzed} is not
renormalisable, requiring in general an infinite number of $Z_i$.
However, as long as we do not consider a perturbative expansion in
$G_F$, we can maintain predictability by renormalising $G_F$ as just
another coupling through a new $Z_{G_F}$ which would usually be
assumed in the \ac{msbar} scheme. However, it turns out that we do
not even have to do that as $Z_{G_F}=1$ to all orders in \ac{QED}.

To see this, we first note that $\mathcal{L}$ is invariant under the
exchange $\psi_e \to \gamma^5\psi_e$ and
$m_e\to-m_e$~\cite{BERMAN196220}. However, because this exchanges the
vector and axial-vector current, we only really need to consider a
vectorial coupling. Further, because the neutrinos are uncharged,
there is no difference between $G_F$ and the normal \ac{QED} coupling
from a renormalisation aspect. Hence, the \ac{QED} Ward identity
$Z_1=Z_2$ still holds. The only contribution left to influence
$Z_{G_F}$ is the equivalent of $Z_3$. The \ac{QED} contribution to
this quantity can be fixed by considering \ac{QED} corrections to
$\nu\nu\to\nu\nu$.  Because the neutrinos are uncharged under
\ac{QED}, these vanish exactly. Of course, terms that are higher order
in $G_F$ exist in principle.

To summarise, we will be using~\eqref{eq:fierzed} for all calculations
involving the muon decay. As long as we only consider LO in $G_F$, the
results will be UV finite after \ac{QED} renormalisation. Higher-order
corrections in $G_F$ have been considered in~\cite{Fael:2013pja}.

\section{Infrared safety}\label{sec:irsafety}
\newcommand{\mat}[1]{\mathcal{M}^{(#1)}}

After the \ac{UV} renormalisation, our \term{virtual} matrix element
is unfortunately still \ac{IR} divergent. This is in so-far physical
that \ac{IR} singularities cannot just be absorbed through
redefinition of quantities. Instead, such fully \term{exclusive}
quantities are just not physical until they are combined with
\term{real} matrix elements involving extra radiation. While it is of
course possible to distinguish events with extra hard radiation in an
appropriate detector, there always exist a physical cut-off $\Delta$
below which radiation cannot be detected any more. As cross sections
usually scale like $\log\Delta$, the cross section would diverge when
integrating over the entire phase space including $\Delta\to0$. This
\term{soft} divergence is exactly cancelled by the \ac{IR} divergence
of the virtual matrix element. Observables for which this is true are
called \term{IR safe}. Totally inclusive cross sections like
\begin{align}
\sigma(a+b\to c+d+\text{any number of}\ \gamma)
\quad\text{where }c,d\neq \gamma
\end{align}
are examples for IR safe observables. The existence of these
observables is guaranteed by the Kinoshita-Lee-Nauenberg theorem
(\ac{KLN}) that states that any sufficiently inclusive observable
(such as the total cross section) will always be finite.  The
condition imposed by the \ac{KLN} theorem can be translated into a
condition on the measurement function as we will see
later~\cite{Kunszt:1992tn}.

As mentioned above, we would very much like to integrate over the
phase space numerically. However, we cannot do that in $d$ dimensions.
Instead, we need special methods to treat these divergences in $d$
dimensions without spoiling our ability to integrate numerically. We
will discuss one such method in detail in Chapter~\ref{ch:fks}.

In a theory with massless fermions there is an additional source of
singularities due to (hard) radiation becoming \term{collinear} with a
massless fermion. This is not an immediate problem as we will mostly be
dealing with massive particles where the mass $m$ serves as a
regulator, giving rise to $\log m$. However, these
\aterm{pseudo-collinear singularities}{PCS} cause a lot of numerical
instabilities making them difficult to integrate over as we will
discuss in Section~\ref{sec:pcs} and again in Section~\ref{sec:ps}.

An unfortunate aspect of perturbative calculations is that, for
processes with very different scales $\mu_i$, logarithms of the form
$L\sim\log\mu_1^2/\mu_2^2$ become very large. Hence, each new loop
order not just brings a new power of $\alpha$ but also often two
powers of $L$ -- one due to soft and one due to collinear emission. At
least in \ac{QCD}, this can easily become large enough s.t. $\alpha_s
L^2\sim 1$, spoiling the expansion completely. But even in \ac{QED}
this is troublesome as it would require computations to an infeasibly
high order.

This means that we have to revise our counting~\eqref{eq:pcount},
assuming that we get two powers of $L$ per loop order
\begin{align}
\def\term#1#2{\alphapi^#1L^#2\sigma^{(#1)}_{#2}}
\begin{split}
\sigma = \sigma^{(0)}_0 &
                                           + \term12 + \term11 + \term10
\\&                    + \term24 + \term23 + \term22 + \term21 + \term10
\\&+\term36 + \term35  + \term34 + \cdots\,.
\end{split}
\end{align}
The rows of this equation correspond to the \term{fixed-order} results
obtained above. However, we can use the fact that the terms
$\sigma^{(i)}_{2i}$ usually follow a predictable pattern. Hence, if we use
$\alpha L^2/\pi$ as the expansion parameter instead of $\alpha/\pi$ we
can get control over these logarithms. This process is known as
\term{resummation}. The first column is known as
\aterm{leading-logarithm}{LL}, the second as \aterm{next-to-leading
logarithm}{NLL} and so on. 

A particularly efficient way to calculate the \ac{LL} contribution is
a \aterm{parton shower}{PS}. This involves including a cascade of soft
and collinear radiation to all involved particles. This is
particularly interesting because \ac{PS} can be constructed
independent of the measurement function. Unfortunately, at the time of
this writing, no \ac{NLL} \ac{PS} has been presented though work is
ongoing towards a construction of such a method. Until then, \ac{NLL}
resummation must be done anew for each observable.  However, much work
has been dedicated to obtaining results that are almost \ac{NLL}
accurate.

\section{Infrared prediction}
\label{sec:irpred}

When performing multi-loop calculations, an important cross-check is
the cancellation of \ac{IR} singularities. However, to use this as a
practical tool, it is necessary to predict the \ac{IR} poles without
having to calculate the (potentially very difficult) real corrections. 

For this discussion we assume that we work in \ac{QCD} with (some)
massless flavours instead as the \ac{IR} structure will be much
richer. We will come back to massive \ac{QED} later.

Infrared predictions have been worked out for massless \ac{QCD} in
dimensional regularisation~\cite{Gardi:2009qi, Gardi:2009zv,
Becher:2009cu, Becher:2009qa}. This was extended to gauge theories
with massive fermions~\cite{Becher:2009kw}.

To predict the \ac{IR} structure of \ac{QCD} we remember that in an
\ac{EFT}, the Wilson coefficients need to be renormalised. However,
the \ac{UV} singularities removed this way were not present in the
full theory. This implies that the part of the calculation entering
the Wilson coefficient is \ac{IR} divergent. We now need to construct
a low-energy theory s.t. its \ac{UV} divergences match the \ac{IR}
poles of \ac{QCD} because we can predict \ac{UV} singularities using
renormalisation theory. The \ac{EFT} in question is
\aterm{soft-collinear effective theory}{SCET}~\cite{Bauer:2000yr,
Bauer:2001yt, Beneke:2002ph} (for a pedagogical introduction, for
example cf.~\cite{Becher:2014oda}) that splits soft and collinear
modes off from the full underlying theory, be it \ac{QED} or \ac{QCD}.

While a full derivation of the \ac{IR} prediction is well beyond the
scope of this work, we can sketch the necessary concepts, especially
because we will encounter some of them later.

\def\scetz{{\bf Z}} 
Let us define the, in principle, all-order renormalised\footnote{We
will assume that the coupling is renormalised in the \ac{msbar} scheme
to be consistent with the literature} matrix element $\mathcal{M}$ for
an arbitrary process as the sum of $\ell$-loop contributions
$\mathcal{M}^{(\ell)}$
\begin{align}
\mathcal{M} = \sum_{\ell=0}^\infty \mathcal{M}^{(\ell)} =
\mathcal{M}^{(0)} + \mathcal{M}^{(1)} + \mathcal{M}^{(2)} + ...\,,
\end{align}
where each $\mathcal{M}^{(\ell)}$ contains one power more of
$\bar\alpha$.  We now define the corresponding $\scetz$ s.t.
\begin{align}
\mathcal{M}_\text{sub} = \big(\scetz\big)^{-1} \mathcal{M} 
\quad\text{with}\quad
\scetz=1+\delta\scetz^{(1)}+\delta\scetz^{(2)}+...
\label{eq:scetzdefnomu}
\end{align}
is finite in the limits $\epsilon\to0$. We call $\mathcal{M}_\text{sub}$
\term{MSlikesubtracted}, because $\scetz$ is constructed to contain no
finite parts, up to trivial terms induced by the loop measure.
However, just like \ac{msbar} renormalisation introduces a
renormalisation scale, the factorisation into \ac{IR} finite and
\ac{IR} divergent quantities of~\eqref{eq:scetzdefnomu} introduces a
new \term{factorisation scale}.

It is important to note, that, while important for what follows, there
is nothing wrong with defining a different $\scetz'$ that contains
finite parts but no factorisation scale (cf.  Chapter~\ref{ch:fks}).
For now, however, we will stick to \ac{msbar}-like subtraction and
re-write \eqref{eq:scetzdefnomu} to account for the new scale $\mu$
\begin{align}
\mathcal{M}_\text{sub}(\mu) = \big(\scetz(\mu)\big)^{-1} \mathcal{M}
\,.
\label{eq:scetzdef}
\end{align}
Next, we note that, even though $\mathcal{M}_\text{sub}$ and $\scetz$
depend on the factorisation scale, the original matrix element
$\mathcal{M}$ does not. Hence, we can a obtain a \ac{RGE} for
$\mathcal{M}_\text{sub}(\mu)$ by differentiating \eqref{eq:scetzdef}
w.r.t. $\mu$, resulting in
\begin{subequations}
\begin{align}
\frac\D{\D\log\mu} \mathcal{M}_\text{sub}(\mu) 
   = {\bf\Gamma}(\mu) \mathcal{M}_\text{sub}(\mu)\,.
\label{eq:zdef}
\end{align}
with
\begin{align}
{\bf\Gamma}(\mu) = -\frac{\D\log\scetz}{\D\log\mu}\,.
\label{eq:zrge}
\end{align}
\end{subequations}
Here, $\bf\Gamma(\mu)$ is the \term{anomalous dimension} of the
process. This is very similar to how the anomalous dimension of, for
example, the fermion that is obtained by
\begin{align}
\gamma_f = \frac{\D\log \bar Z_2}{\D\log\mu_F}\,,
\end{align}
with the \ac{msbar} fermion wave function renormalisation $\bar Z_2$.

The formal solution of \eqref{eq:zrge} is~\cite{Becher:2009qa}
\begin{align}
\log\scetz(\mu) = \int_\mu^\infty \frac{\D\mu'}{\mu'}{\bf\Gamma}(\mu')
                = \int_{\log\mu}^\infty \D(\log\mu'){\bf\Gamma}(\mu')
\,.
\end{align}
Unfortunately, integrating \eqref{eq:zrge} is complicated by the fact
that $\bf\Gamma$ is not just a function of $\mu$ but also of the
\ac{msbar} coupling $\bar\alpha(\mu)$ that has its own
RGE~\eqref{eq:betafunc}\footnote{In~\cite{Becher:2009qa}, $\beta$ is
defined as $\beta_\text{\cite{Becher:2009qa}}=2\beta$.}
\begin{align}
\frac{\partial\alpha(\mu)}{\partial\log\mu} = 2\beta(\alpha(\mu))
\,.
\end{align}
Hence, we need to distinguish the explicit scale dependence from the
one induced by the running of $\bar\alpha$. We substitute
$\mu'\to\alpha'(\mu')$ and write schematically
\begin{align}
\log\scetz(\mu) =\int_0^{\bar\alpha} \frac{\D\alpha'}{-2\beta(\alpha')}
    \Bigg( {\bf\Gamma}(\alpha') +
    \int_0^\alpha\frac{\D\alpha''}{-2\beta(\alpha'')}
    \frac{\partial{\bf\Gamma}(\alpha'')}{\partial(\log\mu)}
\Bigg)\,,
\end{align}
where we have used that the only explicit dependency of $\log\mu$ in
${\bf\Gamma}$ is linear as we will see below.  By identifying
${\bf\Gamma}'$ as
\begin{align}
{\bf\Gamma}'= \frac{\partial{\bf\Gamma}}{\partial\log\mu}\,,
\end{align}
we can solve this order-by-order~\cite{Becher:2009cu,Becher:2009qa}
\begin{align}
\log\scetz = \Big(\frac{\bar\alpha}{4\pi}\Big)\Bigg(
   \frac{{\bf\Gamma}_1'}{4\epsilon^2}
  +\frac{{\bf\Gamma}_1 }{2\epsilon  }
\Bigg)
+ \Big(\frac{\bar\alpha}{4\pi}\Big)^2\Bigg(
  -\frac{3\beta\cdot {\bf\Gamma}_1'}{16\epsilon^3}
  -\frac{ \beta\cdot {\bf\Gamma}_1 }{ 4\epsilon^2}
  +\frac{{\bf\Gamma}_2'}{16\epsilon^2}
  +\frac{{\bf\Gamma}_2 }{4\epsilon  }
\Bigg)
+\mathcal{O}(\alpha^3)\,,
\label{eq:logZ}
\end{align}
where ${\bf\Gamma}_i$ (${\bf\Gamma}_i'$) is the $\mathcal{O}
(\alpha^i)$ coefficient of ${\bf\Gamma}$ (${\bf\Gamma}'$) and
$\beta\cdot{\bf\Gamma}_1 = \beta_0\cdot{\bf\Gamma}_1$ in the notation
of~\cite{Broggio:2015dga} and Appendix~\ref{ch:fdhconst}.

\def\gcusp{{\gamma}_\text{cusp}}
\def\gq   {{\gamma}_i          }
\def\gQ   {{\gamma}_I          }
It has been conjectured by~\cite{Becher:2009cu} that, assuming a
theory without massive flavours, the anomalous dimension ${\bf\Gamma}$
can be constructed to all orders by just considering two-particle
correlations. This ceases to be true in a theory with massive
particles~\cite{Mitov:2009sv}, requiring a more complicated
structure~\cite{Becher:2009kw} that we will not reproduce here.

For the two-parton case ${\bf\Gamma}$ is constructed from a \term{cusp
anomalous dimension} $\gcusp$ relating two partons and quark anomalous
dimensions $\gq$ (or $\gQ$ for massive quarks) that has to do with
just one parton. Assuming trivial colour-flow (as in $t\to Wb$ or of
course any \ac{QED} calculation)
\begin{align}\begin{split}
{\bf\Gamma}(\mu) &= 
  \sum_{i,j} \gcusp \log\frac{\mu^2}{-{\rm sign}_{ij} 2p_i\cdot p_j} + \sum_i\gq\\
&-\sum_{I,J} \gcusp(\chi_{IJ})               + \sum_I\gQ\\
&+\sum_{I,j} \gcusp \log\frac{m_I\mu}{-{\rm sign}_{Ij} 2p_I\cdot p_j}\,.
\label{eq:gammair}
\end{split}\end{align}
We use capital letters $I$ to indicate massive particles and
lower-case letters for massless particles. The signs in front of the
scalar product depend on the types of spinors
involved~\cite{Becher:2009cu}. To be precise, ${\rm sign}_{ij} =
(-1)^{n_{ij}+1}$, where $n_{ij}$ is the number of incoming particles
or outgoing antiparticles among the particles $i$ and $j$.

In a theory without massive particles, the first line of
\eqref{eq:gammair} describes the anomalous dimension of any number of
particles with the sum going over all possible unordered pairs as
conjectured by~\cite{Becher:2009cu}.

The angle $\chi_{IJ}$ of the fully massive case is sometimes called
\term{cusp angle}
\begin{align}
\chi_{IJ} = {\rm arcosh}\frac{-{\rm sign}_{IJ} p_I\cdot p_J}{m_Im_J}\,.
\end{align}
A comprehensive list of the anomalous dimensions required at the
two-loop level can be found in Appendix~\ref{sec:const:scet}.

The procedure to cross-check \ac{IR} poles is now:
\begin{enumerate}

    \item
    Calculate the \ac{msbar}-renormalised matrix element.

    \item
    In a theory with massive flavours, perform a \term{decoupling
    transformation} relating \ac{SCET} parameters, in which heavy
    fermions have been integrated out, and fields such as
    $\alpha_\text{SCET}$ to those of the full
    theory~\cite{Chetyrkin:1997un}
    \begin{align}
        \alpha_\text{full} = \zeta_\alpha\times \alpha_\text{SCET}\,,
    \end{align}
    where $\zeta_\alpha$ is given in Appendix~\ref{sec:const:scet}.
    In a theory without massive flavours there is no need for
    decoupling.

    \item
    Calculate the anomalous dimension ${\bf\Gamma}$ for the
    process under consideration.

    \item
    Use~\eqref{eq:logZ} to calculate $\scetz$ and use~\eqref{eq:scetzdef}
    to check whether $\mathcal{M}_\text{sub}$ is finite.
    
\end{enumerate}
We will see an example of this in Chapter~\ref{ch:reg}.

Even though the above discussion holds in \ac{QED}, there is a much
simpler way to predict \ac{IR} singularities in massive \ac{QED}.
This is done by noting that soft singularities exponentiate. This
means that $\log\scetz$ vanishes at all orders, except the first.

This can be re-formulated to all orders as
\begin{align}
\sum_{\ell=0}^\infty \mat\ell = e^{-\alpha S} \times
\text{finite}\,.\label{eq:yfs}
\end{align}
This was shown by Yennie, Frautschi, and Suura
(\ac{YFS})~\cite{Yennie:1961ad}. Only the pole of $S$ is fixed by
this equation; its finite and $\mathcal{O}(\epsilon)$ contribution can
be chosen at will. In Section~\ref{sec:fks} we will find a
particularly helpful choice of $S$.


\chapter{Regularisation schemes}\label{ch:reg}

As mentioned before, loop integrals are usually divergent and require
regularisation. The most common way to achieve this is to formally
shift the space-time dimension~\cite{Bollini:1972ui, tHooft:1972fi,
Wilson:1970ag, Wilson:1971dc, Ashmore:1972uj} (dimensional
regularisation, \dreg) to
\begin{align}
d=4-2\epsilon\,.
\end{align}
Correspondingly, we change the loop integration to\footnote{In many of
the original references, $\hat k$, $\tilde k$, $\bar k$ etc. were used
with different meanings depending on paper, scheme, and context.  We
avoid that by instead using the notation developed
in~\cite{Gnendiger:2017pys}.}
\begin{align}
\int\frac{\D^4k}{(2\pi)^4} \to \mu^{4-d}\int\frac{\D^d\sd k}{(2\pi)^d}
 \equiv \int[\D k]\,,
\label{eq:d}
\end{align}
where we have defined a convenient integral measure $[\D k]$ (cf.
\eqref{eq:measure}). We use $\s{\rm dim}k$ to indicate a vector of
(quasi-)dimension $\rm dim$. We will specify what precisely is meant
by this in Section~\ref{sec:formal}.

\ac{UV} and \ac{IR} singularities now manifest as poles of the form
$1/\epsilon^n$. \dreg{} is indeed a consistent prescription and
the resulting integrals still fulfil properties like linearity and
invariance under shifts~\cite{Wilson:1972cf,Collins:1984xc}.

Note that~\eqref{eq:d} only specifies the dimensionality of the
integration momentum $k$. The dimensionality of other objects such as
$\gamma$ matrices are not yet constrained. In order to systematically
classify different approaches, one has to consider two questions
\begin{itemize}
    \item
    are all parts of a diagram regularised or only those leading to
    divergences?

    \item
    are algebraic objects like metric tensors or $\gamma$ matrices
    regularised in $d$ dimensions or in a different dimensionality?
\end{itemize}

In Section~\ref{sec:formal} we will introduce a unified framework for
the discussion of (dimensional) \emph{regularisation schemes} (\rs).
Using this, we will briefly discuss $\gamma_5$ in \dreg{} in
Section~\ref{sec:gamma5}.  In Section~\ref{sec:mu} we will use our
unified framework to discuss the muon decay in the common schemes
\hv{}~\cite{tHooft:1972fi}, \cdr{}~\cite{Collins:1984xc}, and
\fdh{}~\cite{Bern:1991aq,Signer:2008va}. In particular, in
Section~\ref{sec:fdf} we will provide a practitioner's guide to a
particularly simple formulation of \fdh{}, the \fdf{}
scheme~\cite{Fazio:2014xea} by once again calculating the muon decay.
Finally, we will discuss how we can use IR prediction (cf.
Section~\ref{sec:irpred}) to predict the regularisation scheme
dependence, both generally and on the example of the muon decay.

\section{Formal aspects}\label{sec:formal}
To study the questions asked above and to elegantly unify all common
variations of \dreg{}, we need to introduce a series of vector
spaces~\cite{Wilson:1972cf, Collins:1984xc, Stockinger:2005gx}: the
strictly four-dimensional Minkowski space $\Sf$ as well as the
infinite-dimensional spaces $\QSs$, $\QSd$, and $\QSe$. The
infinite-dimensional spaces are equipped with the correct
quasi-dimensionality, s.t. the metric tensor for each space fulfils
\begin{align}
\big(\sg{g}^{\mu\nu}\big)^2 = {\rm dim}\,.
\end{align}

Most aspects of \dreg{} can be understood from
the hierarchy between these spaces
\begin{align}
\Sf\subset \QSd
\quad\text{and}\quad
\QSs=\QSd\oplus\QSe\,.\label{eq:split}
\end{align}
The space $\QSd$ is the space in which $\sd{k}$ exist. It is
\emph{enlarged} to the bigger space $\QSs$ by the orthogonal sum with
$\QSe$. The dimensionality $d_s$ is
\begin{align}
d_s = d+\neps = 4-2\epsilon+\neps\,.
\end{align}
Note that for many actual calculations we will be setting
$d_s=4$ and $\neps=2\epsilon$. For now, however, we will keep all
values independent.

Using \eqref{eq:split} we can now construct all necessary
$d_s$-dimensional objects
\begin{align}
\sds{g}^{\mu\nu} = \sd{g}^{\mu\nu} + \se{g}^{\mu\nu}
\quad\text{and}\quad
\sds{\gamma}^\mu = \sd{\gamma}^\mu + \se{\gamma}^\mu\,.
\end{align}
Of course these objects have no finite-dimensional representation. To
practically work with them, we rely on their algebraic properties
\begin{align}
(\sd{g}\se{g})^{\mu\nu} = 0
\,,\quad
\{\s{\rm dim}\gamma^\mu, \s{\rm dim}\gamma^\nu\} 
    = 2\s{\rm dim}g^{\mu\nu}
\,,\quad
\{\sd\gamma^\mu,\se\gamma^\nu\} = 0\,.
\end{align}

Finally, we need to distinguish between two types of vector
fields\footnote{Again, we refer to the notation of~\cite{
Gnendiger:2017pys} instead of using the old names internal and
external, respectively.}:
\begin{itemize}

    \item
    Fields associated with particles in one-particle irreducible
    (\ac{1PI}) diagrams or with soft and collinear radiation are
    called \term{singular},

    \item
    all other fields are \term{regular}.

\end{itemize}
In general, there is no need to regularise regular fields so that
there is some freedom regarding their treatment. We can now identify
the four flavours of \dreg{} through their answers
to the questions considered above, i.e. what particles are treated in
which of the three spaces.
\begin{center}
\begin{tabular}{l|cc|cc}
                & \cdr   &  \hv   & \fdh   & \dred\\\hline
Singular fields & $\QSd$ & $\QSd$ & $\QSs$ &$\QSs$\\
Regular fields  & $\QSd$ & $ \Sf$ & $ \Sf$ &$\QSs$\\\hline
&\multicolumn{2}{c|}{`dim. reg.'}
&\multicolumn{2}{c }{`dim. red.'}
\end{tabular}
\end{center}
\cdr{} and \hv{} belong to a class of schemes that used to be referred
to as `dimensional regularisation' while \fdh{} and \dred{} belong to
what was called `dimensional reduction'. We will not be using these
terms further to avoid confusion and refer to all four schemes as
\dreg{}.

This seems to suggest that \fdh{} is ideal for the calculation of
multi-loop contributions because all quantities are either strictly or
quasi four-dimensional, keeping the algebra simple without introducing
too many new problems (cf. Section~\ref{sec:fdh}). Similarly, \dred{}
is ideal for any type of real corrections because it does not
distinguish between singular and regular fields while still minimising
the nightmare that are the $\epsilon/\epsilon$
contributions~\cite{Gnendiger:2019vnp}.

\section{\texorpdfstring{$\gamma_5$}{gamma5} in dimensional schemes}
\label{sec:gamma5}
In $\Sf$, $\gamma_5$ is defined through two equivalent relations
\begin{align}
\{\s4\gamma^\mu,\gamma_5\} = 0
\quad\text{or}\quad
\tr\big(\underbrace{
    \s4\gamma^\mu\s4\gamma^\nu\s4\gamma^\rho\s4\gamma^\sigma}_{\s4\Gamma}\gamma_5\big)
 = 4\I\s4\varepsilon^{\mu\nu\rho\sigma}
 \equiv
   4\I\varepsilon^{\mu\nu\rho\sigma}\,,
\end{align}
where we have defined
\begin{align}
\sg\Gamma = \sg\gamma^\mu\sg\gamma^\nu\sg\gamma^\rho\sg\gamma^\sigma
\end{align}
in an arbitrary dimension $\rm dim$. However, these definitions are
equivalent only on $\Sf$. In any of the other spaces, they are
mutually exclusive if we want to keep the cyclicity of traces. The
proof of this is simple but lengthy~\cite{Jegerlehner:2000dz} but
results in
\begin{align}
2\Big(\big(\sd g^{\alpha\beta}\big)^2-4\Big)\,\tr\big(\sd\Gamma\gamma_5\big)
 + \tr\big(\sd\Gamma\,\sd\gamma^\alpha\{\gamma_5,\sd\gamma^\alpha\}\big)=0\,,
\end{align}
which is only valid if $d=\big(\sd g^{\alpha\beta}\big)^2=4$. A
similar proof can be found in \cite{Siegel:1980qs}, resulting in
\begin{align}
0=(d-4)(d-3)^2(d-2)^2(d-1)^2d\,.
\end{align}
This suggests that the two definitions are only equivalent in for
integer $d$.

There are two commonly used solutions to this problem that change the
definition of $\gamma_5$ that maintain cyclic traces
\begin{itemize}
\item
    $\gamma_5$ is constructed to fulfil the trace relation as done in
    the original \hv{} scheme~\cite{tHooft:1972fi} and later picked up
    by Breitenlohner and Maison (\BM)~\cite{Breitenlohner:1977hr}
    \begin{align}
        \gamma_5^\BM=\frac\I{4!}\s4{\Big(\varepsilon^{\mu\nu\rho\sigma}   \Gamma\Big)}
                    =\frac\I{4!}      \s4\varepsilon^{\mu\nu\rho\sigma}\sd\Gamma
                    \,.
    \end{align}
    This way, we still have the $\gamma$-algebra in $\QSd$ but also
    generate many more $\gamma$-matrices, complicating traces.

    In this scheme, we find for \fdh{} and \dred{}
    \begin{align}
        \{\gamma_5^\BM,\s4\gamma^\mu\} = 0
        \quad\text{and}\quad
         [\gamma_5^\BM,\se\gamma^\mu ] = [\gamma_5^\BM, \sm\gamma^\mu] = 0\,.
    \end{align}
    This also implies that~\cite{Breitenlohner:1977hr}
    \begin{align}
        \{\gamma_5^\BM,\sd\gamma^\mu\} = 2\sm\gamma^\mu\gamma_5^\BM
    \end{align}
    and similarly for $\{\gamma_5^\BM,\sds\gamma^\mu\}$.
    
    Unfortunately, in combination with \dreg{} \BM{} breaks the chiral
    symmetry because for $P_{L,R} = \tfrac12(1\pm\gamma_5)$ to be a
    chiral projectors for both $\psi$ and $\bar\psi$,
    $\{\gamma^0,\gamma^5\}=0$ is required. This no longer naively
    works meaning that chiral symmetry is broken. Hence, we require
    another finite renormalisation~\cite{Larin:1993tq,Gnendiger:2017rfh}
    \begin{align}
        \gamma_5^\BM\to Z_5\gamma_5^\BM
        \quad\text{with}\quad
        Z_5 = 1+\frac{\alpha}{4\pi}\Big(\frac{\neps}\epsilon-4\Big)\,.
        \label{eq:z5}
    \end{align}
    We will later see an explicit example of this.

\item
    Alternatively we can define $\gamma_5$ algebraically s.t. the
    anti-commutator vanishes~\cite{Gnendiger:2017rfh,Korner:1991sx,
    Kreimer:1993bh}
    \begin{align}
        \{\gamma_5^\AC,\sd \gamma^\mu\}
       =\{\gamma_5^\AC,\se \gamma^\mu\}
       =\{\gamma_5^\AC,\sds \gamma^\mu\}
       =0\,.
    \end{align}
    This scheme is workable but not in the strict sense consistent as
    it fails to reproduce the Adler-Bell-Jackiw or triangle
    anomaly~\cite{Adler:1969gk,Bell:1969ts,Adler:1969er}. Despite
    this, it was proposed by~\cite{Jegerlehner:2000dz} that we can use
    \AC{} if we restore the anomaly by hand afterwards wherever
    necessary (though it often is not). One way to make $\AC$
    consistent is by giving up the cyclicity of the
    trace~\cite{Korner:1991sx,Kreimer:1993bh}.

\end{itemize}
Both methods, if used properly, lead to consistent results. A more
complete review of $\gamma_5$ in \fdh{} and \dred{} can be found
in~\cite{Gnendiger:2017rfh}.

\section{The muon decay in all schemes}\label{sec:mu}

\begin{figure}
\centering
\input{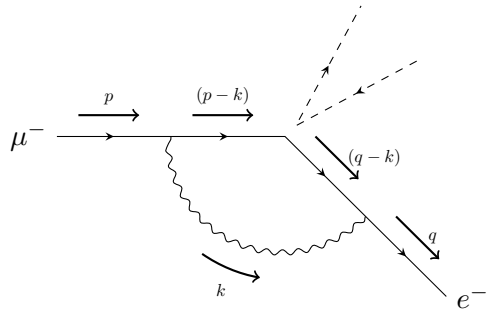}
\caption{The Feynman diagram contributing to the muon decay at the
one-loop level with momentum routing. The dimensionality of the
momenta will depend on the scheme used.}
\label{fig:muonfeynman}
\end{figure}

As an illustration of the various aspects of the different schemes, we
will calculate the muon decay
\begin{align}
\mu(p) \to \nu(q_3)\bar\nu(q_4)  e(q)\,,
\end{align}
with $p^2=M^2\equiv m_\mu^2$ and $q^2=0$ in the three schemes \cdr,
\hv{}, and \fdh{}. In particular, we will set the electron mass to
zero in this section because it will result in a more interesting
singularity structure -- helping us to understand the different
schemes better.  Note that we keep all mass-effects in the
phenomenological discussion later.

To simplify this discussion, we will perform the computation in the
Fierz rearranged effective theory of the muon
decay~\eqref{eq:fierzed}.  As we have already discussed in
Section~\ref{sec:eft}, $\mathcal{L}$ is invariant under the exchange
$\psi_e \to \gamma^5\psi_e$ (and $m_e\to-m_e$ had we not assumed
$m_e=0$)~\cite{BERMAN196220}, allowing us to relate the axial-vector
current to the vector current. Hence, we can calculate the matrix
elements without needing to worry about $\gamma_5$.

When considering this process, we need to compute one diagram at tree
level and one diagram at one-loop, the latter is shown in
Figure~\ref{fig:muonfeynman}. As aids, we will be using the
Mathematica programs {\tt TRACER}~\cite{Jamin:1991dp} for the Dirac
algebra and Package-X~\cite{Patel:2015tea} for the one-loop calculus.

All renormalisation constants necessary for this calculation can be
found in Appendix~\ref{sec:const:ren}.

\subsection{Neutrino average}\label{sec:neutrinoavg}
As a first step that is universal to all schemes, we will deal with
the neutrinos, realising that we cannot actually measure them. Hence,
we would like to remove them as much from the calculation as possible.
To do this we note that when we calculate any observable
using~\eqref{eq:fierzed}, a term corresponding to the neutrino current
\begin{align}
\mathcal{N}^\mu = \bar u(q_3) \gamma^\mu u(q_4)
\end{align}
will be present in the amplitudes. We can factor out the neutrino
tensor $\mathcal{N}^\mu \mathcal{N}^{*\nu}$ that appears in the
squared amplitude by averaging over all possible neutrino momenta. To
do this, we note that
\begin{align}
\mathcal{N}^\mu \mathcal{N}^{*\nu} = 
  4q_3^\mu q_4^\nu
 +4q_3^\nu q_4^\mu
 +4q_3\cdot q_4 g^{\mu\nu}\,,
\end{align}
where $q_3$ and $q_4$ are the momenta of the neutrinos. Here and
henceforth, the sum over spin states is implicit.
Had we not removed the $\gamma_5$ earlier, there would also be
anti-symmetric terms that would not change the discussion below. We
now define the average of an arbitrary function $f(q_3, q_4)$ as the
normalised $1\to2$ phase space integration
\begin{align}
\langle f(q_3, q_4) \rangle = 
  \frac{\int \D\Phi\ f(q_3, q_4)}
       {\int\D\Phi\ 1} = 8\pi{\int \D\Phi\ f(q_3, q_4)}\,.
\end{align}
However, it turns out that, as long as $\langle1\rangle=1$, it does
not matter how the phase space is defined as long as it is Lorentz
invariant and integrates over the neutrinos.

When we average over $q_3$ and $q_4$, the result can only depend on
$Q=q_3+q_4$. Hence, the most general ansatz for $\langle q_3^\mu
q_4^\nu \rangle$ is
\begin{align}
\langle q_3^\mu q_4^\nu \rangle = A \frac{Q^\mu Q^\nu}{Q^2} + B
g^{\mu\nu}\,.
\end{align}
By applying the projectors $g^{\mu\nu}$ and $Q^\mu Q^\nu$ we find
\begin{align}\begin{split}
g^{\mu   \nu}\langle q_3^\mu q_4^\nu \rangle &= \langle q_3\cdot q_4\rangle = A+B\,d\\
Q^ \mu Q^\nu \langle q_3^\mu q_4^\nu \rangle &= \langle Q\cdot q_3\,Q\cdot q_4\rangle = Q^2(A+B)
\,.
\end{split}\end{align}
Using that $q_3^2=q_4^2=0$, i.e. $Q^2=2q_3\cdot q_4$, we can re-write
these equations
\begin{align}
\frac12 Q^2\langle1\rangle = A+B\,d
\quad\text{and}\quad
\frac14 \big(Q^2\big)^2 \langle1\rangle= Q^2(A+B)
\,,
\end{align}
allowing us to determine $A$ and $B$
\begin{align}
A=\frac{d-2}{d-1} \frac{Q^2}{4} \langle 1 \rangle
\quad\text{and}\quad
B=\frac{ 1 }{d-1} \frac{Q^2}{4} \langle 1 \rangle\,.
\end{align}
And hence with $\langle1\rangle=1$
\begin{align}
\mathcal{N}^{\mu\nu} = \langle\mathcal{N}^\mu \mathcal{N}^{*\nu}\rangle 
 = 2\frac{d-2}{d-1} Q^2\bigg( \frac{Q^\mu Q^\nu}{Q^2} - g^{\mu\nu}
 \bigg)\,.
\end{align}
Note that the neutrino tensor $\mathcal{N}^{\mu\nu}$ will be the same
in all parts of the calculation (both real and virtual) as a global
pre-factor. While the dimensionality of $\mathcal{N}$ certainly
influences intermediary results, any physical quantity must be
independent of its dimensionality as it will only influence terms
$\mathcal{O}(\epsilon)$ that vanish in the limit $\epsilon\to0$.  This
means that we could choose its dimensionality independently of the
scheme under consideration.

\subsection{Conventional dimensional regularisation (\textsc{cdr})}
In \cdr{} all quantities are considered $d$-dimensional, even the
external momenta. However, because $\Sf\subset \QSd$ the dimension of
the external fermion momenta does not matter and they could in
principle be chosen from either space. Nevertheless, for consistency
we will still keep them in the space they would be in if they were
internal momenta.

The tree-level amplitude is
\begin{align}
\mathcal{A}_\cdr^{(0)} = \frac12\times\frac{4G_F}{\sqrt2}
  \Big(-\bar u(\sd q) \sd\gamma^\mu u(\sd p)\
\sd{\mathcal{N}}^\mu\Big)\,.
\end{align}
Here, the factor $1/2$ arises from the projector $P_L=(1-\gamma^5)/2$.
The matrix element\footnote{By `matrix element' we denote the result
of squaring the amplitude. In particular, we refrain from calling it
the matrix element squared} is
\begin{align}\begin{split}
\mat0_\cdr &= \frac12\Big|\mathcal{A}_\cdr^{(0)}\Big|^2 
  = \frac12\times\frac{4G_F^2}2\tr \Big[
  \sd{\slashed{q}}\ \sd\gamma^\mu
 \ \big(\sd{\slashed{p}}+M\big)\ \sd\gamma^\nu 
\Big] \times \sd{\mathcal{N}}^{\mu\nu}
\\& = \frac83G_F^2 M^4 x(3+2x(\epsilon-1)-2\epsilon)\,,
\end{split}\end{align}
with $x=2p\cdot q/M^2$ the dimensionless quantity describing the
process.  To obtain this result we have used standard $d$-dimensional
trace techniques as implemented in {\tt TRACER}.

At the one-loop level, we have to calculate
\begin{align}\begin{split}
\mat1_\cdr\Big|_\text{bare}
    &= \frac12\times2\Re\big(
        \mathcal{A}_\cdr^{(1)}\times
        \mathcal{A}_\cdr^{(0)}
    \big) \\&= 
\frac{\alpha_0G_F^2}\pi \int[\D k] \frac{
\tr\Big[
   \sd{\slashed{q}}\,\,
   \sd\gamma^\sigma
   \big(\sd{\slashed{k}}-\sd{\slashed{q}}  \big)
   \sd\gamma^\mu
   \big(\sd{\slashed{k}}-\sd{\slashed{p}}-M\big)
   \sd\gamma^\sigma\,\,
   \big(\sd{\slashed{p}}+M\big)\,\,
   \sd\gamma^\nu
\Big]}{\den1\den2\den3}\sd{\mathcal{N}}^{\mu\nu}\,,
\end{split}\label{eq:cdrbare}\end{align}
with
\begin{align}
\den1 = \sd k^2
\,,\quad
\den2 = (\sd k+\sd p)^2-M^2
\,,\quad
\den3 = (\sd k-\sd q)^2\,.
\end{align}
\eqref{eq:cdrbare} can be evaluated using standard techniques,
obtaining the unrenormalised \cdr{} result
\begin{align}\begin{split}
\mat1_\cdr\Big|_\text{bare} =
\frac{\alpha_0}\pi\bigg(\frac{\mu^2}{M^2x^2}\bigg)^\epsilon \mat0_\cdr \Bigg( 
    &-\frac1{2\epsilon^2}
    -\frac1{2\epsilon  }
    - 2 - \frac32\zeta_2 + \frac{x}{2x-3}\log(x) \\&\qquad+ \log(1-x)\log(x)
      + {\rm Li}_2(x)
\Bigg)\,,
\end{split}\end{align}
with the dilogarithm ${\rm Li}_2(x)$, the first polylogarithm with
order $o=2$, and $\zeta(2)=\pi^2/6$ the Riemann $\zeta$-function.
This and many more expressions we will encounter can be compactly
written by using so-called \aterm{harmonic polylogarithms}{HPL},
introduced in~\cite{Remiddi:1999ew} and implemented for Mathematica
in~\cite{Maitre:2005uu}. These functions extend the notion of
polylogarithms by generalising the order $o$ to a weight vector $\vec
w$. In particular, introducing $\vec w=\{0\}$ and $\vec w=\{1,0\}$ 
\begin{align}
H_0(x)=\log(x)
\quad\text{and}\quad
H_{1,0}(x) = -\log(1-x)\log(x)-{\rm Li}_2(x)\,,
\end{align}
we find
\begin{align}
\mat1_\cdr\Big|_\text{bare} =
\frac{\alpha_0}\pi\bigg(\frac{\mu^2}{M^2x^2}\bigg)^\epsilon \mat0_\cdr \Bigg( 
    -\frac1{2\epsilon^2}
    -\frac1{2\epsilon  }
    - 2 - \frac32\zeta_2 + \frac{x}{2x-3}H_0(x) - H_{1,0}(x)
\Bigg)\,,
\end{align}

We now need to renormalise this quantity. No mass renormalisation is
necessary because $\mathcal{A}^{(0)}$ does not contain $M$. Masses
from the spin-sum are taken care through the renormalisation of the
wave function in the \ac{OS} scheme as mandated by the \ac{LSZ}
formula.  For the muon, this means we have to multiply with (in
Feynman gauge)
\begin{align}
\sqrt{Z_{2,\cdr}^\mu} = 1+\frac{\alpha_0}{2\pi}\bigg( -\frac3{2\epsilon} +
\frac32\log\frac{M^2}{\mu^2} - \frac42 \bigg)\,.
\end{align}
Technically, the $\alpha_0$ here is unrenormalised. However, at the
current loop order there is no difference because
$\alpha=\alpha_0+\mathcal{O}(\alpha^2)$.  The corresponding factor
$Z_2^e=1+\mathcal{O}(\alpha^2)$ because we treat the electron
massless. Hence, we have
\begin{align}
\mat1_\cdr = \frac\alpha\pi\bigg(\frac{\mu^2}{M^2x^2}\bigg)^\epsilon \mat0_\cdr \Bigg(
    -\frac1{2\epsilon^2}
    -\frac5{4\epsilon  }
    - 3 - \frac32 \zeta_2 - \frac{9-4x}{6-4x}H_0(x) - H_{1,0}(x)
\Bigg)\,.
\end{align}
In particular we do not renormalise the electromagnetic coupling
$\alpha$ because no \ac{QED} vertex is present at \ac{LO} (cf.
Section~\ref{sec:renorm:practical}).

\subsection{The original scheme (\textsc{hv})}
In \hv, we treat the regular fields four dimensionally. This means
that at tree level we do a strictly four-dimensional calculation
\begin{align}\begin{split}
\mat0_\hv &= \frac12\times \frac{4G_F^2}2\tr \Big[
  \s4{\slashed{q}}\ \s4\gamma^\mu
 \ \big(\s4{\slashed{p}}+M\big)\ \s4\gamma^\nu 
\Big] \times \s4{\mathcal{N}}^{\mu\nu}
 = \frac{8G_F^2}3 M^4 x(3-2x) \\&= \mat0_\cdr\Big|_{d=4}\,.
\end{split}\end{align}

At one-loop we have to be more careful as we have objects of
different dimensions in one trace
\begin{align}
\mat1_\hv\Big|_\text{bare} = \frac{\alpha_0G_F^2}\pi \int[\D k] \frac{
\tr\Big[
   \s4{\slashed{q}}\,\,
   \sd\gamma^\sigma
   \big(\sd{\slashed{k}}-\sd{\slashed{q}}  \big)
   \s4\gamma^\mu
   \big(\sd{\slashed{k}}-\sd{\slashed{p}}-M\big)
   \sd\gamma^\sigma\,\,
   \big(\s4{\slashed{p}}+M\big)\,\,
   \s4\gamma^\nu
\Big]}{\den1\den2\den3}\s4{\mathcal{N}}^{\mu\nu}\,.
\label{eq:numeratorHV}
\end{align}
Note that the only meaningful difference to the \cdr{} discussion is
the dimensionality of the $\gamma^\mu$ and $\gamma^\sigma$.  When
calculating in \hv{}, we need to utilise that $\Sf\subset \QSd$. This
is a rather powerful statement because it allows us to calculate the
product of two vectors in different spaces as
\begin{align}
 \sd a\cdot \s4 b = \s4 a\cdot \s4b\,.
\end{align}
This relation goes both ways. After using standard trace techniques
(taking care of the dimensionality of each $\gamma$ matrix) we can use
it the other way around to write the numerator again in terms of the
familiar $\sd k\cdot \sd p$, $\sd k\cdot\sd q$, and $\sd k\cdot \sd
k$. However, we also have a new type numerator with $\s4 k\cdot \s4k$
from
\begin{align}
\tr\Big[ \sd k\,\, \s4\gamma^\mu\,\,\sd k\,\,\s4\gamma^\mu\Big]
= 8\s4k\cdot\s4k - 16\sd k\cdot\sd k\,.
\end{align}
Using these relations, we find
\begin{align}\begin{split}
\mat1_\hv\Big|_\text{bare} = \frac{\alpha_0}\pi\frac{8G_F^2}3 \int[\D k]\ &\frac{1}{
    \den1\den2\den3}
\bigg(
  2(3M^2 - 2s)s^2 \\&
+ 4(3M^4 - 4M^2s + 2s^2) (\sd k\cdot \sd q) 
- 4(3M^2-2s)s            (\sd k\cdot \sd p) \\&
+ 4(d-2)M^2              (\sd k\cdot \sd q)^2
+ 4(d-2)(M^2-2s)         (\sd k\cdot \sd p)(\sd k\cdot \sd q) \\&
- (d-2)(2s-3M^2)s        (\sd k\cdot \sd k)
- 2(M^2 - s)s(d-2)       (\s4 k\cdot \s4 k)
\bigg)\,.\end{split}
\end{align}
With the $(\s4 k\cdot \s4 k)$ in the numerator, we have in principle a
new class of integrals to discuss. These will be related to the
$\mu$-integrals of Section~\ref{sec:fdf}. However, for now we can just
solve these integrals using Passarino-Veltman
decomposition~\cite{Passarino:1978jh} (for a didactic introduction
cf.~\cite{Ellis2011One-loop})
\begin{align}
\int[\D k]\frac{\s4k\cdot \s4k}{\den1\den2\den3} = 
\s4g^{\mu\nu} \int[\D k]\frac{\sd k^\mu\sd k^\nu}{\den1\den2\den3}\,,
\label{eq:hv:s4k}
\end{align}
because $\Sf\subset\QSd$.

Solving the loop integral and renormalising with $Z_{2,\hv}^\mu =
Z_{2,\cdr}^\mu$ we find a familiar result
\begin{align}
\mat1_\hv = \frac\alpha\pi\bigg(\frac{\mu^2}{M^2x^2}\bigg)^\epsilon \mat0_\hv \Bigg(
    -\frac1{2\epsilon^2}
    -\frac5{4\epsilon  }
    - 3 - \frac32\zeta_2 - \frac{9-4x}{6-4x}H_0(x) - H_{1,0}(x)
\Bigg)\,.
\label{eq:hvres}
\end{align}
Note that here we pulled out a factor $\mat0_\hv$ instead of
$\mat0_\cdr$. This makes the fully expanded result simpler in
comparison with the fully expanded result of $\mat1_\cdr$ as there are
no terms $\epsilon/\epsilon$ from the poles with the linear parts of
the tree level result. We refer to this as \term{trivial scheme
dependence} because nothing relevant has changed. This is similar to
the scheme dependence due to the neutrino tensor. The $\mat0_\rs$ term
will appear in all parts of the calculations, i.e. both real and
virtual. Intermediary results will have to be different due to the
trivial scheme dependence but any physical, i.e. finite, result will
be independent because $\mat0_\rs$ acts as a pre-factor.

\subsection{The four-dimensional helicity scheme (\textsc{fdh})}
\label{sec:fdh}
The goal of \fdh{} is to treat as many objects in $d_s\equiv4$
dimensions as possible. While this simplifies things a lot, `there
ain't no such thing as a free lunch'. This popular saying manifests
itself in the existence of so-called $\epsilon$-scalars.

We treat singular vector fields in $\QSs=\QSd\oplus\QSe$ which means
that we have to write the covariant derivative as
\begin{align}
\sds{D^\mu}\psi_0 = \sd\partial^\mu\psi_0 + 
    \I\big(e_0 \sd{(A_0)^\mu}+e_{e,0} \se{(A_0)^\mu}\big)\psi_0\,,
    \label{eq:splita}
\end{align}
with a bare $\epsilon$-scalar $\se{(A_0)^\mu}$ with an evanescent
coupling $e_{e,0}$ to fermions. This split, that spoils \fdh's
simplicity, is necessary as $e_{e,0}$ is not protected by the
$d$-dimensional gauge symmetry and is renormalised differently. In
\ac{QED}, the corresponding $\beta$-functions
are~\cite{Gnendiger:2014nxa} (cf. Appendix~\ref{sec:const:ren})
\begin{align}\begin{split}
\beta   &= -\alpha \Bigg(\Big(\frac{\alpha}{4\pi}\Big)
    \underbrace{\Big[-\frac43 N_F\Big]}_{\beta_0\equiv\beta_{20}}
    +\mathcal{O}(\alpha^2)
    \Bigg)
\,,\\
\beta_e &= -\alpha_e\Bigg(
           \Big(\frac{\alpha_e}{4\pi}\Big)\underbrace{\Big[-4-2N_F\Big]}_{\beta_{02}}
        +  \Big(\frac{\alpha  }{4\pi}\Big)\underbrace{\Big[+6\Big]}_{\beta_{11}}
 +\mathcal{O}(\alpha_i^2) \Bigg) \,.
\end{split}\label{eq:beta}\end{align}
Therefore, one would have to perform any $\ell$-loop calculation with
both $\epsilon$-scalars and normal gauge bosons, keeping the couplings
$e$ and $e_e$ different. After renormalisation one can safely set
$e_e\to e$. This increases the number of diagrams by
$\mathcal{O}(2^\ell)$.

Fortunately, there is a silver lining: because the effect of
$\epsilon$-scalars is limited the their coupling's renormalisation,
there is actually no need to use~\eqref{eq:splita} at the $\ell$-loop
level.  This gives us the following prescription for an $\ell$-loop
calculation

\begin{itemize}

    \item
    Use~\eqref{eq:splita} for anything up to the $(\ell-1)$-level and
    renormalise correctly.

    \item
    At the $\ell$-loop level, perform the calculation using only
    quasi-four-dimensional objects.

    \item
    Add everything up and set $e_e\to e$ and $\neps\to2\epsilon$. If
    necessary, convert to a different renormalisation scheme for the
    coupling now.

\end{itemize}

Especially for one-loop $\epsilon$-scalars are not needed at all
because the renormalisation could only influence tree level
$\epsilon$-scalars that do not exist because there are no singular
fields at leading order. Hence, one-loop calculations can be performed
without worrying about $\epsilon$-scalars (for a particular efficient
way to exploit this, cf. Section~\ref{sec:fdf}). This would still work
in QCD when including \ac{PDF}s~\cite{Signer:2008va}.

At tree level in \fdh{}, we obtain the same result as in \hv{} because
both schemes treat regular fields in $\Sf$. For illustration, we will
calculate the one-loop \fdh{} result twice: once carefully
differentiating $\epsilon$-scalar contributions with
$\epsilon$-scalars and once ignoring $\epsilon$-scalars at one-loop,
while instead working in $d_s\equiv4$ dimensions.

\subsubsection{Calculation with $\epsilon$-scalars}
\begin{figure}[t]
\centering
\input{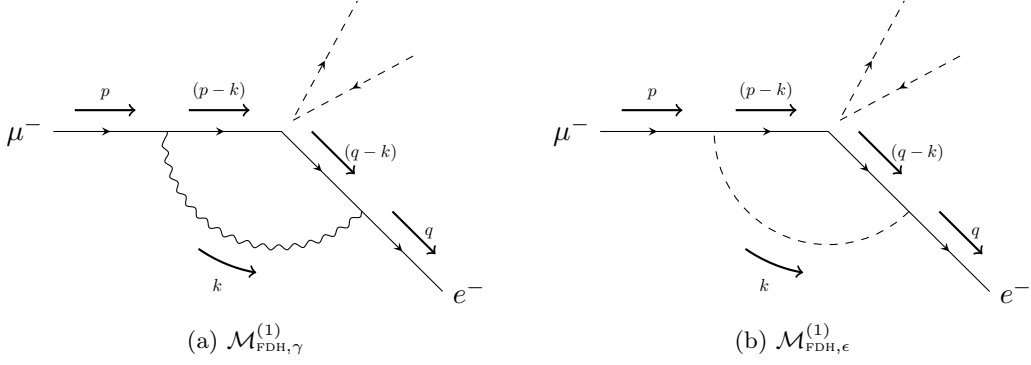}
\caption{The two diagrams in full \fdh{}. The first diagram is
identical to Figure~\ref{fig:muonfeynman}.}
\label{fig:muonfdh}
\end{figure}

We have to calculate the two diagrams of Figure~\ref{fig:muonfdh}: one
diagram with a virtual photon, Figure~\ref{fig:muonfdh:photon},
\begin{subequations}
\begin{align}
 \mat1_{\fdh,\gamma} = \frac{\alpha_0G_F^2}\pi \int[\D k] \frac{
\tr\Big[
   \s4{\slashed{q}}\,\,
   \sds\gamma^\sigma
   \big(\sd{\slashed{k}}-\sd{\slashed{q}}  \big)
   \s4\gamma^\mu
   \big(\sd{\slashed{k}}-\sd{\slashed{p}}-M\big)
   \sds\gamma^\sigma\,\,
   \big(\s4{\slashed{p}}+M\big)\,\,
   \s4\gamma^\nu
\Big]}{\den1\den2\den3}\s4{\mathcal{N}}^{\mu\nu}\,,
\label{eq:fdhgammapre}
\end{align}
and one with a virtual $\epsilon$-scalar,
Figure~\ref{fig:muonfdh:scalar},
\begin{align}
 \mat1_{\fdh,\epsilon} = \frac{\alpha_{e,0}G_F^2}\pi \int[\D k] \frac{
\tr\Big[
   \s4{\slashed{q}}\,\,
   \se\gamma^\sigma
   \big(\sd{\slashed{k}}-\sd{\slashed{q}}  \big)
   \s4\gamma^\mu
   \big(\sd{\slashed{k}}-\sd{\slashed{p}}-M\big)
   \se\gamma^\sigma\,\,
   \big(\s4{\slashed{p}}+M\big)\,\,
   \s4\gamma^\nu
\Big]}{\den1\den2\den3}\s4{\mathcal{N}}^{\mu\nu}\,,
\end{align}
\end{subequations}
where $\alpha_{e,0}=e_{e,0}^2/(4\pi)$ in accordance with the
definition of $\alpha_0$.  Because $\sd{\slashed{k}} = \sd\gamma^\mu
\sd k^\mu = \sds\gamma^\mu \sd k^\mu$, we can do most of the algebra
of $\mat1_{\fdh,\gamma}$ in $d_s$ dimensions as long as we keep track
of the dimensionality of the $k$.

By employing the same tricks as above for the Dirac algebra, we find
without specifying $d_s$ for the bare matrix element
\begin{subequations}
\begin{align}
\begin{split}
\mat1_{\fdh,\gamma}\Big|_\text{bare}
  &= \frac{\alpha_0}\pi\bigg(\frac{\mu^2}{M^2x^2}\bigg)^\epsilon \mat0_\fdh \Bigg(
    -\frac1{2\epsilon^2}
    -\frac{6-d_s}{4\epsilon  }
    - \frac{10-d_s}{4} - \frac32\zeta_2 \\&\qquad\qquad+ \frac{(d_s-4)(3-x)-2x}{6-4x}H_0(x) - H_{1,0}(x)
\Bigg)
\end{split}\label{eq:fdhgamma}\,,\\
\mat1_{\fdh,\epsilon}\Big|_\text{bare}
  &=
  \frac{\alpha_{e,0}}\pi\neps\bigg(\frac{\mu^2}{M^2x^2}\bigg)^\epsilon \mat0_\fdh \Bigg(
     \frac{1}{4\epsilon  }
    +\frac14 + \frac{3-x}{6-4x} H_0(x)
\Bigg)\,.
\end{align}
\end{subequations}
For the renormalisation we need to consider the effect of
$\epsilon$-scalars to $Z_2^\mu$
\begin{align}
 \sqrt{Z_{2,\fdh}^\mu} = 1+
 \frac{\alpha_0  }{2\pi}\bigg( -\frac3{2\epsilon} + \frac32\log\frac{M^2}{\mu^2} - \frac42 \bigg)
+\frac{\alpha_{e,0}}{2\pi}\neps\bigg( -\frac1{4\epsilon} + \frac14\log\frac{M^2}{\mu^2} - \frac14 \bigg)
\,.
\end{align}
Our renormalised \fdh{} result is therefore
\begin{align}
\begin{split}
\mat1_{\fdh}&=
\mat1_{\fdh,\gamma}+\mat1_{\fdh,\epsilon}\\&
  = \bigg(\frac{\mu^2}{M^2x^2}\bigg)^\epsilon \mat0_\fdh \Bigg[
  \frac{\alpha}{\pi}
  \Bigg(
    -\frac1{2\epsilon^2}
    +\frac{d_s-9}{4\epsilon}
    - \frac{10+(4-d_s)}{4} - \frac32\zeta_2
      \\&\hspace{4.5cm}
     + \frac{(d_s-4)(3-x)-9+4x}{6-4x}H_0(x) - H_{1,0}(x)
\Bigg)
\\&\hspace{3cm}
 +\frac{\alpha_e}{\pi}\frac\neps4
  \Bigg(
    \frac1{2\epsilon}+\frac12-\frac3{2x-3}H_0(x)
\Bigg)\Bigg]\,.
\end{split}
\label{eq:resfdh}
\end{align}
This result is what we refer to as \term{two-loop ready}. It has the
explicit dependence on the $\epsilon$-scalars so that we could -- and
in Chapter~\ref{ch:twoloop} will -- perform the two-loop calculation
in \fdh{} with the correct renormalisation of $\alpha_e$.

\subsubsection{Calculation without $\epsilon$-scalars}
Assuming we do not actually want to perform a two-loop calculation, we
can simplify the calculation by just setting $d_s\equiv4$ and
$\neps=0$ in the one-loop calculation from the get-go with
$\alpha_e=\alpha$. However, we still need to keep in mind that
$\Sf\subset \QSs\equiv{\rm QS}_{[4]}$, i.e.  that $\sd a\cdot \sds
b=\sd a\cdot \sd b$, allowing us to perform the algebra $d_s\equiv 4$
dimensionally. For complicate processes this can simplify the algebra
massively as there is no need to keep track of $\epsilon$-terms
induced by the algebra. We of course lose the generality of two-loop
readiness.

Our one-loop bare result is just~\eqref{eq:fdhgamma} with $d_s=4$.
However, we need to keep the $\neps$ in $Z_2^\mu$ resulting in
\begin{subequations}
\begin{align}
\mat1_{\fdh}\Big|_\text{bare}
  &= \frac\alpha\pi\bigg(\frac{\mu^2}{M^2x^2}\bigg)^\epsilon \mat0_\fdh \Bigg(
    -\frac1{2\epsilon^2}
    -\frac{1}{2\epsilon  }
    - \frac{3}{2} - \frac32\zeta_2- \frac{2x}{6-4x}H_0(x) - H_{1,0}(x)
\Bigg)
\,,\\
Z_{2,\fdh}^\mu&= 1+
 \frac{\alpha  }{2\pi}\bigg( -\frac3{2\epsilon} +
 \frac32\log\frac{M^2}{\mu^2} - \frac52 \bigg)\,.
\end{align}\end{subequations}
This directly results in what we find if we set $d_s=4-2\epsilon$ and
$\neps=2\epsilon$ in~\eqref{eq:resfdh}
\begin{align}
\mat1_\fdh = \frac\alpha\pi\bigg(\frac{\mu^2}{M^2x^2}\bigg)^\epsilon \mat0_\fdh \Bigg(
    -\frac1{2\epsilon^2}
    -\frac5{4\epsilon  }
    - \frac{11}4 - \frac32\zeta_2 - \frac{9-4x}{6-4x}H_0(x) - H_{1,0}(x)
\Bigg)\,.
\label{eq:fdh4res}
\end{align}
Compare this result to the \hv{} result from~\eqref{eq:hvres}. The
only difference is the rational number in the finite part that changes
from $-3\to -11/4$. This the first time we have encountered
\term{non-trivial scheme dependence} (this is in contrast to the
trivial scheme dependence between \cdr{} and \hv{}).  As we will see
in Section~\ref{sec:schemedep}, we can predict the scheme dependence
without having to calculate the different contributions.

It is important that all non-trivial scheme dependence will cancel as
soon as $\mat1_\rs$ is combined with the real correction in the same
scheme $\rs$. This is crucial as otherwise the different schemes would
not be consistent.

\subsubsection{Renormalisation of \textsc{fdh} beyond leading order}
Beyond what we have discussed in Section~\ref{sec:renorm}, the
renormalisation in \fdh{} is complicated by the presence of
$\epsilon$-scalars at the one-loop level. Additionally to the issue of
the different coupling we have already discussed, there is one more
subtlety in a theory with massive flavours. In contrast to the vector
boson propagator, there is no symmetry that protects the propagator of
the $\epsilon$-scalar from acquiring a mass term $\propto m^2
\se{g^{\mu\nu}}$~\cite{ Gnendiger:2016cpg, Jack:1994rk}. This
effectively shifts the scalar's mass from zero, requiring an
appropriate counter term to restore a vanishing $\epsilon$-scalar
mass.  Hence, we add the following steps in the discussion of
Section~\ref{sec:renorm:practical}
\begin{enumerate}
\setcounter{enumi}{4}
     \item
     Maintain the masslessness of the $\epsilon$-scalars by
     substituting
     \begin{align}
         \frac{-\I\se{g^{\mu\nu}}}{p^2} \to
         \delta m_\epsilon^{(l)}\ \frac{-\I\se{g^{\mu\nu}}}{p^2}
     \end{align}
     at tree level and one-loop.
 
     \item
     Perform the coupling renormalisation by shifting $\alpha_{0,i}\to
     (1+\dZ[\alpha_i]1 + \dZ[\alpha_i]2) \times \alpha_i$ and sorting
     terms according to the now renormalised coupling, dropping every
     term with too high a power in $\alpha_i$.
 
     \item
     Identify $\alpha_e\equiv\alpha$ and set $\neps=2\epsilon$.

\end{enumerate}

\subsection{Four-dimensional formulation of \textsc{fdh} (\textsc{fdf})}
\label{sec:fdf}

At the one-loop level, \fdh{} is seemingly complicated by the presence
of objects with different dimensions in the traces and the need to
include $\epsilon$-scalars at tree level. There is also still the
problem of $\gamma^5$ which we have ignored so-far. The
four-dimensional formulation of \fdh{} (\fdf) solves both problems at
one-loop~\cite{Fazio:2014xea}.

Originally, the \fdf{} scheme was constructed to best use unitarity
methods for one-loop calculation. This is done by essentially
constructing one-loop amplitudes by sewing together tree-level
amplitudes, allowing for extremely efficent numerical evaluation of
one-loop amplitudes as done by e.g.  GoSam~\cite{Cullen:2014yla}.
However, for this to work all momenta including the loop momentum must
be in $\Sf$. We will not be discussing these methods further as \fdf{}
is for us just a particularly efficient way to calculate complicated
one-loop amplitudes.

We can simplify \fdh{} by realising that all our problems arise
because of the $\s4k^2$-terms in numerators of integrals that we have
postponed in our discussion of \hv. To understand these terms better,
we will introduce another space $\QSm$
\begin{subequations}
\begin{align}
\QSd=\Sf\oplus\QSm\,.
\end{align}
We can now write $\s4k^2$ as
\begin{align}
\sd k^2 = \s4k^2-\mu^2
\,,\label{eq:fdfsplit:k}
\end{align}
\label{eq:fdfsplit}
\end{subequations}
where $\mu^2$ is the remnant of the part of the loop momentum from
$\QSm$.  However, we can realise this already at the level of $\gamma$
matrices by setting
\begin{align}
\sd{\slashed k} = \s4{\slashed k} + \I \gamma^5 \mu\,.\label{eq:rule1}
\tag{Rule I}
\end{align}
One can easily verify that this definition
satisfies~\eqref{eq:fdfsplit:k}. We can now perform the entire
calculation in $\Sf$, up to terms $\mu\in\QSm$ for which we use
\begin{align}
\text{Odd powers of $\mu$ are set to zero}\,.
\tag{Rule II}
\end{align}
In our case, we write (cf.~\eqref{eq:numeratorHV})
\begin{align}\begin{split}
\mat1_\fdf\Big|_\text{bare} = \frac{\alpha}{\pi}\frac{8G_F^2}3 \Bigg[&\int[\D k] \frac{1}{\sd
k^2\,\den2\den3}\Bigg(2 (3M^2 - 2s)s^2
\\&\qquad
+ 4(3M^4 - 4M^2s + 2s^2)(\s4k\cdot \s4q)
- 4(3M^2  - 2s)s        (\s4k\cdot \s4p)
\\&\qquad
+ 8M^2                  (\s4k\cdot \s4q)^2
+ 8(M^2 - 2s)           (\s4k\cdot \s4p)(\s4k\cdot \s4q) 
\\&\qquad
+ 2M^2s \s4k^2 
\Bigg)
-2\int[\D k]\frac{\mu^2}{\sd k^2\,\den2\den3}s(3M^2-2s)\Bigg] \,.
\end{split}
\end{align}
Obtaining a result such as this was the original goal of \fdf{}. All
momenta are strictly four-dimensional, allowing to use (numerical)
unitarity. We, however, are not interested in unitarity, instead
wanting to use standard tools to calculate these integrals. For this
we need to be able to cancel the $\s4k^2$-terms against denominators
containing $\sd k^2$.  Hence, we reverse~\eqref{eq:fdfsplit:k},
properly implementing~\eqref{eq:hv:s4k}
\begin{align}
\begin{split}
&\text{After completing the algebra, use}\\
&\int[\D k]\frac{\s4k^2}{\sd k^2\den{2}\cdots\den{n}} = 
 \int[\D k]\frac{1}{\den{2}\cdots\den{n}}
+\int[\D k]\frac{\mu^2}{\sd k^2\den{2}\cdots\den{n}}
\,.
\end{split}
\tag{Rule III}
\end{align}
Now only terms of the form $\s4k\cdot\s4p$ remain in the numerators.
These terms map directly to $\sd k\cdot\s4p$ because external momenta
have no contribution in $\QSm$. Hence, we can now safely set $\s4k\to
\sd k$ and solve the $\mu^2$-independent integrals using standard
one-loop calculus as implemented in Package-X. For the
$\mu$-integrals, one can show that~\cite{Bern:1995db}
\begin{align}
\int[\D^dk] \frac{(\mu^2)^r}{\den1\cdots\den n} = -\epsilon (4\pi)^r 
\frac{\Gamma(r-\epsilon)}{\Gamma(1-\epsilon)}
  \int[\D^{d+2r}] \frac1{\den1\cdots\den n}\,.
\tag{Rule IV}
\end{align}
As these dimensionally shifted integrals are at most
\ac{UV}-divergent, we only ever need their \ac{UV} pole which is
generally very simple to obtain.  In our case,
\begin{align}
\mat1_\fdf\Big|_\text{$\mu$-integral} = -\frac{16G_F^2}3M^4(x-1) x \epsilon
\times \Big(-\frac1{2\epsilon} + \mathcal{O}(\epsilon^0)\Big)\,.
\end{align}
Thus, we have reproduced the bare \fdh{} result~\eqref{eq:fdhgamma}
with techniques more amenable for automated calculations without
introducing unnecessary terms such as in \cdr. Unfortunately, the
\fdf{} scheme has only been shown to work at one-loop. Worse yet, the
simplest extension to the two-loop level is known to be incorrect.

\subsubsection{$\gamma^5$ in \textsc{fdf}}
We have avoided the $\gamma^5$ problem by using the
$\psi_e\to\gamma^5\psi_e$ symmetry. However, a side effect of
\ref{eq:rule1} is that \fdf{} comes with a hard-coded $\gamma^5$
scheme~\cite{Gnendiger:2017rfh}. Because all objects except $\mu$ are
in $\Sf$, our $\gamma^5$ is anti-commuting in practice. However, due
to the implementation of the $\QSm$ terms as $\sim\gamma^5\mu$, these
terms effectively commute with $\gamma^5$. This means that we have
implemented the \BM{} scheme. Hence, \fdf{} requires the additional
$Z_5$ renormalisation.

In practice the Born matrix element is
\begin{align}
\mat0_\fdf &= \frac{4G_F^2}2\tr \Big[
   \big(\s4{\slashed{p}}+M\big)\ P_L\s4\gamma^\mu
\  \s4{\slashed{q}}\ P_L\s4\gamma^\nu 
\Big] \times \s4{\mathcal{N}}^{\mu\nu}\,,
\end{align}
where we set
\begin{align}
 P_L = \frac12(1-Z_5\gamma^5)\,,
\end{align}
with $Z_5$ as in~\eqref{eq:z5}.

\section{Regularisation-scheme dependence and IR prediction}\label{sec:schemedep}
As mentioned in the prologue to this chapter, we ideally want to
compute every part of the calculation in the best suited scheme. For
this it is important to understand how to convert from one scheme to
another.

For this we distinguish two different types of scheme dependence: a
trivial scheme dependence that is due to the dimensionality of the
Born matrix element and non-trivial scheme dependence. The trivial
scheme dependence is best described between \cdr{} and \hv{}. As
suggested in~\eqref{eq:hvres} this is just
\begin{align}
 \mat\ell_\cdr = \frac{\mat0_\cdr}{\mat0_\hv} \mat\ell_\hv\,.
\end{align}
The scheme dependence between \hv{} and \fdh{} is more interesting.
It is encapsulated by divergent diagrams involving $\epsilon$-scalars.
At one-loop we have
\begin{align}
\mat1_\fdh - \mat1_\hv = 2\epsilon \times
\mat1_\fdh\Big|_\text{$\neps$ coefficient} = \Delta_{\textsc{rs}}\,.
\end{align}
In our case, $\Delta_\textsc{rs} = M^4x(3-2x)/6 +
\mathcal{O}(\epsilon)$. Had we assumed a finite electron mass instead,
we would have found $\Delta_\textsc{rs}^{m_e>0}\propto \epsilon$ and
hence no non-trivial scheme dependence. At \ac{NLO}, scheme
dependence is induced by collinear
singularities~\cite{Kunszt1993One-loop} of which there are non for
finite electron masses. Beyond \ac{NLO}, also soft singularities
contribute to the scheme dependence, i.e. even in the case of
non-vanishing electron mass there is a non-trivial scheme dependence.

To formalise the scheme dependence, we need to extend our discussion
of \ac{IR} predictions from Section~\ref{sec:irpred} to \fdh{} in
\ac{QCD}. The relevant results are given in
Appendix~\ref{ch:fdhconst}. We begin by noting that the
\term{light-quark}\footnote{The terms light and heavy quark are
universally used to describe massless and massive quark,
respectively.} anomalous dimension $\gq$ has terms proportional to
$\neps$
\begin{align}
    \gq &= 
     \Big(\frac{\alpha  }{4\pi}\Big)(-3C_F)
    +\Big(\frac{\alpha_e}{4\pi}\Big)\neps\frac{C_F}2
    +\mathcal{O}(\alpha_i^2)\,.
\end{align}
The \term{heavy-quark} anomalous dimension $\gQ$ on the other hand has
no scheme dependence
\begin{align}
\gQ 
  &= \Big(\frac{\alpha  }{4\pi}\Big)(-2C_F)\,,
\end{align}
meaning that non-trivial scheme dependence is due collinear
singularities at one-loop. The cusp-anomalous dimension
\begin{align}
    \gcusp &= \Big(\frac{\alpha  }{4\pi}\Big)(4)
\end{align}
also has no scheme dependence at one-loop but develops a term $\propto
\alpha^2 C_A\neps$ at two-loop. Up to at least two-loop, $\gcusp$ has
\term{Casimir scaling}, i.e. $\gcusp^q/C_F=\gcusp^g/C_A=\gcusp$.

A further source of scheme dependence in $\scetz$ is that we now have
to include terms induced by the $\epsilon$-scalar coupling (cf.
Appendix~\ref{sec:const:scet}). This makes $\log\scetz_\rs$
regularisation scheme dependent as indicated by the subscript $\rs$.
Note that, because $\scetz_\rs$ matches the \ac{IR} poles exactly, the
$\MS$-like subtracted matrix element $\mathcal{M}_\text{sub}(\mu)$ is
regularisation scheme independent in the limit $\neps\to2\epsilon$ and
$\epsilon\to0$. This allows us to predict the regularisation scheme
dependence of any matrix element by predicting the scheme dependence
$\log\scetz_\rs$ using the \ac{IR} prediction discussed in
Section~\ref{sec:irpred}.

To illustrate this method, let us predict the \ac{IR} pole of
$\mat0_\rs$. For this, we first need to write down $\bf\Gamma$ as
\begin{align}\begin{split}
{\bf\Gamma}_1(\mu) &= \gq + \gQ - \gcusp \log\frac{M\mu}{2p\cdot q}
\\&
= -\frac{5\alpha}{4\pi} + \frac{\alpha_e}{8\pi}\neps -
\frac\alpha\pi\log\frac{\mu}{M x}\,.
\end{split}\end{align}
We now can construct $\log\scetz_\rs$ as in \eqref{eq:scetzfdh}
\begin{align}\begin{split}
\log\scetz_\rs &= 
   \frac{{\bf\Gamma}_1'}{2\epsilon^2}
  +\frac{{\bf\Gamma}_1 }{ \epsilon  }
+ \mathcal{O}(\alpha_i^2)
\\&
= \frac{\alpha}{\pi}\Bigg(-\frac1{2\epsilon^2} -
\frac1\epsilon\bigg[\frac54+2\log\frac{\mu^2}{M^2}-\log x\bigg]\Bigg)
+ \frac{\alpha_e}{\pi}\neps\Bigg(\frac1{8\epsilon}\Bigg)
+ \mathcal{O}(\alpha_i^2)\,,
\end{split}\end{align}
where we have used that ${\bf\Gamma}_1' =
\partial_{\log\mu}{\bf\Gamma} = \gcusp$. By exponentiating
$\log\scetz_\rs$ we can obtain $\scetz_\rs^{-1}$ as required
by~\eqref{eq:scetzdef}
\begin{align}
\scetz_\rs^{-1}
= 1
+ \frac{\alpha}{\pi}\Bigg(-\frac1{2\epsilon^2} -
\frac1\epsilon\bigg[\frac54+2\log\frac{\mu^2}{M^2}-\log x\bigg]\Bigg)
+ \frac{\alpha_e}{\pi}\neps\Bigg(\frac1{8\epsilon}\Bigg)
+ \mathcal{O}(\alpha_i^2)\,.
\end{align}
This, once multiplied with $\mat0_\rs+\mat1_\rs$, produces a finite,
scheme independent result in the limit $d_s\to4-2\epsilon$,
$\neps\to2\epsilon$, $\alpha_e\to\alpha$, and finally $\epsilon\to0$
\begin{align}\begin{split}
\scetz_{\rs}^{-1}\Big(\mat0_{\rs}+\mat1_{\rs}\Big) =
\mat0_{\rs}\Bigg(1+\frac\alpha\pi\bigg[&
    \Big\{3\frac{x-1}{2x-3}+L\Big\}
    -\frac54L - \frac14L^2-3\\&-\frac32\zeta_2
    -2H_{0,0}(x)-H_{1,0}(x)
    \Bigg]
    +\mathcal{O}(\alpha^2)\Bigg)\,,
\end{split}
\end{align}
with $L=\log(\mu^2/M^2)$ for all schemes $\rs\in\{\cdr,\hv,\fdh\}$.

After calculating $\scetz_{\rs}^{-1}\Big(\mat0_{\rs}+\mat1_{\rs}\Big)$
in any scheme $\rs$ we can obtain $\mat1_{\rs'}$ in any other scheme
$\rs'$ by multiplying with the corresponding $\scetz_{\rs'}$. This is
a very powerful statement, allowing us to perform any part of any one-
and two-loop calculation in any scheme we wish and convert to any
other scheme. In calculations for the LHC this is particularly
important because the parton distribution functions (\ac{PDF}s) are
usually only available in \cdr{}.

\chapter{The \texorpdfstring{FKS$^2$}{FKS2} scheme}\label{ch:fks}

As already discussed in Section~\ref{sec:irsafety}, cross sections
beyond \ac{LO} are constructed of several \ac{IR} divergent parts.
In this chapter, we will focus on the \ac{IR} divergences arising
during the phase-space integration of real corrections. As already
mentioned, we would like to do this integration numerically. However,
we cannot do this in $d$ dimensions. A common way to circumvent this
problem is called a \term{subtraction scheme}. The basic idea is to
write the divergent integrand over the extra emission as
\begin{align}
 \int_{n+1}\D\sigma_{n+1} =
 \int_{n+1}\big(\D\sigma_{n+1}-\D{\rm CT}\big)
+\int_{n  }\int_1\D{\rm CT}\,,
\end{align}
where the subscript refers to the number of particles integrated over.
$\D{\rm CT}$ is constructed to ensure that the first integral is
finite while being easy enough so that the integral over
the one-particle phase space can be done analytically in $d$
dimensions.

In the case of massive \ac{QED} the only \ac{IR} singularity is due
to soft photon emission; collinear divergences are regulated by the
presence of fermion masses. Hence, $\D{\rm CT}$ can be quite simple as
we will see below.

In this chapter we will review one of the central pieces of this
project, the \ac{FKS2} subtraction scheme (Section~\ref{sec:fks2}),
as well as its predecessor, the \ac{FKS} scheme
(Section~\ref{sec:fks}). Next, we will comment on the possibility of
extending the scheme beyond \ac{NNLO} in Section~\ref{sec:beyond}.
Finally, we will comment on properties of \ac{FKS2} in
Section~\ref{sec:comments}.

\section{FKS for soft singularities at NLO}\label{sec:fks}

In this section we will briefly summarise the necessary aspects of the
\ac{FKS} scheme at \ac{NLO}. Because we only treat soft
singularities, \ac{FKS} is dramatically simplified. The \ac{NLO}
correction to a cross section is split into virtual and real parts
\begin{align}
\label{eq:sigmanlo}
    \sigma^{(1)} = \int \Big(\bbit{1}{v} + \bbit{1}{r}  \Big)
= \int\D\Phi_n\,\M n1 +\int\D\Phi_{n+1}\, \M{n+1}0\,.
\end{align}
In \eqref{eq:sigmanlo} we implicitly assume the presence of the flux
factor (or the analogous factor for a decay rate) as well as a
measurement function that defines the observable in terms of the
particle momenta. The measurement function has to respect infrared
safety, i.e. the observable it defines must not depend on whether or
not one or more additional soft photons are present as arguments of
this function.

The real corrections
\begin{align}
\label{eq:nloreal}
\bbit{1}{r} = \D\Phi_{n+1}\, \M{n+1}0\, 
\end{align}
are obtained by integrating the tree-level matrix element $\M{n+1}0$
over the phase space $\D\Phi_{n+1}$. To simplify the discussion we
assume that in the tree-level process described by $\M{n}{0}$ no
final-state photons are present. Hence, in $\M{n+1}0$ only the
particle (photon) with label $n+1$ can potentially become soft. If
there are additional photons (i.e. photons in the \ac{LO} process)
the measurement function and combinatorics become slightly more
involved, but the essential part of the discussion is not affected.

When computing a cross section in the centre-of-mass frame, we choose
coordinates where the beam axis is in $z$ direction. Further, we
denote the (partonic) centre-of-mass energy by $\sqrt s$.  When
computing a decay width we instead parametrise one of the outgoing
particles in $z$ direction and, if necessary, rotate the coordinate
system afterwards.

Following~\cite{Frixione1995Three-jet} we parametrise the momentum of
the additionally radiated particle $n+1$ as\footnote{Note that this
parametrisation could also tackle initial-state collinear
singularities because $y_1$ corresponds to the angle between the photon
and the incoming particles. However, a different parametrisation may
be sensible (and is allowed here) to better account for \ac{PCS} from
light particles (cf.  Section~\ref{sec:pcs} and Section~\ref{sec:ps}).
What is important in the following is that the scaled energy $\xi_1$
is chosen as a variable in the parametrisation to ensure a consistent
implementation of the distributions defined in \eqref{eq:xidist}. }
\begin{align}
\label{eq:kdef}
k_1 = p_{n+1} = \frac{\sqrt s}2\xi_1   (1,\sqrt{1-y_1  ^2}\vec e_\perp,y_1  )\,,
\end{align}
where $\vec e_\perp$ is a $(d-2)$ dimensional unit vector and the
ranges of $y_1$ (the cosine of the angle) and $\xi_1$ (the scaled
energy) are $-1\le y_1 \le 1$ and $0\le\xi_1 \le\xi_\text{max}$,
respectively. The upper bound $\xi_\text{max}$ depends on the masses
of the outgoing particles.  Following \cite{Frederix2009Automation} we
find
\begin{align}
\xi_\text{max} = 1-\frac{\Big(\sum_i m_i\Big)^2}{s}\,.
\end{align}
Further kinematic constraints are assumed to be implemented through
the measurement function.  We write the single-particle phase-space
measure for particle $n+1$ as
\begin{align}
\D\phi_1 &\equiv
   \mu^{4-d}\frac{\D^{d-1}k_1}{(2\pi)^{d-1}\,2k_1^0}
   = \frac{\mu^{2\epsilon}}{2(2\pi)^{d-1}}
   \left(\frac{\sqrt{s}}2\right)^{d-2}
  \xi_1^{1-2\epsilon}(1-y_1^2)^{-\epsilon}\,
      \D\xi_1\,\D y_1\,\D\Omega_1^{(d-2)} =
      \D\Upsilon_1 \D\xi_1 \ \xi_1^{1-2\epsilon}
      \, ,
\label{eq:para}
\end{align}
where the angular integrations and other trivial factors are collected
in $\D\Upsilon_1$.  Denoting by $\D \Phi_{n,1}$ the remainder of the
$(n+1)$-parton phase space, i.e. $\D \Phi_{n+1} = \D \Phi_{n,1}
\D\phi_1$, we write the real part of the \ac{NLO} differential cross
section as
\begin{align}
\bbit{1}{r} &= \D \Phi_{n,1}\,  \D\phi_1\,\M{n+1}0
= \pref1\,  \D\xi_1\ \xi_1^2\M{n+1}0 \xi_1^{-1-2\epsilon} \,.
\label{eq:realnlo}
\end{align}
To isolate the soft singularities in the phase-space integration we
use the identity
\begin{align}\label{eq:xidist}\begin{split}
\xi^{-1-2\epsilon} &=
  -\frac{\xc^{-2\epsilon}}{2\epsilon}\delta(\xi) +
  \cdis{\xi^{1+2\epsilon}}
\,,\\
\left\langle \cdis{\xi^n}, f\right\rangle
&=
\int_0^1\D\xi\,\frac{f(\xi)-f(0)\theta(\xc-\xi)}{\xi^n}
\,,
\end{split}\end{align}
to expand $\xi_1^{-1-2\epsilon}$ in terms of a
\term{$c$-distribution}.  Here we have introduced an unphysical free
parameter $\xc$ that can be chosen
arbitrarily~\cite{Frixione1995Three-jet,Frederix2009Automation} as
long as
\begin{align}
0<\xc\le\xi_\text{max}\,.
\end{align}
The dependence of $\xc$ has to drop out exactly since no approximation
was made.  Therefore, any fixed value could be chosen. However,
keeping it variable is useful to test the implementation of the
scheme.

Using~\eqref{eq:xidist} we split the real cross section into a hard
and a soft part\footnote{In~\cite{Frixione1995Three-jet} the second
  term is called $\bit{ns}$ for `non-soft'. We will label it $h$ (for
  `hard') instead to avoid confusion when we need more than one such
  label later.}
\begin{subequations}
\begin{align}
\bbit{1}{r}\phantom{(\xc)} &= \bbit{1}{s}(\xc) + \bbit{1}{h}(\xc)
\,,\\
\bbit{1}{s}(\xc)
 &=
  -\pref1\ \frac{\xc^{-2\epsilon}}{2\epsilon}\ \delta(\xi_1)\,
      \D\xi_1\,
  \Big(\xi_1^2\M{n+1}0\Big)
\,,\\
\label{eq:nloh}
\bbit{1}{h}(\xc)
 &=
 +\pref1\ \cdis{\xi_1^{1+2\epsilon}}
      \D\xi_1
      \Big(\xi_1^2\M{n+1}0\Big)\,.
\end{align}
\end{subequations}
In $\bbit{1}{s}$ we can now (trivially) perform the $\xi_1$
integration.  To do this systematically, we define for photons the
general soft limit $\mathcal{S}_i$ of the $i$-th particle
\begin{align}
\mathcal{S}_i\M m0\equiv\lim_{\xi_i\to0}\xi_i^2\M m0
    =\eik_i\M{m-1}0
\qquad\text{with}\qquad
\xi_i = \frac{2E_i}{\sqrt{s}}\,,
\end{align}
where $\M{m-1}0$ is the matrix element for the process without
particle $i$. The \term{eikonal factor}
\begin{align}
\label{eq:eikonal}
    \eik_{i} \equiv 4\pi \alpha \, \sum_{j,k} \frac{p_j\cdot
      p_k}{p_j\cdot n_i\,p_k\cdot n_i}\ {\rm sign}_{jk}
    \qquad\text{with}\qquad p_i=\xi_i n_i\,,
\end{align}
is assembled from self- and mixed-eikonals. ${\rm sign}_{jk} =
(-1)^{n_{jk}+1}$ as in Section~\ref{sec:irpred}, where $n_{jk}$ is the
number of incoming particles or outgoing antiparticles among the
particles $j$ and $l$. Further, we define the \term{integrated
eikonal}
\begin{align}
\label{eq:inteik}
\ieik(\xc)
\equiv -\frac{\xc^{-2\epsilon}}{2\epsilon} \int\D\Upsilon_i\ \eik_i
= \xc^{-2\epsilon}\ \ieik(1)
= \sum_{j,k} \ieik_{jk}(\xc)\,.
\end{align}
$\ieik$ has been computed for example in~\cite{Frixione1995Three-jet,
Frederix2009Automation} and can be found in Appendix~\ref{ch:eik}.
This definition of $\ieik$ completes the definition of the \ac{YFS}
split~\eqref{eq:yfs} with $\alpha S=\ieik$.  After $\D \Upsilon_1$ and
$\D\xi_1$ integration (under which $\D\Phi_{n,1} \to \D \Phi_n$) we
obtain
\begin{align}
\bbit{1}{s}(\xc)\
    & \stackrel{\int\D\Upsilon_1\D\xi_1}{\longrightarrow}
\
\D \Phi_{n}\ \ieik(\xc)\,\M n0\,.
\label{eq:nlo:s}
\end{align}
This part now contains explicit $1/\epsilon$ poles that cancel against
poles in the virtual cross section. The second term of the real
corrections, $\bbit{1}{h}$ given in \eqref{eq:nloh}, is finite and can
be integrated numerically after setting $d=4$.  Combining the real and
virtual corrections, the \ac{NLO} correction is given by
\begin{subequations}
\label{eq:nlo:4d}
\begin{align}
\sigma^{(1)} &=
\sigma^{(1)}_n(\xc) + \sigma^{(1)}_{n+1}(\xc) \, , \\
\sigma^{(1)}_n(\xc) &= \int
\ \D\Phi_n^{d=4}\,\Bigg(
    \M n1
   +\ieik(\xc)\,\M n0
\Bigg) = 
\int \ \D\Phi_n^{d=4}\, \fM n1
\,,
\label{eq:nlo:n}
\\
\sigma^{(1)}_{n+1}(\xc) &= \int 
\ \D\Phi^{d=4}_{n+1}
  \cdis{\xi_1} \big(\xi_1\, \fM{n+1}0 \big)
\label{eq:nlo:n1}
\, .
\end{align}
\end{subequations}
We have defined $\fM{n+1}0 = \M{n+1}0$ and absorbed one of the $\xi_1$
factors multiplying $\M{n+1}0$ in~\eqref{eq:nloh} in the phase space
$\D\Phi^{d=4}_{n+1}$. Contrary to \eqref{eq:sigmanlo}, there are no
soft singularities present in \eqref{eq:nlo:4d}. According to
\eqref{eq:yfs} the explicit $1/\epsilon$ poles cancel between the two
terms in the integrand of \eqref{eq:nlo:n} and the phase-space
integration in \eqref{eq:nlo:n1} is also manifestly finite.

In \eqref{eq:nlo:4d} we see first terms of the build-up of the YFS
split~\eqref{eq:yfs}
\begin{align}
e^{\alpha \ieik}\, \sum_{\ell = 0}^\infty \M{n}{\ell} = 
\sum_{\ell = 0}^\infty \fM{n}{\ell}
= 
    \M n1
   +\ieik(\xc)\,\M n0 + \mathcal{O}(\alpha^2)\,.
\label{eq:yfsnew}
\end{align}

Finally, we note that $\mathcal{S}_i$ is invariant under rotations,
but not Lorentz invariant, because it contains the explicit energy
$E_i$. Hence, also $\eik_i$ and $\ieik$ are only invariant under
rotations but not under general Lorentz transformations. The
integrated eikonal $\ieik_{jk}$ has been computed
in~\cite{Frederix2009Automation}, dropping terms of
$\mathcal{O}(\epsilon)$. As we will see this is sufficient even beyond
\ac{NLO}. The expression is given in Appendix~\ref{ch:eik}, using our
conventions.

\section{\texorpdfstring{FKS$^2$}{FKS2}: NNLO extension}
\label{sec:fks2}

In the following, we discuss the extension of \ac{FKS} to \ac{NNLO},
while still limiting ourselves to massive \ac{QED}.  To simplify the
discussion in this section, we assume that all (suitably renormalised)
matrix elements are known to sufficient order in the coupling and
expansion in $\epsilon$. In Section~\ref{sec:fksscheme} we will state
what precisely is needed for a \ac{NNLO} computation.

We write the \ac{NNLO} cross section $\sigma^{(2)}$ as
\begin{align}
\label{eq:sigmannlo}
    \sigma^{(2)} = \int \Big(\bbit{2}{vv} + \bbit{2}{rv} +
    \bbit{2}{rr} \Big)
=
\int\D\Phi_n\,\M n2
+\int\D\Phi_{n+1}\, \M{n+1}1 +\int\D\Phi_{n+2}\, \M{n+2}0\,.
\end{align}
The double-virtual corrections are obtained by integrating $\M n2$
over the Born phase space $\D\Phi_n$. Here $\M n2$ contains all terms
of the $n$-particle (renormalised) matrix element with two
additional powers of the coupling $\alpha$. This includes the
interference term of the two-loop amplitude with the tree-level
amplitude as well as the one-loop amplitude squared. Similarly, the
real-virtual contribution is obtained by integration of $\M{n+1}1$,
the interference of the (renormalised) $(n+1)$-particle one-loop
amplitude with the corresponding tree-level amplitude, over the
$(n+1)$-particle phase space $\D\Phi_{n+1}$. Finally, for the
double-real contribution the tree-level matrix element with two
additional particles, $\M{n+2}0$, is integrated over the corresponding
phase space. 

\subsection{Real-virtual correction}

The treatment of the real-virtual contribution
\begin{align}
\label{eq:nnlorv}
\bbit{2}{rv} = \D\Phi_{n+1}\, \M{n+1}1
\end{align}
proceeds along the lines of normal \ac{FKS} because it is a
$(n+1)$-particle contribution. Again we assume that there is only one
external particle, with label $n+1$, that can potentially become
soft. We use \eqref{eq:xidist} with another unphysical cut-parameter
$\xi_{c_A}$ to split the real-virtual cross section into a soft and a
hard part
\begin{align}
\label{eq:shnnlo}
\bbit{2}{rv} = \bbit{2}{s}(\xi_{c_A}) + \bbit{2}{h}(\xi_{c_A}) \, .
\end{align}
For $\bbit{2}{s}$ the analogy to the \ac{NLO} case is particularly
strong because there is no genuine one-loop eikonal
contribution~\cite{Bierenbaum:2011gg,Catani:2000pi}, i.e. the soft
limit of the real-virtual matrix element is
\begin{align}
\mathcal{S}_{n+1}\M{n+1}1=\eik_{n+1} \M n1\,,
\end{align}
with the same $\eik_{n+1}$ as in~\eqref{eq:eikonal}. Therefore,
compared to \eqref{eq:nlo:s} the definition of the soft part remains
essentially unchanged
\begin{align}
\bbit{2}{s}(\xi_{c_A})\
    & \stackrel{\int\D\Upsilon_1\D\xi_1}{\longrightarrow}
  \ \D \Phi_{n}\ \ieik(\xi_{c_A})\,\M n1\,.
\label{eq:nnlo:s}
\end{align}
However, $\bbit{2}{s}$ has a double-soft $1/\epsilon^2$ pole from the
overlap of the soft $1/\epsilon$ poles of $\ieik$ and $\M n1$.

Unfortunately, $\bbit{2}{h}$ is not yet finite as it contains an
explicit $1/\epsilon$ pole from the loop integration. With the
$\MS$-like \ac{IR} subtraction of Section~\ref{sec:irpred}, we
already found one way to remove this pole by defining
$\scetz=1+\alpha\delta\scetz$ s.t.
\begin{align}
\mathcal{M}_\text{sub}(\mu) = \M{n+1}1 - \delta\scetz(\mu)\M{n+1}0
\end{align}
is finite.  This is \eqref{eq:scetzdef} expanded in $\alpha$ and
applied to our discussion.  However, it turns out that a different
subtraction, called \term{eikonal subtraction}, is more advantageous.
We split the real-virtual matrix element according to
\begin{align}
 \label{eq:polemrv}
  \fM{n+1}1 = \M{n+1}1(\xi_{c_B}) + \ieik(\xi_{c_B})\,\M{n+1}0
\end{align}
into a finite and a divergent piece.  The pole of $ \M{n+1}1$ is now
contained in the integrated eikonal of $\ieik(\xi_{c_B})\,\M{n+1}0$,
whereas the eikonal-subtracted matrix element $\fM{n+1}1$ is free from
poles. This is again the \ac{YFS} split, mentioned in \eqref{eq:yfs}
and \eqref{eq:yfsnew}.  In \eqref{eq:polemrv} we have introduced yet
another initially independent cut-parameter $\xi_{c_B}$.

\begin{subequations}
With the help of \eqref{eq:polemrv} we can now write
\begin{align}
\label{eq:nnloh}
\begin{split}
\bbit{2}{h}(\xi_{c_A}) &=
  \pref1 \D\xi_1\,
  \cdis[c_A]{\xi_1^{1+2\epsilon}} \big(\xi_1^2 \M{n+1}1\big)
\\&=
\bbit{2}{f}(\xi_{c_A},\xi_{c_B}) + \bbit{2}{d}(\xi_{c_A},\xi_{c_B}) \, ,
\end{split}
\end{align}
where $c_A$ indicates that the subtraction should be performed with
the cut parameter $\xi_{c_A}$. The finite piece
\begin{align}
\bbit{2}{f}(\xi_{c_A},\xi_{c_B}) &=
  \pref1 \D\xi_1\,
  \cdis[c_A]{\xi_1^{1+2\epsilon}} \big(\xi_1^2\fM{n+1}1(\xi_{c_B})\big)
\label{eq:nnlo:fin}
\end{align}
can be integrated numerically with $\epsilon=0$. Integrating the
divergent piece, $\bbit{2}{d}$, over the complete phase space we
obtain
\begin{align}
\begin{split}
\int\bbit{2}{d}(\xi_{c_A},\xi_{c_B}) &=
  -\int\pref1 \D\xi_1\, 
  \cdis[c_A]{\xi_1^{1+2\epsilon}}   \big(
      \ieik(\xi_{c_B})\, \xi_1^2\,\M{n+1}0
  \big)
\equiv -\mathcal{I}(\xi_{c_A},\xi_{c_B})
\,,
\end{split}\label{eq:nnlo:sin}
\end{align}
where in $\mathcal{I}$ the first argument refers to the cut-parameter
of the $\xi$ integration and the second to the argument of
$\hat{\mathcal{E}}$. This process- and observable-dependent function
is not finite and generally very tedious to compute. Even for the
simplest cases such as the muon decay it gives rise to complicated
analytic expressions including for example Appell's $F_i$ functions.
However, as we will see it is possible to cancel its contribution
exactly with the double-real emission.
\end{subequations}

To summarise, the real-virtual corrections are given by
\begin{align}
\label{eq:rv}
\bbit{2}{rv} &= \bbit{2}{s}(\xi_{c_A})
+ \bbit{2}{f}(\xi_{c_A},\xi_{c_B})
+ \bbit{2}{d}(\xi_{c_A},\xi_{c_B})
\, ,
\end{align}
where the expressions for $\bbit{2}{s}$, $\bbit{2}{f}$, and
$\bbit{2}{d}$ can be read off from \eqref{eq:nnlo:s},
\eqref{eq:nnlo:fin}, and \eqref{eq:nnlo:sin}, respectively. We point
out that $\bbit{2}{rv}$ is independent of both $\xi_{c_A}$ and
$\xi_{c_B}$.

\subsection{Double-real correction}

For the double-real contribution
\begin{align}
\label{eq:nnlorr}
\bbit{2}{rr} = \D\Phi_{n+2}\, \M{n+2}0
\end{align} 
we have to consider $\M{n+2}{0}$, the matrix element for the process
with two additional photons (with labels $n+1$ and $n+2$) w.r.t. the
tree-level process. We extend the parametrisation~\eqref{eq:kdef}
accordingly to
\begin{align}
k_1 = p_{n+1} = \frac{\sqrt s}2\xi_1 (1,\sqrt{1-y_1^2}\vec e_\perp,y_1)\,,
&\qquad
k_2 = p_{n+2} = \frac{\sqrt s}2\xi_2 R_\phi(1,\sqrt{1-y_2^2}\vec e_\perp,y_2)\,,
\end{align}
with $-1\le y_i\le 1$, $0\le\xi_i\le\xi_\text{max}$ and a
$(d-2)$-dimensional rotation matrix $R_\phi$. Writing the phase space
as $\D \Phi_{n+2} = \D \Phi_{n,2} \D\phi_1 \D\phi_2$, the double-real
contribution becomes
\begin{align}
\begin{split}
\bbit{2}{rr} &= \D \Phi_{n,2} \D\phi_1 \D\phi_2\,\ \frac1{2!} \M{n+2}0
\\&=
\pref2\ \D\xi_1\,\D\xi_2\, \frac1{2!} \big(\xi_1^2\xi_2^2\M{n+2}0\big)\ 
  \xi_1^{-1-2\epsilon}\,\xi_2^{-1-2\epsilon}\,,
\end{split}\label{eq:nnlo:sym}
\end{align}
where we have used analogous definitions as in \eqref{eq:para} and
\eqref{eq:realnlo}. The only difference between $\D \Phi_{n,1}$ and
$\D \Phi_{n,2}$ is in the argument of the $\delta$~function that
ensures momentum conservation.  Note that the factor $1/2!$ is 
the symmetry factor due to two identical particles.

Again, we use \eqref{eq:xidist} with two new cut parameters
$\xi_{c_1}$ and $\xi_{c_2}$ to expand $\bbit{2}{rr}$ in terms of
distributions as
\begin{align}
\label{eq:rr}
&\qquad  \bbit{2}{rr} = \bbit{2}{ss}(\xi_{c_1},\xi_{c_2}) +
\bbit{2}{sh}(\xi_{c_1},\xi_{c_2}) + \bbit{2}{hs}(\xi_{c_1},\xi_{c_2})
+ \bbit{2}{hh}(\xi_{c_1},\xi_{c_2})
\,,\\[10pt] \nonumber
& \left\{\def\arraystretch{1.6}\begin{array}{c}
   \bbit{2}{ss}(\xi_{c_1},\xi_{c_2}) \\[5pt] 
   \bbit{2}{hs}(\xi_{c_1},\xi_{c_2}) \\[5pt]
   \bbit{2}{sh}(\xi_{c_1},\xi_{c_2}) \\[5pt] 
   \bbit{2}{hh}(\xi_{c_1},\xi_{c_2})
\end{array}\right\} =
    \pref2\ \frac1{2!}\ 
\left\{\def\arraystretch{1.9}\begin{array}{c}
   \frac{\xi_{c_1}^{-2\epsilon}}{2\epsilon}\delta(\xi_1) \,
   \frac{\xi_{c_2}^{-2\epsilon}}{2\epsilon}\delta(\xi_2)
   \\
  -\frac{\xi_{c_2}^{-2\epsilon}}{2\epsilon}\delta(\xi_2) \,
   \cdis[c_1]{\xi_1^{1+2\epsilon}}
   \\
  -\frac{\xi_{c_1}^{-2\epsilon}}{2\epsilon}\delta(\xi_1) \,
   \cdis[c_2]{\xi_2^{1+2\epsilon}}
   \\
   \cdis[c_1]{\xi_1^{1+2\epsilon}}\,
   \cdis[c_2]{\xi_2^{1+2\epsilon}}
\end{array}\right\}\, \D\xi_1\,\D\xi_2
\ \xi_1^2\xi_2^2\M{n+2}0\,.
\end{align}
We note that for $\xi_{c_1} = \xi_{c_2} \equiv \xi_c$ we have
$\int\bbit{2}{sh}(\xi_{c},\xi_{c}) =
\int\bbit{2}{hs}(\xi_{c},\xi_{c})$.

The contribution from $\bbit{2}{hh}$ can be integrated numerically
with $\epsilon=0$ because it is finite everywhere.

For the mixed contributions  $\bbit{2}{hs}$ and $\bbit{2}{sh}$ we use
\begin{align}
\mathcal{S}_{i}\M{n+2}0 = \eik_i \M{n+1}0
\qquad\text{with}\qquad
i\in\{n+1,n+2\}\, .
\end{align}
Considering first $\bbit{2}{hs}$, we perform the $\xi_2$ integration
(under which $\D \Phi_{n,2} \to \D \Phi_{n,1}$) and use
\eqref{eq:inteik} to do the $\D\Upsilon_2$ integration to obtain
\begin{subequations}
\begin{align}
\int\bbit{2}{hs}(\xi_{c_1},\xi_{c_2}) &=
    \int \D\Upsilon_1 \D \Phi_{n,1} \ \frac1{2!}\ \int\D\xi_1\ \cdis[c_1]{\xi_1^{1+2\epsilon}}
    \big(\xi_1^2\M{n+1}0\big)
    \ieik(\xi_{c_2})
 = \frac1{2!}\mathcal{I}(\xi_{c_1},\xi_{c_2})\,.
\end{align}
Similarly, we get
\begin{align}
\int\bbit{2}{sh}(\xi_{c_1},\xi_{c_2}) &=
  \frac1{2!}\mathcal{I}(\xi_{c_2},\xi_{c_1})
\,.
\end{align}
\end{subequations}
Thus, we find again the integral $\mathcal{I}$ of \eqref{eq:nnlo:sin}.

Finally, we turn to the double-soft contribution $\bbit{2}{ss}$. Since
\begin{align}
\Big(\mathcal{S}_i\circ\mathcal{S}_j\Big)\M{n+2}0 
&= \Big(\mathcal{S}_j\circ\mathcal{S}_i\Big)\M{n+2}0
 = \eik_i\eik_j\, \M{n}0
\qquad\text{with}\qquad
i\neq j\in\{n+1,n+2\}\,,
\end{align}
the $\xi$ integrals in $\bbit{2}{ss}$ factorise. Therefore, we can do
the $\D\xi_1\D\Upsilon_1$ integrations independently from the
$\D\xi_2\D\Upsilon_2$ integrations and obtain
\begin{align}
\begin{split}
\bbit{2}{ss}(\xi_{c_1},\xi_{c_2}) \ 
&\stackrel{\int\D\Upsilon_{1,2}\D\xi_{1,2}}{\longrightarrow} \
\D \Phi_{n}\ \frac1{2!}\
\ieik(\xi_{c_1})\ieik(\xi_{c_2})\,\M{n}0\, .
\end{split}\label{eq:nnlo:ss}
\end{align}
It is clear that the simplicity of the infrared structure of \ac{QED}
with massive fermions is crucial for reducing the complexity of the
procedure described in the steps above.

\subsection{Combination}

At this stage we have introduced four different cutting parameters
$\xi_{c_A}$ and $\xi_{c_B}$ as well as $\xi_{c_1}$ and $\xi_{c_2}$.
All of these are unphysical, arbitrary parameters that can take any
value $0<\xi_{c_i}\le\xi_\text{max}$.  In total we have to deal with
seven different contributions. Two of them, $\bbit{2}{s}$ and
$\bbit{2}{ss}$, are very simple as they just depend on the
eikonal. Another two contributions $\bbit{2}{f}$ and $\bbit{2}{hh}$
can be calculated numerically with $\epsilon=0$.

The sum of the three remaining \term{auxiliary contributions}
$\bbit{2}{d}$, $\bbit{2}{sh}$, and $\bbit{2}{hs}$, only depend on the
function $\mathcal{I}$ defined above
\begin{align}
\nonumber
\int \bbit{2}{aux}(\{\xi_{c_i}\}) &\equiv
\int \Big( \bbit{2}{d}(\xi_{c_A},\xi_{c_B})
+\bbit{2}{hs}(\xi_{c_1},\xi_{c_2})+\bbit{2}{sh}(\xi_{c_1},\xi_{c_2}) \Big)
\\
\label{eq:aux}
&=
-\mathcal{I}(\xi_{c_A},\xi_{c_B})
+\frac1{2!}\mathcal{I}(\xi_{c_1},\xi_{c_2})
+\frac1{2!}\mathcal{I}(\xi_{c_2},\xi_{c_1})
\,.
\end{align}
Note that, due to the sign difference and the symmetry factor,
$\bbit{2}{aux}$ vanishes if we choose 
\begin{align}
\label{eq:allxi}
\xc \equiv \xi_{c_A} = \xi_{c_B} = \xi_{c_1} = \xi_{c_2}\,.
\end{align}
This cancellation will not be affected by the measurement
function. Thus, in what follows we will make the choice
\eqref{eq:allxi}, avoiding the computation of the potentially
difficult $\mathcal{I}$ function. 

It is possible to compute the auxiliary contribution $\bbit{2}{aux}$
numerically keeping all $\xi_{c_i}$ different by implementing the
$d$-dimensional phase space mapping explicitly. While this complicates
the implementation of the scheme it can be helpful to validate the
code by confirming that physical quantities are in fact $\xi_c$
independent. We have indeed done that by calculating
$\mathcal{I}(\xi_{c_1},\xi_{c_2})$ for the muon decay.

We can now collect the non-vanishing contributions, sorted by
remaining integrations
\begin{subequations}
\begin{align}
\sigma^{(2)}\phantom{(\xc)} &= \sigma^{(2)}_n    (\xc) + 
          \sigma^{(2)}_{n+1} (\xc) + 
          \sigma^{(2)}_{n+2} (\xc)
\,,\\
\label{eq:nnloind0}
\sigma^{(2)}_n (\xc) &=
\int\Big(
\D\Phi_n\, \M n2  +
    \bbit{2}{s}+\bbit{2}{ss} \Big)
\,,\\
\label{eq:nnloind1}
\sigma^{(2)}_{n+1} (\xc) &= \int \bbit{2}{f} =\int
 \pref1 \D\xi\, 
  \cdis{\xi^{1+2\epsilon}} \, \xi^2\fM{n+1}1(\xc)
\,,\\
\label{eq:nnloind2}
\sigma^{(2)}_{n+2} (\xc) &= \int \bbit{2}{hh}
     = \int \pref2 \D\xi_1 \D\xi_2  \frac1{2!}
   \cdis{\xi_1^{1+2\epsilon}}\,
   \cdis{\xi_2^{1+2\epsilon}}\,
     \Big(\xi_1^2\xi_2^2\M{n+2}0\Big)\, .
\end{align}\label{eq:nnloint}
\end{subequations}
The three terms of the integrand of $\sigma^{(2)}_n$ are separately
divergent. However, in the sum the $1/\epsilon$ poles cancel. The
other parts, $\sigma^{(2)}_{n+1}$ and $\sigma^{(2)}_{n+2}$, are finite
by construction. Hence, we can set $d=4$ everywhere (except in the
individual pieces of the integrand of $\sigma^{(2)}_n$) and obtain
\begin{subequations}
\label{eq:nnlo:4d}
\begin{align}
\begin{split}
\sigma^{(2)}_n(\xc) &= \int
\ \D\Phi_n^{d=4}\,\bigg(
    \M n2
   +\ieik(\xc)\,\M n1
   +\frac1{2!}\M n0 \ieik(\xc)^2
\bigg) = 
\int \ \D\Phi_n^{d=4}\, \fM n2
\,,
\end{split}\label{eq:nnlo:n}
\\
\sigma^{(2)}_{n+1}(\xc) &= \int 
\ \D\Phi^{d=4}_{n+1}
  \cdis\xi \Big(\xi\, \fM{n+1}1(\xc)\Big)
\,,\\
\sigma^{(2)}_{n+2}(\xc) &= \int
\ \D\Phi_{n+2}^{d=4}
   \cdis{\xi_1}\,
   \cdis{\xi_2}\,
     \Big(\xi_1\xi_2\, \fM{n+2}0\Big) \, .
\end{align}
\end{subequations}
This is the generalisation of \eqref{eq:nlo:4d} to \ac{NNLO}. In the
integrand of \eqref{eq:nnlo:n} the build-up of the exponentiated
singular part $e^{\ieik}$ is recognisable (cf. \eqref{eq:yfs} and
\eqref{eq:yfsnew}). For $\fM{n}\ell$ to be finite, $\ieik$ has to
contain the soft $1/\epsilon$ pole. However, any choice of the finite
part is possible in principle. We have chosen to define the finite
matrix elements through eikonal subtraction, \eqref{eq:polemrv}. This
ensures that the auxiliary contributions cancel and the remaining
parts $\sigma^{(2)}_{n+1}$ and $\sigma^{(2)}_{n+2}$ have a very simple
form. Terms of $\mathcal{O}(\epsilon)$ in $\ieik$ have no effect since
they do not modify $\fM{n}\ell$ after setting $d=4$. This means we can
set them to zero and there is no need to compute the integral
\eqref{eq:inteik} beyond finite terms.

\section{Beyond NNLO} \label{sec:beyond}

\subsection{\texorpdfstring{FKS$^3$: extension to N$^3$LO}{FKS3: extension to N3LO}}
\label{sec:fks3}

First steps towards extending universal schemes beyond \ac{NNLO} have
been made in \ac{QCD}~\cite{Currie:2018fgr}.  The simplicity of
\ac{FKS2} suggests that this paradigm is a promising starting point
for further extension to \ac{n3lo} in massive \ac{QED}, provided
that all matrix elements are known.

At \ac{n3lo}, we have four terms
\begin{align}
\sigma^{(3)} = \int\D\Phi_{n  } \M{n  }3
       + \int\D\Phi_{n+1} \M{n+1}2
       + \int\D\Phi_{n+2} \M{n+2}1
       + \int\D\Phi_{n+3} \M{n+3}0\,,
\end{align}
which are separately divergent. In order to reorganise these four
terms into individually finite terms, we repeatedly use
\eqref{eq:xidist} to split the phase-space integrations into hard and
soft and \eqref{eq:polemrv} to split the matrix element into finite
and divergent parts.  In principle we could choose many different
$\xc$ parameters. However, from the experience of \ac{FKS2} we expect
decisive simplifications if we choose them all to be the same. Indeed,
as is detailed in Appendix~\ref{ch:fks3}, there are now at least three
different auxiliary integrals that enter in intermediate
steps. However, if all $\xc$ parameters are chosen to be equal, their
contributions cancel for any cross section, similar to \eqref{eq:aux}.
Hence, writing 
\begin{align}
\label{eq:nnnlocomb}
\D\sigma^{(3)} &= \bbit{3}{n}(\xc) + \bbit{3}{n+1}(\xc) 
+ \bbit{3}{n+2}(\xc) + \bbit{3}{n+3}(\xc)\,,
\end{align}
all terms are separately finite and, as discussed in detail in
Appendix~\ref{ch:fks3}, given by
\begin{subequations}
\label{eq:nnnloparts}
\begin{align}
\bbit{3}{n}(\xc) &=
  \D\Phi_n^{d=4}   \fM{n}3 
\,,\\
\bbit{3}{n+1}(\xc) &= \D\Phi_{n+1}\, \cdis{\xi_1}
                  \Big(\xi_1\, \fM{n+1}2(\xc)\Big)
\,,\\
\bbit{3}{n+2}(\xc) &=  \frac1{2!}\, \D\Phi_{n+2}\, 
   \cdis{\xi_1}\, \cdis{\xi_2}\,
   \Big( \xi_1 \xi_2 \, \fM{n+2}1(\xc) \Big)
\,,\\
\bbit{3}{n+3}(\xc) &= \frac1{3!}\,
  \D\Phi_{n+3}\, 
   \cdis{\xi_1}\,
   \cdis{\xi_2}\,
   \cdis{\xi_3} \ \Big(
   \xi_1\xi_2\xi_3\, \fM{n+3}0(\xc)\Big) \,.
\end{align}
\end{subequations}
Once more we have used the fact that for tree-level amplitudes
$\M{n+3}0 = \fM{n+3}0$.  As always, the $\xc$ dependence cancels
between the various parts s.t. $\bit{3}$ is independent of this
unphysical parameter.

\subsection{\texorpdfstring{FKS$^\ell$: extension to N$^\ell$LO}{FKSl: extension to NlLO}}
\label{sec:nllo}

The pattern that has emerged in the previous cases leads to the 
following extension to an arbitrary order $\ell$ in perturbation
theory:
\begin{subequations}
\begin{align}
\D\sigma^{(\ell)} &= \sum_{j=0}^\ell \bbit{\ell}{n+j}(\xc)\, ,
\label{eq:nellocomb:a}
\\
\bbit{\ell}{n+j}(\xc) &=  \D\Phi_{n+j}^{d=4}\,\frac{1}{j!} \, 
\bigg( \prod_{i=1}^j \cdis{\xi_i} \xi_i \bigg)\,
 \fM{n+j}{\ell-j}(\xc)\,.
\label{eq:nellocomb:b}
\end{align}
\label{eq:nellocomb}
\end{subequations}
The eikonal subtracted matrix elements
\begin{align}
\fM{m}\ell &= \sum_{j=0}^\ell\frac{\ieik^j}{j!} \M{m}{\ell-j}\,,
\end{align}
(with the special case $\fM{m}0 = \M{m}0$ included) are free from
$1/\epsilon$ poles, as indicated in \eqref{eq:yfs}. Furthermore, the
phase-space integrations are manifestly finite.

\section{Comments on and properties of \texorpdfstring{FKS$^\ell$}{FKSl}}
\label{sec:comments}

With the scheme now established, let us discuss a few non-trivial
properties that are helpful during implementation and testing.

\subsection{Regularisation-scheme and scale dependence}
\label{sec:fksscheme}

As we have explained in Section~\ref{sec:renorm}, it is advantageous
to calculate the matrix elements $\M{n}{\ell}$ in the on-shell scheme
for $\alpha$ (and the masses) because it best exploits the Ward
identity. This way the only $\mu$ dependence is in a global prefactor
$\mu^{2\epsilon}$ induced through the integral measure. The same holds
for the integrated eikonal. Hence, for the finite matrix elements
$\fM{n}{\ell}$ there is no $\mu$ dependence after setting $d=4$.  

A similar argument can be made for the regularisation-scheme
dependence. As discussed in Section~\ref{sec:schemedep}, the
renormalised and $\MS$-like \ac{IR} subtracted
$\mathcal{M}_\text{sub}$ is scheme independent for $\epsilon\to0$
because it is free of terms $\propto\neps/\epsilon$. The same argument
can also be made for the eikonal subtracted matrix element $\fM n\ell$
because the integrated eikonal $\ieik$ is scheme independent, dealing
only with singular vector fields. Of course, this hinges on there
being no collinear singularities.

\subsection{Ingredients required at NNLO}

To be concrete, we list the input that is required for a computation
of a physical cross section at \ac{NNLO} in \ac{QED}. The important
point is that once the final expressions for a \ac{NNLO} cross
section, \eqref{eq:nnlo:4d}, or beyond, \eqref{eq:nellocomb}, are
obtained, we can set $d=4$ everywhere.

\begin{itemize}

    \item
    The two-loop matrix element $\M n2$ is known with non-vanishing
    masses up to $\mathcal{O}(\epsilon^0)$. In general this is a
    bottleneck because the necessary master integrals are only known
    for a very select class of processes, not to mention the algebraic
    complexity. However, it is possible to approximate $\bit{2}$ using
    `massification' of $\M n2$~\cite{Engel:2018fsb, Mitov:2006xs,
      Becher:2007cu} (see Section~\ref{sec:massification}).

    \item
    The renormalised one-loop matrix element $\M n1$ of the
    $n$-particle process is known including $\mathcal{O}(\epsilon^1)$
    terms. This is usually the case for \ac{NNLO} calculation as it
    is needed for the sub-renormalisation $\M n2 \supset \delta
    Z\times \M n1$ as well as the one-loop amplitude squared, which is
    part of $\M n2$. Once these pieces are assembled to $\fM{n}2$, the
    $\mathcal{O}(\epsilon)$ terms can be dropped.  

  \item
    The renormalised real-virtual matrix element $\M{n+1}1$ is known
    with non-vanishing masses. Terms $\mathcal{O}(\epsilon)$ are not
    required.

  \item
    $\M{n+2}0$ is known in four dimensions. In intermediate steps, the
    matrix elements $\M{n}0$ and $\M{n+1}0$ are required to
    $\mathcal{O}(\epsilon^2)$ and $\mathcal{O}(\epsilon)$,
    respectively. However, depending on the regularisation scheme,
    such terms might actually be absent. In any scheme, once $\fM{n}2$
    and $\fM{n+1}1$ is assembled, the $\mathcal{O}(\epsilon)$ terms
    can be dropped.

\end{itemize}

\subsection{Phase-space parametrisation}\label{sec:pcs}

A further issue in connection with small lepton masses is related to
the phase-space parametrisation.  The phase space has to be
constructed in any way that allows the distributions to be
implemented. The easiest way to do this is to ensure that $\xi_i$ is
as an integration variable of the numerical integrator. In addition,
for small $m$ there are potentially numerical problems due to
\ac{PCS}. In fact, these regions produce precisely the $\log(m)$ terms
that correspond to the collinear `singularities' of the real part.
These $\log(m)$ terms will cancel the virtual collinear
`singularities' of similar origin.  Hence, for small $m$ there is a
numerically delicate cancellation.  This requires a dedicated tuning
of the phase-space parametrisation. We will discuss this in detail in
Section~\ref{sec:ps}.

\chapter{Two-loop calculation}\label{ch:twoloop}
\setlength\parskip{0pt}
The major bottleneck in most higher-order calculations is the
evaluation of the $n$-particle amplitude to the required number of
loops. In our case of \ac{NNLO} calculations these are two-loop
diagrams.  The difficulty originates in part from the algebraic
complexity, though this can sometimes be reduced by the choice of
the regularisation scheme. The biggest problem is in any case the lack of
analytic results for the so-called master integrals. This is because
-- especially in massive \ac{QED} -- we have often a lot of active
scales some of which are internal and external masses.

There is a traditional procedure for multi-loop calculations that was
developed over the last decade for the analytic \ac{QCD} calculations
for the \ac{LHC} which we have adopted for massive \ac{QED}:
\begin{enumerate}
 \item
 Generate all Feynman diagrams contributing to the process. While this
 is straightforward to do by hand for amplitudes with few external
 particles and few loops, it quickly becomes a daunting task as the
 number of diagrams grows factorially at higher loops or
 multiplicities. Hence, this step is usually performed with a
 dedicated computer program such as {\sc Qgraf}~\cite{Nogueira:1991ex}.

 \item
 Apply the Feynman rules and perform algebraic simplification. This
 also includes simplifying the Dirac and Lorentz structure of the
 expression to obtain scalar quantities that can be treated later.
 Common ways to do this include

 \begin{itemize}
    \item
    The reduction to helicity amplitudes, i.e. fixing the helicity and
    polarisation of every external particle and then employ
    completeness relations and Fierz identities to simplify the
    result. This works very well for tree and one-loop diagrams
    involving few massless particles as the number of helicity
    combinations is small. For massive particles, this is still
    possible but in practice a lot more involved.
    
    We will not be using helicity amplitudes in this project.

    \item
    The standard approach of squaring the amplitude $\mathcal{A}$ to
    obtain the matrix element $\mathcal{M}=|\mathcal{A}|^2$ and using
    the completeness relation of spinors to convert the expression of
    $\mathcal{M}$ to traces.

    \item
    The projection onto form factors. This is done by writing the most
    general expression that satisfies all symmetries of the theory
    with arbitrary coefficients and fixing those by applying
    projectors onto the amplitude. This procedure has the advantage of
    being completely general and independent of the observable to be
    calculated. However, this mechanism also falls short when too many
    particles are involved because too many form factors need to be
    defined.

 \end{itemize}

 \item
 Take stock of all integrals appearing and try to find relations
 between them.

 \item
 Calculate all remaining master integrals using various methods.

\end{enumerate}

In this chapter, we will briefly discuss all these steps as a short
tutorial on two-loop calculations. As an example we will be using the
calculations in \mcmule{} but the discussion is far more general. In
fact, the relevant calculations were performed in \ac{QCD} first and
only later was the abelian limit taken. We begin by discussing ways of
organising a calculation in a gauge-invariant fashion in
Section~\ref{sec:colour}. Next, we will discuss various aspects of the
actual loop integration, focussing on reductions to scalar
(Section~\ref{sec:scalar}) and master integrals
(Section~\ref{sec:ibp}) as well as the eventual calculation of these
integrals in Section~\ref{sec:intergals}. Finally, we will discuss a
method of coping with massive fermions in
Section~\ref{sec:massification}.

\section{Gauge-invariant splitting of amplitudes}\label{sec:colour}
Often we would like to be able to decompose our expressions into
simpler contributions. If we do not want to break gauge invariance, we
cannot use Feynman diagrams for this. Instead, we are forced to find a
different strategy to classify the contributions. 

A natural strategy in \ac{QCD} is to sort the expression by
\term{colour factors}, the result of solving the colour algebra.
While public codes are available for this (for
example~\cite{Sjodahl:2012nk}), it is often easier to just implement
the colour algebra directly. After solving the traces in colour space,
one is left with combinations of the different colour factors of
Table~\ref{tab:colour}.

When adapting results from \ac{QCD} for \ac{QED}, the limit of the
Casimir operators $C_F\to1$ and $C_A\to0$ is trivial. However, the
factor associated to closed fermion loops $T_Rn_i$ is often written
assuming $T_R=1/2$ s.t. adapting amplitudes may require $n_i\to2n_i$
before use.

Note that even for \ac{QED} without colour structure it may make
sense to separate purely photonic contributions ($\propto C_F^\ell$)
from fermionic contributions with their $n_i$.

\begin{figure}[h]
\centering
\scalebox{0.99}{
\newcommand{\cell}[2]{
\parbox[c][1.75cm][c]{#1}{#2}
}

\def\dcf{\scalebox{0.7}{
\begin{tikzpicture}
 \centerarc[decorate, draw=black,decoration={coil,amplitude=4pt,segment length=4pt}](0,0)(0:180:0.7)
 \draw [mfermion] (-1.3,0) -- (-0.6,0) -- (0.6,0) -- (1.3,0);
\end{tikzpicture}}
}
\def\dca{\scalebox{0.7}{
\begin{tikzpicture}
 \centerarc[decorate, draw=black,decoration={coil,amplitude=4pt,segment length=4pt}](0,0)(0:180:0.7)
 \draw [gluon] (-1.3,0) -- (-0.6,0) -- (0.6,0) -- (1.3,0);
\end{tikzpicture}}
}

\newcommand\df[2]{\scalebox{0.7}{
\begin{tikzpicture}
 \draw [gluon] (-1.3,0) -- (-0.4,0);
 \centerarc[#1,fermion](0,0)(0:180:0.4);
 \centerarc[#1,antifermion](0,0)(0:-180:0.4);
 \draw [gluon] (0.4,0) -- (1.3,0);
 \node at (0,0) {#2};
\end{tikzpicture}}
}

\begin{tabular}{c|c|c|c|l}

     & example diagram & QCD value & QED value & \\\hline

\cell{1.0cm}{$C_F$} & \cell{2cm}{\dcf} 
  & \cell{3cm}{\centering$(N^2-1)/(2N)$\\$=4/3$} 
  & \cell{1cm}{\centering$1$} 
  & \cell{4cm}{Casimir operator of the fundamental repr.}

\\\hline

\cell{1.0cm}{$C_A$} & \cell{2cm}{\dca}
  & \cell{3cm}{\centering$N=3$} 
  & \cell{1cm}{\centering$0$}
  & \cell{4cm}{Casimir operator of the adjoint repr.}

\\\hline

\cell{1.0cm}{$T_Rn_f$} & \cell{2cm}{\df{thin}{$0$}} 
  & \cell{3cm}{\centering$\tfrac12\times4$} 
  & \cell{1cm}{\centering$1\times0$} 
  & \cell{4cm}{no. of massless fermions}

\\\hline

\cell{1.0cm}{$T_Rn_m$} & \cell{2cm}{\df{thick}{$m$}} 
  & \cell{3cm}{\centering$\tfrac12\times1$} 
  & \cell{1cm}{\centering$1\times1$} 
  & \cell{4cm}{no. of light fermions}

\\\hline

\cell{1.0cm}{$T_Rn_h$} & \cell{2cm}{\df{thick}{$M$}} 
  & \cell{3cm}{\centering$\tfrac12\times1$} 
  & \cell{1cm}{\centering$1\times1$} 
  & \cell{4cm}{no. of heavy fermions}
\end{tabular}
}
\renewcommand{\figurename}{Table}
\caption{Colour factors in \ac{QCD} and \ac{QED}. The values for the
$n_i$ refer to calculation of $t\to W^\pm b$ and $\mu\to\nu\bar\nu e$,
respectively.}
\label{tab:colour}
\end{figure}

Unfortunately, sorting the contributions by colour factor is often not
sufficient in \ac{QED} as there are just not enough different colour
factors. Assuming we have multiple flavours of leptons, as is the case
in most \mcmule{} processes, we can exploit this as a new
strategy~\cite{Alacevich:2018vez, MUonEwriteup} by assigning
(formally) different charges to each flavour. For example, the
amplitude for $\mu$-$e$ scattering at \ac{LO} can be written as
\begin{subequations}

\newcommand{\muoneLO}{
\begin{tikzpicture}[scale=0.1, anchor=base, baseline=6]
\draw[mfermion] (-6,0) -- (0,0) -- (6,0);
\draw[mfermion] (-6,6) -- (0,6) -- (6,6);
\draw[photon] (0,0) -- (0,6);
\end{tikzpicture}
}
\newcommand{\muoneel}{
\begin{tikzpicture}[scale=0.1, anchor=base, baseline=6]
\draw[mfermion] (-6,6) -- (0,6) -- (6,6);
\draw[mfermion] (-6,0) -- (-3,0) -- (0,0) -- (3,0) -- (6,0);
\draw[photon] (0,0) -- (0,6);
\centerarc[photon](0,0)(0:180:-3);
\end{tikzpicture}
}
\newcommand{\muonemu}{
\begin{tikzpicture}[scale=0.1, anchor=base, baseline=6]
\draw[mfermion] (-6,0) -- (0,0) -- (6,0);
\draw[mfermion] (-6,6) -- (-3,6) -- (0,6) -- (3,6) -- (6,6);
\draw[photon] (0,0) -- (0,6);
\centerarc[photon](0,6)(0:180:3);
\end{tikzpicture}
}
\newcommand{\muoneint}{
\begin{tikzpicture}[scale=0.1, anchor=base, baseline=6]
\draw[mfermion] (-6,0) -- (-2,0) -- (2,0) -- (6,0);
\draw[mfermion] (-6,6) -- (-2,6) -- (2,6) -- (6,6);
\draw[photon] (-2,0) -- (-2,6);
\draw[photon] (+2,0) -- (+2,6);
\end{tikzpicture}
}

\begin{align}
  \cA^{(0)}(\mu e\to\mu e) = q Q\  \Bigg(\muoneLO\Bigg)\, ,
\end{align}
where $q$ ($Q$) is the charge of electron (muon). The one-loop
amplitude can now be split as
\begin{align}
 \cA^{(1)}   &= q^3 Q  \  \Bigg(\muoneel \Bigg)
              + q   Q^3\  \Bigg(\muonemu \Bigg)
              + q^2 Q^2\  \Bigg(\muoneint\Bigg)\, .\label{eq:gsplit}
\end{align}
\end{subequations}
Hence, we now can consider the terms separately without breaking gauge
invariance. 

A counting that include logarithms as well as powers of $\alpha$
suggests that the $q^3Q$ ($qQ^3$) term would be $\alpha^3
\log\tfrac{m^2}{s}$ ($\alpha^3\log\tfrac{M^2}{s}$). Hence, the
splitting~\eqref{eq:gsplit} allows us to use the large hierarchy
between the lepton masses to prioritise the $q^3Q$ term over the much
more complicated $q^2Q^2$ term because we expect the former to be much
larger than the latter. This was indeed found at
\ac{NLO}~\cite{Alacevich:2018vez}.

This decomposition is very similar to the colour decomposition used in
connection with helicity amplitudes (for a review,
cf.~\cite{Mangano2005Multi-Parton}). When using this method to
calculate $\cA$ directly, one collects the different colour structures
at the amplitude level to split that into different gauge-invariant
subparts.

\section{Scalar integrals}\label{sec:scalar}

Beyond the one-loop level, it is generally advisable to work with
scalar objects. Unfortunately, calculations involving fermions will
eventually require the manipulation of Dirac matrices. While computer
algebra programs can certainly handle this, it is generally a good
idea to reduce the expression to only contain scalar quantities as
soon as possible. As discussed above, this could be achieved by
interfering the amplitude with the corresponding Born amplitude.
However, one stays more flexible in the calculation when instead
decomposing the amplitude into form factors using appropriate
projectors. This has the added advantage of reducing the amount of
algebra necessary if the number of form factors is not unreasonably
large. This is due to the simple fact that projectors can often be
written with less objects than the physical Born amplitude.

The next step is the removal of scalar products involving one (or
more) loop momenta from the numerator. For this, we need to identify
one of the propagators of the diagram and use it to cancel the scalar
product. If no propagator of the diagram contains this scalar product,
we need to add a fictitious propagator that does. We call a set of
propagators that is guaranteed to achieve this for any scalar product
a \term{family}. For a process with $\rho$ external momenta (after
applying momentum conservation) and $\ell$ loop momenta there are
\begin{align}
p=\mbinom{\rho+\ell}{2} - \mbinom{\rho}{2} = \ell\frac{1+\ell+2\rho}2
\end{align}
possible scalar products, requiring a family of that size. Here, we
have defined the multichoose function
\begin{align}
\mbinom nk = \binom{n+k-1}{k} = \frac{(n+k-1)!}{k!(n-1)!}\,,
\end{align}
that counts the number of ways one can pick $k$ unordered elements
from a set of $n$ elements, allowing for repetition.

Once the families are fixed, the next step is to bring the expression
into the following form
\newcommand{\prop}[1]{\mathcal{P}_{#1}^{\alpha_{#1}}}
\begin{align}
\sum_n C_n \times 
\int \prod_{j=1}^{\ell}[\D k_j] 
    \frac{1}{\prop{1,n}\cdots\prop{p,n}}\,.
\end{align}
The powers $\alpha_i$ of the propagators $\mathcal{P}_i$ may be
negative or zero and the $C_n$ are functions of the external
kinematics and the dimension $d$. These integrals are referred to as
\term{reducible scalar integrals}.

In virtually no case are all $\alpha_i>0$. Hence, it makes sense to
group integrals into so-called \term{sectors} by looking at the
propagators $\mathcal{P}$ that are present as denominators (with
whatever power). As this is a boolean decision for every $i$, each
group can be represented as a binary number. An integral that has $t$
propagators with $\alpha_i>0$ shall have sector
ID~\cite{vonManteuffel:2012np}
\begin{align}
{\rm ID} = \sum_{k=1}^t 2^{i_k-1}
  \quad\text{with}\quad
  \alpha_{i_1},...,\alpha_{i_t} > 0\,.
\end{align}
This serves to organise integrals because as soon as one integral in
a sector can be calculated all integrals of the sector can be
calculated, at least in principle. We call the easiest integral of
each sector, i.e. the one with $\alpha_{i_1}=...=\alpha_{i_t}=1$, the
\term{corner integral} of this sector. To further categorise
integrals, we also define
\begin{align}
  r = \alpha_{i_1} + ... + \alpha_{i_t}
  \quad\text{and}\quad
  s = -\sum_{\alpha_i<0} \alpha_i\,,
\end{align}
as the sum of positive and negative propagator indices, respectively.
Obviously $s\ge0$ and $r\ge t$ with $r=t$ for the corner integral.

\section{Integration-by-parts reduction}
\label{sec:ibp}

At this stage, one has a large number of scalar loop integrals. As the
calculation of any one of them can be incredibly difficult, one would
like to reduce them to a minimal set of so-called \term{master
integral}. We will mostly follow~\cite{Smirnov:2012gma} in the
discussion below.

It turns out that such a reduction to master integrals can indeed be
achieved with the use of \aterm{integration-by-parts}{IBP}
identities~\cite{Chetyrkin:1981qh}.  In contrast to the standard
integration-by-parts theorem
\begin{align*}
\int\D x\ u\,v' = u\,v - \int\D x\ u'\,v\,,
\end{align*}
where one chooses $u$ and $v$ s.t. $u'\,v$ is simpler, we now use
that the surface term $u\,v$ vanishes in dimensional regularisation.
In particular, we have
\begin{align}\label{eq:IBPIdentities}
	\int \prod_{j=1}^{\ell}[\D k_j] 
        \frac{\partial}{\partial k_i} \cdot\Bigg( q
    \ \frac{1}{\prop1\cdots\prop{t}}
    \Bigg)=0\,,
    \quad 		i=1,...,\ell\, ,
\end{align}
where $q$ represents either a loop or an external momentum. Note that
$q$ is inside the derivative s.t. if $q=k_i$ the product rule has
to used on the integrand with
\begin{align}
\frac\partial{\partial k}\cdot k
   \equiv \frac{\partial}{\partial k_\mu}k^\mu
   = d\,.
\end{align}
\ac{IBP} relations now allow us to get identities between different
integrals.

To illustrate the usefulness of these \ac{IBP} identities in the
calculation of a large number of loop integrals, we consider as a toy
example a simple class of one-loop integrals
\begin{align}
	 I(a,b) = \int [\D k] 
        \frac{1}{\big[k^2-m^2\big]^{a}\big[(k-p)^2-M^2\big]^{b}}
\quad\text{with}\quad
p^2=M^2\,.
\end{align}
For now we will only consider the case where $m=0$. Of course, this
integral could be trivially calculated for arbitrary powers of $a$ and
$b$.  However, even if this is possible in practice, it is often not
very helpful because the resulting functions of $a$ and $b$ might be
very complicated.

Setting $q=k$, we apply~\eqref{eq:IBPIdentities}
\begin{align}\begin{split}
i(a,b) &= \frac{\partial}{\partial k_\mu}\Bigg( k^\mu
    \ \frac{1}{\big[k^2\big]^{a}\big[(k-p)^2-M^2\big]^{b}}
\Bigg)\\&
= k^\mu\frac{\partial}{\partial k_\mu}\Bigg(
    \frac{1}{\big[k^2\big]^{a}\big[(k-p)^2-M^2\big]^{b}}
\Bigg)
+ \Bigg(\frac{\partial}{\partial k_\mu}k^\mu\Bigg)
    \frac{1}{\big[k^2\big]^{a}\big[(k-p)^2-M^2\big]^{b}}
\\&
= -2k^\mu\Bigg(
 \frac{a\, k^\mu       }{\big[k^2\big]^{1+a}\big[(k-p)^2-M^2\big]^{  b}}
+\frac{b\,(k^\mu-p^\mu)}{\big[k^2\big]^{  a}\big[(k-p)^2-M^2\big]^{1+b}}
\Bigg)\\&\qquad\qquad
+ \frac{d}{\big[k^2\big]^{a}\big[(k-p)^2-M^2\big]^{b}}
\\&
= \frac{(d-2a-2b)k^2 + 2(2a+b-d)k\cdot p}{\big[k^2\big]^{a}\big[(k-p)^2-M^2\big]^{b+1}}
\,.
\end{split}\end{align}
We now again use the algorithm described in Section~\ref{sec:scalar}
to turn this expression back into scalar integrals of the form
$I(a',b')$. After loop integration and setting $\int[\D k]\ i(a,b)=0$ we
finally have our first \term{seed identity}
\begin{align}
0 = -b I(a-1,b+1) + (d - 2 a - b) I(a,b)\,.
\end{align}
We can write this and the seed identity from $q=p$ using the
short-hand notation of~\cite{Smirnov:2012gma}.  ${\bf n}^\pm$
indicates that the power of the $n$-th propagator is raised (lowered)
by one.
\begin{align}
\begin{split}
0&=         d -        2 a - b - b\ {\bf 1}^- {\bf 2}^+\,, \\
0&=\phantom{d}-\phantom2 a + b - b\ {\bf 1}^- {\bf 2}^+
       + a\ {\bf 2}^-\, {\bf 1}^+ + 2 b M^2\ {\bf 2}^+\,.
\end{split}
\end{align}

It is a good idea to consider an integral family with $p$ propagators
as an element of a $p$-dimensional vector space of the
$\{\alpha_i\}$. The \ac{IBP} relations then provide linear dependences
between vectors of this space.

Let us now choose $b=1$ and use the fact that
\begin{align}
{\bf 2}^- I(a,1) = I(a,0) = 0
\end{align}
is scaleless. We now have
\begin{subequations}
\begin{align}
\begin{split}
0&=(         d -1 -          2 a)\ I(a, 1) - I(a-1 , 2) \\
0&=(\phantom{d}+1 - \phantom{2}a)\ I(a, 1) - I(a-1 , 2) + 2 M^2\ I(a, 2)
\,,
\end{split}
\end{align}
which is usually expressed as a matrix equation
\begin{align}
\begin{pmatrix}
d-1-2a & 0  & -1\\
  1- a &2M^2& -1
\end{pmatrix}\cdot\begin{pmatrix}
I(a,1) \\ I(a,2) \\ I(a-1,2)
\end{pmatrix}
=0\,.
\end{align}\label{eq:matrixibp}
\end{subequations}
This system of equations is under-determined, but we could make it
over-determined by variing $a$, a common feature of \ac{IBP}
relations.

For now, we will eliminate $I(a,1)$ and obtain a recursion relation
(shifting $a\to a+1$)
\begin{subequations}
\begin{align}
I(a+1,2) = \frac1{2M^2}\frac{3 + a - d}{3 + 2 a - d} I(a,2)
\,.
\end{align}
Alternatively, we eliminate $I(a-1,2)$ and obtain
\begin{align}
I(a, 2) = -\frac{2+a-d}{2M^2} I(a,1)\,.
\end{align}
This is again classic behaviour for \ac{IBP} reduction. We have
multiple ways of solving the system and have do make decisions on what
integrals are more complicated. In our case we arrive at
\begin{align}
I(a+1,2) = -\frac{2+a-d}{4M^4}\frac{3 + a - d}{3 + 2 a - d} I(a,1)\,,
\end{align}
\label{eq:IBPrec}
\end{subequations}
where $I(1,1)$ is a \term{master integral} that has to be computed.

In real-world calculations it is often not possible to write down a
simple recursion relation as~\eqref{eq:IBPrec}. Instead, one writes
down the linear system \eqref{eq:matrixibp} for $b=1,...,b_\text{max}$
for whatever $b_\text{max}$ the problem under consideration mandates.
One would naively assume that this system grows out of control rapidly
as more and more integrals are added. However, assuming a cut-off
point such as $b_\text{max}$, the system is naturally over-determined
because the number of new integrals grows slower than the number of
equations. We now define what is called a \term{lexicographic
ordering} that, given two integrals, determines which is more
complicated. The exact specification of this ordering does not matter
as long as it is consistent. We can keep generating seed identities
and solve the resulting matrix through Gaussian elimination, favouring
integrals that were deemed simpler by the lexicographic ordering. This
is called Laporta's algorithm~\cite{Laporta:2001dd, Laporta:1996mq}.

There are many public codes that implement Laporta's algorithm or
other, similar algorithms such as {\tt LiteRed}~\cite{Lee:2012cn}, {\tt
AIR}~\cite{Anastasiou:2004vj}, {\tt FIRE}~\cite{Smirnov:2019qkx}, {\tt
Kira}~\cite{Maierhoefer:2017hyi}, and {\tt reduze}~\cite{
vonManteuffel:2012np}. We will be focusing on the latter two.
Additionally to the \ac{IBP} reduction, {\tt reduze} and {\tt Kira}
are also capable of exploiting shift symmetries, i.e.  shifting the
loop momenta $k_i\to k_i+p$. This means that it can find relations
between sectors of different families which reduces the number of
integrals that need to be manually considered.  More importantly yet,
this feature can also be used to find shift relations between
diagrams, reducing the number of families that need to be considered.

Note that there are other relations that can be used to generate seed
identities that use other properties of loop integrals such as Lorentz
invariance.

\section{Calculation of master integrals}\label{sec:intergals}
We now turn to the calculation of master integrals, beginning with
some general comments on Feynman parametrisation in
Section~\ref{sec:feynman}, followed by a discussion of how mass
hierachies can be best exploited in Section~\ref{subsec:MoRGeneral}
(for heavy scales) and Section~\ref{subsec:LightConeCoordinates} (for
light scales).  A more detailed discussion can be found
in~\cite{Engel:2018}.

\subsection{Feynman parametrisation}\label{sec:feynman}
If we want to calculate loop integrals in whatever form, we often
employ \term{Feynman parametrisation} as some stage. For an
$\ell$-loop integral\footnote{The propagators can be either
physical propagators of the form $(k_j + p)^2-m^2$ or linear
propagators $k_j\cdot p$ that appear in
Section~\ref{subsec:MoRGeneral}}
\begin{align}
I=\int \prod_{j=1}^{\ell}[\D k_j] 
    \frac{1}{\prop1\cdots\prop t}
\quad\text{with}\quad
\alpha_1,...,\alpha_t>0
\end{align}
we could either solve the loops one by one or all in one go. Both
methods are equivalent though difficult to relate in practical
examples. We will be focusing on the latter case as the former can be
viewed as a sub-class. Following~\cite{Smirnov:2013eza}, we write
\begin{align}
\frac{1}{\prop1\cdots\prop t}
= \frac{\Gamma(r)}{\prod_j\Gamma(\alpha_j)}
  \int_0^\infty \prod_{j=1}^t \D x_j\ x_j^{\alpha_j-1}
  \delta\Bigg(\sum_{i\in\nu}x_i-1\Bigg)
 \frac1{\Big(\mathcal{P}_1 x_1+...+\mathcal{P}_tx_t\Big)^r}\,.
\end{align}
where $r=\sum_j\alpha_j$ and $\nu$ a non-empty subset of $\{1,...,t\}$
(Cheng-Wu theorem, \cite{Cheng:1987ga}).  The $x_i$ are called
\term{Feynman parameters}.  Note that most books on \ac{QFT} will
assume $\nu=\{1,...,t\}$, reducing the integration region to
$[0,1]\times[1,1-x_1]\times...$. However, for analytic calculations we
have found that having just one element, say $i=1$, in $\nu=\{1\}$ is
a better choice, setting one $x_1=1$ and keeping the integration
bounds at $[0,\infty]^{t-1}$.

The denominator can be written as
\begin{align}
D=\mathcal{P}_1 x_1+...+\mathcal{P}_tx_t = k^T\cdot M(x_i)\cdot k -
2Q(x_i,q_j)^Tk+J(x_i,s_{jk})\,,
\end{align}
with a $\ell\times\ell$ matrix $M$, $\ell$-vectors
$k=(k_1,...,k_\ell)$ and $Q$, depending on the Feynman parameters
$x_i$, external momenta $q_j$ and invariants $s_{jk}=2q_j\cdot q_k$.
By shifting $k\to k+M^{-1}Q$ we cancel the linear term so that after
diagonalising $M$ (with eigenvalues $\lambda_i$) we have
\begin{align}
D = k^T\cdot \mathrm{diag}(\lambda_i)\cdot k - \Delta
\quad\text{with}\quad
\Delta = Q^TM^{-1}Q-J\,.
\end{align}
After rescaling $k_i\to\lambda_i^{-1/2}k_i$ we have factorised the
loop integrations and can use that
\begin{align}
\int[\D k]\frac1{(k^2-\Delta)^r} = (-1)^r\Gamma(1-\epsilon)
\frac{\Gamma(r-d/2)}{\Gamma(r)} \bigg(\frac1\Delta\bigg)^{r-d/2}
\end{align}
to find the general Feynman-parametrised form of the $\ell$-loop
integral $I$, keeping in mind that we chose to have only $x_i$ in the
$\delta$ function
\begin{subequations}
\begin{align}
I = (-1)^r \Gamma(1-\epsilon)^\ell 
  &\frac{\Gamma(r-\ell d/2)}{\prod_j \Gamma(\alpha_j)}
  \int_0^\infty \prod_{j=1}^t \D x_j\ x_j^{\alpha_j-1}
  \delta\Big(x_i-1\Big)
  \underbrace{
    \frac{\mathcal{U}^{r-(\ell+1)d/2}}{\mathcal{F}^{r-\ell d/2}}
  }_{\mathcal{G}}
\,,\label{eq:feynman:int}\\[1em]
&\mathcal{U}=\det M = \prod_j\lambda_j\,,\
\qquad
\mathcal{F}=\det M\times\Delta\,.\label{eq:feynman:UF}
\end{align}
We dub the polynomials $\mathcal{U}$ and $\mathcal{F}$ \term{graph
polynomials} or Symanzik polynomials because they can be computed
without having to go through the motions of finding and diagonalising
$M$ by instead studying graph theoretical aspects of the Feynman
diagram corresponding to $I$ as implemented in {\tt
UF}~\cite{Smirnov:2013eza}.
\label{eq:feynman}
\end{subequations}

Note that $I$ can now be viewed (up to a pre-factor) as the
$(p-1)$-dimensional \term{Mellin transform} of $\mathcal{G}$ evaluated
at the indices $\vec\alpha=(\alpha_2,...,\alpha_p)$
\begin{align}
\{\mathcal{M} \mathcal{G} \}(\vec\alpha) = 
  \int_0^\infty \D\vec x\, \vec x^{\,\vec \alpha-1}\,\, 
  \mathcal{G}(\vec x)
  = 
  \int_0^\infty \Bigg(\prod_{j=1}^p\D x_j\, x_j^{\alpha_j-1}\Bigg)\,\, 
  \mathcal{G}(\vec x)
  \propto I\,,
\end{align}
where we have assumed that $\nu=\{1\}$ for simplicity.

The actual calculation of the Feynman
integral~\eqref{eq:feynman:int} is naturally quite involved. However,
in most cases, once a solution has been found for the corner integral,
other integrals with the same structure can be found relatively
easily. For methods to compute $I$, see for example~\cite{Engel:2018,
Smirnov:2012gma}.

There is one last subtlety related to numerators. As discussed in
Section~\ref{sec:scalar}, we implement numerators in integrals by
setting some $\alpha_i<0$. However, that would make the Feynman
parametrisation ill-defined because the $\Gamma$ function diverges for
negative integers. To solve this problem~\cite{Smirnov:2013eza}, we
note an identity for Mellin transforms called \term{Ramanujan's master
theorem}. In our language it states that the Mellin transform of a
function $f(x)$ evaluated at negative integers $-n$ can be written as
the $n$-th derivative of $f$
\begin{align}
\{\mathcal{M}f\}(-n)=\int\D x\,x^{-n-1}\,\,f(x) =\Gamma(-n)f^{(n)}(0)
\,.
\end{align}
Now the $\Gamma(-n)$ cancels, finally leading to our master
formula~\cite{Borowka:2015mxa}
\begin{align}
\begin{split}
I = (-1)^r \Gamma(1-\epsilon)^\ell &\Gamma(r-s-\ell d/2)
  \int_0^\infty 
  \delta\Big(x_i-1\Big)
  \Bigg(
    \prod_{j=1}^t \D x_j\ 
       \frac{x_j^{\alpha_j-1}}{\Gamma(\alpha_j)}
  \Bigg)\\&
  \Bigg(
    \prod_{j=t+1}^{p} 
      \frac{\partial^{-\alpha_j}}{\partial x_j^{-\alpha_j}}
  \Bigg)
    \frac{\mathcal{U}^{r-s-(\ell+1)d/2}}{\mathcal{F}^{r-s-\ell d/2}}
  \Bigg|_{x_{t+1}=...=x_p=0}\,,
\label{eq:masterform}
\end{split}
\end{align}
with $r$ ($s$) the sum of positive (negative) indices, $t$ the number
of positive indices and $p$ the length of the family as defined
above and in~\cite{vonManteuffel:2012np}. This implies that
\begin{align}
\alpha_1,...,\alpha_t > 0
\quad\text{and}\quad
\alpha_{t+1},...,\alpha_p \le 0\,.
\end{align}

Let us now use Feynman parametrisation to calculate the integral
$I(a,b)$ we have introduced above. The polynomials $\mathcal{U}$ and
$\mathcal{F}$ can be easily calculated using~\cite{Smirnov:2013eza}
\begin{align}
\mathcal{U} = x_1+x_2\,,
\qquad\text{and}\qquad
\mathcal{F} = M^2x_2^2 + m^2 x_1 (x_1 + x_2)\,.
\end{align}
With our master formula $l=1$, $\alpha_1=a$, $\alpha_2=b$, and $r=a+b$
\begin{align}\begin{split}
I(a,b) = (-1)^{a+b} 
    \frac{\Gamma(a+b-\tfrac d2)\Gamma(1-\epsilon)}{\Gamma(a)\Gamma(b)}
    \int_0^\infty&\D x_1\,\D x_2\ \delta(1-x_2)\ x_1^{a-1}x_2^{b-1}
    \\&
    \frac{(x_1+x_2)^{a+b-d}}{\big(M^2x_2^2 + m^2 x_1 (x_1 + x_2)\big)^{a+b-d/2}}
    \,.
\end{split}\end{align}
We can see that if we can solve this integral for $a=b=1$, we will
most likely be able to solve it for any value of $a$ and $b$.

For now we will again set $m=0$ to simplify this integral. Calculating
the full integral either requires more complicated integration
techniques (for example cf.~\cite{Engel:2018}) or a 
clever substitution\footnote{The substitution in question is
$m\to M\sqrt{-(\chi-1)^2/\chi^2}$ with $\chi\in\mathbb{C}$}, resulting
in complicated hypergeometric functions that could be expanded in
$\epsilon$ with {\tt HypExp}~\cite{Huber:2005yg}. Needless to say,
this goes beyond the scope of this simple example.

The $\delta$-function makes the $x_2$-integral trivial. The $x_1$-integral we are left with is
\begin{align}\begin{split}
I(a,b) &= 
(-1)^{a+b}\big[M^2\big]^{d/2-a-b}
\frac{\Gamma(a+b-d/2)\Gamma(1-\epsilon)}{\Gamma(a)\Gamma(b)}
\int_0^\infty\D x_1\ x_1^{a-1} (1+x_1)^{a+b-d}
\\&
= (-1)^{a+b}\big[M^2\big]^{d/2-a-b}
  \frac{\Gamma(a + b - d/2) \Gamma(d-2 a - b) \Gamma(1 -
  \epsilon)}{\Gamma(b)\Gamma(d-a-b)}\,.
\end{split}\end{align}

\subsection{Method of regions}\label{subsec:MoRGeneral}

In general, the calculation of master integrals with full dependence
of any parameter is very difficult and time consuming. However, in
many cases this is not needed, often because the parameters have a
strong hierarchy such as the electron mass $m$ being much smaller than
the muon mass $M$ or typical momentum transfers $\sqrt{s}$. In this
case, we instead calculate the integrals expanded in $m$. The
technique used to achieve this is the \term{method of
regions}~\cite{Beneke:1997zp}.

We consider a loop integral that contains two or more disparate
scales, as in the case of the muon decay where $m^2 \ll M^2\sim p\cdot q$.
If the integral under consideration is hard to calculate, the obvious
idea is to expand the integrand in the small parameter with the hope
of achieving a simplification. The method that allows to consistently
perform such an expansion is the method of regions. 

To motivate this method, we again consider the toy example from above
\begin{align}\label{eq:ToyExample}
  I \equiv I(1,1) = \int
       [\D k]\ 
        \frac{1}{\big[k^2-m^2+\io\big]^{a}\big[k^2-2k\cdot p+\io\big]^{b}}
        \quad \text{with}\quad m^2 \ll p^2=M^2 \, .
\end{align}
It is useful to keep track of the $\io$ prescription in this
discussion.  We once again ignore that this integral can be calculated
with the full $m$ dependence and instead try to expand it at the level
of the integrand. Note that the naive expansion
\begin{align}
	\frac{1}{k^2-m^2}=\frac{1}{k^2}\Bigg(1+\frac{m^2}{k^2}+...\Bigg)
\end{align}
is not allowed due to the region of the integration domain where $k
\sim m \ll M$. The key idea of the method of regions is therefore to
split the domain into regions of constant order of magnitude. In the
case of our toy example, we introduce as an intermediary step an
additional scale $\Lambda$ with $m \ll \Lambda \ll M$ and
write\footnote{Note that we have not actually defined what is meant by
these integration boundaries w.r.t. the Minkowski metric. Thus, the
arguments are somewhat heuristic but could be formalised.}
\begin{align}
I =
  \underbrace{ 
    \int_0^\Lambda
       [\D k]\ \frac1{\big[ k^2-m^2+\io \big]\big[k^2-2k\cdot p+\io\big]}
  }_{=I_s^\Lambda} 
 +\underbrace{
    \int_\Lambda^\infty
       [\D k]\ \frac1{\big[ k^2-m^2+\io \big]\big[k^2-2k\cdot p+\io\big]}
  }_{=I_h^\Lambda} \, ,
\end{align}
where $I_s^\Lambda$ corresponds to the region where the loop momentum is
\term{soft} and $I_h^\Lambda$ to the one where it is \term{hard}.

We are now able to expand the integrand in each region according to
the respective scaling regime, namely\footnote{We assume as usual that
infinite summation and integration commutes.}
\begin{align}
\begin{split}
  I_s^\Lambda &\stackrel{k\sim m \ll p}{=}
      \sum_{n=0}^\infty
        \int_0^\Lambda[\D k]
         \ \frac{\big(-k^2\big)^n}{
           \big[k^2-m^2+\io\big]\big[-2k\cdot p+\io\big]^{n+1}
         }
\,, \\
  I_h^\Lambda &\stackrel{m\ll p \sim k}{=}
     \sum_{n=0}^\infty
        \int_\Lambda^\infty[\D k]
        \ \frac{\big(m^2\big)^n}{
           \big[ k^2+\io \big]^{n+1} \big[k^2-2k\cdot p+\io\big]}\,.
\end{split}
\end{align}
The $I_s^\Lambda$ become more and more \ac{UV} divergent as $n$ grows while
the $I_h^\Lambda$ becomes \ac{UV} finite as soon as $n>0$ but more \ac{IR}
divergent.  The sum, however, always keeps the same degrees of
divergence.

Before actually calculating $I_s^\Lambda$ and $I_h^\Lambda$, let us see what we can
deduce directly. Clearly, the new integrals are much simpler as they
both only have one scale ($m$ and $p^2$, respectively). Further, the
integrand of $I_s^\Lambda$ is proportional to $p^{-n-1}$ while $I_h^\Lambda\propto
m^{2n}$. Because both integrals have mass dimension zero, we can now
write
\begin{align}
I_s^\Lambda \propto \Big(\frac{m^2}{M^2}\Big)^{(n+1)/2}
\quad\text{and}\quad
I_h^\Lambda \propto \Big(\frac{m^2}{M^2}\Big)^{n}\,.
\end{align}
Note that the symmetry $p\to-p$ is broken by the presence of the $\io$
term. Naively, this would suggest that $I_s^\Lambda=0$ for $n$ even.
This, however, is not the case.  Assuming we want to calculate $I$ to
some order in $m/M$, we already now know how far in $n$ to expand
from these simple considerations without ever calculating
$I_s^\Lambda$ or $I_h^\Lambda$.

Once we actually do calculate the leading term, i.e. $n=0$, we see
that the newly introduced cut-off scale $\Lambda$ drops out in the sum
$I_s^\Lambda+I_h^\Lambda$.  This, of course, is to be expected since
there is no $\Lambda$ present in the original integral, defined
in~\eqref{eq:ToyExample}. 

Ideally, we like for the integration to cover the full domain to avoid
introducing the superfluous scale $\Lambda$, i.e. $\Lambda\to\infty$
for $I_s^\Lambda$ and $\Lambda\to0$ for $I_h^\Lambda$. However, this
potentially introduces additional contributions that need to be
calculated. For example in the first case of $I_s^\Lambda$, the added
term is
\begin{align}
\int_\Lambda^\infty[\D k]
 \ \frac{\big(k^2\big)^n}{
   \big[k^2-m^2\big]\big[-2k\cdot p+\io\big]^{n+1}
}
\stackrel{m\ll\Lambda}=\sum_{i=0}^\infty \big[m^2\big]^i
\int_0^\infty[\D k]
 \ \frac{\big(k^2\big)^n}{
   \big[k^2\big]^{1+i}\big[-2k\cdot p+\io\big]^{n+1}
}\,,
\end{align}
which is scaleless and therefore vanishes in \dreg{}. A similar
argument can also be made for $I_h^\Lambda$. Hence, we are allowed to
remove $\Lambda$ and find
\begin{subequations}
\begin{align}
I_s = \lim_{\Lambda\to\infty}I_s^\Lambda &= 
  \bigg(\frac{m^2}{M^2}\bigg)^{\tfrac{n+1}2}
  \bigg(\frac{\mu^2}{m^2}\bigg)^\epsilon
  \Gamma(1-\epsilon)
  \frac{
    \Gamma\big(\frac{n+1}2\big)
    \Gamma\big(\frac{n-1}2+\epsilon\big)}
  {2\Gamma(1+n)}
\\
I_h = \lim_{\Lambda\to0}I_h^\Lambda &= 
  (-1)^n \bigg(\frac{m^2}{M^2}\bigg)^n
  \bigg(\frac{\mu^2}{M^2}\bigg)^\epsilon
  \Gamma(1-\epsilon)
  \frac{\Gamma(1-2n-2\epsilon)\Gamma(n+\epsilon)}{\Gamma(2-n-2\epsilon)}\,.
\end{align}
\label{eq:morres}
\end{subequations}
Had we calculated $I$ with full mass dependence using appropriate
tricks or referred to a one-loop library such as
Package-X~\cite{Patel:2015tea}, we would have found
\begin{align}
I = \frac1\epsilon +
    2
    +2\beta\sqrt{\beta^2-1}\Big[\I\pi+2\log\big(\beta+\sqrt{\beta^2-1}\big)\Big]
    +2(\beta^2-1)\log\frac{M^2}{m^2}
    +\log\frac{\mu^2}{M^2}\,,
\end{align}
with $\beta=\sqrt{1-m^2/(4M^2)}$. We can expand this in $m/M$,
obtaining
\begin{align}
I = \frac1\epsilon + 2 - \pi\frac{m}{M} + \mathcal{O}(m^2)\,,
\end{align}
which is in agreement with~\eqref{eq:morres} for
$n=0$.

We are now ready to formulate the method of regions in general:
\begin{enumerate}
	\item 
    Identify all momentum regions that yield non-zero contributions.
    Note that in real-life applications more regions than just hard
    ($k\sim M$) and soft ($k\sim m$) may contribute.
	
    \item 
    Expand the integrand in each region and integrate the result over
    the full domain.
	
    \item 
    Sum up the contributions from all regions.
	
\end{enumerate}
When following these steps, one ends up with the expanded solution of
the integral. The method of region is intimately linked with the
concept of \ac{EFT}s as both exploit hierarchies of scales. In fact,
the \ac{EFT} framework can be viewed as a field-theoretical
formulation of the method of region. Assuming only soft and hard
contribution exists, the soft contribution can be viewed as a
calculation in an \ac{EFT} and the hard contribution as a matching
calculation to determine the Wilson coefficients of the \ac{EFT}.
Indeed, the \ac{UV} poles of the soft contributions (\ac{EFT}
calculation) match the \ac{IR} poles of the hard contribution (Wilson
coefficients).

\subsection{Light-cone coordinates and momentum regions}
\label{subsec:LightConeCoordinates}
The discussion of the method of regions above deals mostly with heavy
degrees of freedom that we remove. However, as we have seen in
Section~\ref{sec:irpred} it is also possible to remove light degrees
of freedom. This is particularly interesting if we want to study
(small) mass effects in a hard scattering process using \ac{SCET}, an
\ac{EFT} that splits soft and collinear modes off from the full
underlying theory, be it \ac{QED} or \ac{QCD}. A full review of
\ac{SCET} is well beyond the scope of this review, hence we refer
to~\cite{Becher:2014oda}. Instead, we will just discuss those points
we need in order to extend our previous discussion to also cover
(anti-)collinear and ultrasoft regions. Both of these are relevant for
the muon decay which we will calculate in a \ac{SCET}-inspired way.

Coordinates suitable for the description of the relevant regions (i.e.
hard, soft, collinear) are the light-cone coordinates, which are based
on the light-like momenta $e =(1,\vec{0},1)/\sqrt{2}$ and
$\bar{e}=(1,\vec{0},-1)/\sqrt{2}$.\footnote{Our definition of $e$ and
$\bar{e}$ differs from the standard convention by the normalisation
factor $1/\sqrt{2}$.} These allow to decompose any momentum $l$ into
its light-cone components as
\begin{align}
	l^\mu
       =l_+^\mu + l_-^\mu+l_\perp^\mu
       =(l_+,l_-,l_\perp)
       =(e\cdot l,\bar{e}\cdot l, l_\perp) \, .
\end{align}
Let $r$ be another arbitrary momentum, then these components satisfy
the properties
\begin{align}\label{eq:LightconeProperties}
	l_{\pm}^2=l_{\pm}\cdot l_\perp =0
\,, \quad 
    l \cdot r = l_+ \cdot r_- + l_-\cdot r_++l_\perp \cdot r_\perp
\,, \quad 
    l^2=2l_+\cdot l_-+l_\perp^2 \, .
\end{align}
As an example, let us now write the kinematics of the muon decay in
terms of these coordinates. If the muon is considered at rest, we find
$p=(M,M,0)/\sqrt{2}$ in light-cone coordinates. Furthermore, we choose
the electron momentum as $q=(0,q_-,q_\perp)=(0,\sqrt2E,q_\perp)$. This
yields for the kinematic invariants
\begin{align}\label{eq:InvariantsInLightcone}
	p^2=2 p_+\cdot p_-=M^2 \sim\lambda^0
\,, \quad 
    q^2=q_\perp^2=m^2 \sim\lambda^2
\,, \quad 
    s = 2 p \cdot q = 2 p_+\cdot q_-=2M E \sim\lambda^0 \, .
\end{align}
We use $\lambda$ to indicate the relative size of the parameters and
as a book keeping tool that was not strictly necessary in the
discussion above.  Next, we need a componentwise scaling of the
momenta $p$ and $q$ that reproduces these scalings
\begin{align}
    p\sim(M,M,0) \sim (1,1,0)
\,, \quad
    q\sim(0,E,q_\perp)\sim (0,1,\lambda)\, .
\end{align}
A region of the loop momentum $k$ is then defined as a specific choice
of parameters $a$, $b$, and $c$ where $k \sim (\lambda^a, \lambda^b,
\lambda^c)$. At this point we expect an infinite number of regions
corresponding to the infinite possible choices of $a$, $b$, and $c$.
Fortunately, almost all of the infinite number of regions turn out to
be zero.

From the \ac{SCET} point of view, we expect the following contributing
regions:
\begin{subequations}\label{eq:regions}
  \begin{align}
\mbox{hard:} &\quad k\sim(1,1,1) \label{eq:khard} \\
\mbox{soft:} &\quad k\sim(\lambda,\lambda,\lambda)\label{eq:ksoft} \\
\mbox{anti-collinear:} &\quad k\sim(1,\lambda^2,\lambda) \label{eq:kacoll}\\
\mbox{collinear:} &\quad k\sim(\lambda^2,1,\lambda) \label{eq:kcoll}\\
\mbox{ultrasoft:} &\quad k\sim(\lambda^2,\lambda^2,
\lambda^2)\,, \label{eq:kus}
  \end{align}
\end{subequations}
We have included the anti-collinear region for completeness even
though it does not enter in the muon decay. All regions can appear on
the level of individual integrals and even diagrams.  However, once
all diagrams are summed, we expect all regions except hard, soft, and
collinear to drop out.

Let us discuss a simple one-loop example to illustrate how loop
integrations are performed in light-cone coordinates. The integral
\begin{align}
	I= \int \big[\D k\big] 
        \frac1{\big[-2 k_-\cdot p_+\big]\big[(k-q)^2-m^2\big]}
\end{align}
occurs in the method of regions calculation of the one-loop bubble
master integral. In order to perform this integration in the standard
way, we need to write the integrand as a function of $k$ instead of
its light-cone components. This can be achieved with the identity
$k_-\cdot p_+=k \cdot p_+$. Now we can proceed as usual:
Feynman parametrisation, shift to remove all terms linear in $k$ and
integration over loop momentum. Using~\eqref{eq:LightconeProperties}
and \eqref{eq:InvariantsInLightcone} as well as our master
formula for the Feynman parametrisation~\eqref{eq:masterform}, we find
\begin{align}
\begin{split}
 I&= M^{2\epsilon}\Gamma(1-\epsilon)\Gamma(\epsilon) 
    \int_0^\infty \D x \ x^{-2+\epsilon}(s+m^2 x)^{-\epsilon}
\\&= -\frac{m^2}{s \epsilon} + \frac{m^2\big[-1+2\log(m/M)\big]}{s}
        +\mathcal{O}(\epsilon)\, .
\end{split}
\end{align}

\section{Massification}\label{sec:massification}
\def\soft{\mathcal{S}}
The procedure set out above allows, at least in principle, to expand
any amplitude to whatever power in $m/M$ necessary. And while it is
certainly much simpler than the full computation of the amplitude with
massive electron, it would still be a lot of effort to repeat it anew for
each process. However, if we are only interested in the leading term
$(m/M)^0$, we do not have to because we can view the light mass as an
\ac{IR} regulator of collinear singularities.  This way the terms
$\log m$ we are after can be obtained by considering a regularisation
scheme dependence. This formalises the discussion of
Section~\ref{subsec:LightConeCoordinates} above in the \ac{SCET}
framework. We call the resulting procedure \term{massification}.

Massification has been worked out at \ac{NNLO}.  Initially this was
done for \ac{QED} in the context of Bhabha
scattering~\cite{Penin:2005eh}. Later, a more general approach has
been presented~\cite{Mitov:2006xs,Becher:2007cu} that relies on
factorisation and is also valid in \ac{QCD}. \cite{Engel:2018fsb}~has
extended this to include also heavy flavours.  Very recently, these
considerations have been extended beyond \ac{NNLO}, in particular for
the heavy-quark form factor~\cite{Liu:2017axv, Liu:2018czl,
Blumlein:2018tmz}.

To be concrete, massification allows us to write for example
\begin{subequations}
\begin{align}
\mathcal{A}_{t\to W^\pm b}(m)
    &= \sqrt{\Zjet}\times\soft\times
    \mathcal{A}_{t\to W^\pm b}(0)
    +\mathcal{O}(m/M)\label{eq:massimaster:tb}
\,,\\
\mathcal{A}_{\gamma^*\to e e}
    &= \sqrt{\Zjet\times\Zantijet}\times\soft'\times
    \mathcal{A}_{\gamma^*\to ee}(0)
    +\mathcal{O}(m^2/s)\,,
\label{eq:massif}\end{align}
allowing us to relate the massive amplitude $\mathcal{A}(m)$ to the
(partially) massless amplitude $\mathcal{A}(0)$. For this we need a
process-dependent soft contribution $\soft^{(\prime)}$ as well as a
process-independent collinear contribution $\Zjet$ and an
anti-collinear contribution $\Zantijet$. The latter two are universal
and can be obtained by solving~\eqref{eq:massif} (cf.
Appendix~\ref{sec:const:massify}). For this, we have to calculate the
amplitude expanded in $m$. For the case of the heavy-quark form factor
that only contains \ac{HPL}s, we can just take the full
result~\cite{Bernreuther:2004ih} and expand using the Mathematica
package {\tt HPL}~\cite{Maitre:2005uu}. However, obtaining
$\mathcal{A}$ is more involved for the heavy-to-light form factor
which contains \aterm{generalised polylogarithms}{GPL}.  Hence, we
have to resort to the method of regions discussed in
Section~\ref{subsec:MoRGeneral}.
\end{subequations}

With the expressions for $\Zjet$ and $\Zantijet$, we have now the
following recipe to massify any amplitude. The hard part corresponds
to the corresponding amplitude with $m=0$. For each external collinear
(anti-collinear) fermion of mass~$m$, we multiply by the corresponding
$\Zjet^{1/2}$ ($\Zantijet^{1/2}$). Finally, we add a process dependent
soft function.

In the following we will discuss the soft contributions for both
processes (Section~\ref{sec:softtb} and Section~\ref{sec:softhq}),
commenting on subtleties that, in this context, were first discussed
in~\cite{Engel:2018fsb}. Next in Section~\ref{sec:compjet}, we will
compare $\Zjet$ and $\Zantijet$, the explicit expressions of which can
be found in Appendix~\ref{sec:const:massify}, with the literature,
especially~\cite{Becher:2007cu}.

\subsection{The soft function for \texorpdfstring{$t\to W^\pm b$}{t->Wb}}\label{sec:softtb}

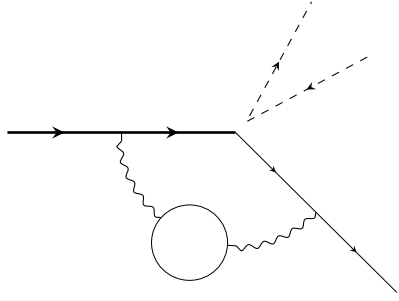
\begin{figure}[t]
\centering
\begin{tikzpicture}

\def\thz{67.5}
\def\R{1.5}
\def\rad{0.5}
 
 \coordinate (muin) at (180:3.0);
 \coordinate (eout) at (-45:3.0);
 \coordinate (vert) at (  0,0.0);
 \coordinate (qed1) at (180:\R);
 \coordinate (qed2) at (-45:\R);

 \draw[mfermion,line width=0.1pt] (vert) -- (qed2) -- (eout);
 \draw[mfermion,line width=1pt] (muin) -- (qed1) -- (vert);
 \draw[mfermion,dashed] (30:2) -- (45:0.2) -- (60:2);

\pgfmathsetmacro\alph{2*atan(\rad/\R)};
\pgfmathsetmacro\tha{\thz+\alph/2};
\pgfmathsetmacro\thb{\thz-\alph/2};
\pgfmathsetmacro\radprime{sqrt(\rad*\rad+\R*\R)};

\centerarc[photon](0,0)(135:\tha:-1.5)
\centerarc[photon](0,0)(\thb:0:-1.5)

\draw [line width=0.1pt](\thz:-\radprime) circle  (\rad);
 
\end{tikzpicture}
\caption{The $n_m$-bubble giving rise to the soft contribution of
$t\to W^\pm b$.}
\label{fig:diagsoft}
\end{figure}

For the soft part $\soft$, we only need to consider diagrams with
internal fermion loops. Indeed, by performing the formal decoupling of
gluon and fermion fields in the \ac{SCET} framework, one can show that
purely gluonic contributions to the soft part vanish to all
orders~\cite{Bauer:2001yt,Becher:2007cu}. A simple counting argument
implies that only the fermion bubble with mass $m$ contributes~(cf.
Figure~\ref{fig:diagsoft}). Therefore, the unrenormalised soft part
$\soft_0$ can easily be calculated from first principle in the
\ac{SCET} framework using \eqref{eq:ksoft}, i.e.
\begin{align}
  \label{eq:soft}
\soft_0 = 1+\bigg(\frac{\alpha_0^2}{4\pi}\bigg)\ C_F \int[\D k]\frac{
\big(-2p^\mu\big)\big(-2q_-^\nu\big)
}{(k^2)^2 (2p\cdot k)(2q_-\cdot k)} \Pi^{\mu\nu}_{(n_m)}(k)\,.
\end{align}
In accordance with \eqref{eq:ksoft}, we only use the large component
of the collinear momentum $q$. Even though the calculation is
performed in \fdh{}, there is no contributions $\propto\neps$, because
$\epsilon$-scalars do not couple to fermions in the eikonal
approximation~\cite{Gnendiger:2016cpg}. The function
$\Pi^{\mu\nu}_{(n_m)}$ is the contribution of $n_m$ fermions with mass
$m$ to the usual tensorial vacuum polarisation~\eqref{eq:pimunu}. When
calculating $\soft_0$, one encounters an anomaly, i.e. the breaking of
naive factorisation~\cite{Becher:2010tm, Becher:2011pf}.
Following~\cite{Beneke:2005}, we call this \term{factorisation
anomaly}\footnote{This is also referred to as collinear anomaly or
rapidity divergence~\cite{Chiu:2011qc}.}. This is a new feature that
is only present due to the large mass $p^2\!=\!M^2$. 

The factorisation anomaly first appears because the
integral~\eqref{eq:soft} is not fully regularised in \dreg{} and hence
requires further \term{analytic regularisation}. We shift the power of
the propagator $p\cdot k$ at the diagrammatic level according
to~\cite{Smirnov:1997gx, Becher:2014oda}
\begin{align}
  \label{eq:reganomaly}
\frac1{(k-p)^2-M^2+\io}\to
\frac1{-2p\cdot k+\io} \to (-\nu^2)^\eta\,
\frac1{(-2p\cdot k+\io)^{1+\eta}}\,,
\end{align}
where the regulator $\eta$ has to be expanded before the dimensional
regulator $\epsilon$. This regularisation also introduces an
associated scale $\nu$ that drops out in the final result. The only
further soft contribution is from $n_m$ terms in the wave-function
renormalisation of the heavy fermion. Including this contribution,
$Z_2^{\soft}$, we obtain
\begin{align}\begin{split}
\soft = \sqrt{Z_2^{\soft}}\times \soft_0 = 1+\Big(a_0(M m)\Big)^2 C_F n_m 
\Bigg[&
  \frac23\frac1\eta\bigg(-\frac1{\epsilon^2}+\frac{5}{3\epsilon}
  -\frac{28}{9}-2\zeta_2\bigg)
\\[10pt]&+
\frac1{2\epsilon^3}
-\frac1{9\epsilon^2}
+\frac1\epsilon\bigg(-\frac{26}{27}+\zeta_2\bigg)
\\[10pt]&
+\frac{11}3-\frac23\zeta_3-\frac29\zeta_2
\Bigg]+\mathcal{O}\Big(a^3,\epsilon,\eta\Big)\,,
\label{eq:softtb}
\end{split}\end{align}
where we define $a_0(x)$ through the bare coupling as
\begin{align}
a_{0,i}(x) = \bigg(\frac{\alpha_{i,0}}{4\pi}\bigg)
      \bigg(\frac{\mu^2}{m^2}\bigg)^{\epsilon}
      (-2+\io)^{\eta/2}
      \bigg(\frac{-\nu^2}{x}\bigg)^{\eta/2}\,,
\qquad
i\in\{s,e\}\,,
\label{eq:acoup}
\end{align}
with an analogous expression for the renormalised couplings.

\subsection{Collinear contribution for \texorpdfstring{$t\to W^\pm b$}{t->Wb}}
Looking at \eqref{eq:softtb}, the pole in $1/\eta$ seems like a
catastrophe as we are required to set $\eta\to0$ in the end. Luckily,
there is still the contribution from collinear region to consider.
$\Zjet$ is much more complicated than $\soft$ with the full result
given in \eqref{eq:zjet}. The relevant $n_m$ bit however is simple
enough
\begin{align}
\sqrt{\Zjet}\Big|_{n_m} &=
 \Big(a_0(s)\Big)^2
    C_F n_m \frac23 \Bigg[
        \frac1\eta\bigg(
            \frac1{\epsilon^2}-\frac5{3\epsilon}+\frac{28}9+2\zeta_2
        \bigg)
        -\frac1{\epsilon^3}+\frac1{2\epsilon^2}
        +\frac1\epsilon\bigg(-\frac{55}{24}-3\zeta_2\bigg)
        \notag\\&\qquad\qquad
        +\frac{1675}{432}-2\zeta_2+\zeta_3
    \Bigg]
+\mathcal{O}(a_s^3,\epsilon,\eta)\,.
\label{eq:zjet:nm}\end{align}
Because $\Zjet$ also has a pole in $\eta$, their sum, as mandated by
\eqref{eq:massimaster:tb} is finite in the analytic regulator $\eta$.
In other words, after finishing the massification, the result is again
free of extra divergences and reproduces the result obtained by
calculating the amplitude directly. 

However, in doing so a new anomalous logarithm is created because the
arguments of $a_0$ in $\Zjet$ and $\soft$ differ. Schematically,
\begin{align}
\sqrt\Zjet + \soft \sim
      \bigg(\frac{-\nu^2}{ s }\bigg)^{\eta/2}\frac1\eta 
     -\bigg(\frac{-\nu^2}{m M}\bigg)^{\eta/2}\frac1\eta 
      = \frac12\log\frac{m M}{s} + \mathcal{O}(\eta)\,.
\end{align}
Combined with the terms $\log(\mu^2/m^2)$ from expanding $a_0$ in
$\epsilon$, this means that the two-loop form factors contain terms
$(\log m)^3$ instead of just $(\log m)^2$ suggested by naive counting.
This extraneous logarithm is cancelled when the process is combined
with the pair-production process $t\to W^\pm b+bb$.

\subsection{The soft function for \texorpdfstring{$\gamma^*\to qq$}{g*->qq}}\label{sec:softhq}
For this process, the soft function was first calculated
in~\cite{Becher:2007cu} as
\begin{align}
  \soft'_{\text{\cite{Becher:2007cu}}}  =
  1+\bigg(\frac{\alpha}{4\pi}\bigg)^2\ C_F \int[\D k]\frac{
\big(-2p^\mu\big)\big(-2q^\nu\big)
}{(k^2)^2 (2p\cdot k)(2q\cdot k)} \Pi^{\mu\nu}_{(n_m)}(k)\,.
\end{align}
This definition is motivated by the eikonal approximation and does not
lead to a factorisation anomaly. Our definition of the soft
contribution to the heavy-quark form factor is motivated by
\ac{SCET}. For consistency with the collinear contribution, one also
has to introduce the same regulator here. Our definition therefore
reads
\begin{align}
  \label{eq:softHQ}
\soft' &= 1+\bigg(\frac{\alpha}{4\pi}\bigg)^2\ C_F \int[\D k]\frac{
    \big(-2p_+^\mu\big)\big(-2q_-^\nu\big)
}{(k^2)^2 (2p_+\cdot k)^{1+\eta}(2q_-\cdot k)} \Pi^{\mu\nu}_{(n_m)}(k)
\,,
\end{align}
where $p$ is assumed to scale anti-collinear and $q$ collinear.
Because any integral of the form
\begin{align}
I(n_1,n_2,n_3) \equiv \int[\D k]\frac{1}{
    (k^2)^{n_1} (2p_+\cdot k)^{n_2} (2q_-\cdot k)^{n_3}}
\Pi^{\mu\nu}_{(n_m)}(k)
\end{align}
depends on $p_+$ and $q_-$ only through $s=2\, p_+\!\cdot q_-$, it is
invariant under simultaneous rescaling $p_+ \to \lambda\ p_+, \ q_-
\to q_-/\lambda$.  This implies $I(n_1,n_2,n_3) =
\lambda^{-n_2}\lambda^{n_3}\ I(n_1,n_2,n_3)$ and, hence, $I=0$ unless
$n_2=n_3$. However, due to the regulator \eqref{eq:reganomaly}, $n_2$
can never be equal to $n_3$. Hence, all occurring integrals vanish and
$\soft'=1$ at two loops.  For the heavy-quark form factor, the
factorisation anomaly in $\Zjet$ is therefore not cancelled by an
anomaly in the soft contribution. In the following we show that,
instead, it is cancelled by an anomaly in $\Zantijet$. This is a
contribution analogous to $\Zjet$, but due to the anti-collinear
fermion.

\subsection{Comparison with the heavy-quark form factor}
\label{sec:compjet}

The collinear contribution, $\Zjet$, agrees with a corresponding
expression obtained in~\cite{Mitov:2006xs} apart from the $n_m$ terms
that were not considered there. However, the different treatment of
the soft function makes a direct comparison with~\cite{Becher:2007cu}
difficult. Instead, we have to include the anti-collinear contribution
$\Zantijet$ whose $n_m$ term (cf.~\eqref{eq:zantijet})
\begin{align}
\sqrt{\Zantijet}\Big|_{n_m} = \Big(a^0_s(m^2)\Big)^2 
    C_F n_m \frac23 \Bigg[&
        -\frac1\eta\bigg(
            \frac1{\epsilon^2}-\frac5{3\epsilon}+\frac{28}9+2\zeta_2
        \bigg)
        +\frac1{2\epsilon^3}
        -\frac5{6\epsilon^2}
        -\frac{253}{72\epsilon}
        \notag\\&\quad
        +\frac{5083}{432}-\frac{14}{3}\zeta_2-\zeta_3
    \Bigg] + \mathcal{O}(a_s^3,\epsilon,\eta)
\end{align}
is different from the one of $\Zjet$, again cancelling the pole in
$\eta$. We find agreement for
\begin{align}
Z_{\text{\cite{Mitov:2006xs}}} =
\sqrt{\Zjet\times\Zantijet}\, \Big|_{n_m\to0}
\qquad\text{and}\qquad 
Z_{\text{\cite{Becher:2007cu}}} \times \soft'_{\text{\cite{Becher:2007cu}}} =
    \sqrt{\Zjet\times\Zantijet}\times\soft'\,.
\end{align}
Hence, our results agree with previous ones but extend them to
processes where additional fermions with a large mass are present.
This agreement as well as the fact that $\Zjet$ is the same for the
heavy-to-light and heavy-quark form factors is a strong indication
that the factorisation presented here is general.

\subsection{Summary}
With massification we have an extremely powerful tool at hand to
calculate the leading mass effects, i.e. the logarithms $\log m$, of
any one- or two-loop matrix element where $m$ is the smallest scale
involved. Unfortunately, this means that massification cannot yet be
used to calculate real-virtual or real-virtual-virtual matrix elements
because those will contain a scale associated to the energy of the
real photon. When integrating over phase space with \ac{FKS2}, this
energy can become arbitrarily small s.t. the assumption that the mass
$m$ is the smallest scale is no longer justified.

For any valid process, we need to write
\begin{align}
\cA(m) = \prod_{\text{inc}}\sqrt{\Zantijet}\times
         \prod_{\text{out}}\sqrt{\Zjet}\times
         \soft\times
         \cA(0) + \mathcal{O}\Big(\frac{m^2}{\{s,t,M^2,...\}}\Big)\,.
\end{align}
This is very similar to the \ac{LSZ} formula except that the products
only run over incoming and outgoing light but non massless flavours
($n_m$).  The function $\soft=1+\delta\soft$ is process dependent but
$\delta\soft\propto n_m$ and hence is relatively simple to calculate.

There is one remaining problem related to the factorisation anomaly.
In~\eqref{eq:acoup}, we are forced to choose a scale of the anomaly
that is different in $\Zjet$, $\Zantijet$, and $\soft$. Presently it
is unclear how this scale must be chosen.

The anomaly also has the unfortunate side effect of giving rise to
logarithms with higher power than suggested by naive counting. For
example at two-loop, the higher power one would naively expect is
$\log^2m$. This is raised to $\log^3m$ due to the anomaly. Hence,
power-suppressed terms $m^i\log^jm$ too might be larger than expected.
Luckily all of this happens only in the $n_m$ part of the amplitude
that is generally easier to obtain with full $m$ dependence than the
remaining amplitude.

This is especially true considering that one might also need to
include contributions from the \ac{HVP} that in any case need to be
done numerically. Some progress has been made to efficiently include
\ac{HVP} effects also in complicated loop
diagrams~\cite{Fael:2018dmz, Fael:2019nsf}. As a side effect, this
also allows the exact numerical calculation of the $n_m$ terms with
just one finite numerical integration.

\chapter{The Monte Carlo code \mcmule{}}\label{ch:mcmule}

\mcmule{} ({\bf M}onte {\bf c}arlo for {\bf Mu}ons and other {\bf
le}ptons) is a generic framework for higher-order \ac{QED}
calculations of scattering and decay processes involving leptons. It
is written in Fortran~95 with two types of users in mind. First,
several processes are implemented, some at \ac{NLO}, some at
\ac{NNLO}. For these processes, the user can define an arbitrary
(infrared safe), fully differential observable and compute cross
sections and distributions. \mcmule{}'s processes, present and, future,
are listed in Table~\ref{tab:mcmuleprocs} together with the relevant
experiments for which the cuts are implemented.  Second, the program
is set up s.t.  additional processes can be implemented by supplying
the relevant matrix elements.

The code can be found at 
\begin{lstlisting}[language=bash]
    (*@\url{https://gitlab.psi.ch/mcmule/mcmule}@*)
\end{lstlisting}
The internal version of the code can be found at
\begin{lstlisting}[language=bash]
    (*@\url{https://gitlab.psi.ch/mcmule/monte-carlo}@*)
\end{lstlisting}
Access will be granted by the \mcmule{} core team (\ac{MMCT}),
usually for new collaborators who wish to extend \mcmule{} in
meaningful ways.

This chapter will often refer to \mcmule{}'s online
manual~\cite{mcmuleman} for specific details. This is to avoid
deprecating this document as new processes are added and technical
details may change. In any case, the online manual will be
authoritative.

\begin{figure}[t]
\centering
\begin{tabular}{l|r|l|p{5cm}|c}
\bf process & \bf order & \bf experiments & \bf comments & \bf
status\\\thickhline
$\mu\to\nu\bar\nu e$ & NNLO & MEG I\&II & polarised, massified \& exact & \cite{Engel:2019nfw}\\\hline

$\mu\to\nu\bar\nu e\gamma$ & NLO & MEG I & polarised &
\cite{Pruna:2017upz}\\
\hline

$\mu\to\nu\bar\nu eee$ & NLO & Mu3e & polarised & \cite{Pruna:2016spf} \\\hline

$\mu\to\nu\bar\nu e\gamma\gamma$ & LO & MEG & polarised  & priv. comm.\\\hline

$\tau\to\nu\bar\nu e\gamma$ & NLO & BaBar & cuts in lab frame &
\cite{Pruna:2017upz}\\\hline

$\tau\to\nu\bar\nu l\ell\ell$ & NLO & Belle II & & $*$ \\\thickhline

\multirow{3}{*}{$e\mu\to e\mu$} & NLO & \multirow{3}{*}{MUonE} & &
complete
\\\cline{2-2}\cline{4-5}
& \multirow{2}{*}{NNLO} &  & purely electronic corrections & 
\cite{Banerjee:2020rww} \\\cline{4-5}
& & & mixed (massified) & $\dag$\\\hline

$\ell p\to\ell p$ & NNLO & P2, MUSE, Prad & only leptonic
corrections & complete\\\thickhline

$e^-e^-\to e^-e^-$ & NNLO & Prad & &$*$\\\hline
$e^+e^-\to e^+e^-$ & NNLO &      & &$\dag$\\\hline
$e^+e^-\to \gamma\gamma$ & NNLO & PADME    & &$\dag$\\\hline
$e^+e^-\to \mu^+\mu^-$ & NNLO & Belle & massified & $\dag$
\end{tabular}

\renewcommand{\figurename}{Table}
\caption{A list of processes that are either already included in
\mcmule{}, almost implemented ($*$), or planned to be implemented
($\dag$).}
\label{tab:mcmuleprocs}
\end{figure}

\mcmule{} consists of several modules with a simple, mostly hierarchic
structure. In this chapter we will describe this structure as follows:
First, we give an overview with a brief description of all modules and
how they are connected in Section~\ref{sec:structure}.  Next, we
discuss in Section~\ref{sec:example} how the code works and how to run
it on the basis of a simple process, the radiative tau decay
$\tau\to\nu\bar\nu e \gamma $. We also discuss tools to analyse the
output of \mcmule.  Technical aspects of \mcmule{} are discussed in
Section~\ref{sec:techno}. Finally, we describe in
Section~\ref{sec:implement} on how to implement additional processes
in \mcmule{}.

\section{Structure of \mcmule{}}
\label{sec:structure} 

\mcmule{} is written in Fortran 95 with helper and analysis tools
written in {\tt python}. To obtain a copy of \mcmule{} we recommend
the following approach
\begin{lstlisting}[language=bash]
$ git clone --recursive https://gitlab.psi.ch/mcmule/mcmule
\end{lstlisting}
\noindent To build \mcmule{}, a Fortran compiler such as {\tt
gfortran} and a python installation is needed. The main executable can
be compiled by running
\begin{lstlisting}[language=bash]
$ ./configure
$ make mcmule
\end{lstlisting}
Alternatively, we provide a Docker container~\cite{Merkel:2014} for
easy deployment and legacy results. In multi-user environments, {\sl
udocker}~\cite{Gomes:2017hct} can be used instead. In either case, a
pre-compiled copy of the code can be obtained by calling
\begin{lstlisting}[language=bash]
$ docker pull yulrich/mcmule  # requires Docker to be installed
$ udocker pull yulrich/mcmule # requires uDocker to be installed
\end{lstlisting}
When started, {\tt mcmule} reads options from {\tt stdin} as specified
in Table~\ref{tab:mcmuleinput} (cf. Section~\ref{sec:example}). The
value and error estimate of the integration is printed to {\tt stdout}
and the full status of the integration is written in a
machine-readable format into a folder called {\tt out/} (see below).

The structure of the code and the relation between the most important
Fortran modules is depicted in Figure~\ref{fig:structure}. A solid
arrow indicates ``using'' the full module, whereas a dashed arrow is
indicative of partial use. In what follows we give a brief description
of the various modules and mention some variables that play a
prominent role in the interplay between the modules.

\begin{figure}
  \centering

\tikzset{
    block/.style={
        rectangle, draw, 
        text width=6em, minimum height=2em, 
        text centered, rounded corners
    },
    pblock/.style={
        rectangle, fill=gray!30, midway
    },
    line/.style={
        decoration={markings,mark=at position \pgfdecoratedpathlength-5pt with {\arrow[scale=3]{>}}},
        postaction={decorate}, shorten >= 5pt
    }
}
\begin{tikzpicture}[x=3cm,y=-2cm]
\node[block       ] (GD)  at (-1,-1) {\tt global\_def};
\node[block       ] (CLL) at ( 1,-1) {\tt collier};
\node[block       ] (FU)  at ( 0, 0) {\tt functions};
\node[block       ] (USR) at ( 1, 3) {\tt user};
\node[block       ] (PS)  at ( 0, 1) {\tt phase\_space};
\node[block       ] (ML)  at (-1, 1) {\tt \{pg\}\_mat\_el};
\node[block       ] (PG)  at (-1, 2) {\tt \{pg\}};
\node[block       ] (MG)  at (-1, 3) {\tt mat\_el};
\node[block       ] (INT) at (-1, 4) {\tt integrands};
\node[block       ] (VEG) at ( 1, 4) {\tt vegas};
\node[block       ] (XS)  at (-1, 5) {\tt mcmule};
\node[block       ] (TST) at ( 1, 5) {\tt test};

\draw [line       ] (CLL) --+ ( 0,0.5) -- (FU) ;
\draw [line       ] (GD)  --+ ( 0,0.5) -- (FU) ;
\draw [line       ] (FU)  --+ ( 1,0.5) -- (USR);
\draw [line       ] (FU)               -- (PS) ;
\draw [line       ] (FU)  --+ (-1,0.5) -- (ML) ;
\draw [line       ] (ML)               -- (PG) ;
\draw [line       ] (PG)               -- (MG) ;
\draw [line       ] (MG)               -- (INT);
\draw [line,dashed] (PS)               -- (PG)   node[pblock] {\tt ksoft};
\draw [line       ] (PS)  --+ (0,2)    -- (INT);
\draw [line       ] (USR)              -- (INT);
\draw [line,dashed] (USR)              -- (VEG) node[pblock] {\tt metadata};
\draw [line,dashed] (VEG)              -- (INT) node[pblock] {\tt bin\_it};
\draw [line       ] (INT)              -- (XS) ;
\draw [line       ] (VEG)              -- (XS) ;
\draw [line       ] (INT)              -- (TST);
\draw [line       ] (VEG)              -- (TST);

\draw [decorate,decoration={brace,amplitude=10pt}] ( -1.5,2.2) --
(-1.5,0.8);
\end{tikzpicture}
  \caption{The structure of \mcmule{}.}
  \label{fig:structure}
\end{figure}

\begin{description}
    \item[{\tt global\_def}:] 
    This module simply provides some parameters such as fermion masses
    that are needed throughout the code. It also defines {\tt prec} as
    a generic type for the precision used.\footnote{For quad precision
    {\tt prec=16} and the compiler flag {\tt -fdefault-real-16} is
    required.} Currently, this simply corresponds to double precision.

    \item[{\tt functions}:] 
    This module is a library of basic functions that are needed at
    various points in the code. This includes dot products, eikonal
    factors, the integrated eikonal, and an interface for scalar
    integral functions among others.

    \item[{\tt collier}:] 
    This is an external module~\cite{Denner:2016kdg, Denner:2010tr,
    Denner:2005nn, Denner:2002ii}. It will be linked to \mcmule{}
    during compilation and provides the numerical evaluations of the
    scalar, and in some cases tensor, integral functions in {\tt
    functions}.

    \item[{\tt phase\_space}:] 
    The routines for generating phase-space points and their weights
    are collected in this module. Phase-space routines ending with
    {\tt FKS} are prepared for the \ac{FKS} subtraction procedure with
    a single unresolved photon. In the weight of such routines a
    factor $\xi_1$ is omitted to allow the implementation of the
    distributions in the \ac{FKS} method. This corresponds to a global
    variable {\tt xiout1}. This factor has to be included in the
    integrand of the module {\tt{integrands}}. Also the variable {\tt
    ksoft1} is provided that corresponds to the photon momentum
    without the (vanishing) energy factor $\xi_1$. Routines ending
    with {\tt FKSS} are routines with two unresolved photons.
    Correspondingly, a factor $\xi_1\,\xi_2$ is missing in the weight
    and {\tt xiout1} and {\tt xiout2}, as well as {\tt ksoft1} and
    {\tt ksoft2} are provided. To ensure numerical stability it is
    often required to tune the phase-space routine to a particular
    kinematic situation.

    \item[{\tt \{pg\}\_mat\_el}]:
    Matrix elements are grouped into \term{process groups} such as
    muon decay ({\tt mudec}) or $\mu$-$e$ and $\mu$-$p$ scattering
    ({\tt mue}). Each process group contains a {\tt mat\_el} module
    that provides all matrix elements for its group.  Simple matrix
    elements are coded directly in this module. More complicated
    results are imported from sub-modules not shown in
    Figure~\ref{fig:structure}. A matrix element starting with {\tt P}
    contains a polarised initial state.  A matrix element ending in
    {\tt av} is averaged over a neutrino pair in the final state (cf.
    Section~\ref{sec:neutrinoavg}).

    \item[{\tt \{pg\}}:] 
    In this module the soft limits of all applicable matrix elements
    of a process group are provided to allow for the soft subtractions
    required in the \ac{FKS} scheme. These limits are simply the
    eikonal factor evaluated with {\tt ksoft} from {\tt phase\_space}
    times the reduced matrix element, provided through {\tt mat\_el}.

    This module also functions as the interface of the process group,
    exposing all necessary functions that are imported by

    \item[{\tt mat\_el},] which collects all matrix elements as well
    as their particle labelling or \aterm{particle
    identification}{PID}.
    
    \item[{\tt user}:]
    For a user of the code who wants to run for an already implemented
    process, this is the only relevant module.  At the beginning of
    the module, the user has to specify the number of quantities to be
    computed, {\tt nr\_q}, the number of bins in the histogram, {\tt
    nr\_bins}, as well as their lower and upper boundaries, {\tt
    min\_val} and {\tt max\_val}. The last three quantities are arrays
    of length {\tt nr\_q}. The quantities themselves, i.e. the
    measurement function, is to be defined by the user in terms of the
    momenta of the particles in {\tt quant}.  Cuts can be applied by
    setting the logical variable {\tt pass\_cut} to
    false\footnote{Technically, {\tt pass\_cut} is a list of length
    {\tt nr\_q}, allowing to decide whether to cut for each histogram
    separately.}. Some auxiliary functions like (pseudo)rapidity,
    transverse momentum etc. are predefined in {\tt functions}. Each
    quantity has to be given a name through the array {\tt names}.

    Further, {\tt user} contains a subroutine called {\tt inituser}.
    This allows the user to read additional input at runtime, for
    example which of multiple cuts should be calculated. It also
    allows the user to print some information on the configuration
    implemented. Needless to say that it is good idea to do this for
    documentation purposes.
    
    \item[{\tt vegas}:] 
    As the name suggests this module contains the adaptive Monte Carlo
    routine {\tt vegas}~\cite{Lepage:1980jk}.  The binning routine
    {\tt bin\_it} is also in this module, hence the need for the
    binning metadata, i.e. the number of bins and histograms ({\tt
    nr\_bins} and {\tt nr\_q}, respectively) as well as their bounds
    ({\tt min\_val} and {\tt max\_val}) and names, from {\tt user}.

    \item[{\tt integrands}:] 
    In this module the functions that are to be integrated by {\tt
    vegas} are coded. There are three types of integrands:
    non-subtracted, single-subtracted, and double-subtracted
    integrands, corresponding to, for example, the three parts
    of~\eqref{eq:nnloint}. The matrix elements to be evaluated and the
    phase-space routines used are set using function pointers through
    a subroutine {\tt initpiece}. The factors $\xi_i$ that were
    omitted in the phase-space weight have to be included here for the
    single- and double-subtracted integrands.
    
    \item[{\tt mcmule}:]
    This is the main program, but actually does little else than read
    the inputs and call {\tt vegas} with a function provided by {\tt
    integrands}.

    \item[{\tt test}:]
    For developing purposes, a separate main program exists that is
    used to validate the code after each change. Reference values for
    matrix elements and results of short integrations are stored here
    and compared against.

\end{description}

\begin{figure}
\input{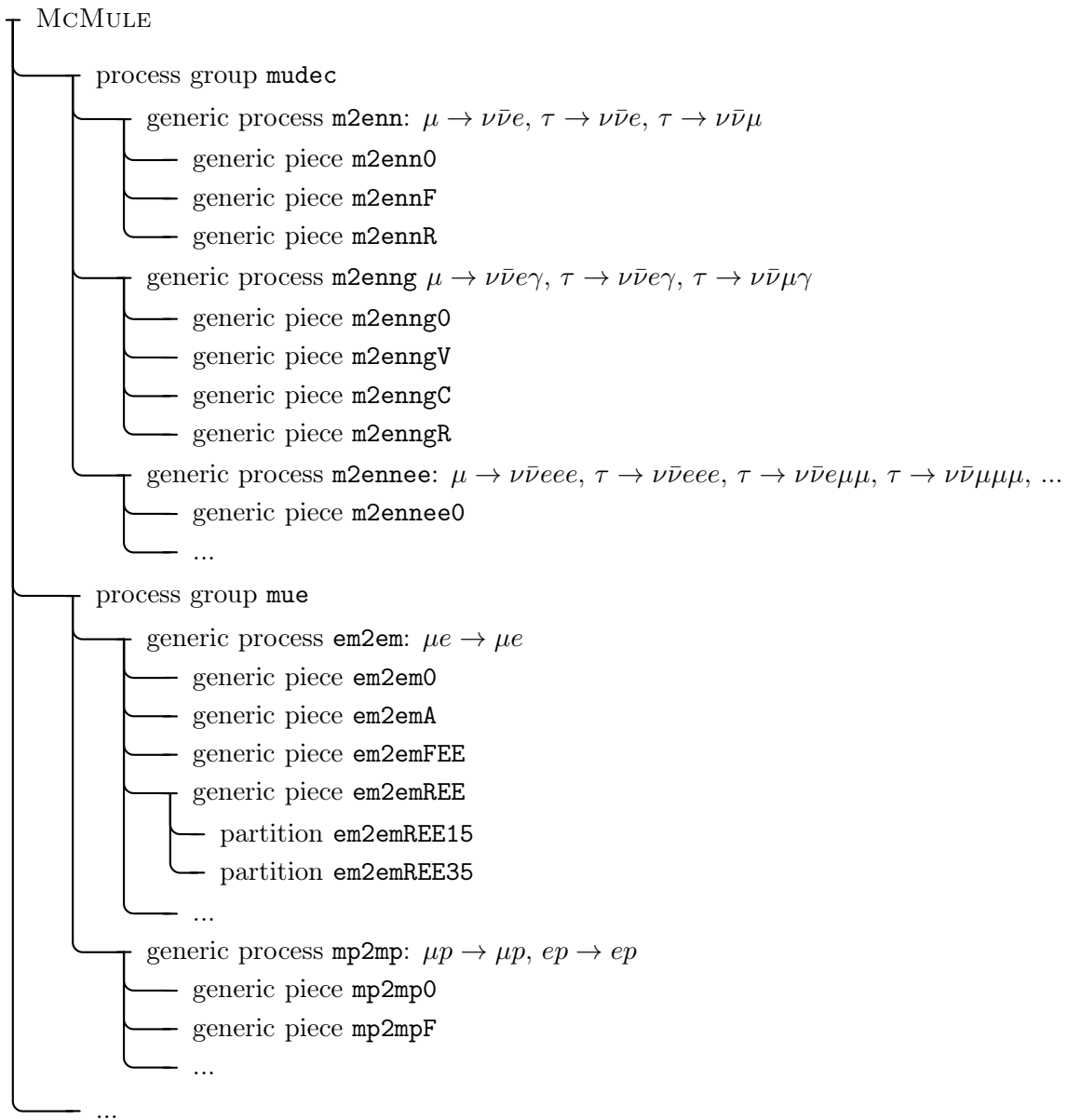}
\caption{The structure of process group, generic process, and generic
piece as used by \mcmule{}.}
\label{fig:processtree}
\end{figure}

The library of matrix elements deserves a few comments. As matrix
elements quickly become very large, we store them separately from the
main code. This makes it also easy to extend the program by minimising
the code that needs to be changed.  We group matrix elements into
process groups, \term{generic processes}, and \term{generic pieces} as
shown in Figure~\ref{fig:processtree}.  The generic process is a
prototype for the physical process such as $\ell\to\nu\bar\nu l\gamma$
where the flavour of the leptons $\ell$ and $l$ is left open. The
generic piece describes a part of the calculation such as the real or
virtual corrections, i.e. the different pieces of~\eqref{eq:nlo:4d}
(or correspondingly~\eqref{eq:nnlo:4d} at NNLO), that themselves may
be further subdivided as is convenient.  In particular, in some cases
a generic piece is split into various partitions (cf.
Section~\ref{sec:ps} for details on why that is important). The
example shown concerns the real part of \ac{NLO} contributions to the
electronic corrections to $\mu$-$e$ scattering.

When running {\tt mcmule}, the code generates a statefile from which
the full state of the integrator can be reconstructed should the
integration be interrupted (cf. Section~\ref{sec:vegasff} for
details). This makes the statefile ideal to also store results in a
compact format.  To analyse these results, we provide a python tool
{\tt pymule}, additionally to the main code for \mcmule{}. {\tt
pymule} uses {\tt numpy}~\cite{Walt:2011np} for data storage and {\tt
matplotlib} for plotting~\cite{Hunter:2007mp}. While {\tt pymule} works with
any python interpreter, {\tt IPython}~\cite{Perez:2007ip} is
recommended. We will encounter {\tt pymule} in
Section~\ref{sec:analyse} when we discuss how to use it to analyse
results. A full list of functions provided can be found in the online
manual of {\tt pymule}~\cite{mcmuleman}.

\section{Running \mcmule{}: an example}
\label{sec:example}

In order to provide a simple example with concrete instructions on how
to run the code and to illustrate how it works, we consider the
radiative decay of the tau $\tau\to e[\nu\bar\nu] \gamma $. Since the
neutrinos are not detected, we average over them, indicated by the
brackets (cf.  Section~\ref{sec:neutrinoavg}). Hence, we have to be
fully inclusive w.r.t. the neutrinos. Still, the code allows to make
any cut on the other final-state particles. As we will see, the
\ac{BR} for this process, as measured by
{\sc BaBar}~\cite{Lees:2015gea, Oberhof:2015snl} has a discrepancy of
more than $3\,\sigma$ from the \ac{SM} value. This will illustrate the
importance of fully differential \ac{NLO} corrections in \ac{QED}.

\subsection{Preparations}

To be concrete let us assume that we want to compute two distributions, the
invariant mass of the $e\gamma$ pair, $m_{\gamma e}\equiv
\sqrt{(p_e+p_\gamma)^2}$, and the energy of the electron, $E_e$, in the
rest frame of the tau. To avoid an \ac{IR} singularity in the \ac{BR},
we have to require a minimum energy of the photon. We choose this to
be $E_\gamma \ge 10\,{\rm MeV}$ as used in~\cite{Lees:2015gea,
Oberhof:2015snl}.

As mentioned in Section~\ref{sec:structure} the quantities are defined
in the module {\tt user} ({\tt src/user.f95}). At the beginning of the module we set
\begin{lstlisting}
  nr_q = 2
  nr_bins = 90
  min_val = (/ 0._prec, 0._prec /)
  max_val = (/ 1800._prec, 900._prec /)
\end{lstlisting}
where we have decided to have 90 bins for both distributions and {\tt
nr\_q} determines the number of distributions. The boundaries for the
distributions are set as $0 < m_{\gamma e} < 1800\,{\rm MeV}$ and $0
\le E_e \le 900\,{\rm MeV}$.

The quantities themselves are defined in the function {\tt quant} of
the module {\tt user}. This function takes arguments, {\tt q1} to {\tt
q7}. These are the momenta of the particles, arrays of length 4 with
the fourth entry the energy.  Depending on the process though not all
momenta are needed and may be zero.

The \ac{PID}, i.e. which momentum corresponds to which particle,
can be looked up in the online documentation of \mcmule{} as well as
the file {\tt mat\_el.f95}, as it may change as new processes are
added or modified. In our case we have {\tt q1} for the incoming
$\tau$, {\tt q2} for the outgoing $e$, and {\tt q5} for the outgoing
$\gamma$.  At \ac{NLO}, we will also need {\tt q6} for the second
$\gamma$. The momenta of the neutrinos do not enter, as we average
over them.

Schematically, the function {\tt quant} is shown in
Listing~\ref{lst:quantlo}.  Here we have used {\tt sq} provided by
{\tt functions} to compute the square of a four-vector. We have also
specified the polarisation vector {\tt pol1} s.t. the initial tau is
considered unpolarised.  The variable {\tt pass\_cut} controls the
cuts. Initially it is set to true, to indicate that the event is kept.
Applying a cut amounts to setting {\tt pass\_cut} to false. The version
of {\tt quant} in Listing~\ref{lst:quantlo} will work for a \ac{LO}
calculation, but will need to be adapted for the presence of a second
photon in an \ac{NLO} computation. Being content with \ac{LO} for the
moment, all that remains to be done is prepare the input read by {\tt
mcmule} from {\tt stdin}, as specified in Table~\ref{tab:mcmuleinput}.

\begin{figure}
\centering
\input{figures/lst/quantlo.tex}
\renewcommand{\figurename}{Listing}
\caption{An example for the function {\tt quant} to calculate the
radiative $\tau$ decay. Note that this is only valid at \ac{LO} and
should not be used in any actual calculation.}
\label{lst:quantlo}
\vspace{0.5cm}
\begin{tabular}{l|l|l}
\bf Variable name& \bf Data type  & \bf Comment
 \\\hline

\tt nenter\_ad   & \tt integer    & calls / iteration during pre-conditioning  \\
\tt itmx\_ad     & \tt integer    & iterations during pre-conditioning         \\
\tt nenter       & \tt integer    & calls / iteration during main run          \\
\tt itmx         & \tt integer    & iterations during main run                 \\
\tt ran\_seed    & \tt integer    & random seed $z_1$                          \\
\tt xinormcut    & \tt real(prec) & the $0<\xc\le1$ parameter                  \\
\tt delcut       & \tt real(prec) & the $\delta_{\text{cut}}$ parameter   
                                    (or at NNLO the second $\xc$)              \\
\tt which\_piece & \tt char(10)   & the part of the calculation to perform     \\
\tt flavour      & \tt char(8)    & the particles involved                     \\
(opt)            & unknown        & the user can request further input during 
                                    {\tt userinit}
\end{tabular}

\renewcommand{\figurename}{Table}
\caption{The options read from {\tt stdin} by \mcmule{}. The calls are
multiplied by 1000.}
\label{tab:mcmuleinput}
\input{figures/lst/mcmuleread.tex}
\renewcommand{\figurename}{Listing}
\caption{Methods to enter configuration into \mcmule{}. All four
invocations will result in the same run assuming the text file {\tt
r1.in} contains the correct data.}
\label{lst:mcmuleinput}
\end{figure}

To be concrete let us assume we want to use 10 iterations with
$1000\times 10^3$ points each for pre-conditioning and 50 iterations
with $1000\times 10^3$ points each for the actual numerical evaluation
(cf. Section~\ref{sec:stat} for some heuristics to determine the
statistics needed). We pick a \term{random seed} between 0 and $2^{31}-1$
(cf. Section~\ref{sec:rng}), say $70\,998$, and for the input
variable {\tt which\_piece} we enter {\tt m2enng0}. This stands for
the generic process $\mu\to\nu\bar\nu e\gamma$ and 0 for tree level.
The {\tt flavour} variable is now set to {\tt tau-e} to change from
the generic process $\mu\to\nu\bar\nu e\gamma$ to the process we are
actually interested in, $\tau\to\nu\bar\nu e\gamma$.  This system is
used for other processes as well. The input variable {\tt
which\_piece} determines the generic process and the part of it that
is to be computed (i.e. tree level, real, double virtual etc.).  In a
second step, the input {\tt flavour} associates actual numbers to the
parameters entering the matrix elements and phase-space generation.

Obviously, in practice the input will typically not be given by typing
in by hand. In Listing~\ref{lst:mcmuleinput}, we have listed four
equivalent ways to input this data into {\tt mcmule}. The two
variables {\tt xinormcut1} and {\tt xinormcut2} have no effect at all
for a tree-level calculation and will be discussed below in the
context of the \ac{NLO} run. We also ignore the optional input for the
moment. 

Now the mule is ready to trot. The first step it does in {\tt mcmule}
is to associate the numerical values of the masses, as specified
through {\tt flavour}. In particular, we set the generic masses {\tt
Mm} and {\tt Me} to {\tt Mtau} and {\tt Mel}. This is done in {\tt
initflavour}, defined in {\tt global\_def}. For other processes this
might also involve setting e.g. centre-of-mass energies {\tt scms} to
default values.

Next, the function to be integrated by {\tt vegas} is determined. This
is a function stored in {\tt integrands}. There are basically three
types of integrands: a standard, non-subtracted integrand {\tt
sigma\_0}, a single-subtracted integrand needed beyond \ac{LO} {\tt
sigma\_1}, and a double-subtracted integrand needed beyond \ac{NLO}
{\tt sigma\_2}. Which integrand is needed and what matrix elements and
phase-space it depends on is determined by calling the function {\tt
initpiece} which uses the variable {\tt which\_piece} to point
function pointers at the necessary procedures.  For our \ac{LO} case,
{\tt initpiece} sets the integrand to {\tt sigma\_0} and fixes the
dimension of the integration to ${\tt ndim}=8$. The matrix element
pointer is assigned to the matrix element that needs to be called,
{\tt Pm2enngAV(q1,n1,q2,q3,q4,q5)}. The name of the function suggests
we compute $\mu(q_1,n_1)\to [\nu(q3) \bar\nu(q4)]e(q_2)  \gamma(q_5)$
with the polarisation vector {\tt n1} of the initial lepton, and the
neutrinos are averaged over. Note that the momenta of the neutrinos
are given as arguments, even if they are redundant. This simplifies
the code a lot because it means that all matrix elements have the same
calling convention.

The interplay between the function {\tt sigma\_0(x,wgt,ndim)} and {\tt
vegas} is as usual, through an array of random numbers {\tt x} of
length {\tt ndim}. In addition there is the {\tt vegas} weight of the
event, {\tt wgt} due to the Jacobian introduced by the importance
sampling.  The function {\tt sigma\_0} simply evaluates the complete
weight {\tt wg} of a particular event by combining {\tt wgt} with the
matrix element supplemented by symmetry, flux, and phase-space
factors. In a first step a phase-space routine of {\tt phase\_space}
is called. For our \ac{LO} calculation, {\tt initpiece} pointed a
pointer to the phase-space routine {\tt psd5\_25(x, p1,Mm, p2,Me,
p3,0., p4,0., p5,0., weight)}. The {\tt d} in the name of the
phase-space routine indicates that we are considering a decay process
(one initial state particle), the {\tt 5} indicates the total number
of momenta generated and the meaning of {\tt fks} will be explained
below. The other labels indicate the particular tuning and partition
which are irrelevant in this case (cf. Section~\ref{sec:ps}).  With
these momenta the observables to be computed are evaluated with a call
to {\tt quant}.  If one of them passes the cuts, the variable {\tt
cuts} is set to true. This triggers the computation of the matrix
element and the assembly of the full weight. In a last step, the
routine {\tt bin\_it}, stored in {\tt vegas}, is called to put the
weight into the correct bins of the various distributions. If the
variable under- or overshoots the bounds specified by {\tt min\_val}
and {\tt max\_val}, the event is placed into dedicated, infinitely big
under- and overflow bins.  These steps are done for all events and
those after pre-conditioning are used to obtain the final
distributions.

For a corresponding computation at \ac{NLO}, a few things need to be
modified. First, the observables have to be specified more carefully.
In particular, we need to decide how we treat the additional photon
due to real radiation. In our example we will consider the exclusive
radiative decay, i.e. we request precisely one photon with energy
$E_\gamma>10\,\mev$. The function {\tt quant} will have to take this
into account with the additional argument {\tt q6}, the momentum of
the second photon. 

An example of how this could be done is shown in
Listing~\ref{lst:quantnlo}.  Here we have just defined the harder and
softer photon {\tt gah} and {\tt gas}, respectively, and require that
the former (latter) has energy larger (smaller) than $10\,{\rm MeV}$.
This version of {\tt quant} is also suitable for the \ac{LO}
calculation, and to ensure infrared-safety, it is generally advisable
to use a single {\tt quant} function for all parts of a computation.
This is also mandatory if \ac{LO} and \ac{NLO} runs are done in one
go, as discussed below.

With this version of {\tt quant} we evaluate the virtual and real
corrections, as well as the infrared counterterm (i.e. the integrated
eikonal times the tree-level matrix element.) The latter is often
combined with the virtual corrections.  The corresponding {\tt
which\_piece} are {\tt m2enngV}, {\tt m2enngR}, and {\tt m2enngC},
respectively. If the counterterm is combined with the virtual part, we
would use {\tt m2enngF} which is not implemented.

\begin{figure}
\centering
\input{figures/lst/quantnlo.tex}
\renewcommand{\figurename}{Listing}
\caption{An example for the function {\tt quant} to calculate the
radiative $\tau$ decay at \ac{NLO}. This implementation is \ac{IR}
safe and exclusive w.r.t. extra photons.}
\label{lst:quantnlo}
\end{figure}

\subsection{Running and analysing}
\label{sec:analyse}

When we run \mcmule{}, we will want to choose various random seeds and
different values for the unphysical parameter $\xc$.  Checking the
independence of physical results on the latter serves as a consistency
check. To do this, it helps to disentangle {\tt{m2enngF}} into
{\tt{m2enngV}} and {\tt{m2enngC}}. Only the latter depends on $\xi_c$
and this part is typically much faster in the numerical evaluation.
However, this can quickly lead to a rather large number of runs that
need to be taken care of. We often also disentangle the \aterm{vacuum
polarisation}{VP} contributions. In this case they would be called
{\tt m2enngA} though this particular piece does not exist.

A particularly convenient way to run \mcmule{} is using \term{menu
files}\footnote{The name menu was originally used by the cryptanalysts
at Bletchley Park to describe a particular set of configurations for
the `computer' to try}. A menu file contains a list of jobs to be
computed s.t. the user will only have to vary the random seed and
$\xc$ by hand as the statistical requirements are defined globally in
a \term{config file}. This is completed by a \term{submission script},
usually called {\tt submit.sh}. The submit script is what will need to
be launched. It will take care of the starting of different jobs. It
can be run on a normal computer or on a Slurm
cluster~\cite{Yoo:2003slurm}.

To prepare the run in this way we can use {\tt pymule} as shown in
Listing~\ref{lst:menubabar}. When using the tool, we are asked various
questions, most of which have a default answer in square brackets. In
the end {\tt pymule} will create a directory that the user decided to
call {\tt babar-tau-e}, where
all results will be stored. The menu and config files generated by
{\tt pymule} are shown in Figure~\ref{lst:menu}

\begin{figure}
\centering
\input{figures/lst/menubabar.tex}
\renewcommand{\figurename}{Listing}
\caption{The steps necessary to use {\tt pymule} to prepare running
\mcmule{}. Input by the user is shown in bold. A red arrow indicates a
graphical line wrap. Note that numbers listed as {\tt seeds} are
random and hence not reproducible.}
\label{lst:menubabar}
\end{figure}
\begin{figure}
\centering
\input{figures/lst/menu.tex}
\renewcommand{\figurename}{Listing}
\caption{The files required for the present calculations as generated
in Listing~\ref{lst:menubabar}. The file has been massively shortened
for presentation. The online manual of \mcmule{} has the full file.}
\label{lst:menu}
\end{figure}

To start {\tt mcmule}, we now just need to execute the created {\tt
babar-tau-e/submit.sh}. Note that per default this will spawn at most
as many jobs as the computer {\tt pymule} ran on had CPU cores. If the
user wishes a different number of parallel jobs, change the fifth line
of {\tt babar-tau-e/submit.sh} to
\begin{lstlisting}[language=bash]
#SBATCH --ntasks=<number of cores>
\end{lstlisting}

After running the code, we need to combine the various {\tt
which\_pieces} into physical results that we will want to use to
create plots. For this purpose, we provide the python tool {\tt
pymule}, though of course other tools can be used as well. Here, we
will only cover the aspects of {\tt pymule} required for the present
analysis as shown in Listing~\ref{lst:pymule}; a full documentation
can be found in the {\tt docstrings} used in {\tt pymule} as well as
the online manual~\cite{mcmuleman}. First, we import {\tt pymule}.
Next, we need to point {\tt pymule} to the output directory of {\tt
mcmule} with the {\tt setup} command. In our example this is {\tt
babar-tau-e/out}.
\begin{figure}
\centering
\begin{lstlisting}[language=python,firstline=10,lastline=55]
from pymule import *

# To normalise branching ratios, we need the tau lifetime
lifetime = 1/(1000*(6.582119e-25)/(2.903e-13))

# The folder where McMule has stored the statefiles
setup(folder='babar-tau-e/out/')

# Import LO data and re-scale to branching ratio
LO = scaleset(mergefks(sigma('m2enng0')), GF**2*lifetime*alpha)

# Import NLO corrections from the three pieces
NLO = scaleset(mergefks(
    sigma('m2enngR'),      # real corrections
    sigma('m2enngCT'),     # counter term
    anyxi=sigma('m2enngV') # virtual corrections
), GF**2*lifetime*alpha**2)

# The branching ratio at NLO = LO + correction
fullNLO = plusnumbers(LO['value'], NLO['value'])

# Print results
print "BR_0 = ", printnumber(LO['value'])
print "dBR  = ", printnumber(NLO['value'])
print "BR_1 = ", printnumber(fullNLO)

# Produce energy plot
fig1, (ax1, ax2) = kplot(
    {'lo': LO['Ee'], 'nlo': NLO['Ee']},
    labelx=r"$E_e\,/\,{\rm MeV}$",
    labelsigma=r"$\D\mathcal{B}/\D E_e$"
)
ax2.set_ylim(-0.2,0.01)

# Produce visible mass plot
fig2, (ax1, ax2) = kplot(
    {'lo': LO['minv'], 'nlo': NLO['minv']},
    labelx=r"$m_{e\gamma}\,/\,{\rm MeV}$",
    labelsigma=r"$\D\mathcal{B}/\D m_{e\gamma}$"
)

ax1.set_yscale('log')
ax1.set_xlim(1000,0)
ax1.set_ylim(5e-9,1e-3)
ax2.set_ylim(-0.2,0.)
\end{lstlisting}
\renewcommand{\figurename}{Listing}
\caption{An example code to analyse the results for $\tau\to\nu\bar\nu
e\gamma$ in {\tt pymule}. Note that, in the Fortran code
$G_F=\alpha=1$. In {\tt pymule} they are at their physical
values~\cite{PDG}.}
\label{lst:pymule}
\end{figure}

As a next step, we import the \ac{LO} and \ac{NLO} {\tt
which\_pieces} and combine them using two central {\tt pymule}
commands: {\tt sigma} and {\tt mergefks}. {\tt sigma} takes the {\tt
which\_piece} as an argument and imports matching results, already
merging different random seeds. {\tt mergefks} takes the results of
(multiple) {\tt sigma} invocations, adds results with matching $\xc$
values and combines the result. In the present case, $\sigma_n^{(1)}$
is split into multiple contributions, namely {\tt m2enngV} and {\tt
m2enngC}.  This is indicated by the {\tt anyxi} argument.

Users should keep in mind that \mcmule{} ships with a version of {\tt
global\_def} where the couplings $G_F={\tt GF}$ and $\alpha={\tt
alpha}$ are set to $G_F=\alpha=1$. Hence, we use {\tt pymule}'s
function {\tt scaleset} to multiply the result with the correct values
of $G_F$ (in ${\rm MeV}^{-1}$) and $\alpha$ (in the \ac{OS} scheme).

Next, we can use some of {\tt pymule}'s tools (cf.
Listing~\ref{lst:pymule}) to calculate the full \ac{NLO} \ac{BR}s from
the corrections and the \ac{LO} results
\begin{align}
\mathcal{B}|_\text{LO} &= 1.8339(1) \times 10^{-2}\,, &
\mathcal{B}|_\text{NLO} &= 1.6451(1) \times 10^{-2}\,,
\end{align}
which agree with~\cite{Fael:2015gua,Pruna:2017upz}, but
$\mathcal{B}|_\text{NLO}$ is in tension with the value
$\mathcal{B}|_\text{exp} = 1.847(54)\times 10^{-2}$ reported by
{\sc BaBar}~\cite{Lees:2015gea, Oberhof:2015snl}. As discussed
in~\cite{Pruna:2017upz, Ulrich:2017adq} it is very likely that this
tension would be removed if a full \ac{NLO} result was used to take
into account the effects of the stringent experimental cuts to extract
the signal. We will come back to this issue in
Section~\ref{sec:babarsol}.

\begin{figure}
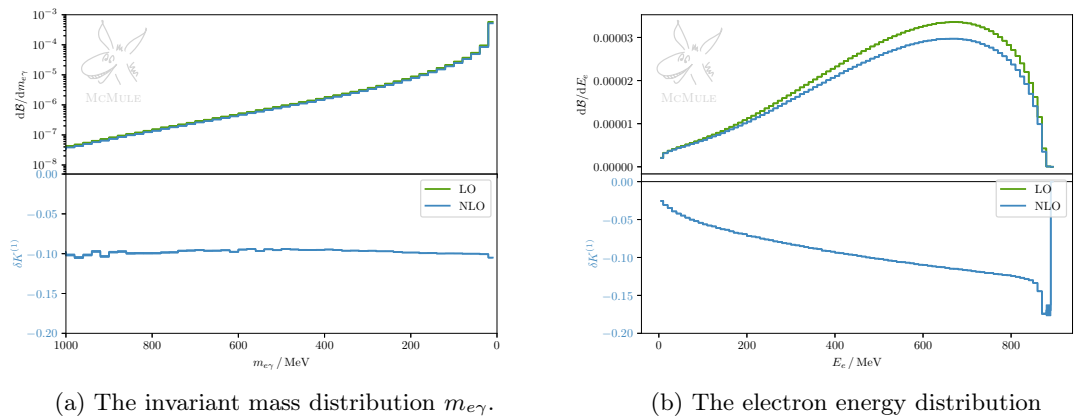

\centering
\subfloat[The invariant mass distribution $m_{e\gamma}$.\label{fig:babares:minv}
]{
    \scalebox{0.45}{\input{figures/mule/tauminv.pgf}}
}
\subfloat[The electron energy distribution
    \label{fig:babares:el}
]{
    \scalebox{0.45}{\input{figures/mule/tauenergy.pgf}}
}
\caption{Results of the toy run to compute $m_{e\gamma}$ (left) and
$E_e$ (right) for $\tau\to\nu\bar\nu e\gamma$. Upper panels show the
\ac{LO} (green) and \ac{NLO} (blue) results, the lower panels show
the \ac{NLO} $K$ factor.}
\label{fig:babares}
\end{figure}

As a last step, we can use the {\tt matplotlib}-backed {\tt kplot}
command to present the results for the distributions (logarithmic for
$m_{e\gamma}$ and linear for $E_e$). The results are shown in
Figure~\ref{fig:babares}. The upper panel of
Figure~\ref{fig:babares:minv} shows the results for the invariant mass
$m_{e\gamma}$ at \ac{LO} (green) and \ac{NLO} (blue) in the range
$0\le m_{e\gamma} \le 1\,\gev$. Note that this, for the purposes of
the demonstration, does not correspond to the boundaries given in the
run.

The distribution falls sharply for large $m_{e\gamma}$.  Consequently,
there are only few events generated in the tail and the statistical
error becomes large. This can be seen clearly in the lower panel,
where the \ac{NLO} $K$ factor is shown.  It is defined as
\begin{align}
    K^{(1)} = 1 + \delta K^{(1)} 
    =1+\frac{\D\sigma^{(1)}/\D x}{\D\sigma^{(0)}/\D x}\,,
\end{align}
and the band represents the statistical error of the Monte Carlo
integration. To obtain a reliable prediction for larger values of
$m_{e\gamma}$, i.e. the tail of the distribution, we would have to
perform tailored runs. To this end, we should introduce a cut
$m_\text{cut}\ll m_\tau$ on $m_{e\gamma}$ to eliminate events with
larger invariant mass. Due to the adaption in the numerical
integration, we then obtain reliable and precise results for values of
$m_{e\gamma} \lesssim m_\text{cut}$.

Figure~\ref{fig:babares:el} shows the electron energy distribution,
again at \ac{LO} (green) and \ac{NLO} (blue). As for $m_{e\gamma}$
the corrections are negative and amount to roughly $10\%$. Since this
plot is linear, they can be clearly seen by comparing \ac{LO} and
\ac{NLO}. In the lower panel once more the $K$ factor is depicted.
Unsurprisingly, at the very end of the distribution, $E_e\sim
900\,{\rm MeV}$, the statistics is out of control.

\section{General aspects of using \mcmule}
\label{sec:general}
In this section, we will collect a few general points of interest
regarding \mcmule{}. In particular, we will discuss heuristics on how
much statistics is necessary for different contributions in
Section~\ref{sec:stat}. This is followed by a more in-depth discussion
of the analysis strategy in Section~\ref{sec:analysis}.

\subsection{Statistics}
\label{sec:stat}
\mcmule{} is a Monte Carlo program. This means it samples the
integrand at $N$ (pseudo-)random points to get an estimate for the
integral. However, because it uses the adaptive Monte Carlo
integration routine {\tt vegas}~\cite{Lepage:1980jk}, we split
$N=i\times n$ into $i$ iterations ({\tt itmx}), each with $n$ points
({\tt nenter}). After each iteration, {\tt vegas} changes the way it
will sample the next iteration based on the results of the previous
one. Hence, the performance of the integration is a subtle interplay
between $i$ and $n$ -- it is not sufficient any more to consider their
product $N$.

Further, we always perform the integration in two steps: a
pre-conditioning with $i_\text{ad}\times n_\text{ad}$ ({\tt
nenter\_ad} and {\tt itmx\_ad}, respectively), that is used to
optimise the integration strategy and after which the result is
discarded, and a main integration that benefits from the integrator's
understanding of the integrand.

Of course there are no one-size-fits-all rules of how to choose the
$i$ and $n$ for pre-conditioning and main run. However, the following
heuristics have proven helpful:
\begin{itemize}
    \item
    $n$ is always much larger than $i$. For very simple integrands,
    $n=\mathcal{O}(10\cdot 10^3)$ and $i=\mathcal{O}(10)$.

    \item
    Increasing $n$ reduces errors that can be thought of as systematic
    because it allows the integrator to `discover' new features of the
    integrand. Increasing $i$ on the other hand will rarely have that
    effect and only improves the statistical error. This is especially
    true for distributions.

    \item
    There is no real limit on $n$, except that it has to fit into the
    datatype used -- integrations with $n=\mathcal{O}(2^{31}-1)$ are
    not too uncommon -- while $i$ is rarely (much) larger than 100.

    \item
    For very stringent cuts it can happen that that typical values of
    $n_\text{ad}$ are too small for any point to pass the cuts.
    In this case {\tt vegas} will return {\tt NaN}, indicating that no
    events were found. Barring mistakes in the definition of the cuts,
    a pre-pre-conditioning with extremely large $n$ but $i=1\!-\!2$
    can be helpful.

    \item
    $n$ also needs to be large enough for {\tt vegas} to reliably find
    all features of the integrand. It is rarely obvious that it did,
    though sometimes it becomes clear when increasing $n$ or looking at
    intermediary results as a function of the already-completed
    iterations.

    \item
    The main run should always have larger $i$ and $n$ than the
    pre-conditioning. Judging how much more is a delicate game though
    $i/i_\text{ad} = \mathcal{O}(5)$ and $n/n_\text{ad} =
    \mathcal{O}(10\!-\!50)$ have been proven helpful.

    \item
    If, once the integration is completed, the result is
    unsatisfactory, take into account the following strategies
    \begin{itemize}
        \item
        A large $\chi^2/\rm{d.o.f.}$ indicates a too small $n$. Try to
        increase $n_\text{ad}$ and, to a perhaps lesser extent, $n$.

        \item
        Increase $i$. Often it is a good idea to consciously set $i$
        to a value so large that the integrator will never reach it
        and to keep looking at `intermediary' results.

        \item
        If the error is small enough for the application but the
        result seems incorrect (for example because the $\xc$
        dependence does not vanish), massively increase $n$.

    \end{itemize}
    \item
    Real corrections need much more statistics in both $i$ and $n$
    ($\mathcal{O}(10)$ times more for $n$, $\mathcal{O}(2)$ for $i$)
    than the corresponding \ac{LO} calculations because of the
    higher-dimensional phase-space.

    \item
    Virtual corrections have the same number of dimensions as the
    \ac{LO} calculation and can go by with only a modest increase to
    account for the added functional complexity.

    \item
    {\tt vegas} tends to underestimate the numerical error.
\end{itemize}
These guidelines are often helpful but should not be considered
infallible as they are just that -- guidelines.

\mcmule{} is not parallelised; however, because Monte Carlo
integrations require a random seed anyway, it is possible to calculate
multiple estimates of the same integral using different random seeds
$z_1$ and combining the results obtained this way. This also allows to
for a better, more reliable understanding of the error estimate.

\subsection{Analysis}
\label{sec:analysis}\setcounter{enumi}{4}
Once the Monte Carlo has run, an offline analysis of the results is
required. This entails loading, averaging, and combining the data. This
is automatised in {\tt pymule} but the basic steps are
\begin{enumerate}
\setcounter{enumi}{-1}
    \item
    Load the data into a suitable analysis framework such as {\tt
    python}.

    \item
    Combine the different random seeds into one result per
    contribution and $\xc$. The $\chi^2/{\rm d.o.f.}$ of this merging
    must be small. Otherwise, try to increase the statistics or choose
    of different phase-space parametrisation.

    \item
    Add all contributions that combine into one of the physical
    contributions~\eqref{eq:nellocomb:b}. This includes any
    partitioning done in Section~\ref{sec:ps}.

    \item
    (optional) At N$^\ell$LO, perform a fit\footnote{Note that it is
    important to perform the fit after combining the phase-space
    partitionings (cf. Section~\ref{sec:ps}) but before
    adding~\eqref{eq:nellocomb:a} as this model is only valid for the
    terms of~\eqref{eq:nellocomb:b}}
    \begin{align}
        \sigma_{n+j}^{(\ell)} = c_0^{(j)} + c_1^{(j)} \log\xc +
        c_2^{(j)} \log^2\xc + \cdots + c_\ell^{(j)} \log^\ell
        = \sum_{i=0}^\ell c_i^{(j)}\log^i\xc\,.
    \label{eq:xifit}
    \end{align}
    This has the advantage that it very clearly quantifies any
    residual $\xc$ dependence. We will come back to this issue in
    Section~\ref{sec:xicut}.

    \item
    Combine all physical contributions of~\eqref{eq:nellocomb:a} into
    $\sigma^{(\ell)}(\xc)$ which has to be $\xc$ independent.

    \item
    Perform detailed checks on $\xc$ independence. This is especially
    important on the first time a particular configuration is run.
    Beyond \ac{NLO}, it is also extremely helpful to check whether
    the sum of the fits~\eqref{eq:xifit} is compatible with a
    constant. In case it is not, try to run the Monte Carlo again with
    an increased $n$. {\tt pymule}'s {\tt mergefkswithplot} can be
    helpful here.

    \item
    Merge the different estimates of~\eqref{eq:nellocomb:a} from the
    different $\xc$ into one final number $\sigma^{(\ell)}$. The
    $\chi^2/{\rm d.o.f.}$ of this merging must be small.

    \item
    Repeat the above for any distributions produced, though often
    bin-wise fitting as in Point 3 is rarely necessary or helpful.

    If a total cross section is $\xc$ independent but the
    distributions (or a cross section obtained after applying cuts)
    are not, this is a hint that the distribution (or the applied
    cuts) is not IR safe.

\end{enumerate}
These steps have been almost completely automatised in {\tt pymule}
and Mathematica. Though all steps of this pipeline could be easily
implemented in any other language by following the specification of
the file format below (Section~\ref{sec:vegasff}).

\section{Technical aspects of \mcmule{}}
\label{sec:techno} 

In this section, we will review the very technical details of the
implementation. This is meant for those readers, who wish to truly
understand the nuts and bolts holding the code together. We begin by
discussing the phase-space generation and potential pitfalls in
Section~\ref{sec:ps}. Next, in Section~\ref{sec:fksfor}, we discuss
how the \ac{FKS} scheme of Chapter~\ref{ch:fks} is implemented in
Fortran code. This is followed by a brief review of the random number
generator used in \mcmule{} in Section~\ref{sec:rng}. Finally, we give
an account of how the statefiles work and how they are used to store
distributions in Section~\ref{sec:vegasff}.

\subsection{Phase-space generation}\label{sec:ps}
We use the {\tt vegas} algorithm for numerical
integration~\cite{Lepage:1980jk}.  As {\tt vegas} only works on the
hypercube, we need a routine that maps $[0,1]^{3n-4}$ to the momenta
of an $n$-particle final state, including the corresponding Jacobian.
The simplest way to do this uses iterative two-particle phase-spaces
and boosting the generated momenta all back into the frame under
consideration. An example of how this is done is shown in
Listing~\ref{lst:psn}.

As soon as we start using \ac{FKS}, we cannot use this simplistic
approach any longer. The $c$-distributions of \ac{FKS} require the
photon energies $\xi_i$ to be variables of the integration. We can fix
this by first generating the photon explicitly as~\eqref{eq:kdef}
and~\eqref{eq:para} and then generate the remaining particles
iteratively again. This can always be done and is guaranteed to work.

For processes with one or more \ac{PCS} this approach is suboptimal.
The numerical integration can be improved by orders of magnitude by
aligning the pseudo-singular contribution to one of the variables of
the integration, as this allows {\tt vegas} to optimise the
integration procedure accordingly. As an example, consider once again
$\mu\to\nu\bar\nu e\gamma$. The \ac{PCS} comes from
\begin{align}
\M{n+1}\ell \propto \frac{1}{(q\cdot k)^2} =\frac1{\xi^2}\frac1{(1-y\beta)^2}
\,,
\end{align}
where $y$ is the angle between photon ($k$) and electron ($q$). For
large velocities $\beta$ (or equivalently small masses), this becomes
almost singular as $y\to1$. If now $y$ is a variable of the
integration this can be mediated. An example implementation is shown
in Listing~\ref{lst:psmu}. 

The approach outlined above is very easy to do in the case of the muon
decay as the neutrinos can absorb any timelike four-momentum. This is
because the $\delta$ function of the phase-space was solved through
the neutrino's {\tt pair\_dec}. However, for scattering processes
where all final state leptons could be measured, this fails. Writing a
routine for $\mu$-$e$-scattering
\begin{align}
e(p_1)+\mu(p_2) \to e(p_3)+\mu(p_4) + \gamma(p_5)\,,
\end{align}
that optimises on the incoming electron is rather trivial because its
direction stays fixed s.t. the photon just needs to be generated
according to~\eqref{eq:kdef}. The outgoing electron $p_3$ is more
complicated.  Writing the $p_4$-phase-space four- instead of
three-dimensional
\begin{align}
\D\Phi_5 &= \delta^{(4)}(p_1+p_2-p_3-p_4-p_5)
 \delta(p_4^2-M^2) \Theta(E_4)
 \frac{\D^4\vec p_4}{(2\pi)^4}
 \frac{\D^3\vec p_3}{(2\pi)^32E_3}
 \frac{\D^3\vec p_5}{(2\pi)^32E_5}\,,
\end{align}
we can solve the four-dimensional $\delta$ function for $p_4$ and
proceed for the generation $p_3$ and $p_5$ almost as for the muon
decay above. Doing this we obtain for the final $\delta$ function
\begin{align}
\delta(p_4^2-M^2) =
\delta\bigg(m^2-M^2+s(1-\xi)+E_3\sqrt{s}\Big[\xi-2-y
\xi\beta_3(E_3)\Big]\bigg)\,.
\label{eq:psdel}
\end{align}
When solving this for $E_3$, we need to take care to avoid extraneous
solutions of this radical equation~\cite{Gkioulekas:2018}. We have now
obtained our phase-space parametrisation, albeit with one caveat: for
anti-collinear photons, i.e. $-1<y<0$ with energies
\begin{align}
\xi_1 = 1-\frac{m}{\sqrt{s}}+\frac{M^2}{\sqrt{s}(m-\sqrt{s}} < \xi <
\xi_\text{max}=1-\frac{(m+M)^2}{s}
\end{align}
there are still two solutions. One of these corresponds to very
low-energy electron that are almost produced at rest. This is rather
fortunate as most experiments will have an electron detection
threshold higher that this. Otherwise, phase-spaces optimised this way
also define a {\tt which\_piece} for this \term{corner region}.

There is one last subtlety when it comes to these type of phase-space
optimisations. Optimising the phase-space for emission from one leg
often has adverse effects on terms with dominant emission from another
leg. In other words, the numerical integration works best if there is
only one \ac{PCS} on which the phase-space is tuned. As most
processes have more than one \ac{PCS} we need to resort to something
that was already discussed in the original \ac{FKS}
paper~\cite{Frixione1995Three-jet}.  Scattering processes that involve
multiple massless particles have overlapping singular regions. The
\ac{FKS} scheme now mandates that the phase-space is partitioned in
such a way as to isolate at most one singularity per region with each
region having its own phase-space parametrisation. Similarly we have
to split the phase-space to contain at most one \ac{PCS} as well as
the soft singularity. In \mcmule{} $\mu$-$e$ scattering for instance
is split as follows\footnote{When implementing this, care must be
taken to ensure that the split is also well defined if the photon is
soft, i.e. if $\xi=0$.}
\begin{align}
1 = \theta\big( s_{15} > s_{35} \big) 
  + \theta\big( s_{15} < s_{35} \big)\,,
\end{align}
with $s_{ij} = 2p_i\cdot p_j$ as usual. The integrand of the first
$\theta$ function has a final-state \ac{PCS} and hence we use the
parametrisation obtained by solving~\eqref{eq:psdel}. The second
$\theta$ function, on the other hand, has an initial-state \ac{PCS}
which can be treated by just directly parametrising the photon in the
centre-of-mass frame as per~\eqref{eq:kdef}. This automatically makes
$s_{15}\propto(1-\beta_\text{in}y_1)$ a variable of the integration.

For the double-real corrections of $\mu$-$e$ scattering, we proceed
along the same lines except now the argument of the $\delta$ function
is more complicated.

\begin{figure}
\input{figures/lst/psn.tex}
\renewcommand{\figurename}{Listing}
\caption{Example implementation of iterative phase-space. Not shown
are the checks to make sure that all particles have at least enough
energy for their mass, i.e. that $E_i\ge m_i$.}
\label{lst:psn}
\end{figure}

\begin{figure}
\input{figures/lst/psmu.tex}
\renewcommand{\figurename}{Listing}
\caption{Example implementation of a so-called \ac{FKS} phase-space
where the fifth particle is an \ac{FKS} photon that may becomes soft.
Not shown are checks whether $E_i\ge m_i$.}
\label{lst:psmu}
\end{figure}

\begin{figure}
\input{figures/lst/fks.tex}
\renewcommand{\figurename}{Listing}
\caption{An example implementation of the \ac{FKS} scheme in Fortran.
Not shown are various checks performed, the binning as well as
initialisation blocks.}
\label{lst:fks}
\end{figure}

\subsection{Implementation of FKS schemes}
\label{sec:fksfor}
Now that we have a phase-space routine that has $\xi_i$ as variables
of the integration, we can start implementing the relevant
$c$-distributions~\eqref{eq:xidist}
\begin{align}\begin{split}
\bbit{1}{h}(\xc)
 &=
 \pref1\ \cdis{\xi_1}
      \D\xi_1
      \Big(\xi_1^2\M{n+1}0\Big)
\\&
 = \D\xi_1\frac1{\xi_1}\Bigg(
   \pref1\ \Big(\xi_1^2\M{n+1}0\Big) 
  -\pref1\ \Big(\eik\M{n}0\Big) \ \theta(\xc-\xi_1)
\Bigg)\,.
\end{split}\end{align}
We refer to the first term as the \term{event} and the second as the
\term{counter-event}.

Note that, due to the presence of $\delta(\xi_1)$ in the counter-event
(that is implemented through the eikonal factor $\eik$,
cf.~\eqref{eq:eikonal}) the momenta generated by the phase-space
$\pref1$ are different. Thus, it is possible that the momenta of the
event pass the cuts or on-shell conditions, while those of the counter
event fail, or vice versa.  This subtlety is extremely important to
properly implement the \ac{FKS} scheme and many problems fundamentally
trace back to this.

Finally, we should note that, in order to increase numerical
stability, we introduce cuts on $\xi$ and sometimes also on a
parameter that encodes the \ac{PCS} such as $y={\tt y2}$
in~\eqref{eq:kdef} and Listing~\ref{lst:psmu}. Events that have values
of $\xi$ smaller than this \term{soft cut} are discarded immediately
and no subtraction is considered. The dependence on this slicing
parameter is not expected to drop out completely and hence, the soft
cut has to be chosen small enough to not influence the result.

An example implementation can be found in Listing~\ref{lst:fks}.

\subsection{Random number generation}
\label{sec:rng}
A Monte Carlo integrator relies on a (pseudo) \emph{random number
generator} (\ac{RNG} or PRNG) to work. The pseudo-random numbers need
to be of high enough quality, i.e. have no discernible pattern and a
long period, to consider each point of the integration independent but
the \ac{RNG} needs to be simple enough to be called many billion
times without being a significant source of runtime. \ac{RNG}s used
in Monte Carlo applications are generally poor in quality and often
predictable s.t. they could not be used for cryptographic
applications.

A commonly used trade-off between unpredictability and simplicity,
both in speed and implementation, is the Park-Miller \ac{RNG}, also
known as {\tt minstd}~\cite{Park:1988RNG}. As a linear congruential
generator, its $(k+1)$th output $x_{k+1}$ can be found as
\begin{align}
z_{k+1} = a\cdot z_k \ \text{mod}\ m = a^{k+1} z_1 \ \text{mod}\ m
\qquad\text{and}\qquad
x_k = z_k / m \in (0,1)\,,
\end{align}
where $m$ is a large, preferably prime, number and $2<a<m-1$ an
integer.  The initial value $z_1$ is called the random seed and
is chosen integer between 1 and $m-1$. It can easily be seen that any
such \ac{RNG} has a fixed period\footnote{Note that, because of the
simple recursion the \ac{RNG} will not repeat any number until the
full period is complete} $p<m$ s.t. $z_{k+p} = z_k$ because any
$z_{k+1}$ only depends on $z_k$ and there are finitely many possible
$z_k$.  We call the \ac{RNG} attached to $(m,a)$ to be of \term{full
period} if $p=m-1$, i.e. all integers between 1 and $m-1$ appear in
the sequence $z_k$.

Assuming $z_1=1$ then the existence of $p$ s.t. $z_{p+1}=1$ is
guaranteed by Fermat's Theorem\footnote{If $p$ is prime, for any
integer $a$, $a^p-a$ is a multiple of $p$.}. This means that the
\ac{RNG} is of full period iff $a$ is a primitive root modulo $m$,
i.e. 
\begin{align}
\forall g\ \text{co-prime to $m$}\quad
\exists k\in\mathbb{Z}
\quad\text{s.t.}\quad
a^k\equiv g\ (\text{mod}\ m)\,.
\end{align}
Park and Miller suggest to use the Mersenne prime $m=2^{31}-1$, noting
that there are 534,600,000 primitive roots of which 7 is the smallest.
Because $7^b\ \text{mod}\ m$ is also a primitive root as long as $b$
is co-prime to $(m\!-\!1)$, \cite{Park:1988RNG} settled on $b=5$, i.e.
$a=16807$ as a good choice for the multiplier that, per construction,
has full period and passes certain tests of randomness.

The points generated by any such \ac{RNG} will fall into
$\sqrt[n]{n!\cdot m}$ hyperplanes if scattered in an $n$ dimensional
space~\cite{Marsaglia25}. However, for bad choices of the multiplier
$a$ the number of planes can be a lot smaller\footnote{An infamous
example is {\tt randu} that used $a=2^{16}+3$ and $m=2^{31}$ that in
three dimension produces only 15 planes instead of the maximum 2344.}.

Presently, the period length of $p=m-1=2^{31}-2$ is believed to be
sufficient though detailed studies quantifying this would be welcome.

\subsection{Differential distributions and intermediary state files}
\label{sec:vegasff}

Distributions are always calculated as histograms by binning each
event according to its value for the observable $S$. This is done by
having an $(n_b\times n_q)$-dimensional array\footnote{To be precise,
the actual dimensions are $(n_b+2)\times n_q$ to accommodate under-
and overflow bins} {\tt quant} where $n_q$ is the number of histograms
to be calculated ({\tt nr\_q}) and $n_b$ is the number of bins used
({\tt nr\_bins}). The weight of each event $\D\Phi\times\mathcal{M}
\times w$ is added to the correct entry in {\tt bit\_it} where $w={\tt
wgt}$ is the event weight assigned by {\tt vegas}.

After each iteration of {\tt vegas} we add {\tt quant} (${\tt
quant}^2$) to an accumulator of the same dimensions called {\tt
quantsum} ({\tt quantsumsq}). After $i$ iterations, we can calculate
the value and error as
\begin{align}
\frac{\D\sigma}{\D S} \approx \frac{\tt quantsum}{\Delta\times i}
\qquad\text{and}\qquad
\delta\bigg(\frac{\D\sigma}{\D S}\bigg)\approx \frac1\Delta \sqrt{
    \frac{{\tt quantsumsq}-{\tt quantsum}^2/i}{i(i-1)}
}\,,
\end{align}
where $\Delta$ is the bin-size.

Related to this discussion is the concept of intermediary state
files. Their purpose is to record the complete state of the integrator
after every iteration in order to recover should the program crash --
or more likely be interrupted by a batch system. \mcmule{} uses a
custom file format {\tt .vegas} for this purpose which uses
Fortran's record-based (instead of stream- or byte-based) format. This
means that each entry starts with 32bit unsigned integer, i.e. 4 byte,
indicating the record's size and ends with the same 32bit integer. As
this is automatically done for each record, it minimises the amount of
metadata that have to be written.

The current version ({\tt v3}) must begin with the magic header and
version self-identification shown in Figure~\ref{fig:vegf:head}. The
latter includes file version information and the first five characters
the source tree's \ac{SHA1} hash, obtained using {\tt make hash}.

The header is followed by records describing the state of the
integrator as shown in Figure~\ref{fig:vegf:body}. Additionally to
information required to continue integration such as the current value
and grid information, this file also has 300 bytes for a message. This
is usually set by the routine to store information on the fate of the
integration such as whether it was so-far uninterrupted or whether
there is reason to believe it to be inconsistent. 

The latter point is particularly important. While \mcmule{} cannot
read intermediary files from a different version of the file format,
it will continue any integration for which it can read the state file.
This also includes cases where the source tree has been changed. In
this case \mcmule{} prints a warning but continues the integration
deriving potentially inconsistent results.

\begin{figure}
\begin{center}
\def\at{\textbackslash t}
\def\an{\textbackslash n}
\def\af{}
\def\v#1{$v_#1$}
\def\s#1{$s_#1$}
\small\begin{tabular}{l||c|c|c|c|c|c|c|c|c|c|c|c|c|c|c|c}
\bf offset  &00 &01 &02 &03 &04 &05 &06 &07 &08 &09 &0A &0B &0C &0D &0E &0F\\\hline
\bf hex     &09 &00 &00 &00 &20 &4D &63 &4D &75 &6C &65 &20 &20 &09 &00 &00\\
\bf ASCII   &\at&   &   &   &' '&M  &c  &M  &u  &l  &e  &' '&' '&\at&   &   
\\\hline\hline
\bf offset  &10 &11 &12 &13 &14 &15 &16 &17 &18 &19 &1A &1B &1C &1D &1E &1F\\\hline
\bf hex     &00 &0A &00 &00 &00 &76 &xx &xx &20 &20 &20 &20 &20 &20 &20 &0A\\
\bf ASCII   &   &\an&   &   &   &v  &\v1&\v2&' '&' '&' '&' '&' '&' '&' '&\an
\\\hline\hline
\bf offset  &20 &21 &22 &23 &24 &25 &26 &27 &28 &29 &2A &2B &2C &2D &2E &2F\\\hline
\bf hex     &00 &00 &00 &05 &00 &00 &00 &xx &xx &xx &xx &xx &05 &00 &00 &00\\
\bf ASCII   &   &   &   &\af&   &   &   &\s1&\s2&\s3&\s4&\s5&\af&   &   &   
\end{tabular}

\end{center}
\caption{The magic header and version information used by {\tt v3}.
$v_1$ indicates the current version number and $v_2$ whether long
integers are used ({\tt L}) or not ({\tt N}). $s_1$-$s_5$ indicate the
first five characters of the \ac{SHA1} hash produced by the source
code at compile time ({\tt make hash}).}
\label{fig:vegf:head}

\begin{center}
\small
\begin{tabular}{r|l|l||l|p{5cm}}
\bf Off  & \bf Len        & \bf Type                    & \bf Var.      & \bf Comment\\\hline
\tt 0030 &\tt 000C        & \tt integer                 & \tt it        & the current iteration\\\hline
\tt 003C &\tt 000C        & \tt integer                 & \tt ndo       & subdiv. on an axis\\\hline
\tt 0048 &\tt 0010        & \tt real                    & \tt si        &$\sigma/(\delta\sigma)^2$\\\hline
\tt 0058 &\tt 0010        & \tt real                    & \tt swgt      &$1/(\delta\sigma)^2$\\\hline
\tt 0068 &\tt 0010        & \tt real                    & \tt schi      &$(1-{\tt it})\chi + \sigma^2/(\delta\sigma)^2$\\\hline
\tt 0078 &\tt 1A98        & \tt real(50,17)             & \tt xi        & the integration grid\\\hline
\tt 1B10 &\tt 000C        & \tt integer                 & \tt randy     & the current random number seed\\\hline
\tt 1B1C &\tt 0014        & \tt integer                 & $n_q$         & number of histograms\\
\tt      &                & \tt integer                 & $n_b$         & number of bins\\
\tt      &                & \tt integer                 & $n_s$         & len. histogram name\\\hline
\tt 1B30 &$10n_q+8$       & \tt real($n_q$)             & \tt minv      & lower bounds \\
\tt      &                & \tt real($n_q$)             & \tt maxv      & upper bounds \\\hline
\tt      &$n_sn_q+8$      & \tt character($n_s$,$n_q$)  & \tt names     & names of $S$\\\hline
\tt      &$10n_q(n_b+2)+8$& \tt real($n_q$,$n_b$+2)     & \tt quantsum  & accu. histograms\\
\tt      &                & \tt real($n_q$,$n_b$+2)     & \tt quantsumsq& accu. histograms squared\\\hline
\tt-0144 &\tt 0010        & \tt real                    & \tt time      & current runtime in seconds\\\hline
\tt-0134 &\tt 0134        & \tt character(300)          & \tt msg       & any message\\\hline\hline
\tt-0000 &\multicolumn{4}{c}{\tt EOF}
\end{tabular}

\end{center}
\caption{The body of a {\tt.vegas} file storing all important
information. Each horizontal line indicates as dressed record. In the
offset and length columns, all integers are in hexadecimal notation.
Negative numbers count from the end of file ({\tt EOF}).}
\label{fig:vegf:body}
\end{figure}

\section{Implementing new processes in \mcmule{}}
\label{sec:implement} 

In this section we will discuss how new processes can be added to
\mcmule{}. Not all of the points below might be applicable to any
particular process. Further, all points are merely guidelines that
could be deviated from if necessary as long as proper precautions are
taken.

\begin{figure}
\input{figures/lst/mollerlo.tex}
\renewcommand{\figurename}{Listing}
\caption{An example implementation of $\M n0$ for M{\o}ller
scattering. Note that the electron mass and the centre-of-mass energy
are calculated locally. A global factor of $8e^4=128\pi^2\alpha^2$ is
included at the end.}
\label{lst:mollerlo}
\end{figure}

As an example, we will discuss how M{\o}ller scattering $e^-e^-\to
e^-e^-$ could be implemented.
\begin{enumerate}
    \item
    A new process group may need to be created if the process does not
    fit any of the presently implemented groups. This requires a new
    folder with a makefile as well as modifications to the main
    makefile as discussed in the online manual.

    In our case, $ee\to ee$ does not fit any of the groups, so we
    create a new group that we shall call {\tt ee}. 

    \item
    Calculate the tree-level matrix elements needed at \ac{LO} and
    \ac{NLO}: $\M{n}0$ and $\M{n+1}0$. This is relatively
    straightforward and -- crucially -- unambiguous as both are finite
    in $d=4$. We will come back to an example calculation in
    Section~\ref{sec:matel}.
    
    \item
    A generic matrix element file is needed to store `simple' matrix
    elements as well as importing more complicated matrix elements.
    Usually, this file should not contain matrix elements that are
    longer than a few dozen or so lines. In most cases, this applies
    to $\M n0$. 

    After each matrix element, the \ac{PID} needs to be denoted in a
    comment. Further, all required masses as well as the
    centre-of-mass energy, called {\tt scms} to avoid collisions with
    the function ${\tt s(pi,pj)}=2{\tt pi}\cdot{\tt pj}$, need to be
    calculated in the matrix element to be as localised as possible.

    In the case of M{\o}ller scattering, a file {\tt
    ee/ee\_mat\_el.f95} will contain $\M n0$. For example, $\M n0$ is
    implemented there as shown in Listing~\ref{lst:mollerlo}.
    
    \item
    Further, we need an interface file that also contains the soft
    limits. In our case this is called {\tt ee/ee.f95}.

    \item
    Because $\M{n+1}0$ is border-line large, we will assume that it
    will be stored in an extra file, {\tt ee/ee2eeg.f95}. The required
    functions are to be imported in {\tt ee/ee\_mat\_el.f95}.

    \item
    Calculate the one-loop virtual matrix element $\M n1$,
    renormalised in the \ac{OS} scheme. In particular \ac{VP}
    contributions should not be included but implemented in a separate
    function. Of course, this could be done in any regularisation
    scheme. However, results in \mcmule{} shall be in the \fdh{} (or
    equivalently the \fdf{}) scheme. Divergent matrix elements in
    \mcmule{} are implemented as $c_{-1}$, $c_0$, and $c_1$
    \begin{align}
    \M n1 = \frac{(4\pi)^\epsilon}{\Gamma(1-\epsilon)}\Bigg(
        \frac{c_{-1}}\epsilon + c_0 + c_1\epsilon+\mathcal{O}(\epsilon^2)
    \Bigg)\,.
    \end{align}
    For $c_{-1}$ and $c_0$ this is equivalent to the conventions
    employed by Package-X~\cite{Patel:2015tea} up to a factor
    $1/16\pi^2$. While not strictly necessary, it is generally
    advisable to also include $c_{-1}$ in the Fortran code.

    For \ac{NLO} calculations, $c_1$ does not enter. However, we wish
    to include M{\o}ller scattering up to \ac{NNLO} and hence will
    need it sooner rather than later anyway.

    In our case, we will create a file {\tt ee/ee\_ee2eel.f95}, which
    defines a function
    \begin{lstlisting}
  FUNCTION EE2EEl(p1, p2, p3, p4, sing, lin)
    !! e-(p1) e-(p2) -> e-(p3) e-(p4)
    !! for massive electrons
  implicit none
  real(kind=prec), intent(in) :: p1(4), p2(4), p3(p4), p4(4)
  real(kind=prec) :: ee2eel
  real(kind=prec), intent(out), optional :: sing, lin
  ...
  END FUNCTION
    \end{lstlisting}
    The function shall return $c_0$ in {\tt ee2eel} and, if {\tt
    present} $c_{-1}$ and $c_1$ in {\tt sing} and {\tt lin}.

    \item
    At this stage, a new subroutine in the program {\tt test} with
    reference values for all three matrix elements should be written
    to test the Fortran implementation. This is done by generating a
    few points using an appropriate phase-space routine and comparing
    to as many digits as possible using the routine {\tt check}.

    In our case, we would construct a subroutine {\tt TESTEEMATEL} as
    shown in Listing~\ref{lst:mollertest}

    \item
    Define a default observable in {\tt user} for this process. This
    observable must be defined for any {\tt which\_piece} that might
    have been defined and test all relevant features of the
    implementation such as polarisation if applicable.

    \item
    Add the matrix elements to the integrands defined in {\tt
    integrands.f95} as discussed above. A second test routine should
    be written that runs short integrations against a reference value.
    Because {\tt test\_INT} uses a fixed random seed, this is expected
    to be possible very precisely. To guarantee reproducibility, the
    reference values for these tests need to be obtained by running
    \mcmule{} in a Docker container.

    \item
    After some short test runs, it should be clear whether new
    phase-space routines are required. Add those, if need be, to {\tt
    phase\_space} as described in Section~\ref{sec:ps}.

    \item
    Per default the stringent soft cut, that may be required to
    stabilise the numerical integration (cf.
    Section~\ref{sec:fksfor}), is set to zero. Study what the smallest
    value is that still permits integration.

    \item
    Perform very precise $\xc$ independence studies. Tips on how to do
    this can be found in Section~\ref{sec:xicut}.

\end{enumerate}

At this stage, the \ac{NLO} calculation is complete and may, after
proper integration into \mcmule{} and adherence to coding style has
been confirmed, be added to the list of \mcmule{} processes in a new
release. Should \ac{NNLO} precision be required, the following steps
should be taken

\begin{enumerate}
\setcounter{enumi}{12}

    \item
    Calculate the real-virtual and double-real matrix elements
    $\M{n+1}1$ and $\M{n+2}0$ and add them to the test routines as
    well as integrands.

    \item
    Prepare the $n$-particle contribution $\sigma_n^{(2)}$. In a
    pinch, massified results can be used also for $\ieik(\xc)\M n1$
    though of course one should default to the fully massive results.

    \item
    Study whether the pre-defined phase-space routines are sufficient.
    Even if it was possible to use an old phase-space at \ac{NLO},
    this might no longer work at \ac{NNLO} due to the added
    complexity. Adapt and partition further if necessary, adding more
    test integrations in the process.

    \item
    Perform yet more detailed $\xc$ and soft cut analyses.

\end{enumerate}

\begin{figure}
\centering
\input{figures/lst/mollertest.tex}
\renewcommand{\figurename}{Listing}
\caption{Test routine for $ee\to ee$ matrix elements and integrands.
The reference values for the integration are yet to be determined.}
\label{lst:mollertest}
\end{figure}

In the following we comment on a few aspects of this procedure such as
the $\xc$ study (Section~\ref{sec:xicut}), the calculation of matrix
elements (Section~\ref{sec:matel}), and a brief style guide for
\mcmule{} code (Section~\ref{sec:style}).

\subsection{Study of \texorpdfstring{$\xc$}{xic} dependence}\label{sec:xicut}

When performing calculations with \mcmule{}, we need to check that the
dependence of the unphysical $\xc$ parameter introduced in
Chapter~\ref{ch:fks} actually drops out at \ac{NLO} and \ac{NNLO}.
In principle it is sufficient to do this once during the development
phase.  However, we consider it good practice to also do this (albeit
with a reduced range of $\xc$) for production runs.

Because the $\xc$ dependence is induced through terms as
$\xc^{-2\epsilon}/\epsilon$, we know the functional dependence
of $\sigma^{(\ell)}_{n+j}$. For example, at \ac{NLO} we have
\begin{subequations}
\begin{align}\begin{split}
\sigma^{(1)}_{n  }(\xc) &= a_{0,0} + a_{0,1}\log(\xc)\,,\\
\sigma^{(1)}_{n+1}(\xc) &= a_{1,0} + a_{1,1}\log(\xc)\,,
\end{split}\end{align}
where $\xc$ independence of $\sigma^{(1)}$ of course requires
\begin{align}
a_{0,1}+ a_{1,1} = 0\,.
\end{align}
\label{eq:xinlo}
\end{subequations}
\begin{subequations}
At \ac{NNLO} we have
\begin{align}\begin{split}
\sigma^{(2)}_{n  }(\xc) &= a_{0,0} + a_{0,1}\log(\xc) + a_{0,2}\log(\xc)^2\,,\\
\sigma^{(2)}_{n+1}(\xc) &= a_{1,0} + a_{1,1}\log(\xc) + a_{1,2}\log(\xc)^2\,,\\
\sigma^{(2)}_{n+2}(\xc) &= a_{2,0} + a_{2,1}\log(\xc) + a_{2,2}\log(\xc)^2\,.
\end{split}\end{align}
We require 
\begin{align}
a_{0,i} + a_{1,i} + a_{2,i} = 0
\end{align}
for $i=1,2$. However, the \ac{IR} structure allows for an even
stronger statement for the $a_{j,2}$ terms
\begin{align}
a_{0,2} = a_{2,2} = -\frac{a_{1,2}}2\,.
\end{align}
\label{eq:xinnlo}
\end{subequations}
Of course we cannot directly calculate any of the $a_{1,i}$ or
$a_{2,i}$ because we use numerical integration to obtain the
$\sigma^{(\ell)}_{n+j}$. Still, knowing the coefficients can be
extremely helpful when debugging the code or to just quantify how well
the $\xc$ dependence vanishes. Hence, we use a fitting routine to fit
the Monte Carlo results \emph{after} any phase-space partitioning has
been undone. Sometimes non of this is sufficient to pin-point the
source of a problem to any one integrand. However, if the goodness of,
for example, $\sigma^{(2)}_{n+2}(\xc)$ is much worse than the one for
$\sigma^{(2)}_{n+1}(\xc)$, a problem in the double-real corrections
can be expected.

A worked example can be found in the next chapter in
Section~\ref{sec:muone}.

\subsection{Example calculations in Mathematica}
\label{sec:matel}
A thorough understanding of one-loop matrix elements is crucial for
any higher-order calculation. In \mcmule{}, one-loop matrix elements
either enter as the virtual contribution to \ac{NLO} corrections or
the real-virtual contribution in \ac{NNLO} calculations. In any case,
a fast numerical routine is required that computes the matrix element.

We perform all one-loop calculations in \fdf{} as this is arguably the
simplest scheme available. For theoretical background, we refer to
Section~\ref{sec:fdf} and references therein.

As already discussed in Section~\ref{ch:twoloop}, we use {\sc Qgraf}
for the diagram generation. Using the in-house Mathematica package
{\tt qgraf.wl} we convert {\sc Qgraf}'s output for manipulation with
Package-X~\cite{Patel:2015tea}. This package is available on request
through the \ac{MMCT}
\begin{lstlisting}[language=bash]
    (*@\url{https://gitlab.psi.ch/mcmule/qgraf}@*)
\end{lstlisting}

An example calculation for the one-loop calculation of
$\mu\to\nu\bar\nu e\gamma$ can be found in Listing~\ref{lst:pkgx}. Of
course this example can be made more efficient by, for example,
feeding the minimal amount of algebra to the loop integration routine.

When using {\tt qgraf.wl} for \fdf{} some attention needs to be paid
when considering diagrams with closed fermion loops. By default, {\tt
qgraf.wl} evaluates these traces in $d$ dimensions. {\tt RunQGraf} has
an option to keep this from happening.

\begin{figure}
\input{figures/lst/pkgx.tex}
\renewcommand{\figurename}{Listing}
\caption{An example on how to calculate the renormalised one-loop
matrix element for $\mu\to\nu\bar\nu e$ in \fdf.}
\label{lst:pkgx}
\end{figure}

There is a subtlety here that only arise for complicated matrix
elements. Because the function Package-X uses for box integrals, {\tt
ScalarD0IR6}, is so complicated, no native Fortran implementation
exists in \mcmule{}.  Instead, we are defaulting to
COLLIER~\cite{Denner:2016kdg} and should directly evaluate the finite
part of the {\tt PVD} function above.  The same holds true for the
more complicated triangle functions. In fact, only the simple {\tt
DiscB} and {\tt ScalarC0IR6} are natively implemented without need for
external libraries. For any other functions, a judgement call is
necessary of whether one should {\tt LoopRefine} the finite part in
the first place. In general, if an integral can be written through
logarithms and dilogs of simple arguments (resulting in real answers)
or {\tt DiscB} and {\tt ScalarC0IR6}, it makes sense to do so.
Otherwise, it is often easier to directly link to COLLIER.

\subsection{Coding style and best practice}
\label{sec:style}
A large-scale code base like \mcmule{} cannot live without some basic
agreements regarding coding style and operational best practice. These
range from a (recommended but not enforced) style guide over the
management of the git repository to how to best run \mcmule{} in
development scenarios. All aspects have been discussed within the
\ac{MMCT}.

Fortran code in \mcmule{} is (mostly) written in accordance with the
following style guide. If new code is added, compliance would be
appreciated but deviation is allowed if necessary. If in doubt,
contact any member of the \ac{MMCT}. 
\begin{itemize}
    \item
    Indentation width is two spaces. In Vim this could be implemented
    by adding the following to {\tt.vimrc}
    \begin{lstlisting}[language=vim]
autocmd FileType fortran set tabstop=8 softtabstop=0 expandtab shiftwidth=2 smarttab
    \end{lstlisting}

    \item
    Function and subroutine names are in all-upper case.

    \item
    A function body is not indented beyond its definition.

    \item
    When specifying floating point literals specify the precision when
    possible, i.e. \lstinline{1._prec}.

    \item
    Integrands should have \lstinline{ndim} specified.

    \item
    Internal functions should be used where available.

    \item
    Masses and other kinematic parameters must be calculated in the
    matrix elements as local variables; using the global parameters
    {\tt Mm} and {\tt Me} is strictly forbidden.

    \item
    These rules also hold for matrix elements.

\end{itemize}
For python code, i.e. {\tt pymule} as well as the analysis code, PEP8
compliance is strongly encouraged with the exception of {\tt E231}
(Missing whitespace after {\tt ,}, {\tt;}, and {\tt:}), {\tt E731} (Do
not assign a lambda expression, use a {\tt def}) as well, in justified
cases, i.e. if required by the visual layout, {\tt E272} (Multiple
spaces before keyword), and {\tt E131} (Continuation line unaligned for
hanging indent).

\mcmule{} uses two git repositories for version management. One
internal repository and one public-facing one. Releasing to the latter
is the responsibility of the \ac{MMCT} after sufficient vetting was
performed by squashing commits to avoid the accidental release of
embarrassing or wrong code to the public. However, even the internal
repository has certain rules attached. In general, developers are
encouraged to not commit wrong or unvetted code though this can
obviously not be completely avoided in practice. To avoid
uncontrollable growth of the git repository, large files movements are
strongly discouraged. This also means that matrix elements should not
be completely overhauled barring unanimous agreement. Instead,
developers are encouraged to add a new matrix element file and link to
that instead.

Even when running \mcmule{} for development purposes the usage of menu
files is strongly encouraged because the code will do its utmost to
automatically document the run by storing the git version as well as
any modification thereof. This allows for easy and unique
reconstruction of what was running. For production runs this is not
optional; these must be conducted with menu files after which the run
folder must be stored with an analysis script and all data on the AFS
as well as the user file library to ensure data retention.

\chapter{Phenomenology}\label{ch:pheno}

In this chapter we will demonstrate example calculations with
\mcmule{}. We will come back to the list of processes presented in
Chapter~\ref{ch:intro}, reviewing various scattering processes and
muon decay modes sorted by experimental situation. We will begin by
discussing the scattering experiments MUonE (Section~\ref{sec:muone})
and Muse (Section~\ref{sec:muse}). Next, we will review MEG in
Section~\ref{sec:meg}. Afterwards in Section~\ref{sec:babarsol}, we
review the $3\,\sigma$ discrepancy in the radiative $\tau$ decays we
observed earlier. This is followed by a discussion of the Michel
decay $\mu\to\nu\bar\nu e$ that is independent of any particular
experiment in Section~\ref{sec:michel}. Finally, we briefly present
results for the Mu3e experiment in Section~\ref{sec:mu3e}.

For the present discussion we will only provide examples that show
\mcmule's capabilities. This list is not meant to be exhaustive of all
results that have ever been produced. Such a list is in
preparation~\cite{Ulrich:legacy}.

All results presented here use the following input parameters
\begin{align*}
m_\mu &= 105.6583715\,{\rm MeV}\,,\qquad&
m_e   &= 0.510998928\,{\rm MeV}\,,
\\
m_\tau&= 1776.82\,{\rm MeV}\,,\qquad&
m_p   &= 938.2720813\,{\rm MeV}\,,
\\
\alpha &= \frac{1}{137.03599907}\,,\qquad&
G_F &= 1.16637\cdot 10^{-11}\,{\rm MeV}^{-2}
\\
{\tt conv} &= (c\hbar)^2 = 3.8937936\cdot 10^8\ {{\rm MeV}^2}{\rm \upmu b}\,,
\end{align*}
where the masses and the coupling is understood to be in the on-shell
scheme. Here, {\tt conv} is the factor used to convert cross sections
from ${\rm MeV}^{-2}$ to $\rm \upmu b$.

\section{MUonE (\texorpdfstring{$\mu^- e^-\to\mu^- e^-$}{mu- e- -> mu- e-})}\label{sec:muone}
Following the renewed interest into $\mu$-$e$ scattering, previous
\ac{NLO} calculations~\cite{Bardin:1997nc,Kaiser:2010zz} have been
redone in a fully differential Monte Carlo~\cite{Alacevich:2018vez,
NLOmfmp} as well as \mcmule{}~\cite{Banerjee:2020rww}. However, to match the
required experimental accuracy a \ac{NNLO} calculation is required
(for a review cf.~\cite{MUonEwriteup}). 

The full \ac{NNLO} is currently under investigation though impressive
progress has been made. The required master integrals are known for
vanishing electron masses~\cite{Mastrolia:2017pfy, DiVita:2018nnh,
Mastrolia:2018sso, Ronca:2019kcw}. Similarly, the real-virtual
diagrams have been calculated both for $m=0$ and
$m>0$~\cite{MUonEwriteup}. The signal, i.e. the \ac{HVP} contribution,
has been studied at \ac{NNLO}~\cite{Fael:2018dmz,Fael:2019nsf} and an
integration of these results into \mcmule{} is being validated.
Finally, the impact of \ac{BSM} physics has been found to be
negligible~\cite{Masiero:2020vxk,Dev:2020drf}.

For all calculations we will assume a muon beam with a fixed energy
$E_{\rm{beam}}=150\ \rm GeV$, consistent with the M2 beam line at CERN
North Area~\cite{MUonE:LOI}. Let us further remark that the total
cross section is ill-defined due to the behaviour $\D\sigma/\D t \sim
t^{-2}$ with $t_{\rm min}\leq t\leq 0$. We therefore have to apply a
cut on the maximal value of $t$ or equivalently on the minimal energy
of the outgoing electron. In the results below we have chosen $E_{\rm
min}=1\ \rm{GeV}$ (`Setup 2' of~\cite{Alacevich:2018vez}). Further,
to demonstrate the versatility of \mcmule{}, we apply a cut
restricting photon emission in a way that could be measured by MUonE.
To be precise, we require that the acoplanarity is
\begin{align}
\big|\pi - |\phi_e-\phi_\mu|\big| < 3.5\,{\rm mrad}\,,
\label{eq:acocut}
\end{align}
in correspondence with `Setup 4' of~\cite{Alacevich:2018vez}.

We will be more verbose in the discussion of $\mu$-$e$ scattering than
in the other calculations presented in this chapter, as these results
have not been presented elsewhere yet. However, all of them follow the
same procedure.

In the following, we will present selected results for $\mu$-$e$
scattering in the context of MUonE. In particular, we will compare
\mcmule{}'s \ac{NLO} calculation~\cite{Banerjee:2020rww}
with~\cite{Alacevich:2018vez} putting special emphasis on the gauge
invariant split into \term{electronic corrections} (emission only from
the electron line), \term{muonic corrections} (emission from the muon
line), and \term{mixed corrections} (cf. Section~\ref{sec:colour}).
With the splitting properly motivated, we present the electronic
corrections without any \ac{VP} contribution at \ac{NNLO} and compare
with~\cite{CarloniCalame:2020yoz}. This is of course much simpler
because the muon becomes a spectator, reducing the number of scales.

As discussed many times before, detailed $\xc$ studies are crucial. In
Section~\ref{sec:xicut}, we have outlined a procedure on how to best
do this by fitting the \mcmule{} data. In Figure~\ref{fig:xicut}, the
result of the fitting procedure as well as the final combination can
be seen for the full \ac{NLO} corrections as well as the electronic
\ac{NNLO} corrections. Note that while it is of course possible to use
small $\xc$ values such as $10^{-3}$ for production runs, this rarely
is a good idea due to the large cancellation between the different
contributions. At \ac{NLO}, the ideal spot for running is the
intersection between $\sigma^{(1)}_n(\xc)$ and
$\sigma^{(1)}_{n+1}(\xc)$. In this case this is $\xc\approx0.15$. At
\ac{NNLO} this is less clear cut because the three parabolas might not
conveniently intersect. Here, $\xc\approx0.1$ might be a good idea.

We split the total cross section $\sigma$ into different contributions
by order in perturbation theory and origin (either $e$ for electronic,
$\mu$ for muonic, or $m$ for mixed)
\begin{align}
\sigma = \sigma^{(0)}
   +\Big(\sigma_e^{(1)}+\sigma_\mu^{(1)}+\sigma_m^{(1)}\Big)
   +\Big(\sigma_e^{(2)}+\cdots\Big) + \mathcal{O}(\alpha^5)\,.
   \label{eq:muexplit}
\end{align}
The different contributions are shown in Table~\ref{tab:xsmue} for
$\mu^-$-$e^-$ scattering. To obtain results for $\mu^+$-$e^-$
scattering, the sign of $\sigma_m^{(1)}\propto (q^2Q^2)Q$ needs to be
flipped (cf. Section~\ref{sec:colour}). Results are compared with the
results from~\cite{Alacevich:2018vez,CarloniCalame:2020yoz}, finding excellent agreement.
All errors given are purely statistical. Especially, parametric
uncertainties and those arising from the uncomputed \ac{n3lo} are not
considered. It is clearly visible that the electronic corrections at
\ac{NLO} are by far the largest contributor to the full \ac{NLO}.  The
high precision to which the cross sections were calculated is a side
effect of wanting to obtain precise histograms. The present dataset
corresponds to roughly $1.4\times 10^8\,{\rm CPU\,s} \approx 4.6\,{\rm
CPU\,years}$ on \ac{PSI}'s Slurm system.
\begin{figure}
\centering
\subfloat[The $\xc$ dependence at \ac{NLO}]{
\scalebox{0.55}{\input{figures/mule/xi-nlo.pgf}}
}\\
\subfloat[The $\xc$ dependence at \ac{NNLO}]{
\scalebox{0.55}{\input{figures/mule/xi-nnlo.pgf}}
}
\caption{The $\xc$ dependence at \ac{NLO} and \ac{NNLO}, split into
different contributions and fitted to \eqref{eq:xinlo} and
\eqref{eq:xinnlo}, respectively. The upper panels show all different
contributions in $\rm\upmu b$. The lower panels show the sum
normalised to the averaged value. The fit shows a $1\sigma$ band.}
\label{fig:xicut}
\vspace{5mm}
\centering
\scalebox{0.9}{
\begin{tabular}{l|r|r||r|r||r|r}
 & \multicolumn{2}{c||}{\cite{Alacevich:2018vez}}  & \multicolumn{2}{c||}{\mcmule{}} & \multicolumn{2}{c}{$K$} \\\cline{2-7}
 & \multicolumn{1}{c|}{Setup 2} & \multicolumn{1}{c||}{Setup 4}          
 & \multicolumn{1}{c|}{Setup 2} & \multicolumn{1}{c||}{Setup 4}          
 & \multicolumn{1}{c|}{Setup 2} & \multicolumn{1}{c  }{Setup 4}  \\\hline
$\sigma  ^{(0)}$ & \multicolumn{2}{c||}{\tt245.038906(3)}  & \multicolumn{2}{c||}{\tt245.038910(1)}\\
\hline
$\sigma_e  ^{(1)}$ & \tt     10.510(2) &\tt     -21.605(2)  &\tt    10.51037(5) &\tt   -21.60054(3) &\tt   0.0429  &\tt -0.0882 \\
$\sigma_\mu^{(1)}$ & \tt     -0.069(2) &\tt      -0.627(2)  &\tt  -0.06824902(5) &\tt    -0.62546(4) &\tt  -0.0003  &\tt -0.0026 \\
$\sigma_m  ^{(1)}$ & \tt     -0.360(5) &\tt       0.042(5)  &\tt  -0.3599420(3) &\tt     0.04113(1) &\tt  -0.0015  &\tt  0.0002 \\
$\sigma    ^{(1)}$ & \tt     10.081(2) &\tt     -22.188(2)  &\tt    10.08218(5) &\tt   -22.18488(3) &\tt   0.0411  &\tt -0.0905 \\
\hline
$\sigma_e  ^{(2)}$ & \tt    10.5793(7) &\tt      1.0409(7)  &\tt    0.02277(2) &    1.04118(2) &\tt  0.0023 &\tt -0.0469 \\
\hline
\hline
$\sigma          $ & \multicolumn{2}{c||}{}         &\tt  255.14385(5) &  223.89521(4) &    &    \\
\end{tabular}}
\renewcommand{\figurename}{Table}
\caption{The cross section for $\mu$-$e$ scattering at MUonE in Setup
2 ($E_e>1\,{\rm GeV}$) and Setup 4  (also
including~\eqref{eq:acocut}). The results are split into the gauge
invariant subsets introduced in Section~\ref{sec:colour} and again in
\eqref{eq:muexplit}. It is apparent that at \ac{NLO}, the electronic
contributions are by far the largest as discussed before. To compare
with~\cite{CarloniCalame:2020yoz} no \ac{VP} contributions were
included.}
\label{tab:xsmue}
\end{figure}

\begin{figure}
\centering
\subfloat[$\D\sigma/\D\theta_e$ without the acoplanarity
cut~\eqref{eq:acocut} (Setup 2)]{
\scalebox{0.8}{\input{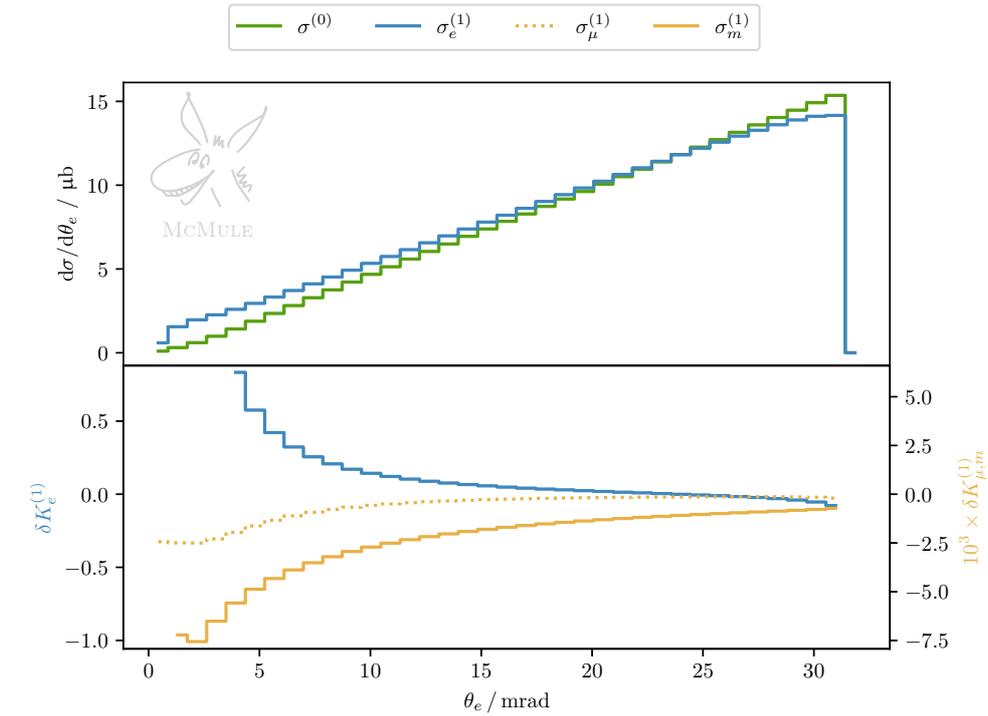}}
}

\subfloat[$\D\sigma/\D\theta_e$ with the acoplanarity
cut~\eqref{eq:acocut} (Setup 4)]{
\scalebox{0.8}{\input{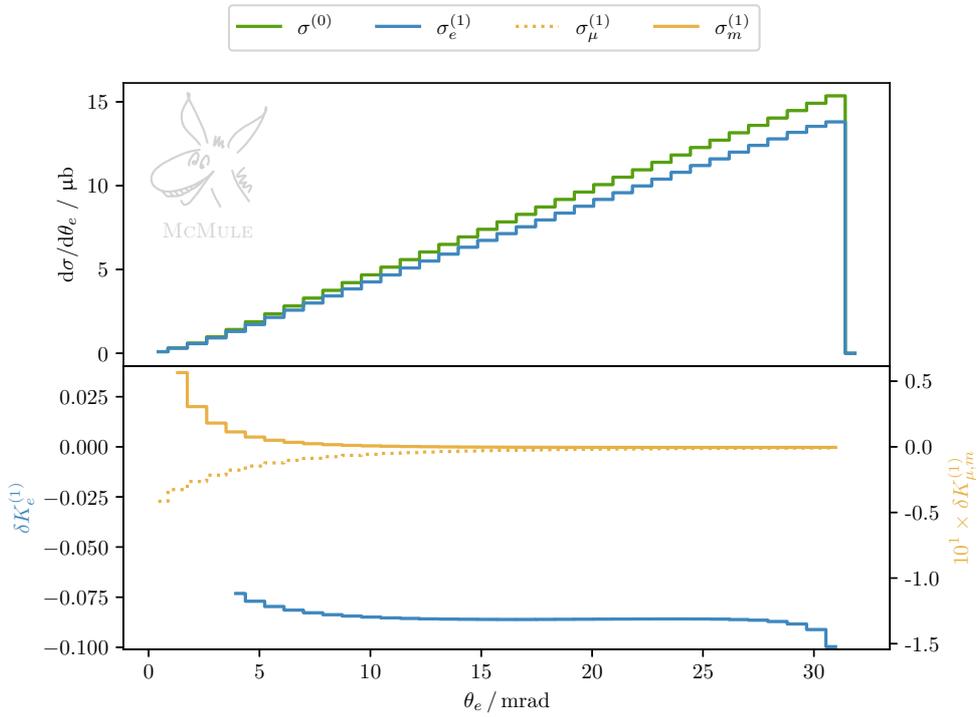}}
}
\caption{The angular distribution $\D\sigma/\D\theta_e$ at \ac{NLO}.
The $K$-factors are presented split into different
contributions~\eqref{eq:muexplit}. The electronic corrections (blue,
left axis) are a lot larger than the muonic (orange, dashed, right
axis) or mixed contributions (orange, right axis).}
\label{fig:dsigma:split}
\end{figure}

After we have justified the split into contributions for the cross
section, let us now look at a differential distribution such as
$\D\sigma/\D\theta_e$. In Figure~\ref{fig:dsigma:split}, this
distribution is shown, once without the acoplanarity
cut~\eqref{eq:acocut} and with it. The $K$-factor is shown split into
the different classes~\eqref{eq:muexplit}
\begin{align}
    K_j^{(i)}
    =\frac{\D\sigma_{j,i}/\D\theta_e}{\D\sigma_{j,i-1}/\D\theta_e} = 1+\delta K_j^{(i)}
    \quad
    \text{with}
    \quad
    j\in\{e,\mu,m\}\,.
\end{align}
For now, we only do this at \ac{NLO}, i.e. $i=1$, because the mixed
contributions at \ac{NNLO} are not yet available. It is clearly
visible that even for differential spectra, the electronic corrections
are by far the largest. Indeed, considering e.g. $\theta_e=5\,{\rm
mrad}$, the fixed-order \ac{NLO} electronic correction $K_e$ amount to
nearly $50\%$ (see scale on the left) whereas the muonic and mixed
corrections ($K_\mu$ and $K_m$, respectively) are less than half a
percent (see scale on the right).

\begin{figure}
\centering
\subfloat[$\D\sigma/\D\theta_e$ with the acoplanarity cut~\eqref{eq:acocut}]{
\scalebox{0.8}{\input{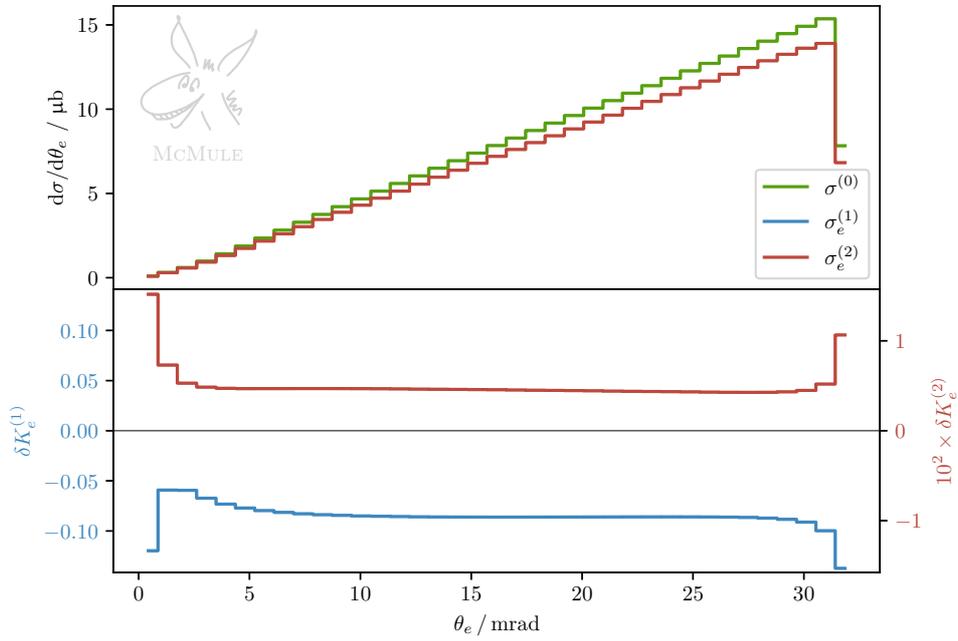}}
}

\subfloat[$\D\sigma/\D\theta_e$ without the acoplanarity cut~\eqref{eq:acocut}]{
\scalebox{0.8}{\input{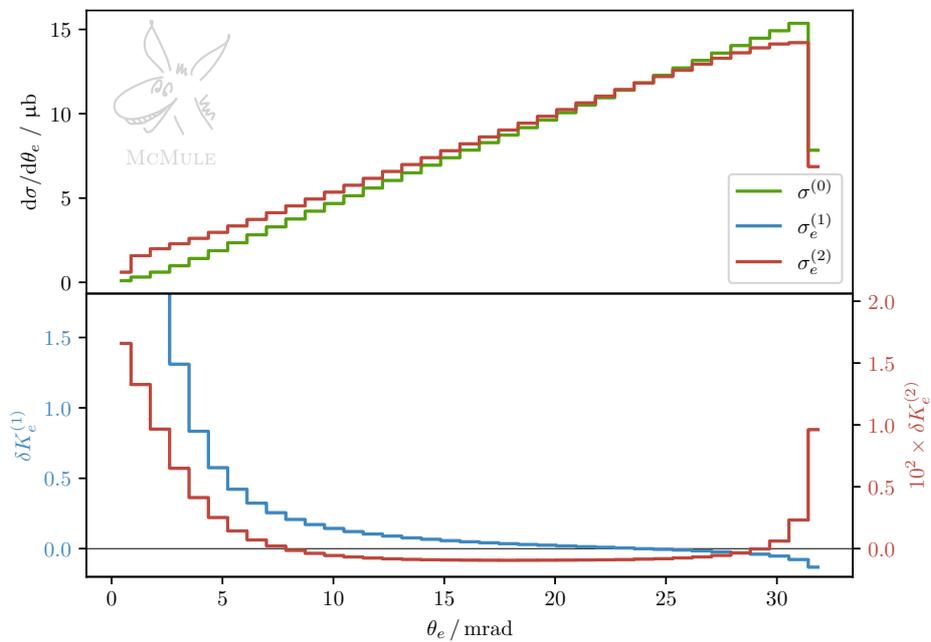}}
}
\caption{The distribution of the outgoing electron's angle relative to
the beam axis at \ac{LO}, \ac{NLO}, and \ac{NNLO}. The upper panels
only show the \ac{LO} and \ac{NNLO} curves.}
\label{fig:dsigma:nnlo}
\end{figure}

Before we discuss Figure~\ref{fig:dsigma:split} in detail, let us add
the electronic corrections at \ac{NNLO} ignoring contributions due to the \ac{VP} in
Figure~\ref{fig:dsigma:nnlo}. It is clearly visible that for small
scattering angles $\theta_e\to0$ the \ac{NLO} corrections become
extremely large ($K>2$ for $\theta_e < 5\,{\rm mrad}$). As this is the
region of interest for MUonE, these corrections are especially
troubling. However, they can be almost entirely accounted to the new
process $\mu e\to\mu e\gamma$. Hard photon radiation can knock the
electron back towards the beam axis resulting in more small-angle
electrons, i.e. positive corrections.  At \ac{NNLO}, the radiative
process $\mu e\to\mu e\gamma$ has now essentially been included at
\ac{NLO}. This results in reduced, but still large
($\mathcal{O}(10\%)$), corrections for small-angle electrons. As soon
as the acoplanarity cut~\eqref{eq:acocut} is applied, the corrections
dramatically decrease in size because the cut restricts hard photon
emission.

Still, the \ac{NNLO} corrections are very large in the relevant
regions ($\mathcal{O}(0.5\%)$). A naive extrapolation to \ac{n3lo}
would suggest $\mathcal{O}(5\times 10^{-4})$ corrections. This is a
long way from the requirement that all systematic uncertainties need
to be below $10^{-5}$. Fortunately, the corrections are almost
exclusively driven by large logarithms for $\theta_e\to0$ that can be
resummed.  However, only \ac{LL} resummation is feasible due to the
complexity of the cuts because those can be obtained using a \ac{PS}.
It is unlikely that for example a cross section with the acoplanarity
cut~\eqref{eq:acocut} could ever be resummed analytically to \ac{NLL}.
While it is of course possible to construct other observables that
could be resummed to \ac{NLL} or even beyond, most of those could not
be measured at MUonE due to the lack of precise energy measurements of
the outgoing particle.

\section{MUSE \texorpdfstring{($ep\to ep$)}{ep -> ep}}\label{sec:muse}
Lepton proton scattering has an extremely long history in particle
physics that we will not recount in full here. On the theoretical
side, higher-order corrections have been calculated long
ago~\cite{Mo:1968cg,Tsai:1961zz} and later revisited~\cite{
deCalan:1990eb,Kwiatkowski:1990es, Arbuzov:1995id, Maximon:2000hm,
Ent:2001hm, Afanasev:2001nn, Weissbach:2004ij, Weissbach:2008nx,
Akushevich:2015toa, Gakh:2016xby}. Unfortunately, these calculations
can often not be directly reused as they typically rely on assumptions
on the energy scales involved or what particles are and are not measured. While
this helps to arrive at concise formulas, these assumptions are not
universally valid. Hence, we need a fully-differential \ac{NNLO}
calculation to best exploit past, present and future data.

Results similar to those shown here and~\cite{Banerjee:2020rww} have
been presented in~\cite{Bucoveanu:2018soy}, not including \ac{VP}
contributions.  Our NLO results (without \ac{VP}) agree with these results.
However, we disagree substantially with the NNLO corrections
of~\cite{Bucoveanu:2018soy}, even if we adapt to their calculation and
include the electron loop in the two-loop vertex diagram.  With
respect to the results presented in Section~\ref{sec:muone} that have
been verified independently by~\cite{CarloniCalame:2020yoz}, the only
new ingredients are the matrix elements. They have been compared
pointwise with~\cite{Bucoveanu:2018soy} and agree.

We can use \mcmule{} to calculate $e$-$p$ scattering by repurposing
the electronic corrections to $\mu$-$e$ scattering. However, unlike
the muon, the proton is not point-like. For small virtualities $Q^2$
of the $t$-channel photon, we change the proton's interaction with the
photon to
\begin{align}
\bar u(m_\mu) \gamma_\mu u(m_\mu)
\to
      \bar u(m_p) \Big( 
        F_1(Q^2)\gamma^\mu+F_2(Q^2)\tfrac{{\rm i}\sigma^{\mu\nu}Q_\nu}{2m_p}
      \Big)u(m_p)\,,
\end{align}
where $F_1$ and $F_2$ are determined through measurements. For
simplicity, we assume a dipole parametrisation of a proton with charge
radius $R_p$
\begin{align}
F_1(Q^2) =
\frac{1+\kappa\tau}{1+\tau}\Big(1+\frac{Q^2}{\Lambda^2}\Big)^{-2}
\quad\text{and}\quad
F_2(Q^2) =
\frac{-1+\kappa}{1+\tau}\Big(1+\frac{Q^2}{\Lambda^2}\Big)^{-2}
\,,
\end{align}
with $\tau=Q^2/(4m_p^2)$, $\Lambda^2=12/R_p^2=0.71\,{\rm GeV}^2$ and
$\kappa=2.7928$ the proton's magnetic moment.  However, the specific
values and parametrisation used have no large
influence~\cite{Bucoveanu:2018soy}.

As an example of $e$-$p$ scattering we will calculate the process for
the Muse experiment~\cite{Gilman:2013eiv}. It is situated on the
$\pi{\rm M}1$ beam line at \ac{PSI}, measuring both $e$-$p$ and
$\mu$-$p$ scattering with different momenta. The geometric acceptance
of the detector for outgoing electrons is
\begin{align}
20^\circ < \theta_e < 100^\circ\,.
\end{align}
For now, we will only consider one value of the beam momentum
$p_\text{in} = 115\,{\rm MeV}$. 

For this calculation, we have included only electronic effects. Just
as for the discussion of $\mu$-$e$ scattering, we have not included
\ac{HVP} or leptonic vacuum polarisation effects as they are being
still vetted.  Hence, these results should not be considered definite.

\begin{figure}
    \centering
    \scalebox{0.8}{\input{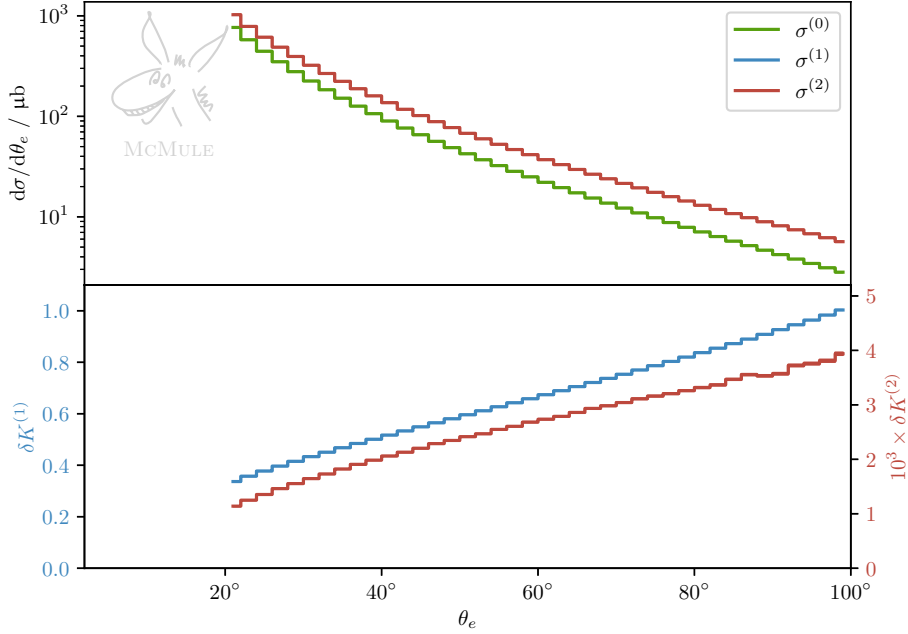}}
\caption{The angular distribution of the outgoing electron for a
$p_\text{in} = 115\,{\rm MeV}$ electron beam in Muse.}
\label{fig:muse}
\end{figure}

The resulting distribution $\D\sigma/\D\theta_e$ is shown in
Figure~\ref{fig:muse}. Just as for MUonE, the large corrections are
due to unrestricted photon emission in $ep\to ep\gamma$ that shifts
the entire spectrum towards larger angles such that eventually more
electrons hit the detector. This view is reaffirmed by the fact that
the \ac{NNLO} corrections are very small, especially compared to the
\ac{NLO} corrections.

\section{MEG and MEG II}\label{sec:meg}
The MEG experiment and its successor MEG II are designed to search for
the \ac{LFV} process $\mu\to e\gamma$. However, they are also
interested in the single ($\mu\to\nu\bar\nu e\gamma$) and double
($\mu\to\nu\bar\nu e\gamma\gamma$) radiative muon decays as they serve
as backgrounds to searches like $\mu\to e\gamma$ and $\mu\to
eJ(\to\gamma\gamma)$.

MEG and MEG II are running on the $\pi{\rm E}5$ beam line that
delivers (partially) polarised $\mu^+$ that are stopped in the
detector.  We define the $z$-axis against the polarisation axis s.t.
the muon polarisation $\vec P_\mu = -0.85\,\vec z$~\cite{Baldini:2015lwl}. 
The geometric acceptance of the MEG detector is then simulated as
\begin{subequations}
\begin{align}
\big|\cos\sphericalangle(\vec p_\gamma, \vec z)\big| \equiv
\big|\cos\theta_\gamma\big| &< 0.35\,,
&\big|\phi_\gamma\big| > \frac{2\pi}3\,,\label{eq:meg:geo:g}
\\
\big|\cos\sphericalangle(\vec p_e, \vec z)\big| \equiv
\big|\cos\theta_e\big| &< 0.5\,,
&\big|\phi_e\big| > \frac{\pi}3\,.\label{eq:meg:geo:e}
\end{align}\label{eq:meg:geo}
\end{subequations}
Further cuts may be applied, depending on the physics search.

\subsection{Single radiative muon decay
(\texorpdfstring{$\mu\to\nu\bar\nu e\gamma$}{mu -> nu anti-nu e gamma})}
The radiative muon decay $\mu\to\nu\bar\nu e\gamma$ is an important
background to the \ac{LFV} searches for $\mu\to e\gamma$ in MEG.
Hence, precise understanding of this decay is crucial.

We begin by noting that the \ac{BR} for this process depends on the
energy cut on the photon, that is required to make the
quantity well defined. For the standard choice of $E_\gamma > 10\,{\rm
MeV}$ the \ac{BR} is roughly 1\%. Given the vast number of muons that
can be produced, it should be possible to study radiative muon decays
with very good precision. Apart from measuring the \ac{BR} and as a
background to \ac{LFV} searches, the \ac{SM} could in principle also be
tested by measuring Michel parameters of a general formula for muon
decays~\cite{Eichenberger:1984gi, Pocanic:2014mya, Arbuzov:2016ywn}.
Unfortunately, the cuts employed by MEG are far too restrictive to do
this. Hence, we have to rely on other experiments for these
measurements.

Corrections beyond the Fermi theory due to the $W$-boson
propagator~\cite{Ferroglia:2013dga, Fael:2013pja} turn out to be much
smaller than the \ac{NLO} corrections. The tree-level calculation within
the Fermi theory has been considered by several authors a long time
ago~\cite{PhysRev.101.866, Fronsdal:1959zzb, Eckstein1959297,
Kinoshita:1959uwa}. Due to the photon bremsstrahlung
the helicity of the final-state lepton does not have to be
left-handed~\cite{Falk:1993tf,Sehgal:2003mu,Schulz:2004xd,Gabrielli:2005ek}.
After some partial results~\cite{Fischer:1994pn, Arbuzov:2004wr} a
full \ac{NLO} calculation for the \ac{BR} was presented
in~\cite{Fael:2015gua,Fael:2016hnz}. As for a related calculation of
the rare decays of leptons~\cite{Fael:2016yle}, the results presented
in~\cite{Fael:2015gua,Fael:2016hnz} allow to obtain the differential
decay width at \ac{NLO} with cuts on the photon and electron energy and
angles between them. With \mcmule{}, we generalise these results
because it allows us to implement arbitrary cuts, allowing to mirror
the experimental situation more closely.

When searching for $\mu\to e\gamma$, MEG applies, in addition
to~\eqref{eq:meg:geo}, energy cuts requiring
\begin{subequations}
\begin{align}
E_\gamma>40\,\mev
\qquad\text{and}\qquad
E_e>45\,\mev\,,
\end{align}
This reduces the amount of data taken without infringing on the signal
which is at $E_e\approx E_\gamma=M/2$. Also, MEG will veto any event
with multiple visible photons. We simulate this by requiring for the
second photon (if present)
\begin{align}
E_{\gamma_2} < \begin{cases}2\,\mev & \text{if \eqref{eq:meg:geo:g} is
satisfied} \\ \infty & \text{otherwise}\end{cases}\,.
\end{align}
Of course this is rather simplistic because it assumes that the
detector could tell two photons apart regardless of how closely
clustered they are. In contrast to \ac{QCD}, there is fortunately no
mechanism driving the two photons collinear. Hence, this model is
sufficient for current purposes. In Section~\ref{sec:drmd}, we will
discuss a more detailed model that does require spatial separation in
the detector.
\label{eq:meg:energy}
\end{subequations}

\begin{figure}
    \centering
    \scalebox{0.8}{\input{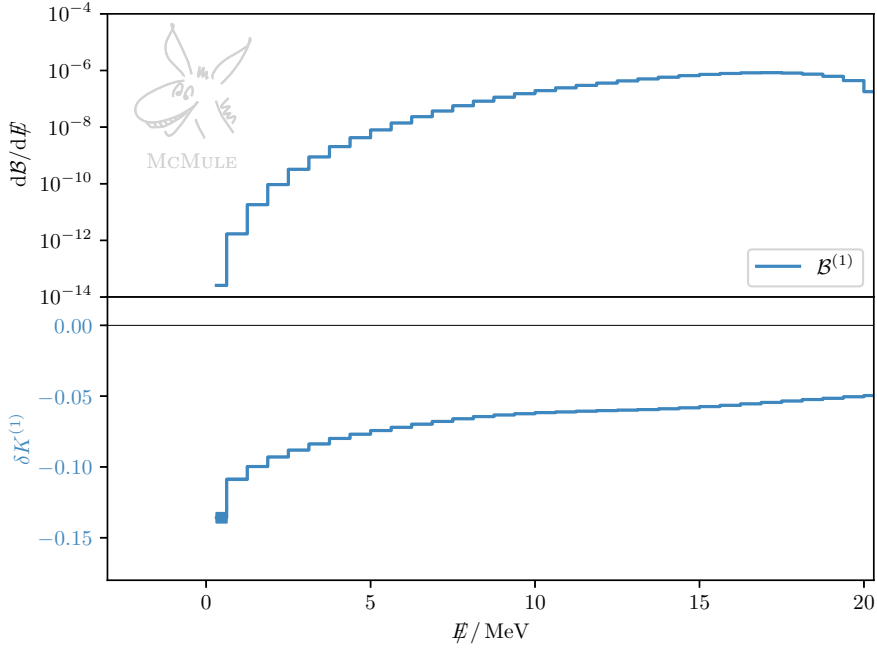}}
    \caption{The missing energy spectrum with full MEG
    cuts~\eqref{eq:meg:geo} and~\eqref{eq:meg:energy} at \ac{NLO}.}

    \label{fig:meg:emiss}
\end{figure}

We now can use the cuts~\eqref{eq:meg:geo} and~\eqref{eq:meg:energy}
to calculate the missing energy spectrum $\D\mathcal{B}/\D\Einv$ in
Figure~\ref{fig:meg:emiss} where the missing energy is \emph{defined}
as
\begin{align}
\Einv = M-E_e-E_\gamma\,,
\end{align}
which includes both the neutrinos as well as a potential second 
photon. To obtain precise results in the region of small $\Einv$, we
perform two runs, one with the full range of $\Einv$ allowed
by~\eqref{eq:meg:energy} (roughly $0<\Einv\lesssim20\,{\rm MeV}$) and
a tailored run where $\Einv < 6\,{\rm MeV}$ is enforced (cf.
Section~\ref{sec:analyse}).

For the bulk of the distribution, the corrections are of the order of
$-5\%$, but in the tail they increase substantially. We also note that
the distribution itself falls rapidly towards zero for $\Einv \to
20\,{\rm MeV}$, due to the kinematic constraints.

\begin{figure}
    \centering
    \scalebox{0.8}{\input{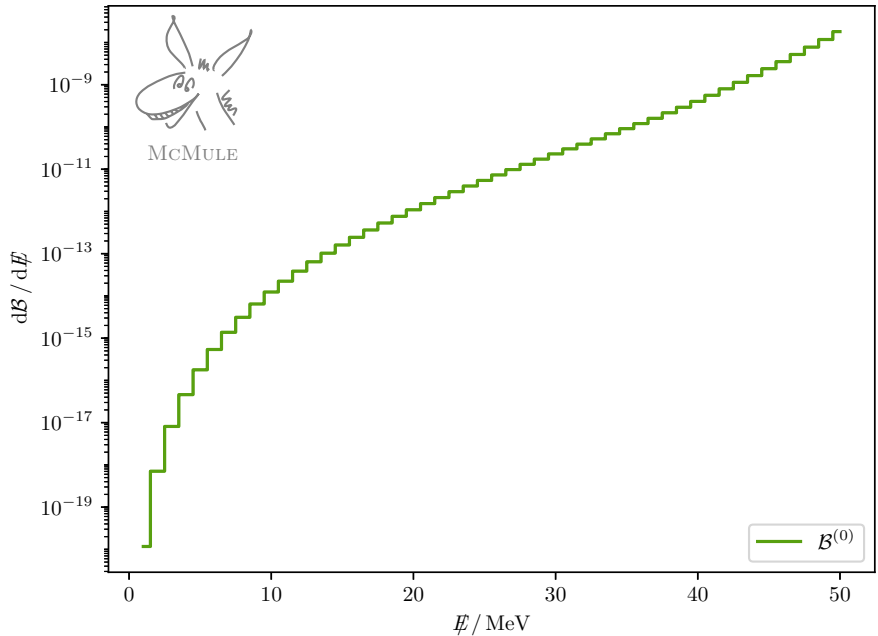}}
    \caption{The missing energy spectrum for the double-radiative muon
    decay at \ac{LO}.}
    \label{fig:meg:drmdmiss}
\end{figure}

\subsection{Double radiative muon decay
(\texorpdfstring{$\mu\to\nu\bar\nu e\gamma\gamma$}{mu -> nu anti-nu e
gamma gamma})}
\label{sec:drmd}
The double radiative muon decay is a background for searches of light
New Physics that induces the \ac{LFV} muon decay $\mu\to eJ$ where
the $J$ is a Majoron, a light but not massless new particle, that
could promptly decays into $J\to\gamma\gamma$. 

\begin{subequations}
For this study we only apply the cuts on photon
geometry~\eqref{eq:meg:geo:g} as well as
\begin{align}
E_{\gamma_i} \ge 10\,\mev\,,
\end{align}
which is necessary for \ac{IR} safety. Further, we require that the
two photons can be separated in the calorimeter. This is implemented
by specifying them to be $\delta x=20\,{\rm cm}$ apart on the detector
surface which is at a radius of $R=67.85\,{\rm cm}$ resulting in
\begin{align}
\sphericalangle(\vec p_{\gamma_1}, \vec p_{\gamma_2})
= \theta_{\gamma\gamma}
> \tan^{-1}\bigg(\frac{\delta x}{R}\bigg)
\approx 16.4^\circ\,.
\end{align}
\end{subequations}
With these cuts, we can now again calculate the missing energy
spectrum at \ac{LO}, depicted in Figure~\ref{fig:meg:drmdmiss}.

\section{{\sc BaBar} (\texorpdfstring{$\tau\to\nu\bar\nu e\gamma$}{tau -> nu
anti-nu e gamma})}\label{sec:babarsol}

The example calculation in Section~\ref{sec:example} already indicated
a discrepancy between the \ac{NLO} results and the experimental
measurement~\cite{Oberhof2015Measurement,Lees:2015gea} for the
radiative $\tau$ decay. As argued in~\cite{Fael:2015gua} these
measurements are to be compared with the exclusive \ac{BR} we have
calculated before
\begin{align}\begin{split}
\mathcal{B}|_\text{NLO} &= 1.6451(1) \times 10^{-2}\,\\
\mathcal{B}|_\text{exp} &= 1.847(54)\times 10^{-2}\,.
\end{split}\end{align}
We will now use \mcmule{} to revisit this  $3.5\,\sigma$ discrepancy,
making use of our fully differential \ac{NLO} computation to match the
actual measurement as closely as possible.

For the \textsc{BaBar} measurement, tau pairs are produced through
$e^+ e^-$ collisions at $\sqrt{s} = M_{\Upsilon(4S)} = 10.58\,{\rm
GeV}$. The event is then divided into a signal- and tag-hemisphere. In
order to reduce background events, rather stringent cuts on the
kinematics of the decay products $e$ and $\gamma$ in the signal
hemisphere are applied. In particular, the following requirements are
made:
\begin{align}
\cos\theta^*_{e\gamma} &\ge 0.97, &
0.22\,{\rm GeV} &\le E^*_{\gamma} \le 2.0\,{\rm GeV}, &
M_{e\gamma} &\ge 0.14\,{\rm GeV}\, .
\label{cut:taue}
\end{align}
All the quantities are given in the centre-of-mass frame. These cuts
can be easily implemented in our code. To this end, we generate taus
in their rest frame, boost them to a frame such that they have energy
$\sqrt{s}/2$ and then apply the cuts \eqref{cut:taue} in this boosted
frame. As we will see, the \ac{NLO} corrections will have an important
effect when `undoing' the cuts, i.e. when extracting the exclusive
\ac{BR} (with only the cut $E_\gamma > 10\,{\rm MeV}$ in the tau
rest frame).

\begin{figure}[t]
\centering
\begin{tabular}{c||c|c}
                          & $\tau\to \bar\nu\nu e\gamma$& $\tau\to \bar\nu\nu\mu\gamma$\\\hline
\ac{LO}                   & $1.834(1) \cdot 10^{-2}$    & $3.662(1)\cdot 10^{-3}$      \\
exclusive \ac{NLO}        & $1.645(1) \cdot 10^{-2}$    & $3.571(1)\cdot 10^{-3}$      \\
inclusive \ac{NLO}        & $1.727(3) \cdot 10^{-2}$    & $3.604(1)\cdot 10^{-3}$      \\\hline
$\mathcal{B}_{\text{exp}}$& $1.847(54)\cdot 10^{-2}$    & $3.69(10)\cdot 10^{-3}$      \\\hline\hline
$\epsilon_{\mathrm{ LO}} $ & $48.55(1)$                 & $4.966(1)$                   \\
$\epsilon_{\mathrm{NLO}} $ & $44.80(1)$                 & $4.911(1)$                   \\
$\epsilon' =   \epsilon_{\mathrm{NLO}}/\epsilon_{\mathrm{LO}}$ 
                           & $0.923(1)$                 & $0.989(1)$                   \\
$\epsilon' \cdot
 \mathcal{B}_{\text{exp}}$ & $1.704(50)\cdot 10^{-2}$   & $3.65(10) \cdot 10^{-3}$     \\
\end{tabular}

\renewcommand{\figurename}{Table}
\caption{Branching ratios for the radiative decays of the $\tau$. The
minimum photon energy is $10\,{\rm MeV}$. For the theoretical results
only the numerical error due to the Monte Carlo integration is given.
The errors on the experimental results are combined statistical and
systematic errors, as given by~\cite{Lees:2015gea}.
}
\label{tab:branching}
\end{figure}

In order to illustrate this we have devised the following simplified
scheme: let $N_{\text{obs}}$ be the measured number of events
including all cuts. To obtain the \ac{BR} this is multiplied by a
factor $\epsilon_{\text{(N)LO}}^{\text{exp}}$
\begin{align}
\mathcal{B}^{\mathrm{(N)LO}}_{\text{exp}} = \epsilon_{\mathrm{(N)LO}}^{\text{exp}} \cdot
N_{\text{obs}} = 
\epsilon_{\text{det}}\cdot
\epsilon_{\mathrm{(N)LO}} \cdot N_{\text{obs}} \,.
\end{align}
$\epsilon_{\text{det}}$ contains detector efficiencies needed to
compute the fiducial \ac{BR}. On the other hand,
$\epsilon_{\mathrm{(N)LO}}$ is a theoretical correction factor that is
needed to convert the actually measured \ac{BR} with the
cuts~\eqref{cut:taue} to the desired \ac{BR} with $E_{\gamma}\ge
10\,{\rm MeV}$.  This factor can be computed easily at \ac{LO} and
\ac{NLO}\footnote{Note that, to remain consistent with the discussion
above, we will denote the decay rate by $\sigma$ instead of $\Gamma$.}
\begin{align}
\epsilon_{\mathrm{(N)LO}} =
\frac{\sigma_{\mathrm{(N)LO}}^{\text{total}}}
  {\sigma_{\mathrm{(N)LO}}^{\text{with cuts}}} \bigg|_{\rm theory} \,,
\end{align}
where $\sigma_{\mathrm{(N)LO}}^{\text{total}}$ and
$\sigma_{\mathrm{(N)LO}}^{\text{with cuts}}$ again refer to the cut
$E_{\gamma}\ge 10\,{\rm MeV}$ and the cuts~\eqref{cut:taue},
respectively. More precisely, we require that exactly one photon
passes the cuts. To assess the importance of \ac{NLO} corrections when
extracting $\mathcal{B}_{\text{exp}}$ we write
\begin{align}
\mathcal{B}^{\mathrm{NLO}}_{\text{exp}}
= \epsilon_{\mathrm{NLO}}^{\text{exp}} \cdot N_{\text{obs}}
= \epsilon_{\mathrm{NLO}} \cdot \epsilon_{\text{det}}\cdot N_{\text{obs}}
= \frac{\epsilon_{\mathrm{NLO}}}{\epsilon_{\mathrm{LO}}} \cdot 
  \mathcal{B}^{\mathrm{LO}}_{\text{exp}}
= \epsilon' \cdot \mathcal{B}^{\mathrm{LO}}_{\text{exp}} \,,
\end{align}
where we assume that $\epsilon_\text{det}$ remains unchanged by the
inclusion of radiative corrections. Thus, $\epsilon'$ is a purely
theoretical factor that describes the difference of using a \ac{LO} or
\ac{NLO} computation in the determination of
$\mathcal{B}_{\text{exp}}$. 

The results for the various factors described above are given in the
first row of Table~\ref{tab:branching}. The salient feature is that
\ac{NLO} effects are very important in the $\tau\to
e\,\nu\bar\nu\gamma$ case and amount to a correction of 7\%. Since the
corresponding {\sc BaBar} result was obtained using theory at \ac{LO} the
inclusion of the \ac{NLO} corrections changes the result from
$\mathcal{B}_{\text{exp}} = 1.847(54)\cdot 10^{-2}$ to $\epsilon'
\cdot \mathcal{B}_{\text{exp}} = 1.704(50)\cdot 10^{-2}$, in much
better agreement with the theoretical \ac{NLO} result $\mathcal{B} =
1.645(1)\cdot 10^{-2}$.

Of course, the same procedure can be repeated for the
$\tau\to\nu\bar\nu\mu\gamma$ decay. In this case, some of the cuts
applied by {\sc BaBar} are 
\begin{align}
\cos\theta^*_{\mu\gamma} &\ge 0.99, &
0.10\,{\rm GeV} &\le E^*_{\gamma} \le 2.5\,{\rm GeV}, &
M_{\mu\gamma} &\le 0.25\,{\rm GeV}\, .
\label{cut:taumu}
\end{align}
A computation of the $\epsilon'$ factor reveals that the effects here
are more modest and amount only to a correction of about 1\%. The
resulting value $\epsilon' \cdot \mathcal{B}_{\text{exp}} = 3.65(10)
\cdot 10^{-3}$ agrees well with the \ac{NLO} result $\mathcal{B} =
3.571(1)\cdot 10^{-3}$.

Obviously, this is only a simplistic and by far not complete
simulation of the full analysis.  While the cut on $E^*_{\gamma}$ has
the biggest impact, the results for the $\epsilon'$ factor actually
depend quite significantly on all the details of the cuts.  In
particular, in the presence of a second photon it is important to
precisely specify how the cuts are applied.  This can also be seen
from the rather large difference between the exclusive and inclusive
results for $\tau\to \nu\bar\nu e\gamma$.  We do not claim that this
is the conclusive resolution to the apparent $3.5~\sigma$ deviation
for the measured branching ratio of $\tau\to \nu\bar\nu e\gamma$.
However, we do claim that a proper inclusion of \ac{NLO} effects is
mandatory for such a measurement, in particular if stringent cuts on
the decay products are applied.

\section{Michel decay (\texorpdfstring{$\mu\to\nu\bar\nu e$}{mu -> nu
anti-nu e})}
\label{sec:michel}

The conventional Michel decay $\mu\to\nu\bar\nu e$ is used to
determine the Fermi constant $G_F$ by measuring the muon lifetime.
Hence, this process is of high phenomenological relevance. However,
many experiments have measured this and the present analysis is not
connected to any one experiment.

\ac{NLO} corrections to the Michel decay have been known for many
decades~\cite{Kinoshita:1958ru,PhysRev.101.866}. Using the optical
theorem, the \ac{NNLO} \ac{QED} corrections to the decay width were
calculated around the turn of the millennium, assuming vanishing
electron masses~\cite{vanRitbergen:1999fi}. Over the course of the
next decade, the electron energy spectrum, which is not infrared
finite in the limit $m_e\to0$, was calculated. At first, only its
logarithms were known
analytically~\cite{Arbuzov:2002pp,Arbuzov:2002cn}. A few years later,
the full spectrum was calculated with a numerical loop
integration~\cite{Anastasiou2005The-electron} and the original
calculation of \cite{vanRitbergen:1999fi} was extended to include mass
effects~\cite{Pak:2008qt}. It was only recently that the form factors
necessary for a fully differential calculation were
published~\cite{Chen:2018dpt, Engel:2018fsb}.

In what follows, we have included muon and electron loops but neither
tau nor hadronic contributions~\cite{vanRitbergen:1998hn,
Davydychev:2000ee}. We treat the electromagnetic coupling $\alpha$ in
the on-shell scheme, except in Table~\ref{tab:stuart} where, in order
to compare to~\cite{vanRitbergen:1999fi}, we need the \ac{msbar} coupling
$\bar\alpha \equiv \bar\alpha(\mu=M)$.

Apart from the form factors needed for $\bbit{2}{n}$, we also need
matrix elements for $\bbit{2}{n+1}$ and $\bbit{2}{n+2}$ that were
calculated using the strategies detailed in previous
chapters.

\subsection{Results for the decay rate}

The first quantity we consider is the full decay width
\begin{align}
\label{def:resnnlo}
\sigma = \sigma_0  + \frac{\bar\alpha}{\pi}\, \sigma^{(1)}  
+ \Big(\frac{\bar\alpha}{\pi}\Big)^2\, \sigma^{(2)} 
+ \mathcal{O}(\bar\alpha^3) \, ,
\end{align}
where we have pulled out factors of the \ac{msbar} coupling
$\bar\alpha/\pi$.  We compute $\sigma^{(2)}$ using the massified form
factors to obtain the leading terms in $z=m/M$, as well as the form
factor with full $m$ dependence~\cite{Engel:2018fsb, Chen:2018dpt}. We
will label these two results `massified' and `massive', respectively.
In the case of the massified result, we expand all three parts of the
integrand contributing to $\sigma^{(2)}_n$, see \eqref{eq:nnloind0}
and \eqref{eq:nnlo:n}. Of course, the exact mass dependence of
$\bbit{1}{s}$ and $\bbit{2}{ss}$ is usually much easier to obtain than
for $\M n2$. However, the complete cancellation of singularities
requires a consistent expansion in $z$ of all contributions at the
$n$-particle level.

Because the full decay rate does not contain terms $\log z \sim \log
m$ the limit $m\to 0$ exists and we can compare our massified and
massive results with the result for a massless
electron~\cite{vanRitbergen:1999fi}.  We note that in this particular
case (contrary to distributions, where $\log m$ terms exist), the
massified result is not expected to be superior to the massless
computation.

Following \cite{vanRitbergen:1999fi}, we split the result into three
parts: photonic corrections $\sigma_\gamma^{(1)}$ and
$\sigma_\gamma^{(2)}$, corrections due to an electron pair (real or
virtual) $\sigma_e^{(2)}$, and corrections due to a muon pair
(virtual) $\sigma_\mu^{(2)}$. These parts have been defined and their
analytic results in the massless case given in equations (2.11),
(2.13) and (2.15) of \cite{vanRitbergen:1999fi}.  The individual
results for the \ac{NNLO} corrections are shown in Table~\ref{tab:stuart},
where the Monte Carlo error is smaller than the significant
digits. Note that~\cite{vanRitbergen:1999fi} had to include the
`open-lepton production' $\mu\to\nu\bar\nu e\, ee$ into their
calculation of $\sigma_e^{(2)}$ to guarantee finiteness.  We have
included this process as well~\cite{Pruna:2016spf} since it
contributes to $\sigma_e^{(2)}$ (two-trace contribution) and
$\sigma_\gamma^{(2)}$ (one-trace contribution).\footnote{The amplitude
  for $\mu^-\to\nu\bar\nu e^-\,e^+e^-$ has a (anti)symmetry under
  exchange of the two $e^-$.  This gives rise to two types of
  interference terms in the matrix element: first the contribution
  that is also present without this symmetry (two-trace) and one where
  the swapped is interfered with the non-swapped contribution
  (one-trace).}

\begin{figure}
\centering
\begin{tabular}{l|ccc|c}

          & $\sigma_\gamma^{(2)}/\sigma_0$ &$\sigma_\mu^{(2)}/\sigma_0$ &$\sigma_e^{(2)}/\sigma_0$ & total\\\hline
massified & 3.42    & -0.0364 & 3.24    & 6.62 \\
massive   & 3.54    & -0.0364 & 3.16    & 6.66 \\
\hline
massless~\cite{vanRitbergen:1999fi}    & 3.56    & -0.0364 & 3.22    & 6.74 \\
\hline
 $\Delta_{\text{rel}}$ massified & $3.7\times10^{-2}$ & $ 0$
& $6.1\times 10^{-3}$ & $1.6\times10^{-2}$ \\
 $\Delta_{\text{rel}}$ massive      & $5.0\times10^{-3}$ &
$1.9\times10^{- 4}$
& $2.0\times 10^{-2}$ & $1.1\times10^{-2}$ \\
\end{tabular}

\renewcommand{\figurename}{Table}
\caption{The different contributions to $\sigma^{(2)}$. Note that
$\sigma_e^{(2)}$ also includes the process $\mu\to\nu\bar\nu e\, ee$.
See text for interpretation. The coupling $\bar\alpha(\mu=M)$ is
renormalised in the \ac{msbar} scheme. $\Delta_{\text{rel}}$ denotes the
relative difference of our results to the massless result~\cite{vanRitbergen:1999fi}.}
\label{tab:stuart}
\end{figure}

\noindent 
The results of Table~\ref{tab:stuart} merit a few comments:
\begin{itemize}

    \item
    The good agreement for the purely photonic contributions
    $\sigma_\gamma^{(2)}$ between the massive and massless result is
    due to the absence of terms $\log m$ and $m\log m$ as discussed
    by~\cite{vanRitbergen:1999fi}.

    \item
    The massified results differs by about 3\% from the massive (and
    massless) result for $\sigma_\gamma^{(2)}$. This is due to the
    mismatch between the real corrections, that were calculated with
    the full electron mass dependence, and the massified two-loop
    amplitude that only includes logarithmically enhanced mass
    effects. 

    \item
    The massified results agrees perfectly
    with~\cite{vanRitbergen:1999fi} for the $\sigma_\mu^{(2)}$ part
    because the contribution comes purely from one two-loop diagram
    that is free of any soft or collinear logarithms and hence
    effectively massless.

    \item
    The massive and massless results for $\sigma_e^{(2)}$ agree only
    up to two percent. This difference can be accounted for through
    the two-trace contribution of the open-lepton production. In the
    pure electron trace $m$ must not be neglected to lead to finite
    expressions. However, in the other trace the electron mass can be
    set to zero.  Our value of $3.16$ was calculated with full
    electron mass dependence. If we were to set $m\to0$ in the this
    trace, we would obtain $3.23$ in much better agreement
    with~\cite{vanRitbergen:1999fi}.

    \item
    The $\sigma_e^{(2)}$ part contains the factorisation anomaly,
    already discussed in Chapter~\ref{ch:twoloop} and~\cite{Engel:2018fsb}.

\end{itemize}
Note that in any case the `massive' result should be considered the
reference. Our results agree with~\cite{Pak:2008qt}.  For the pure
mass effects of the photonic part, this agreement is only at the 20\%
level. This is due to large numerical cancellations between
$\sigma^{(2)}_n$, $\sigma^{(2)}_{n+1}$ and $\sigma^{(2)}_{n+2}$ which
make the extraction of a few-percent effect on the \ac{NNLO} corrections
numerically challenging. In fact, an efficient numerical evaluation of
the integrals with full mass dependence~\cite{Chen:2018dpt} has only
recently been implemented~\cite{Naterop:2019}.

\subsection{The electron energy spectrum}

\begin{figure}
\centering
\scalebox{0.8}{\input{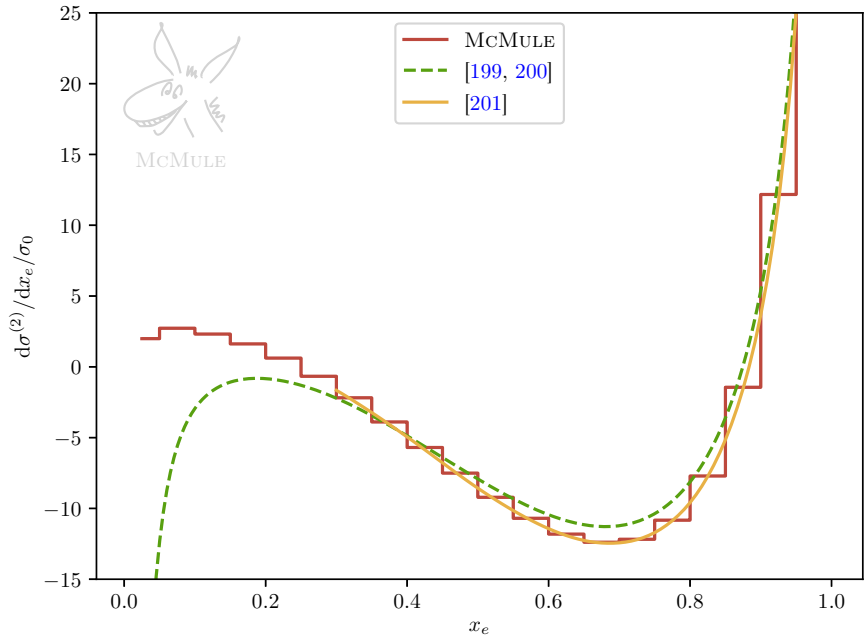}}
\caption{The \ac{NNLO} corrections to the  electron energy
  spectrum, omitting a factor $(\alpha/\pi)^2$. The logarithmic
  contributions of~\cite{Arbuzov:2002pp,Arbuzov:2002cn} (green dashed)
  agree reasonably well with our massive result (red histogram) for large
  $x_e$. Adding the constant terms
  of~\cite{Anastasiou2005The-electron} (orange) we obtain very good
  agreement. }
\label{fig:xe}
\end{figure}

In order to validate our computation, we consider the \ac{NNLO} corrections
to the normalised electron energy spectrum $x_e=2E_e/M$ and compare them
to results available in the literature. If two (negatively charged)
electrons are present in the final state, we include both of them in
the $x_e$ distribution.  The leading and sub-leading logarithmic
contributions for this observable were calculated
in~\cite{Arbuzov:2002pp,Arbuzov:2002cn}.  Because this corresponds to
a strict expansion in $z$, we expect good agreement for large $x_e$ as
noticed in~\cite{Engel:2018fsb}.  In Figure~\ref{fig:xe} we compare
the two results and see that the differences are compatible with the
constant (logarithm-free) terms missing
in~\cite{Arbuzov:2002pp,Arbuzov:2002cn}. These terms were computed
numerically and shown in a plot for $x_e>0.3$
in~\cite{Anastasiou2005The-electron}. If we include these constant
terms of~\cite{Anastasiou2005The-electron}, we obtain perfect
agreement with our result, using the massive form factors.  Note that
the difference between massified and massive result in
Figure~\ref{fig:xe} is at the percent level and only becomes visible
around the zero crossing at $x_e\approx0.21$ and $x_e\approx0.88$,
never changing the overall picture. The on-shell coupling
$(\alpha/\pi)^2$ is omitted in the results shown in the
Figure~\ref{fig:xe}.

With a fully differential Monte Carlo code, we can compute arbitrary
distributions, including cuts. As an example, we consider again the
normalised electron energy spectrum but impose a cut on photon
emission through lepton isolation. Concretely, we restrict the total
energy of all photons within a cone of angle $\theta\equiv
\sphericalangle(\vec p_e, \vec p_\gamma) = 37^\circ$ (i.e. a cone with
$|\cos\theta|>0.8$) around the electron to be less than $10\,{\rm
MeV}$.

\begin{figure}
\centering
\scalebox{0.8}{\input{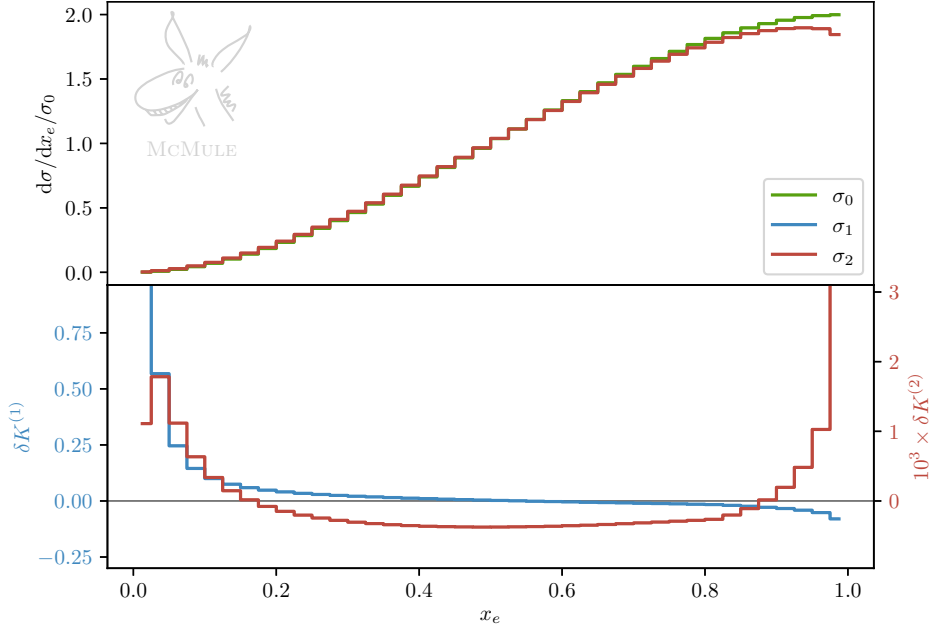}}
\caption{Top panel: The normalised electron energy spectrum at \ac{LO} 
  (green histogram) and \ac{NNLO} (red histogram) with a cut on photon
  emission. The lower panel shows the \ac{NLO} and \ac{NNLO} $K$
  factors.}
\label{fig:xecut}
\end{figure}

The results are shown in Figure~\ref{fig:xecut}. Comparing the
normalised \ac{NNLO} result (red histogram) to the normalised \ac{LO}
result (green histogram) in the top panel reveals that only for large
$x_e$ the corrections to the shape are relevant. This is driven by the
\ac{NLO} corrections. They are large at both ends of the $x_e$
spectrum, as shown by the \ac{NLO} $K^{(1)}$ factor
\begin{align}
K^{(i)}=\frac{\D\sigma_i/\D x_e}{\D\sigma_{i-1}/\D x_e} = 1+\delta K^{(i)}
\,.
\end{align}
Typically, the \ac{NNLO} corrections are below 0.1~\% and even in the
regions of huge \ac{NLO} corrections they are below 0.5\%.

\subsection{Michel decay as a background in MEG}
The Michel decay is not just a signal for the measurement of the muon
lifetime. It also serves as a background to \ac{BSM} searches. The
light new \ac{LFV} particle $J$ -- introduced above as the signal over
$\mu\to\nu\bar\nu e\gamma\gamma$ -- for example may not decay promptly
but actually leave the detector as missing energy, resulting in the
difficult signature $\mu\to e+\text{invisible}$. This process is
indistinguishable from the Michel decay $\mu\to\nu\bar\nu e$. The
experimental searches are now hunting for a miniscule deviation from
the only spectrum available -- $\D\mathcal{B}/\D x_e$.

Unfortunately, the delicate experimental situation requires a full
detector simulation based on the best possible theory prediction.
Hence, we need to generalise the discussion above. MEG will not be
using the photon detector in search for $\mu\to eJ$.  Hence, we can be
inclusive w.r.t. photon emission, simplifying our analysis. We exploit
that the decay of a polarised muon is completely described to all
orders by an \term{isotropic}, i.e. polarisation independent, and an
\term{anisotropic}, i.e. polarisation dependent, part
\begin{align}
\frac{\D^2\sigma}{\D x_e\,\D(\cos\theta)} = \Gamma_0 \big(f(x_e) +
P\cdot \cos\theta g(x_e)\big)\,,
\label{eq:fandg}
\end{align}
where $x_e$ is the electron energy fraction and $\theta$ the angle
between the polarisation axis and the outgoing electron. Neglecting
mass effects, $f$ and $g$ can be written at \ac{LO} as
\begin{align}\begin{split}
f^{(0)}(x_e)&=x_e^2\big(3-2x_e\big)\,,\\
g^{(0)}(x_e)&=x_e^2\big(1-2x_e\big)\,.
\end{split}\end{align}
For the allowed energies $x_e\lesssim 1$, $f$ is always positive.
However, $g$ crosses zero at $x_e'=1/2$.\footnote{Taking into account
mass effects, this happens at $x_e'=1/2 + (3/2)m^2/M^2$}.

\begin{figure}
\centering
\subfloat[The isotropic $f$ function]{
\scalebox{0.8}{\input{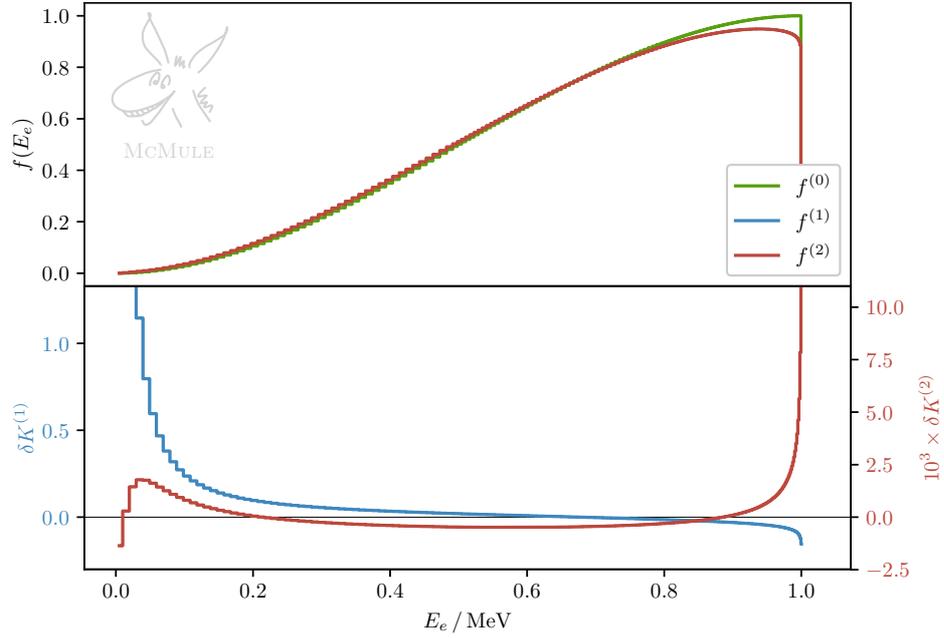}}
}

\subfloat[The anisotropic $g$ function]{
\scalebox{0.8}{\input{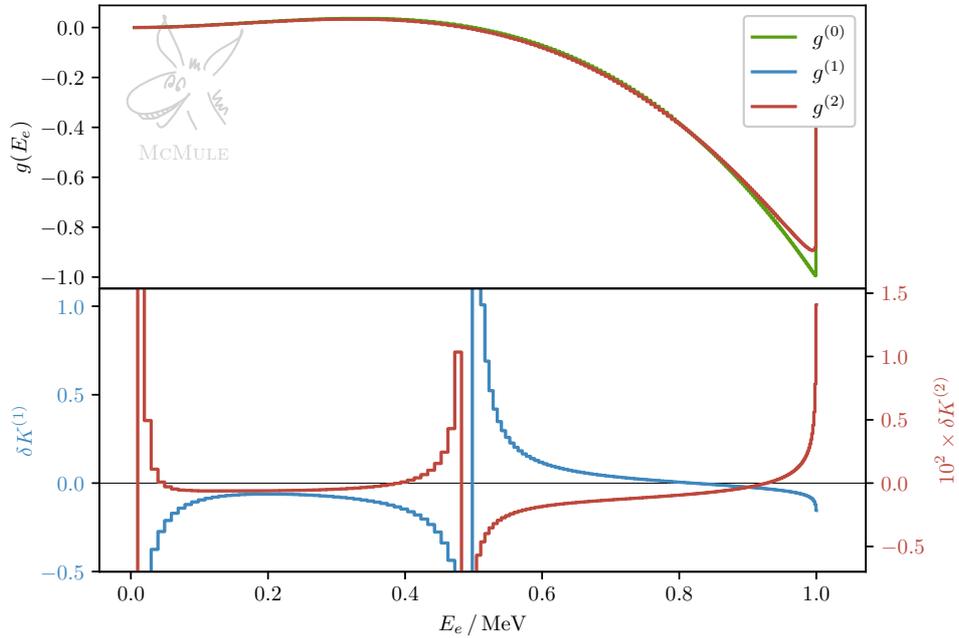}}
}
\caption{The isotropic and anisotropic functions $f(x_e)$ and $g(x_e)$
as defined in~\eqref{eq:fandg}. The upper panel shows the functions
and the lower panel the effect due to radiative corrections as
$\delta K^{(i)} = f^{(i)}/f^{(i-1)}$. The left axis of the lower panel
shows the \ac{NLO} $\delta K^{(1)}$ and the right axis the \ac{NNLO}
$\delta K^{(2)}$. See text for a detailed discussion}.
\label{fig:fandg}
\end{figure}

Beyond \ac{LO}, we can calculate $f(x_e)$ and $g(x_e)$ with two runs
of \mcmule{}\footnote{In reality there are more runs required to
sample the $x$ distributions precise enough} by defining
\begin{align}
\frac{\D\sigma_-}{\D x_e}=\int_{-1}^0\D(\cos\theta)\ \frac{\D^2\sigma}{\D x_e\,\D(\cos\theta)}
\qquad\text{and}\qquad
\frac{\D\sigma_+}{\D x_e}=\int_{ 0}^1\D(\cos\theta)\ \frac{\D^2\sigma}{\D x_e\,\D(\cos\theta)} 
\,.
\end{align}
By combining $\sigma_+$ and $\sigma_-$ we can obtain results for $f$
and $g$ allowing a full detector simulation. The numerical results for
$f$ and $g$ are shown in Figure~\ref{fig:fandg}. 

These results merit a few comments
\begin{itemize}

    \item
    For small and large electron energies the \ac{NLO} $K$ factor
    becomes very large, both for $f$ and $g$. This is a fundamental
    change in the kinematic situation due to extra photon emission.
    However, these large corrections are almost entirely \ac{LL} and
    can be resummed easily enough. Especially for large $x_e$, soft
    photon emission gives rise to logarithms of the form
    $\log(1-x_e)$ that can just be exponentiated. This is currently
    being implemented~\cite{Gurgone:msc}.

    \item
    The corrections are still quite large at \ac{NNLO} towards the
    endpoint though nowhere nearly as large as at \ac{NLO}. The
    build-up of the \ac{LL} tower can hence be clearly seen.

    \item
    Around $x_e'\approx 1/2$ the $K$ factor for $g$ diverges.  This is
    because soft-photon emissions slightly shift the zero crossing of
    $g$ away from $x_e'$ resulting in large relative corrections.

\end{itemize}

\section{Mu3e}\label{sec:mu3e}
The rare muon decay $\mu\to\nu\bar\nu eee$ is a background for
Mu3e, looking for $\mu\to eee$. Mu3e -- just as MEG -- operates on the
$\pi{\rm E}5$ beam line at \ac{PSI} using positive muons and hence, we
again define the $z$ axis s.t. the muon polarisation is $\vec P_\mu =
-0.85\,\vec z$. We model the Mu3e detector with the cuts
\begin{align}
\big|\cos\theta_i\big| < 0.8
\qquad\text{and}\qquad
E_i > 10\,\mev\,.\label{eq:mu3e}
\end{align}
Without special modifications, Mu3e is not sensitive to photons so
that we accept any photon emission.

\begin{figure}
    \centering
    \scalebox{0.8}{\input{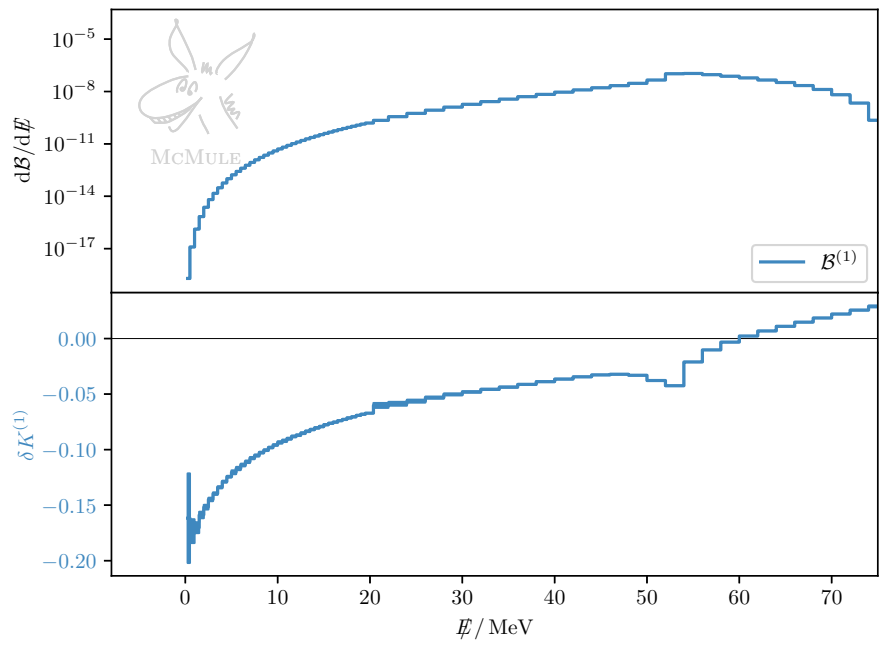}}
    \caption{The invisible energy spectrum with the Mu3e cuts at
    \ac{NLO}. }

    \label{fig:mu3e}
\end{figure}

A simple observable is the invisible energy $\Einv$
\begin{align}
\Einv = M-\sum_i E_i \,,
\end{align}
where the sum runs over all charged tracks, i.e. the electron and the
two positrons. Note that this includes the energy of undetected
photons. The resulting spectrum is shown in Figure~\ref{fig:mu3e}. For
this plot, too, we had to perform dedicated runs with a cut of
$\Einv<20\,{\rm MeV}$ in order to obtain a good enough precision for
this region. This is the reason why the statistical error briefly goes
down again for small $\Einv$.

The \ac{NLO} corrections are negative except for a small region of
maximal $\Einv$. In the low-energy tail, the corrections exceed
$−10\%$, due to the ever-present large logarithms.  Hence, there are
fewer background events to $\mu\to 3e$ from the rare decay than
expected from tree-level simulations. The cuts on the electron and
positrons~\eqref{eq:mu3e} are the reason for the sharp fall of the
distribution at $\Einv = M-30\,{\rm MeV}$. 

The kink in the distribution is at about $M/2$, shifted to somewhat
lower values due to the effects of the non-vanishing electron mass. In
fact, due to the additional real radiation of a photon, the \ac{NLO}
corrections amount to shifting the distribution
$\D\mathcal{B}/\D\Einv$ to higher energies.

\chapter{Outlook}\label{ch:conclusion}

\mcmule{} supports various phenomenologically relevant processes at
\acs{NLO} and \acs{NNLO} in \ac{QED} with massive fermions. The code's
development was driven by and implemented in close cooperation with
the experiments that it will continue to serve. 

When designing \mcmule{} we have two more or less distinct groups of
users in mind: those who just wish to calculate tailored observables
and those who wish to extend it by adding new processes.  The code's
structure serves both groups. 

Defining new observables is as easy as changing a single file; in fact
we will provide a library of legacy results~\cite{Ulrich:legacy} with
the user files to reproduce all previous works by (or related to)
\mcmule{}~\cite{Pruna:2016spf, Pruna:2017upz, Ulrich:2017adq,
Engel:2019nfw, Banerjee:2020rww}.

Thanks to the technical development of massification and \acs{FKS2},
adding new processes at \acs{NNLO} is also relatively painless.
This is good too because the demand for massive \acs{QED} calculation
will not abate. If anything, it will grow as more low-energy
experiments push for higher and higher accuracy.

The bottleneck in the computation of cross sections for massive QED at
\ac{NNLO} is the availability of the matrix element $\M{n}2$. These
computations are usually much simpler if some (or all) fermion masses
$m$ are set to zero.  Unfortunately, this also spoils \ac{FKS2}.
However, if $m$ is small compared to the other kinematic quantities,
an option is to start from the massless case and subsequently massify
$\M{n}{2}$. As we have seen, this converts the collinear $1/\epsilon$
singularities of $\M{n}2(m=0)$ into $\log(m)$ terms that will cancel
against corresponding `singularities' of the real corrections. In
addition, it retains the finite $\log(m)$ terms in $\bbit{2}{n}$ that
are present in differential distributions.  However, terms $m \log(m)$
that vanish in the limit $m\to 0$ will be neglected.  Using full $m$
dependence in $\bbit{2}{n+1}$, but only partial $m$ dependence in
$\bbit{2}{n}$ through a massified $\M{n}{2}$ results in a mismatch in
terms $m \log(m)$.  Since the whole procedure of massification is
anyway only correct up to such terms, the mismatch should not cause
additional problems, as the terms relevant to the $\xc$ independence
can be included exactly.

It should be noted that a similar procedure in $\bbit{2}{n+1}$ is less
straightforward. It is not possible to naively use massification for
$\M{n+1}{1}$. The remaining phase-space integration over the
additional particle requires a non-vanishing $m$ to avoid a collinear
singularity. While this could be patched, massification relies on the
fact that the small mass is the smallest scale of the process.  While
this is certainly often the case, it ceases to be true once we allow
for soft or collinear photon emission. Hence, a crucial step will be
working out the massification for real-emission matrix elements.

The extension of massification is closely connected to the problem of
collinear stabilisation, i.e. finding an efficient numerical treatment
of \acs{PCS}s. So far we have solved this problem by dedicated tuning
of the phase-space. However, this ceases to be feasible for
high-multiplicity processes.  In fact, both problems -- numerical
stabilisation and massification of real-emissions matrix elements
might be solved with the same method.  The idea is to subtract
pseudo-collinear regions from the integrand and add them back in
integrated form~\cite{Dittmaier:1999mb}. However, care must be taken
when integrating these terms analytically to retain a
fully-differential code. This is because, in contrast to \ac{QCD}
where collinear emission cannot be resolved, collinear photon emission
of an electron can very much be resolved experimentally. 

To actually compute $\M{n+1}1$, we use \fdf{} in conjunction with
COLLIER. The usage of \fdf{} over \cdr{} is already a major
simplification. However, maybe a much better strategy exists.
Fundamentally, we are not interested in $\M{n+j}\ell$ but in the
eikonal-subtracted $\fM{n+j}\ell$ that we traditionally obtain from
$\M{n+j}\ell$. However, $\fM{n+j}\ell$ is \emph{finite}. This seems to
suggest that it should be possible to calculate it without ever
leaving $\Sf$ using numerical methods. Schematically, this is similar
to the \mtextsc{fdu} scheme (four-dimensional unsubtraction)~\cite{
Hernandez-Pinto:2015ysa, Sborlini:2016gbr, Sborlini:2016hat,
Rodrigo:2016hqc} that directly combines real and virtual corrections
using only four-dimensional quantities.

Unfortunately, just adding new processes at \acs{NNLO} -- M{\o}ller
scattering and photon pair production come to mind -- is not enough.
Already now, \acs{NNLO} accuracy fails to be good enough for some
applications. To reach the required accuracy for MUonE, it may become
necessary to calculate the electronic \acs{n3lo} corrections. While we
do have a suitable subtraction scheme with FKS$^3$, the relevant
matrix elements are presently not known and most likely will not
become known with the full mass dependence in time. Further, even if
they were available, they would likely be very complicated analytic
functions that are not directly suited for numerical integration. The
big bottleneck here is the real-virtual-virtual contribution $\M{n+1}2$.
The electronic corrections can be constructed from $\gamma^*\to q\bar
qg$~\cite{Gehrmann:2000zt,Gehrmann:2001ck} that are only known for
vanishing quark (or in our case, electron) masses. This makes the need
for massification of real-emission matrix element even more pressing.

Whatever happens with the \acs{n3lo} calculation, it presently seems
exceedingly unlikely that we could go beyond even that to N$^4$LO.
Unfortunately, a naive extrapolation of the trend observed so-far in
the radiative corrections to $\mu$-$e$ scattering seem to suggest that
we need exactly that. Luckily, resummation provides a way out as most
of the corrections come from a single source: large -- and predictable
-- logarithms. In the framework of \mcmule{}, the only way to
implement resummation is adding a \acs{PS} that will resum the full
\acs{LL} tower. This way, we can capture the largest contributions of
all orders without sacrificing our ability to calculate arbitrary
observables.

{\parskip=5pt 
To summarise, \mcmule{} has allowed for relatively easy implementation
of \ac{NNLO} calculations in QED with massive fermions while paving
the way to \ac{n3lo} calculations. In the near future, these will be
matched to \ac{PS} in order to resum the \acs{LL} tower.  Further,
more exotic technical development, such as the numerical and direct
evaluation $\fM{n+j}\ell$, is being proposed.
}

\appendix
\chapter{Conventions}\label{ch:conventions}

When calculating loop integrals one often encounters factors of
$\log(4\pi)$ and $\gamma_E$. These are artefacts of expanding the
$d$-dimensional spherical integration in $\epsilon$ and hence not
physical. Because they will drop out in any physical result, there is
no need to include them to begin with. Hence, we remove the
relevant factor directly by defining the loop measure as~\cite{Bell:2006tz}
\begin{align}
\mu^{2\epsilon}\frac{\D^dk}{(2\pi)^d}
\to
[\D k] = \mu^{2\epsilon} \Gamma(1-\epsilon)\frac{\D^dk}{\I(4\pi)^{-d/2}}
\label{eq:measure}
\end{align}
With this conventions, the tadpole integral reads
\begin{align}
\int[\D k] \frac1{(k^2-\Delta)^\alpha} =
(-1)^\alpha\Delta^{2-\epsilon-\alpha}\times
  \frac{\Gamma(1-\epsilon)\Gamma(\alpha-2+\epsilon)}{\Gamma(\alpha)}
\end{align}
Here we also have removed a factor $16\pi^2$ that is physical but
cumbersome to write. What factors to include is somewhat arbitrary as
long as this is done consistently for virtual and real correction and
all factors that remain for $\epsilon\to0$ are added back. In fact,
Package-X~\cite{Patel:2015tea} uses a different convention
\begin{align}
[\D k]_X = \Bigg(
    \frac{\I\E^{-\gamma_E\epsilon}}{(4\pi)^{d/2}}\Bigg)^{-1}
    \D^dk\,.
\end{align}
Both conventions have the effect of removing unwanted and distracting
constants and are equally valid. However, they do differ at
$\mathcal{O}(\epsilon^2)$ so attention must be paid for massless
calculations where the highest pole is $\epsilon^{-2}$ or when
requiring the $\mathcal{O}(\epsilon)$ terms of a massive one-loop
amplitude. The relevant conversion factor is
\begin{align}
\frac{[\D k]}{[\D k]_X} =
    \frac{\Gamma(1-\epsilon)}{\E^{\epsilon\gamma_E}}
    = 1+\frac{\zeta_2}2\epsilon^2 + \frac{\zeta_3}3 \epsilon^3
        +\mathcal{O}(\epsilon^4)\,.
\end{align}
Of course other conventions exist as well.

\chapter{Constants in \textsc{fdh}}
\label{ch:fdhconst}
\begingroup
\allowdisplaybreaks

For the benefit of the reader, we will collect all relevant constants
for \fdh{} calculations in this chapter.  These include
renormalisation constants (Section~\ref{sec:const:ren}), the anomalous
dimensions required for the \ac{IR} prediction
(Section~\ref{sec:const:scet}) and the massification constants
(Section~\ref{sec:const:massify}). In all cases, the \cdr{} or \hv{}
limit can be obtained by setting $\neps\to0$.

All results present here were previously published in~\cite{
Gnendiger:2014nxa, Gnendiger:2016cpg, Broadhurst:1991fy} for the
renormalisation constants, \cite{Gnendiger:2014nxa, Gnendiger:2016cpg}
for the \ac{SCET} constants and \cite{Engel:2018fsb} for the massification
constants. Some results in different schemes have obviously been
published before.

All results presented use Feynman gauge.

\section{Renormalisation constants}\label{sec:const:ren}
\newcommand{\inclca}[1]{#1}

In this section we present the renormalisation constants $Z_2$ and
$Z_m$ up to $\mathcal{O}(\alpha^2)$ as well as $Z_\alpha$ and
$Z_{m_\epsilon}$ up to $\mathcal{O}(\alpha)$. With this we can
calculate processes to the muon decay as well as the electronic
corrections to $\mu$-$e$ scattering.

All results (except $Z_{\alpha_i}$) in this section are given in the
\emph{unrenormalised} coupling in accordance to the procedure set out
in Section~\ref{sec:renorm:practical}. Further, the scale $\mu$ of the
integration is set to the muon mass $\mu=M$, assuming we are
renormalising the muon field and mass. Obtaining the results for the
electron is straightforward by re-introducing logarithms from
$(m^2/\mu^2)^{\epsilon}$ per coupling.

Using the technology we set out in Chapter~\ref{ch:reg} we find for
$\delta m_\epsilon$
\begin{align}\begin{split}
\delta{m_\epsilon} &= 
\Big(\frac{\alpha_{0,e}}{4\pi}\Big) \big(2 M^2 n_h\big)
\Gamma(1-\epsilon)\Gamma(\epsilon-1)+ \mathcal{O}(\alpha^2)
\\&
=\Big(\frac{\alpha_{0,e}}{4\pi}\Big) \big(-2 M^2 n_h\big)
\Bigg[
    \frac1\epsilon + 1 + (1 + \zeta_2) \epsilon + (1 + \zeta_2) \epsilon^2
\Bigg]
    +\mathcal{O}(\alpha^2,\epsilon^3)
\end{split}\end{align}
A similar term is required for $n_m$ and $m^2$.

The mass renormalisation is
\begin{align}
Z_m = 1 &+ \Big(\frac{\alpha_0}{4\pi}\Big) C_F\left[
    -\frac{3}{\epsilon }-4+(-3 \zeta _2-8) \epsilon +(-4 \zeta _2-16) \epsilon ^2
    +\mathcal{O}(\epsilon ^3)
\right]
\notag\\&+\Big(\frac{\alpha_{0,e}}{4\pi}\Big) C_F \neps\left[
    -\frac{1}{2 \epsilon }-\frac{1}{2}-\frac12(\zeta_2+1) \epsilon-\frac12 (\zeta _2-1) \epsilon ^2
    +\mathcal{O}(\epsilon ^3)
\right]
\notag\\&
+
\Big(\frac{\alpha_0}{4\pi}\Big)^2\Bigg\{
C_F^2\left[
    \frac{9}{2 \epsilon ^2}+\frac{45}{4 \epsilon }
    -21 \zeta _2-12 \zeta _3+48 \zeta _2 \log2+\frac{199}{8}
    +\mathcal{O}(\epsilon)
  \right] 
\notag\\&\quad
\inclca{+ C_AC_F \left[
    -\frac{11}{2 \epsilon ^2}-\frac{91}{4 \epsilon }
    -3 \zeta _2+6 \zeta _3-24 \zeta _2 \log2-\frac{605}{8}
    +\mathcal{O}(\epsilon)
\right] 
\inclca{
\notag\\&\quad 
+C_AC_F\neps \left[
    \frac{1}{4 \epsilon ^2}+\frac{9}{8 \epsilon }
    +\frac{3}{2} \zeta _2+\frac{63}{16}
    +\mathcal{O}(\epsilon)\right]
}
\notag\\&\quad}
+ n_fC_F\left[
    \frac{1}{\epsilon ^2}+\frac{7}{2 \epsilon }+6 \zeta _2+\frac{45}{4}
    +\mathcal{O}(\epsilon)
\right]
+ n_hC_F\left[
    \frac{1}{\epsilon ^2}+\frac{7}{2 \epsilon }-6 \zeta _2+\frac{69}{4}
    +\mathcal{O}(\epsilon)
\right] 
\notag\\&\quad 
+ n_mC_F \Bigg[
    \frac{1}{\epsilon ^2}+\frac{7}{2 \epsilon }+
        \frac{11}{16} - \frac12z + \Big(\frac32 - 2 z^2 + z^4\Big) \zeta_2
        + 8 z^4 H_{0,0}(z) 
        +4 z^2 H_0(z)\notag\\&\qquad\qquad
        + 4 (z-1)^2 (z^2+z+1) H_{1,0}(z)
        - 4 (z^4+z^3+z+1) H_{-1,0}(z) 
    +\mathcal{O}(\epsilon)
\Bigg]\Bigg\}
\notag\\&
+
\Big(\frac{\alpha_0}{4\pi}\Big)\Big(\frac{\alpha_{0,e}}{4\pi}\Big)\neps\Bigg\{
C_F^2\left[
    -\frac{3}{2 \epsilon }+ 6 \zeta_2-\frac{23}{4}+\mathcal{O}(\epsilon)
\right]
\inclca{
+C_AC_F\left[
    -\frac{3}{4 \epsilon }- 3 \zeta_2-\frac{11}{8}+\mathcal{O}(\epsilon)
\right]
}
\Bigg\}
\notag\\&
+
\Big(\frac{\alpha_{0,e}}{4\pi}\Big)^2\neps\Bigg\{
C_F^2\left[
    \frac{1}{\epsilon ^2}+\frac{3}{\epsilon }+6+\mathcal{O}(\epsilon)
\right]
+C_F^2\neps\left[
    -\frac{1}{8 \epsilon ^2}-\frac{13}{16 \epsilon }+\frac{5}{4} \zeta _2-\frac{75}{32}
    +\mathcal{O}(\epsilon)
\right]
\inclca{
\notag\\&\quad
C_AC_F \left[
    -\frac{1}{2 \epsilon ^2}-\frac{3}{2 \epsilon }-3+\mathcal{O}(\epsilon)
\right]
+C_AC_F\neps\left[
    \frac{1}{4 \epsilon ^2}+\frac{3}{4 \epsilon }+\frac{3}{2}+\mathcal{O}(\epsilon)
\right]
}
\notag\\&\quad
+C_Fn_f\left[
    \frac{1}{4 \epsilon ^2}+\frac{5}{8 \epsilon }+\frac{3}{2} \zeta _2+\frac{11}{16}
    +\mathcal{O}(\epsilon)
\right]
+C_Fn_h\left[
    \frac{1}{4 \epsilon ^2}+\frac{5}{8 \epsilon }+\frac{1}{2}\zeta _2+\frac{3}{16}
    +\mathcal{O}(\epsilon)
\right]
\notag\\&\quad
+C_Fn_m\Bigg[
    \frac{1}{4 \epsilon ^2}+\frac{5}{8 \epsilon }
    + \frac{11}{16} - \frac12z^2 + (\frac32 - 2 z^2 + z^4) \zeta_2
    +2 (z^2-2) z^2 H_{0,0}(z)
    -z^2 H_0(z)\notag\\&\qquad\qquad
    -(z^2-1)^2 H_{-1,0}(z)
    +(z^2-1)^2 H_{1,0}(z)
    +\mathcal{O}(\epsilon)
\Bigg]
\Bigg\}
+ \mathcal{O}(\alpha^3)\,.
\end{align}

Similarly, the wave-function is renormalised through
\begin{align}
Z_2 = 1 &+ \Big(\frac{\alpha_0}{4\pi}\Big) C_F\left[
    -\frac{3}{\epsilon }-4+(-3 \zeta _2-8) \epsilon +(-4 \zeta _2-16) \epsilon ^2
    +\mathcal{O}(\epsilon ^3)
\right]
\notag\\&+\Big(\frac{\alpha_{0,e}}{4\pi}\Big) C_F \neps\left[
    -\frac{1}{2 \epsilon }-\frac{1}{2}-\frac12(\zeta_2+1) \epsilon-\frac12 (\zeta _2-1) \epsilon ^2
    +\mathcal{O}(\epsilon ^3)
\right]
\notag\\&
+
\Big(\frac{\alpha_0}{4\pi}\Big)^2\Bigg\{
C_F^2\left[
    \frac{9}{2 \epsilon ^2}+\frac{51}{4 \epsilon }
    -69 \zeta _2-24 \zeta _3+96 \zeta _2 \log2+\frac{433}{8}
    +\mathcal{O}(\epsilon)
\right]
\notag\\&\quad
\inclca{C_AC_F\left[
    -\frac{11}{2 \epsilon ^2}-\frac{101}{4 \epsilon }
    +19 \zeta _2+12 \zeta _3-48 \zeta _2 \log2-\frac{803}{8}
    +\mathcal{O}(\epsilon)
\right]
\notag\\&\quad
+C_AC_F\neps\left[
    \frac{1}{4 \epsilon ^2}+\frac{11}{8 \epsilon }+\frac{3}{2} \zeta _2+\frac{81}{16}
    +\mathcal{O}(\epsilon)
\right]
\notag\\&\quad
}
+C_Fn_f\left[
    \frac{1}{\epsilon ^2}+\frac{9}{2 \epsilon }+6 \zeta _2+\frac{59}{4}
    +\mathcal{O}(\epsilon)
\right]
+C_Fn_h\left[
    \frac{2}{\epsilon ^2}+\frac{19}{6 \epsilon }+\frac{1139}{36}-12 \zeta _2
    +\mathcal{O}(\epsilon)
\right]
\notag\\&\quad
+C_Fn_m\Bigg[
    \frac{2}{\epsilon ^2}
    +\frac1{\epsilon }\Big(\frac{19}{6}-4 H_0(z)\Big)
    +\frac{635}{36} + 14 z^2 + (8 - 18 z - 30 z^3 + 12 z^4) \zeta_2\notag\\&\qquad\qquad
    -2 (6 z^4+5 z^3+3 z+2) H_{-1,0}(z)
    +2 (6 z^4-5 z^3-3 z+2) H_{1,0}(z)\notag\\&\qquad\qquad
    +8 (3 z^4+2) H_{0,0}(z)
    +(8 z^2+\frac{16}{3}) H_0(z)
    +\mathcal{O}(\epsilon)
\Bigg]\Bigg\}
\notag\\&
+
\Big(\frac{\alpha_0}{4\pi}\Big)\Big(\frac{\alpha_{0,e}}{4\pi}\Big)\neps\Bigg\{
+C_F^2\left[
    \frac{3}{2 \epsilon }+\frac{47}{4}-6 \zeta _2
    +\mathcal{O}(\epsilon)
\right]
\inclca{+C_AC_F\left[
    -\frac{9}{4 \epsilon }+\zeta _2-\frac{77}{8}
    +\mathcal{O}(\epsilon)
\right]}
\Bigg\}
\notag\\&
+
\Big(\frac{\alpha_{0,e}}{4\pi}\Big)^2\neps\Bigg\{
C_F^2\left[
    \frac{1}{\epsilon ^2}+\frac{2}{\epsilon }+4 \zeta _2-3
    +\mathcal{O}(\epsilon)
\right]
+C_F^2\neps\left[
    -\frac{1}{8 \epsilon ^2}-\frac{3}{16 \epsilon }+\frac{91}{32}-\frac{7}{4} \zeta _2
    +\mathcal{O}(\epsilon)
\right]
\notag\\&\quad
\inclca{
+C_AC_F\left[
    -\frac{1}{2 \epsilon ^2}-\frac{1}{\epsilon } +\frac{3}{2}-2 \zeta _2
    +\mathcal{O}(\epsilon)
\right]
+C_AC_F\neps\left[
    \frac{1}{4 \epsilon ^2}+\frac{1}{2 \epsilon }+\zeta _2-\frac{3}{4}
    +\mathcal{O}(\epsilon)
\right]
\notag\\&\quad
}
+C_Fn_f\left[
    \frac{1}{4 \epsilon ^2}+\frac{7}{8 \epsilon }
    +\frac{3 \zeta _2}{2}+\frac{21}{16}
    +\mathcal{O}(\epsilon)
\right]
+C_Fn_h\left[
    \frac{1}{4 \epsilon ^2}+\frac{7}{8 \epsilon }
    +\frac{\zeta _2}{2}-\frac{3}{16}+\mathcal{O}(\epsilon)
\right]
\notag\\&\quad
+C_Fn_m\Bigg[
    \frac{1}{4 \epsilon ^2}+\frac{7}{8 \epsilon }
    +\frac{21}{16} - \frac32 z^2 + (\frac32 - 4 z^2 + 3 z^4) \zeta_2
    +(6 z^4-8 z^2) H_{0,0}(z)
    -3 z^2 H_0(z)\notag\\&\qquad\qquad
    +(-3 z^4+4 z^2-1) H_{-1,0}(z)
    +(3 z^4-4 z^2+1) H_{1,0}(z)
    +\mathcal{O}(\epsilon)
\Bigg]\Bigg\}+ \mathcal{O}(\alpha^3)\,.
\end{align}

Note that the one-loop coefficients of $Z_2$ and $Z_m$ match. However,
this is clearly a coincidence as it ceases to be true at two-loop.

If we work in a theory with some massless and some massive flavours we
need a further renormalisation constant that renormalises diagrams
where a heavy-fermion loop is inserted in the light-fermion propagator
corrections. This constant starts at the two-loop level and
is
\begin{align}
Z_{2,l} &= 1
+\Big(\frac{\alpha_0}{2\pi}\Big)^2
C_Fn_h\left[
    \frac{1}{2 \epsilon }-\frac{5}{12}
    +\mathcal{O}(\epsilon)
\right]
+\Big(\frac{\alpha_e}{4\pi}\Big)^2
C_Fn_h\neps\left[
    -\frac{1}{4 \epsilon ^2}+\frac{3}{8 \epsilon }-\frac{\zeta _2}{2}-\frac{13}{16}
    +\mathcal{O}(\epsilon)
\right]\notag\\&\quad+ \mathcal{O}(\alpha^3)\,.
\end{align}

Next, we need the renormalisation constants for the couplings. Those
will be given in the \ac{msbar} scheme. If other schemes, such as the
\ac{OS} scheme for the coupling is desired, this can be fixed
\emph{after} setting $\bar\alpha_e=\bar\alpha$ and $\neps=2\epsilon$. 
\begin{align}\begin{split}
Z_\alpha &= 1 + \frac{\bar\alpha}{4\pi}\Big[
\frac{\beta_{20}}{\epsilon}\Big]+ \mathcal{O}(\bar\alpha_i^2)
\,\\
Z_{\alpha_e} &= 1 + \frac{\bar\alpha_e}{4\pi}\Big[\frac{\beta_{02}}{\epsilon}\Big] 
 + \frac{\bar\alpha}{4\pi}\Big[\frac{\beta_{11}}{\epsilon}\Big]+
 \mathcal{O}(\bar\alpha_i^2)
\,,
\end{split}\end{align}
For obvious reasons, we give $Z_{\alpha_i}$ in the renormalised
couplings. The $\beta$ coefficients are
\begin{align}
\begin{split}
\beta_{20} &= \beta_0 + \neps\Big(-\frac{C_A}6\Big)\,,\\
\beta_{11} &= 6 C_F\,,\\
\beta_{02} &= -4 C_F + 2 C_A - 2T_RN_F + \neps(C_F-C_A)\,,
\end{split}
\label{eq:betafdh}
\end{align}
where $\beta_0 = 11/3C_A - 4/3T_RN_F$ is the normal $\beta_0$ of
\cdr{}.  Here we have defined the shorthand $N_F=n_h+n_m+n_f$ as the
sum of all active flavours, independent of mass. To convert these
results into the \ac{OS} scheme, we set~\cite{Grozin:2005yg}
\begin{align}
\alpha = \bar\alpha\ \frac{\bar Z_\alpha}{Z_\alpha} =
\bar\alpha\Bigg[1-\frac43\frac{\bar\alpha}{4\pi}\log\frac{\mu^2}{M^2}\Bigg]
\,,
\end{align}
for each active fermion with mass $M$.

\section{Infrared prediction in QCD}\label{sec:const:scet}
In this section we will give results for QCD with some massive and
some massless flavours. The QED limit is straightforward by setting
$C_F\to1$, $C_A\to0$, and $n_f\to0$.

As discussed in Section~\ref{sec:irpred}, we use \ac{SCET} to predict the
\ac{IR} structure of an amplitude. For this we need to calculate the
process's anomalous dimension ${\bf\Gamma}$ with~\eqref{eq:gammair}
and determine $\scetz$ by solving the \ac{RGE}~\eqref{eq:zrge}. $\scetz$
then shares the \ac{IR} structure with our amplitude after we have
performed the decoupling transformation~\cite{Chetyrkin:1997un,
Gnendiger:2016cpg}
\begin{align}
\zeta_{\alpha  } &= 1+\Big(\frac{\alpha  }{4\pi}\Big)n_h \frac43\log\frac{\mu^2}{M^2} + \mathcal{O}(\alpha^2)\,,\\
\zeta_{\alpha_e} &= 1+\Big(\frac{\alpha_e}{4\pi}\Big)n_h      2 \log\frac{\mu^2}{M^2} + \mathcal{O}(\alpha^2)\,.
\end{align}

In \fdh{}, the presence of the evanescent coupling makes the \ac{RGE} more
complicated~\cite{Broggio:2015dga}. We restrict ourselves mostly to
\ac{QED} again because three- and four-gauge vertices become very
complicated as soon as they involve $\epsilon$-scalars. We have
\begin{align}\begin{split}
\log\scetz &= \Big(\frac{\vec\alpha}{4\pi}\Big)\Bigg(
   \frac{\vec{\bf\Gamma}_1'}{4\epsilon^2}
  +\frac{\vec{\bf\Gamma}_1 }{2\epsilon  }
\Bigg)\\&
+ \sum_{m+n=2}
    \Big(\frac{\alpha}{4\pi}\Big)^m
    \Big(\frac{\alpha_e}{4\pi}\Big)^n
\Bigg(
  -\frac{3\vec\beta_{mn}\cdot \vec{\bf\Gamma}_1'}{16\epsilon^3}
  -\frac{ \vec\beta_{mn}\cdot \vec{\bf\Gamma}_1 }{ 4\epsilon^2}
  +\frac{{\bf\Gamma}_{mn}'}{16\epsilon^2}
  +\frac{{\bf\Gamma}_{mn} }{4\epsilon  }
\Bigg)+ \mathcal{O}(\alpha^3)\,,
\label{eq:scetzfdh}
\end{split}\end{align}
where we have defined the shorthand notation for terms involving only
one-loop quantities
\begin{align}
\begin{split}
\vec\alpha\cdot \vec{\bf\Gamma}_1&= \alpha\quad\ {\bf\Gamma}_{10}+\alpha_e\ \,{\bf\Gamma}_{01}\,,\\
\vec\beta \cdot \vec{\bf\Gamma}_1&= \beta_{10}\, {\bf\Gamma}_{10}+\beta_{01}\,{\bf\Gamma}_{01}\,.
\end{split}
\end{align}
Here, we have used ${\bf\Gamma}_{mn}$
($\beta_{mn}$) to indicate the $\alpha^m\alpha_e^n$ coefficient of
${\bf\Gamma}$ ($\beta$). $\vec\beta$ is given in \eqref{eq:betafdh} and
$\bf \Gamma$ is constructed as in \eqref{eq:gammair}
\begin{align}\begin{split}
{\bf\Gamma}(\mu) &= 
  \sum_{i,j} \gcusp \log\frac{\mu^2}{-{\rm sign}_{ij} 2p_i\cdot p_j} + \sum_i\gq\\
&-\sum_{I,J} \gcusp(\chi_{I,J})               + \sum_I\gQ\\
&+\sum_{I,j} \gcusp \log\frac{m_I\mu}{-{\rm sign}_{Ij} 2p_I\cdot p_j}\,.
\end{split}\end{align}
Let us go through the four anomalous dimensions appearing here:
\begin{itemize}

\def\lqad{
    \begin{align}\begin{split}
        \gq &= 
         \Big(\frac{\alpha  }{4\pi}\Big)(-3C_F)
        +\Big(\frac{\alpha_e}{4\pi}\Big)\neps\frac{C_F}2\\&
        +\Big(\frac{\alpha  }{4\pi}\Big)^2\Bigg[
            C_F^2\Big(-\frac32+12\zeta_2-24\zeta_3\Big)
           +C_Fn_f\Big(\frac{130}{27}+4\zeta_2\Big)
           \inclca{
           \\&\qquad+C_AC_F\Big(-\frac{961}{54}-11\zeta_2+26\zeta_3\Big)
           }
        \Bigg]\\&
        +\Big(\frac{\alpha  }{4\pi}\Big)\Big(\frac{\alpha_e}{4\pi}\Big)
            \neps\Bigg[
                C_AC_F\frac{11}2 
              - C_F^2\Big(2+2\zeta_2\Big)
            \Bigg]
        +\Big(\frac{\alpha_e}{4\pi}\Big)^2\neps\Bigg[
            - \frac18\neps C_F^2
            -\frac32 n_f
        \Bigg]
    +\mathcal{O}(\alpha^3)\,.
    \end{split}\end{align}
}
\def\hqad{
    \begin{align}\begin{split}
        \gQ 
          &= \Big(\frac{\alpha  }{4\pi}\Big)(-2C_F)
        +\Big(\frac{\alpha  }{4\pi}\Big)^2\Bigg[
            C_AC_F\Bigg(
                -\frac{98}9 + 4\zeta_2 - 4\zeta_3 + \frac89\neps
            \Bigg) + C_Fn_f\frac{40}9
        \Bigg]
    +\mathcal{O}(\alpha^3)\,.
    \end{split}\end{align}
}
\def\cuspad{
    \begin{align}\begin{split}
        \gcusp 
          &= \Big(\frac{\alpha  }{4\pi}\Big)(4)
        +\Big(\frac{\alpha  }{4\pi}\Big)^2\Bigg[
            C_A\Big(\frac{268}9 - 8\zeta_2\Big)
            -\frac{80}9 n_f
            -\neps C_A\frac{16}9
        \Bigg]
    +\mathcal{O}(\alpha^3)\,.
    \end{split}\end{align}
}
\def\velcuspad{
    \begin{align}\begin{split}
        \gcusp(\chi)  &= \gcusp \chi\coth\chi
        +\Big(\frac{\alpha  }{4\pi}\Big)^2\ 8C_A\Bigg\{
            \chi^2+\zeta_2+\zeta_3\\&\qquad
            +\coth\chi\Bigg[
                {\rm Li}_2\big(\E^{-2\chi}\big) - 2\chi\log\big(1-\E^{-2\chi}\big) - \zeta_2(1+\chi)-\chi^2-\frac{\chi^3}{3}
            \Bigg]\\&\qquad
            +\coth^2\chi\Bigg[
                {\rm Li}_3\big(\E^{-2\chi}\big) + \chi {\rm Li}_2\big(\E^{-2\chi}\big) - \zeta_3 + \zeta_2\chi + \frac{\chi^3}3
            \Bigg]
        \Bigg\}
    +\mathcal{O}(\alpha^3)\,.
    \end{split}\end{align}
}

    \item
    The light-quark anomalous dimension $\gq$~\cite{Broggio:2015dga}
    \lqad

    \item
    The heavy-quark anomalous dimension $\gq$~\cite{Gnendiger:2016cpg}
    \hqad

    \item
    The normal cusp anomalous dimension $\gcusp$
    is~\cite{Broggio:2015dga}
    \cuspad

    \item
    Finally, we have the velocity dependent cusp anomalous dimension
    $\gcusp(\chi)$~\cite{Gnendiger:2016cpg}
    \velcuspad
\end{itemize}
Beyond the one-loop level, these anomalous dimensions do not have
$C_F^2$, $C_Fn_m$, or $C_Fn_h$ terms. This implies that soft
singularities associated to these terms exponentiate -- just as
expected.

\section{Massification}\label{sec:const:massify}
Our discussion of massification was still missing the explicit
expression of $\Zjet$
\begin{align}
\sqrt{\Zjet} &= 1
 +a_0 C_F \Bigg\{
    \frac1{\epsilon^2}+\frac1{2\epsilon}+\zeta_2+2
    +\bigg(4+\frac12\zeta_2\bigg)\epsilon
    +\bigg(8+2\zeta_2+\frac74\zeta_4\bigg)\epsilon^2+\mathcal{O}(\epsilon^3)
\Bigg\}\notag\\&\qquad
 +a_{0,e} C_F \frac{\neps}4\Bigg\{
    -\frac1{\epsilon}-1
    -(1+\zeta_2)\epsilon
    -(1+\zeta_2)\epsilon^2+\mathcal{O}(\epsilon^3)
\Bigg\}\notag\\&\qquad
 +\Big(a_0(s)\Big)^2\Bigg\{
    C_F^2\Bigg[
         \frac1{2\epsilon^4} + \frac1{2\epsilon^3}
        +\frac1{\epsilon^2}\bigg(\frac{51}{24}+\zeta_2\bigg)
        +\frac1{\epsilon  }\bigg(\frac{43}{ 8}-2\zeta_2+6\zeta_3\bigg)
        \notag\\&\qquad\qquad
        +\frac{369}{16}+\frac{61}{4}\zeta_2-18\zeta_4-24\zeta_2\log2-3\zeta_3
    \Bigg]\notag\\&\qquad\quad
   +C_FC_A\Bigg[
        \frac{11}{12\epsilon^3}   
       +\frac1{\epsilon^2}\bigg(\frac{25}{9}-\frac12\zeta_2\bigg)
       +\frac1{\epsilon}\bigg(
            \frac{1957}{216}+\frac{13}2\zeta_2-\frac{15}2\zeta_3
        \bigg)
        \notag\\&\qquad\qquad
       +\frac{31885}{1296}+\frac{38}{3}\zeta_2-13\zeta_4+12\zeta_2\log2
       + \frac{13}3\zeta_3
    \Bigg]\notag\\&\qquad\quad
    -C_FC_A\frac{\neps}{12}\Bigg[
        \frac1{2\epsilon^3}+\frac{11}{6\epsilon^2}
       +\frac1{\epsilon}\bigg(\frac{215}{36}+3\zeta_2\bigg)
       +\frac{4559}{216}+11\zeta_2+4\zeta_3
    \Bigg]\notag\\&\qquad\quad
    +C_F n_f\frac16\Bigg[
       -\frac1{\epsilon^3}-\frac{8}{3\epsilon^2}
       -\frac1{\epsilon}\bigg(\frac{149}{18}+6\zeta_2\bigg)
       -\frac{3269}{108}-16\zeta_2-8\zeta_3
    \Bigg]\notag\\&\qquad\quad
    +C_F n_m \frac23 \Bigg[
        \frac1\eta\bigg(
            \frac1{\epsilon^2}-\frac5{3\epsilon}+\frac{28}9+2\zeta_2
        \bigg)
        -\frac1{\epsilon^3}+\frac1{2\epsilon^2}
        +\frac1\epsilon\bigg(-\frac{55}{24}-3\zeta_2\bigg)
        \notag\\&\qquad\qquad
        +\frac{1675}{432}-2\zeta_2+\zeta_3
    \Bigg]\Bigg\}
\notag\\&\qquad
 +a_{0,e}a_0 \Bigg\{
    C_F^2 \frac\neps4 \Bigg[
        -\frac1{\epsilon^3}-\frac9{2\epsilon^2}-\frac{15}{2\epsilon}
        -23\zeta_2-2\zeta_3+1
    \Bigg]\notag\\&\qquad\quad
    +C_AC_F \frac\neps8 \Bigg[
        -\frac{11}\epsilon-\frac{105}2+4\zeta_2+20\zeta_3
    \Bigg]\Bigg\}
\notag\\&\qquad
 +\Big(a_{0,e}\Big)^2 \Bigg\{
    (C_F\neps)^2\frac1{32}\Bigg[
        -\frac3{\epsilon^2}-\frac5\epsilon-30\zeta_2+\frac{85}2
    \Bigg]\notag\\&\qquad\quad
   +C_F n_f \frac{\neps}{8} \Bigg[
        \frac1{\epsilon^2}+\frac7{2\epsilon}+\frac{21}4+6\zeta_2
    \Bigg]\notag\\&\qquad\quad
   +C_F \neps \bigg(\frac{C_F}2+\frac{\neps C_A}{8}-\frac{C_A}4 \bigg)
   \Bigg[
        \frac1{\epsilon^2}+\frac2{\epsilon}-3+4\zeta_2
    \Bigg]\notag\\&\qquad\quad
   +C_F \neps n_m \frac18\Bigg[
        \frac1{\epsilon^2}+\frac{7}{2\epsilon}-\frac34+2\zeta_2
    \Bigg]
\Bigg\}+\mathcal{O}(a_i^3,\epsilon,\eta)\,.
\label{eq:zjet}\end{align} 

Here, we have defined $a^0_i(x)$ as in~\eqref{eq:acoup}
\begin{align}
a^0_i(x) = \bigg(\frac{\alpha_i^0}{4\pi}\bigg)
      \bigg(\frac{\mu^2}{m^2}\bigg)^{\epsilon}
      (-2+\io)^{\eta/2}
      \bigg(\frac{-\nu^2}{x}\bigg)^{\eta/2}\,,
\qquad
i\in\{s,e\}\,,
\end{align}
through the unrenormalised coupling. The factorisation anomaly, i.e.
the pole in $1/\eta$, either cancels with the soft function or with a
similar contribution due to the anti-collinear jet $\Zantijet$ that is
identical to $\Zjet$ except for the $n_m$ term
\begin{align}
\sqrt{\Zantijet}\Big|_{n_m} = \Big(a^0_s(m^2)\Big)^2 
    C_F n_m \frac23 \Bigg[&
        -\frac1\eta\bigg(
            \frac1{\epsilon^2}-\frac5{3\epsilon}+\frac{28}9+2\zeta_2
        \bigg)
        +\frac1{2\epsilon^3}
        -\frac5{6\epsilon^2}
        -\frac{253}{72\epsilon}
        \notag\\&\quad
        +\frac{5083}{432}-\frac{14}{3}\zeta_2-\zeta_3
    \Bigg] + \mathcal{O}(a_s^3,\epsilon,\eta)
    \,.
    \label{eq:zantijet}
\end{align}

\section{Conclusion}
We now have all necessary constants to perform any calculation in
\fdh{} at the two-loop level. These results are three-loop ready
in that they contain $\neps$. However, before we can use even these
two-loop results in any actual three-loop calculations, we need to
expand everything up to at least $\mathcal{O}(\epsilon)$ or even
$\mathcal{O}(\epsilon^2)$ for $\Zjet$. For $Z_m$ and $Z_2$, this is
trivial because their exact $\epsilon$ dependence is known in terms of
hypergeometric functions. This is unfortunately not true for the more
complicated $\Zjet$.

\endgroup

\chapter{Eikonal integrals \texorpdfstring{$\ieik$}{hat E}}
\label{ch:eik}

Here we give the explicit form of integrated eikonal required for
massive \ac{QED}. These expressions have been computed in
\cite{Frederix2009Automation}. As discussed in the text, we do not
need terms $\mathcal{O}(\epsilon)$ or higher: terms of
$\mathcal{O}(\epsilon)$ in $\ieik$ have no effect since they do not
modify $\fM{n}\ell$ after setting $d=4$. This means we can set them to
zero and there is no need to compute $\ieik$ beyond finite terms.

We start with defining a few auxiliary quantities:
\begin{align}
\beta_j &\equiv \sqrt{1-\frac{m_j^2}{E_j^2}}\ , &
v_{kj} &\equiv \sqrt{1-\left(\frac{m_j m_k}{p_j\cdot p_k}\right)^2 }\ , &
a_{kj} &\equiv (1+v_{kj})\frac{p_j\cdot p_k}{m_k^2} \ , &
\nu_{kj} &\equiv \frac{a_{kj}^2 m_k^2 - m_j^2}{2 \, (a_{kj}\, E_k - E_j)}\ . &
\end{align}
Following \cite{Frederix2009Automation}, the integrated eikonal can
then be written as
\begin{align}
\label{app:ieik}
\ieik_{kj} = \frac{\alpha}{2\pi} 
\frac{(4\pi)^\epsilon}{\Gamma(1-\epsilon)}
\left(\frac{\xc^2\, s}{\mu^2}\right)^{-\epsilon} 
\bigg(&-\frac{1}{2\epsilon} \frac{1}{v_{kj}}
\log\frac{1+v_{kj}}{1-v_{kj}} 
\\ \nonumber 
& +
 \frac{a_{kj} (p_j\cdot p_k) }{2\, (a_{kj}^2 m_k^2 - m_j^2)}
\Big( J(a_{kj} E_k, \beta_k,\nu_{kj}) - J(E_j,\beta_j,\nu_{kj}) \Big)
\bigg) + \mathcal{O}(\epsilon)\, ,
\end{align}
where we have used the function
\begin{align}
J(x,y,z) &= 
\Bigg( \log^2\frac{1-y}{1+y} 
+ 4\, \text{Li}_2\left(1-\frac{x(1+y)}{z}\right)
+ 4\, \text{Li}_2\left(1-\frac{x(1-y)}{z}\right)
\Bigg)
\end{align}
For the case $j=k$ this expression simplifies to 
\begin{align}
\label{app:ieikk}
\ieik_{jj} &= \frac{\alpha}{2\pi} 
\frac{(4\pi)^\epsilon}{\Gamma(1-\epsilon)}
\left(\frac{\xc^2\, s}{\mu^2}\right)^{-\epsilon} 
\bigg(-\frac{1}{\epsilon} - \frac{1}{\beta_j}
  \log\frac{1+\beta_j}{1-\beta_j} \bigg) + \mathcal{O}(\epsilon)\,
\end{align}
for the self-eikonals.

\chapter{Explicit derivation of \texorpdfstring{FKS$^3$}{FKS3}}
\label{ch:fks3}

In Section~\ref{sec:fks3}, we have skipped the detailed derivation of
FKS$^3$. While we motivated that all auxiliary integrals can be
avoided by setting all $\xc$ equal, we have not shown this because the
iterative eikonal subtracting and expanding is rather lengthy. In the
following we will go through all contributions and show that indeed
all auxiliary integrals cancel. One concession we will make for
simplicity is to set already those $\xc$ equal that we have set equal
in \ac{FKS2}.

At \ac{n3lo}, we have four terms
\begin{align}
\sigma^{(3)} = \int\D\Phi_{n  } \M{n  }3
       + \int\D\Phi_{n+1} \M{n+1}2
       + \int\D\Phi_{n+2} \M{n+2}1
       + \int\D\Phi_{n+3} \M{n+3}0\,,
\end{align}
which are separately divergent and that we will re-organise according
to the scheme's prescription.

\section{Real-virtual-virtual contribution}

Let us begin with the real-virtual-virtual part that we split again
into a hard and soft contribution
\begin{align}
\label{eq:nnnlorvv}
\bbit{3}{rvv} = \D\Phi_{n+1} \M{n+1}2 = \bbit{3}{s}(\xc) + \bbit{3}{h}(\xc)
\end{align}
as in \eqref{eq:shnnlo}.  Using that even at the two-loop level
\begin{align}
\mathcal{S}_{n+1}\M{n+1}2 = \eik_{n+1}\M{n}2\,,
\end{align}
the soft contribution in analogy to \eqref{eq:nlo:s}
and \eqref{eq:nnlo:s}  is given by
\begin{align}
\bbit{3}{s}(\xc) \to \D \Phi_{n}\ieik(\xc) \M{n}2\,.
\end{align}
The hard contribution  is now
\begin{align}\begin{split}
\bbit{3}{h}(\xc) &= 
  \pref1 \D\xi\, \cdis{\xi^{1+2\epsilon}} \big(\xi^2\M{n+1}2\big)
\\&
 =\pref1 \D\xi\, \cdis{\xi^{1+2\epsilon}} \xi^2\Big(
     \fM{n+1}2 -\ieik(\xc)\M{n+1}1-\frac1{2!}\ieik(\xc)^2\M{n+1}0
\Big)
\\&
 =\bbit{3}{f}(\xc) + \underbrace{\bbit{3}{d1}(\xc) + \bbit{3}{d0}(\xc)}_{\bbit{3}{d}(\xc)}
\end{split}\end{align}
where $\bbit{3}{f}$ is finite and the divergent part $\bbit{3}{d}$
is composed of
\begin{subequations}
\begin{align}
\int \bbit{3}{d1}(\xc) &=
-
\int  \pref1 \D\xi\, \cdis{\xi^{1+2\epsilon}} \xi^2\Big(
      \ieik(\xc)\M{n+1}1
\Big) \equiv -\mathcal{I}^{(1)}(\xc)\,,\\
\int  \bbit{3}{d0}(\xc) &=
- \int \frac1{2!} \pref1 \D\xi\, \cdis{\xi^{1+2\epsilon}} \xi^2\Big(
      \ieik(\xc)^2\M{n+1}0
\Big) \equiv -\frac1{2!}\mathcal{J}(\xc)\,.
\end{align}\end{subequations}
Above we have defined two functions $\mathcal{I}^{(1)}$ and
$\mathcal{J}$ that are potentially tedious to compute. However, as we
will see they cancel in the final result, similar to the function
$\mathcal{I}$ at \ac{NNLO}.

\section{Real-real-virtual contribution}

The real-real-virtual contribution are similar to the double-real
contribution of \ac{FKS2}
\begin{subequations}
\begin{align}
& \qquad
\bbit{3}{rrv} = \D\Phi_{n+2} \M{n+2}{1} = \bbit{3}{ss}(\xc) +
\bbit{3}{sh}(\xc) + \bbit{3}{hs}(\xc) + \bbit{3}{hh}(\xc)
\,,\\[5pt] 
& \left\{\begin{array}{c}
   \bbit{3}{ss}(\xc) \\[5pt] 
   \bbit{3}{hs}(\xc) \\[5pt]  
   \bbit{3}{sh}(\xc) \\[5pt]  
   \bbit{3}{hh}(\xc)
\end{array}\right\} =
    \pref2\ \frac1{2!}\ 
\left\{\begin{array}{c}
   \frac{\xc^{-2\epsilon}}{2\epsilon}\delta(\xi_1) \,
   \frac{\xc^{-2\epsilon}}{2\epsilon}\delta(\xi_2)
   \\[3pt] 
  -\frac{\xc^{-2\epsilon}}{2\epsilon}\delta(\xi_2) \,
   \cdis{\xi_1^{1+2\epsilon}}
   \\[3pt]
  -\frac{\xc^{-2\epsilon}}{2\epsilon}\delta(\xi_1) \,
   \cdis{\xi_2^{1+2\epsilon}}
   \\[3pt]
   \cdis{\xi_1^{1+2\epsilon}}\,
   \cdis{\xi_2^{1+2\epsilon}}
\end{array}\right\}\, \D\xi_1\,\D\xi_2
\ \xi_1^2\xi_2^2\M{n+2}1\,.
\end{align}\end{subequations}
Obviously $\int\bbit{3}{hs}=\int\bbit{3}{sh}$ and
\begin{align}
\int\bbit{3}{hs}(\xc) &=
    \pref2 \ \frac1{2!}\ \int\D\xi_1\ \cdis{\xi_1^{1+2\epsilon}}
    \big(\xi_1^2\M{n+1}1\big)
    \ieik(\xc)
= \frac1{2!}\,\mathcal{I}^{(1)}(\xc) \, .
\end{align}
Furthermore, as for \eqref{eq:nnlo:ss} we find 
\begin{align}
\bbit{3}{ss}(\xc) &\to\ \D\Phi_{n} \frac1{2!} \ieik(\xc)^2 \M{n}1 \,.
\end{align}
The hard contribution is  not yet finite due to the
explicit $1/\epsilon$ pole in $\M{n+2}1$. As is customary by now we
again perform an eikonal subtraction 
\begin{align}
    \M{n+2}1 \equiv \fM{n+2}1(\xc) - \ieik(\xc)\,\M{n+2}0\,.
\end{align}
and write
\begin{subequations}
\begin{align}
\bbit{3}{hh}(\xc) &= \bbit{3}{hf}(\xc) + \bbit{3}{hd}(\xc)\,,\\
\bbit{3}{hf}(\xc) &= \pref2\ \frac1{2!}\ 
   \cdis{\xi_1^{1+2\epsilon}}\,
   \cdis{\xi_2^{1+2\epsilon}}\,
   \xi_1^2\xi_2^2\fM{n+2}1(\xc)\,,\\
\int\bbit{3}{hd}(\xc) &=-\int \pref2\ \frac1{2!}\ 
   \cdis{\xi_1^{1+2\epsilon}}\,
   \cdis{\xi_2^{1+2\epsilon}}\,
   \xi_1^2\xi_2^2\ieik(\xc)\,\M{n+2}0 \equiv -\frac1{2!} \mathcal{K}(\xc)\,.
\end{align}
\end{subequations}
Here we have defined a third auxiliary function $\mathcal{K}$ that
will cancel in the final result. 

\section{Triple-real contributions}

The evaluation of the triple-real contributions proceeds along the
lines of the \ac{FKS2} double-real part, albeit with more (individually
$\xc$ dependent) terms
\begin{align}
\bbit{3}{rrr} &= \D\Phi_{n+3} \M{n+3}0 = 
  \bbit{3}{hhh} 
+ \underbrace{\bbit{3}{hhs} + \bbit{3}{hsh} + \bbit{3}{shh}}_{3\bbit{3}{hhs}}
+ \underbrace{\bbit{3}{hss} + \bbit{3}{shs} + \bbit{3}{ssh}}_{3\bbit{3}{hss}}
+ \bbit{3}{sss}\,.
\end{align}
Because we choose all $\xc$ equal, it does not matter which photon is
soft, just how many. Thus, we are left with four different kinds of
contributions
\begin{align}\begin{split}
\left\{\begin{array}{c}
   \bbit{3}{sss}(\xc) \\[5pt]  
   \bbit{3}{hss}(\xc) \\[5pt]  
   \bbit{3}{hhs}(\xc) \\[5pt]  
   \bbit{3}{hhh}(\xc)
\end{array}\right\} &=
   \prod_{i=1}^3 \Big( \D\Upsilon_i \D\xi_i\, \xi_i^2\Big)\ \D \Phi_{n,3}
   \frac{\M{n+3}0}{3!}\ 
\left\{\begin{array}{c}
   \frac{\xc^{-2\epsilon}}{2\epsilon}\delta(\xi_1) \,
   \frac{\xc^{-2\epsilon}}{2\epsilon}\delta(\xi_2) \,
   \frac{\xc^{-2\epsilon}}{2\epsilon}\delta(\xi_3)
   \\[3pt]
   \frac{\xc^{-2\epsilon}}{2\epsilon}\delta(\xi_2) \,
   \frac{\xc^{-2\epsilon}}{2\epsilon}\delta(\xi_3) \,
   \cdis{\xi_1^{1+2\epsilon}}
   \\[3pt]
  -\frac{\xc^{-2\epsilon}}{2\epsilon}\delta(\xi_3) \,
   \cdis{\xi_1^{1+2\epsilon}}
   \cdis{\xi_2^{1+2\epsilon}}
   \\[3pt]
   \cdis{\xi_1^{1+2\epsilon}}\,
   \cdis{\xi_2^{1+2\epsilon}}\,
   \cdis{\xi_3^{1+2\epsilon}}
\end{array}\right\}\, .
\end{split}\end{align}
The triple-hard $\bbit{3}{hhh}$ contribution is finite and can be
integrated numerically. For the triple-soft $\bbit{3}{sss}$ we get
\begin{align}
\bbit{3}{sss}(\xc) &= \D\Phi_n\,  \frac1{3!} \ieik^3\,\M n0\,.
\end{align}
The double-soft contribution can be expressed in terms of the function
$\mathcal{J}(\xc)$ as
\begin{align}
\int \bbit{3}{hss}(\xc) &=\int 
    \pref1\ \frac1{3!}\ 
   \ieik(\xc)^2\cdis{\xi_1^{1+2\epsilon}}
   \D\xi_1 \ \xi_1^2\M{n+1}0 = \frac1{3!} \mathcal{J}(\xc)\,.
\end{align}
Similarly, the single-soft contribution
\begin{align}
  \int  \bbit{3}{hhs}(\xc) &= \int \pref2\ \frac1{3!}\ \ieik(\xc)
   \cdis{\xi_1^{1+2\epsilon}} \cdis{\xi_2^{1+2\epsilon}}\,
   \D\xi_1\,\D\xi_2 \ \xi_1^2\xi_2^2\M{n+2}0 =
   \frac1{3!}\mathcal{K}(\xc)
\end{align}
involves the auxiliary function  $\mathcal{K}$.

\section{Combination}

Combining all contributions at \ac{n3lo} we need to evaluate
\eqref{eq:nnnlocomb}. Collecting the terms with an $n$-parton phase
space we get 
\begin{align}\begin{split}
\bbit{3}{n}(\xc)  &= \int\D\Phi_n\Big(
     \M{n}3
 + \underbrace{\ieik(\xc) \M{n}2}_{\bbit{3}{s}}
 + \underbrace{\frac1{2!} \ieik(\xc)^2 \M{n}1 }_{\bbit{3}{ss}}
 + \underbrace{1\times\frac1{3!}\ieik(\xc)^3\M{n}0}_{\bbit{3}{sss}} \Big)
\\&\qquad
  \underbrace{ -\mathcal{I}(\xc)
    - \frac1{2!} \mathcal{J}(\xc)}_{\bbit{3}{d}}
 + \underbrace{\frac1{2!}
   \mathcal{I}(\xc)+\frac1{2!}\mathcal{I}(\xc)}_{\bbit{3}{hs}+\bbit{3}{sh}}
   \underbrace{-\frac1{2!} \mathcal{K}(\xc)}_{\bbit{3}{hd}}
 + \underbrace{3\times\frac1{3!} \mathcal{J}(\xc)}_{\bbit{3}{hss}+\cdots}
 + \underbrace{3\times\frac1{3!} \mathcal{K}(\xc)}_{\bbit{3}{hhs}+\cdots}
 \, .
\end{split}\end{align}
The auxiliary integrals $\mathcal{I}^{(1)}$, $\mathcal{J}$ and
$\mathcal{K}$ cancel as do the explicit $1/\epsilon$ poles in the
first line. The other contributions in \eqref{eq:nnnlocomb} are also
separately finite. Thus, after setting $d=4$ the explicit expressions
of the separately finite parts of \eqref{eq:nnnlocomb} are given by
\eqref{eq:nnnloparts} with
\begin{align}
\label{eq:nnnloidiv}
\bbit{3}{n+1}(\xc) &= \bbit{3}{f} \, ; &
\bbit{3}{n+2}(\xc) &= \bbit{3}{hf} \, ; &
\bbit{3}{n+3}(\xc) &= \bbit{3}{hhh} \, .
\end{align}
Comparing \eqref{eq:nnnloidiv} to \eqref{eq:nnloind1} and
\eqref{eq:nnloind2} reveals the pattern of how to extend beyond
\ac{n3lo} as done in Section~\ref{sec:nllo}.


\chapter*{Index}
\addcontentsline{toc}{chapter}{Index}
\markboth{INDEX}{INDEX}

We have used the following acronyms, abbreviations and terminology
\makeatletter
\write\@auxout{%
\unexpanded{\global\@namedef{notfirstrun}{1}}%
}
\makeatother
\ifcsname notfirstrun\endcsname {
\begin{multicols}{2}
\begin{acronym}[MMCT]
\renewcommand*{\aclabelfont}[1]{
    \def\textsc{}
    \textbf{\MakeUppercase{\acsfont{#1}}}
}
\acro{1PI}{one-particle irreducible}
\acro{ac}[\textsc{ac}]{Anti-commuting $\gamma_5$ scheme}
\acro{bm}[\textsc{bm}]{Breitenlohner-Maison scheme}
\acro{BR}{branching ratio}
\acro{BSM}{Beyond the Standard Model}
\acro{cdr}[\textsc{cdr}]{Conventional dimensional regularisation}
\acro{dred}[\textsc{dred}]{Dimensional reduction}
\acro{dreg}[\textsc{dreg}]{Dimensional regularisation}
\acro{EFT}{effective field theory}
\acro{fdf}[\textsc{fdf}]{Four-dimensional formulation of \textsc{fdh}}
\acro{fdh}[\textsc{fdh}]{Four-dimensional helicity scheme}
\acro{fdu}[\textsc{fdu}]{Four-dimensional unsubtraction}
\acro{FKS2}[$\mathrm{FKS}^2$]{FKS double soft}
\acro{FKS}{Frixione-Kunszt-Signer subtraction scheme}
\acro{GPL}{generalised polylogarithm}
\acro{HPL}{harmonic polylogarithm}
\acro{HVP}{hadronic vacuum polarisation}
\acro{hv}[\textsc{hv}]{'t~Hooft-Veltman scheme}
\acro{IBP}{integration-by-parts}
\acro{IR}{infrared}
\acro{KLN}{Kinoshita-Lee-Nauenberg}
\acro{LFV}{lepton-flavour violating}
\acro{LHC}{Large Hadron Collider}
\acro{LL}{leading logarithm}
\acro{LO}{Leading order}
\acro{LSZ}{Lehmann-Symanzik-Zimmermann reduction formula}
\acro{MMCT}{\mcmule{} core team}
\acro{msbar}[$\MS$]{modified minimal subtraction}
\acro{n3lo}[N$^3$LO]{Next-to-next-to-next-to-leading order}
\acro{NLL}{Next-to-leading logarithm}
\acro{NLO}{Next-to-leading order}
\acro{NNLO}{Next-to-next-to-leading order}
\acro{OS}{on-shell}
\acro{PCS}{pseudo-collinear singularity}
\acro{PID}{particle identification}
\acro{PDF}{parton distribution function}
\acro{PSI}{Paul Scherrer Institut}
\acro{PS}{parton shower}
\acro{QCD}{Quantum chromodynamics}
\acro{QED}{Quantum electron dynamics}
\acro{QFT}{Quantum field theory}
\acro{RGE}{Renormalisation-group equation}
\acro{RNG}{Random number generator}
\acro{rs}[\textsc{rs}]{Regularisation scheme}
\acro{SCET}{Soft-collinear effective theory}
\acro{SHA1}{Secure Hashing Algorithm 1}
\acro{SM}{Standard Model}
\acro{UV}{Ultraviolet}
\acro{VP}{vacuum polarisation}
\acro{YFS}{Yennie-Frautschi-Suura}
\end{acronym}
\makeatletter
\AC@withpagetrue
\makeatother
\begin{acronym}[x]
\acro{active scales}{}
\acro{analytic regularisation}{}
\acro{anisotropic}{}
\acro{anomalous dimension}{}
\acro{auxiliary contributions}{}
\acro{$c$-distribution}{}
\acro{Casimir scaling}{}
\acro{collinear}{}
\acro{colour factors}{}
\acro{config file}{}
\acro{corner integral}{}
\acro{corner region}{}
\acro{counter-event}{}
\acro{cusp angle}{}
\acro{cusp anomalous dimension}{}
\acro{decoupling transformation}{}
\acro{dimension-six operator}{}
\acro{effective field theory}{}
\acro{eikonal factor}{}
\acro{eikonal subtraction}{}
\acro{electronic corrections}{}
\acro{exclusive}{}
\acro{event}{}
\acro{factorisation anomaly}{}
\acro{factorisation scale}{}
\acro{family}{}
\acro{Feynman parameters}{}
\acro{Feynman parametrisation}{}
\acro{Fierz identities}{}
\acro{fixed-order}{}
\acro{full period}{}
\acro{fully differential}{}
\acro{generic pieces}{}
\acro{generic processes}{}
\acro{graph polynomials}{}
\acro{hard}{}
\acro{heavy-quark}{}
\acro{integrated eikonal}{}
\acro{integration-by-parts}{}
\acro{IR safe}{}
\acro{isotropic}{}
\acro{lexicographic ordering}{}
\acro{light-quark}{}
\acro{loop diagrams}{}
\acro{loop-induced}{}
\acro{loop integrals}{}
\acro{massification}{}
\acro{master integral}{}
\acro{matching calculation}{}
\acro{measurement function}{}
\acro{Mellin transform}{}
\acro{menu files}{}
\acro{method of regions}{}
\acro{Michel decay}{}
\acro{mixed corrections}{}
\acro{MSlikesubtracted}[$\MS$-like subtracted]{}
\acro{muonic corrections}{}
\acro{non-trivial scheme dependence}{}
\acro{process groups}{}
\acro{radiative muon decay}{}
\acro{Ramanujan's master theorem}{}
\acro{random seed}{}
\acro{rare muon decay}{}
\acro{real}{}
\acro{reducible scalar integrals}{}
\acro{regular}{}
\acro{regularise}{}
\acro{renormalisable}{}
\acro{renormalisation}{}
\acro{renormalisation constants}{}
\acro{renormalisation scale}{}
\acro{renormalisation scheme}{}
\acro{resummation}{}
\acro{running}{}
\acro{sectors}{}
\acro{seed identity}{}
\acro{singular}{}
\acro{soft}{}
\acro{soft cut}{}
\acro{submission script}{}
\acro{sub-renormalisation}{}
\acro{subtraction scheme}{}
\acro{Symanzik polynomials}{}
\acro{tree-level diagrams}{}
\acro{trivial scheme dependence}{}
\acro{two-loop ready}{}
\acro{virtual}{}
\acro{Ward identity}{}
\acro{weak-isospin}{}
\acro{Wilson coefficient}{}
\end{acronym}
\end{multicols}
}
\else yy \fi

\bibliographystyle{JHEP}
\bibliography{../muon_ref}{}

\end{document}